\DocumentMetadata
{
	lang		= en-US, 
	pdfversion  = 1.7,
	pdfstandard = a-2b,
}

\documentclass[twoside]{mitthesis}

\usepackage{amsmath}
\usepackage{amssymb}
\usepackage{xcolor}
\usepackage{graphicx}
\usepackage{empheq}
\usepackage{cancel}
\usepackage{slashed}
\usepackage{yfonts}
\usepackage{float}
\usepackage{tikz}
\usepackage{physics}
\usepackage{pdfpages}
\usepackage{subcaption}
\usepackage{placeins}
\usepackage{enumitem}
\usepackage{setspace}

\usepackage[capitalize,noabbrev]{cleveref}

\hypersetup{%
	pdfsubject={},
	pdfkeywords={Massachusetts Institute of Technology, MIT},
	pdfurl={},
	pdfcontactemail={},
	pdfauthortitle={},
}

\begin{document}

\title{Searches for Non-Standard Neutrino Oscillations\\
with the IceCube Neutrino Observatory}

%
 \Author{Philip Lawrence Rudolf Weigel}{Department of Physics}[B.S. Physics, Drexel University, 2020]
%

%
\Degree{Doctor of Philosophy in Physics}{Department of Physics}

%
 \Supervisor{Janet Conrad}{Professor of Physics}[Department of Physics]

%
\Acceptor{Scott Hughes}{Professor of Physics}{Interim Associate Department Head of Physics} 

\DegreeDate{May}{2026}

\ThesisDate{May 6, 2026}

\maketitle





\begin{abstract}
	The short-baseline neutrino anomalies observed by LSND, MiniBooNE, and the gallium experiments can be explained by additional neutrino states beyond the three established by solar, atmospheric, and reactor oscillation measurements.
The minimal 3+1 sterile neutrino model that parameterizes these anomalies is in significant tension with null results from other oscillation experiments and with cosmological constraints, motivating the exploration of non-minimal extensions in which the additional mass eigenstate is unstable. 
This thesis presents a search for an unstable sterile neutrino model, in which the heavy mass eigenstate $\nu_4$ decays to two invisible particles, using 10.67 years of high-energy atmospheric neutrino data from the IceCube Neutrino Observatory. 
The analysis finds no preference for sterile decay over the no-decay 3+1 hypothesis and excludes most of the parameter space preferred by global fits to short-baseline data at 90\% confidence level, providing a strong constraint on this non-minimal explanation of the anomalies.
The thesis additionally develops the foundation for a future sterile neutrino search at IceCube targeting the resonant disappearance of antineutrinos.
Three contributions are presented: a comprehensive calculation of neutrino-nucleon and neutrino-nucleus deep-inelastic scattering cross sections, new machine-learning reconstruction techniques, and an improved event selection with twice the signal efficiency of the previous iteration. 
Together, these tools enable statistical separation of neutrinos and antineutrinos and establish the infrastructure for the next generation of sterile neutrino searches with IceCube.
\end{abstract}

\doublespacing



\chapter*{Acknowledgments}
\pdfbookmark[0]{Acknowledgments}{acknowledgments}

\begin{singlespace}
I am deeply grateful to my advisor, Janet Conrad. I first started working with Janet in 2017 as a co-op on IsoDAR and returned in 2019 to continue, before beginning graduate school on IceCube in 2020.
She kept inviting me back, so I must have been doing something right. 
Throughout my time as a graduate student, she taught me almost everything I know about being a good physicist and gave me ample intellectual freedom to explore topics that interested me along the way.

Many people shaped my path toward physics. 
John Metroke and Jay Werkheiser fostered my interest in electrical engineering and physics early on in high school. 
Michelle Dolinski, who introduced me to neutrino physics and detector R\&D in my first year at Drexel, is solely responsible for converting me from engineering to physics. 
The real rigor of my hands-on training came from Dan Akerib at SLAC and Daniel Winklehner at MIT, from whom I learned a great deal about hardware and how to build experiments.
I would like to thank the graduate students in my group, past and present: Marjon, Alex, Joe, Loyd, Nick K., Darcy, Josh, Julia, and Claire; and the postdocs: Daniel, Austin, John H., Jarrett, and David. 
Everyone in the Harvard group has been a source of interesting ideas and discussions, and I am very appreciative that they let me use their computing cluster.
Spencer and Miles from the University of Delaware group helped me significantly with several parts of the new MEOWS event selection, and were my go-to contacts for calibration-related topics.

I have had many friends who have supported me along the way.
In particular, I would like to thank Neil, John W., and Aidan, who have been a constant throughout these years.
Many of my friends from Drexel: Alan, Bassem, Brian, Dan$^2$, Nick, Huang, John B., Kar, and Shipper, were indispensable for getting through the COVID times and the lockdown boredom.
In the past year, Owen Jacob, Kyle, Emily, Josh, Igor, Jon, and Preston have been a source of never-ending fun and the closest thing one can call a ``social life'' while finishing up graduate school.

Lastly, I would like to thank Marina for her love and being my biggest supporter throughout these years.
Without her, I doubt I would be where I am today.

\end{singlespace}



\tableofcontents
\listoffigures
\listoftables


\chapter{Introduction}\label{chapter:intro}
This thesis presents a search for physics beyond the Standard Model using high-energy atmospheric neutrinos detected by the IceCube Neutrino Observatory at the geographic South Pole. 
The analysis employs a sample of neutrino interactions with reconstructed energies between 500 GeV and 100 TeV, selected with very high purity, and interprets the observed neutrino energy and zenith angle distributions in the context of non-standard neutrino propagation.

Neutrinos occupy a unique position as probes of physics beyond the Standard Model. 
In the original formulation of the Standard Model, neutrinos were assumed to be massless. 
The observation of evidence for flavor oscillations by Super-Kamiokande \cite{Super-Kamiokande:1998kpq}, SNO \cite{SNO:2002tuh}, and KamLAND \cite{KamLAND:2002uet,KamLAND:2004mhv} established that neutrinos carry non-degenerate masses, constituting the first confirmed experimental evidence for physics beyond the Standard Model. 
The mechanism by which neutrino masses are generated remains unknown, but the mixing angles and squared mass splittings have been measured to good precision \cite{Esteban:2024eli}.
Because neutrinos interact only via the weak force and flavor mixing, they are sensitive to subtle new interactions and propagation effects that would be obscured by hadronic or electromagnetic backgrounds in other experimental contexts. 
Taken together, these properties make the neutrino sector a natural place to search for new physics.

A specific experimental motivation exists in the form of several anomalous results that are difficult to accommodate within the three-flavor oscillation framework. 
The LSND \cite{LSND:1996ubh,LSND:2001aii} and MiniBooNE \cite{MiniBooNE:2012maf,MiniBooNE:2018esg,MiniBooNE:2020pnu} experiments each observed an excess of $\bar{\nu}_e$ and $\nu_e$ appearance events in $\bar{\nu}_\mu$ and $\nu_\mu$ beams at a high statistical significance when combined, at a $\Delta m^2 \sim 1$ eV$^2$ incompatible with solar and atmospheric mass splittings.
The gallium-based experiments SAGE \cite{SAGE:1998fvr, Abdurashitov:2005tb}, GALLEX \cite{GALLEX:1997lja}, and BEST \cite{Barinov:2022wfh} observed a deficit of $\nu_e$ interactions using radioactive sources, also consistent with oscillations at eV-scale $\Delta m^2$ \cite{Elliott:2023cvh}.
These anomalies motivate the hypothesis of at least one additional neutrino mass eigenstate beyond the three established by solar and atmospheric measurements.

The experimental situation does not, however, straightforwardly point to a single theoretical model. 
Because the anomalies are defined by their oscillation-like signatures rather than by a specific underlying mechanism, an experiment-driven approach does not immediately identify the signal to search for.
The natural starting point is the minimal extension of the three-flavor framework: the addition of a fourth (sterile) neutrino mass eigenstate with $\Delta m_{41}^2 \sim 1$ eV$^2$.
This 3+1 model provides a consistent parametrization of the anomalous signals and serves as a benchmark for experimental searches. 
However, the vanilla 3+1 model faces significant tension with null results from other experiments, including reactor and long-baseline disappearance searches, as well as with cosmological constraints on the number of thermalized light neutrino states from measurements of $N_{\mathrm{eff}}$ and $\sum m_\nu$ \cite{Boser:2019rta,Diaz:2019fwt,Hardin:2022muu}.
These tensions indicate that if the short-baseline anomalies reflect genuine new physics, additional interactions or modifications beyond the vanilla 3+1 model are required to render the resulting picture consistent with global oscillation and cosmological data.

This motivates the exploration of non-minimal extensions. 
One well-motivated possibility is that the fourth mass eigenstate is unstable and decays to a lighter sterile state and a light scalar boson \cite{Moss:2017pur, Kopp:2026tnx}. 
This model preserves the oscillation-like phenomenology that explains the short-baseline anomalies while suppressing sterile neutrino thermalization in the early universe, alleviating the cosmological constraints that disfavor the minimal 3+1 model. 
High-energy atmospheric neutrinos detected by IceCube, which traverse baselines of up to $\mathcal{O}(10^4)$ km through the Earth, are sensitive to the effects of sterile decay.

This thesis presents a search for the unstable sterile neutrino model using IceCube data and develops the foundation for the next generation of sterile neutrino searches at IceCube. 
The contributions span three areas. 
First, a comprehensive calculation of the neutrino-nucleon and neutrino-nucleus deep-inelastic scattering cross sections are developed, providing the inelasticity distributions needed to predict the response of a neutrino telescope to sterile-induced antineutrino disappearance.
Second, new machine-learning reconstruction techniques are introduced for IceCube, including dedicated reconstruction of the interaction inelasticity that provides a statistical handle on the neutrino-to-antineutrino composition not directly accessible on an event-by-event basis. 
Third, an improved event selection is developed that doubles the signal efficiency of the previous selection while maintaining high purity, increasing both the statistical power of the analysis and the size of the starting-track sample on which the inelasticity-based analysis depends. 
Together, these tools establish the infrastructure for an improved sterile neutrino search leveraging neutrino-antineutrino separation, and they support broader IceCube physics analyses beyond the sterile neutrino program.

The thesis is structured as follows. 
\cref{chapter:exp_bkg} reviews the experimental and theoretical background of neutrino oscillations and the short-baseline anomalies. 
\cref{chapter:cross_sections} presents the improved cross-section calculations.
\cref{chapter:icecube} describes the IceCube experiment, its systems, and the data it collects. 
\cref{chapter:sterile_signals} investigates the sterile neutrino signals observable at neutrino telescopes.
\cref{chapter:decay_analysis} presents the unstable sterile neutrino search and reports the constraints on the model parameter space.
\cref{chapter:ml_recos} introduces new machine-learning reconstruction techniques.
\cref{chapter:meows2026} develops the next-generation event selection and analysis-framework improvements.
\cref{chapter:conclusions} summarizes the results of this work.
\chapter{Experimental Background}\label{chapter:exp_bkg}

The observation of oscillations implies that neutrinos carry non-degenerate masses: flavor and mass eigenstates are distinct, and a neutrino produced in a definite flavor state propagates as a superposition of mass eigenstates that accumulates a relative phase during propagation. 
It is this phase that gives rise to oscillations.

The oscillation probability depends on the mass squared splitting $\Delta m_{ij}^2 = m_{i}^2 - m_{j}^{2}$ that set the physical length-scale of the oscillations, $L_{ij} \propto \frac{E}{\Delta m_{ij}^2}$.
In the simplified two-neutrino oscillation model, with just a single mixing angle $\theta$ and mass splitting $\Delta m^2$, the survival probability of a neutrino created in flavor state $\alpha$ is,
\begin{equation}
    P_{\alpha\alpha}(L, E) = 1 - \sin^2 (2\theta) \sin^2 \left(1.27\frac{\Delta m^2 L}{E} \right),
\end{equation}
where $\Delta m^2$ is in eV$^2$, $L$ in km, and $E$ in GeV. 
The oscillation probability is a function of $L/E$, meaning that experiments with different baselines and energies are sensitive to different ranges of $\Delta m^2$.
This chapter traces the experimental developments that established the current working model of three-neutrino oscillations, beginning with the first hints of flavor-changing oscillations to the resolution of both the solar and atmospheric neutrino deficits.
During these discoveries, new anomalies that did not fit into this paradigm were observed and hinted at the possible existence of neutrino states that do not interact weakly.

\section{The Solar Neutrino Problem}\label{sec:solar_problem}
\begin{figure}
    \centering
    \includegraphics[width=0.8\linewidth]{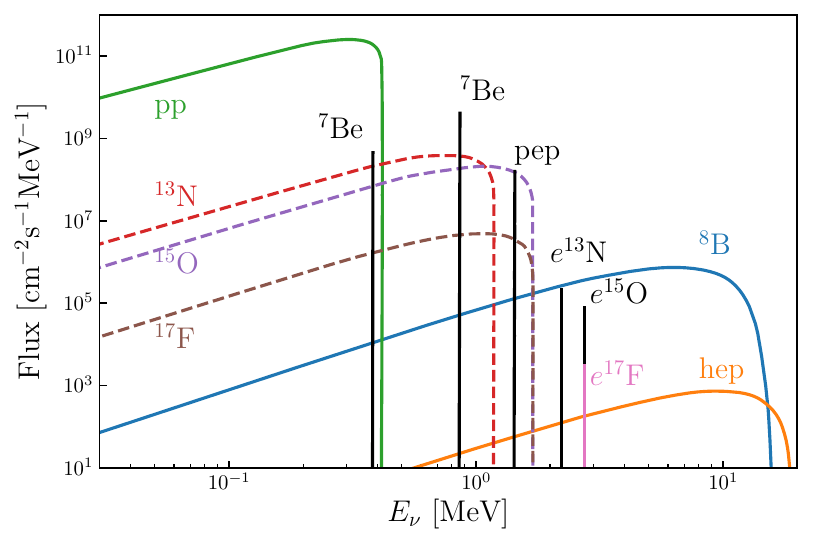}
    \caption{The flux of solar neutrinos as a function of energy separated by the process by which they are produced. Figure from Ref. \cite{Xu:2022wcq}.}
    \label{fig:solar_neutrino_spectra}
\end{figure}

The first hint of neutrino oscillations was observed by the Homestake Experiment \cite{Davis:1968cp}. 
The experiment was designed to be a large counting experiment whereby solar electron neutrinos would be captured on $^{37}$Cl to produce $^{37}$Ar, which could subsequently be counted.
Neutrinos are produced in the sun through several different processes, the dominant one being the $pp$ cycle.
Subdominant processes produce neutrinos at slightly higher energies than the $pp$ process, such as the products of the CNO cycle and the $pep$ process.
The spectra of the solar neutrino processes are shown in \cref{fig:solar_neutrino_spectra}.
A challenge in the detection of solar neutrinos is the energy scale.
Though $pp$ neutrino flux is much larger, the energy threshold of the Homestake experiment was too high to observe these.
The threshold energy for neutrino capture on $^{37}$Cl is 0.81 MeV, whereas the maximum energy of the $pp$ neutrinos is 0.42 MeV \cite{BOREXINO:2014pcl, Nakahata:2022xvq}.
Instead, the experiment measured the neutrinos originating from the more energetic $^8$B decay as well as contributions from $^7$Be electron capture.
The $^8$B decay neutrinos have the additional benefit of being energetic enough to produce an excited state of $^{37}$Ar, increasing the efficiency of chemical detection \cite{Bahcall:1969gd}.

\begin{figure}
    \centering
    \includegraphics[width=0.98\linewidth]{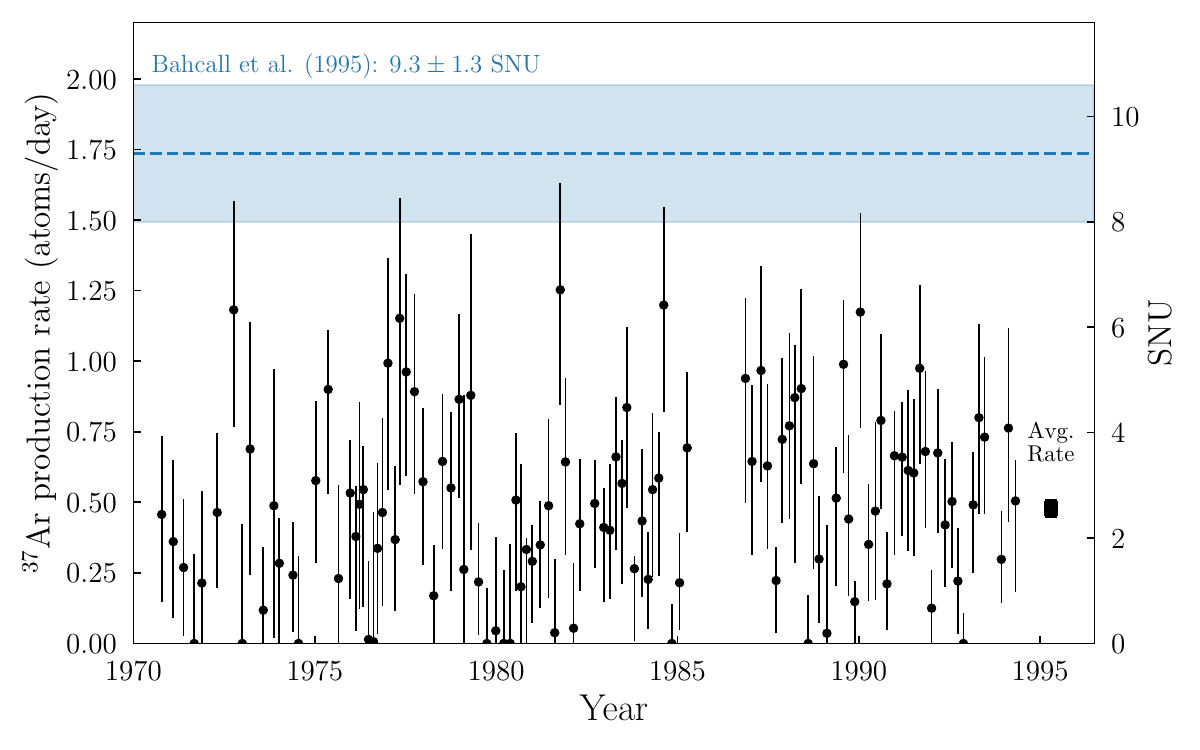}
    \caption{Measured data (black) of the Homestake experiment over 25 years of data taking. The reported average rate of $^{37}$Ar production was $2.56\pm 0.16~(\textrm{stat}.)\pm0.16~(\textrm{syst}.)$ SNU compared to the prediction from the SSM of $9.3\pm1.3$ SNU \cite{Bahcall:1995bt}. This figure is a reproduction of Fig.~13 in Ref. \cite{Cleveland:1998nv} with the addition of standard solar model prediction and uncertainty (blue) from Ref. \cite{Bahcall:1995bt}.}
    \label{fig:homestake_data}
\end{figure}

The Homestake experiment observed solar neutrinos, but with a significantly lower rate than expected. A deficit was observed when comparing the experimental rate, $2.56\pm 0.16~(\textrm{stat}.)\pm0.16~(\textrm{syst}.)$  solar neutrino units (SNU), to the predictions from the standard solar model (SSM), $9.3\pm1.3$ SNU, over nearly 25 years of data taking \cite{Cleveland:1998nv}.
The results of the 25-year Homestake analysis and the SSM prediction are shown in \cref{fig:homestake_data}.
The Kamiokande-II water Cherenkov detector measured $^8$B neutrinos from the Sun via $\nu_e e^- \rightarrow \nu_e e^-$ scattering, which had the benefit of directionality and measurable energy depositions over the radiochemical detector methods \cite{Kamiokande-II:1989hkh}.
However, in 1990, they reported results with over 1,000 days of data and found a measured flux of approximately 46\% of the SSM prediction \cite{Kamiokande-II:1990wrs} consistent with the deficit observed by the Homestake experiment.

A second radiochemical approach used gallium, which offers a lower energy threshold than chlorine.
The Soviet-American-Neutrino-Experiment (SAGE) at the Baksan Neutrino Observatory measured $\nu_{e} +\ ^{71}\mathrm{Ga} \rightarrow e^- +\ ^{71}\mathrm{Ge}$ using 30 tons of liquid gallium \cite{Abazov:1991rx}.
With a lower energy threshold of 0.23 MeV, gallium radiochemical detectors are sensitive to neutrinos from the $pp$ component of the flux in addition to the $^7$Be and $^8$B components.
The $^{71}$Ge atoms were extracted with a $^{nat}$Ge carrier dissolved in the gallium tanks and converted to germane (GeH$_{4}$) gas.
The mixed germane gas was inserted into a proportional counter with a NaI detector for months to accumulate K-shell decays of the $^{71}$Ge atoms.
In 1991, the SAGE upper limit was found to be $<79$ SNU at 90\% C.L. compared to the standard solar model prediction of $125\pm5$ SNU, though consistent with the expected flux from only $pp$ fusion\cite{Abazov:1991rx}.
Importantly, this was the first measurement of the solar neutrino deficit that included contributions beyond the $^{7}$Be and $^{8}$B components, but with large uncertainties.

The GALLEX experiment at the Laboratori Nazionali del Gran Sasso was another gallium-based radiochemical detector with a similar detection mechanism to SAGE \cite{GALLEX:1992gcp}.
Instead of employing metallic liquid gallium, as SAGE did, GALLEX used an aqueous solution of GaCl$_3$ which could be extracted with very high efficiency \cite{GALLEX:1992yvx}.
The measured solar neutrino flux from GALLEX was $83\pm19(\mathrm{stat.})\pm8(\mathrm{syst.})$ SNU, which was the first  measurement of $pp$ solar neutrinos \cite{GALLEX:1992gcp}\footnote{SAGE was only able to place an upper limit on the $pp$ solar neutrino flux \cite{Abazov:1991rx}.}.
This still represented a significant deficit relative to the SSM prediction.

The solar neutrino deficit observed by the Homestake experiment, Kamiokande-II, SAGE, and GALLEX was consistent with neutrino flavor transformations driven by the MSW effect \cite{Wolfenstein:1977ue, Mikheyev:1985zog}.
In the solar interior, where electron density is very high, $\nu_e$ production occurs near or above the MSW resonance condition.
This causes an adiabatic conversion to the heavier mass eigenstate as neutrinos propagate out of the sun through decreasing density.
For $^8$B solar neutrinos, the survival probability is approximately $30\%$, meaning that $70\%$ of the initial $\nu_e$ flux was converted to $\nu_\mu$ and $\nu_\tau$ before reaching the Earth.
Prior to a definitive measurement, the existing data were consistent with a range of mixing solutions \cite{Bahcall:1998jt}.

This definitive measurement came from the Sudbury Neutrino Observatory (SNO), which was a heavy water (D$_2$O) Cherenkov detector at an overburden of 6000 mwe inside the Creighton Mine located in Sudbury, Canada \cite{SNO:1999crp}.
The use of D$_2$O as a target enabled three channels by which solar neutrinos could interact.
Elastic scattering (ES) of solar neutrinos on electrons can occur for all flavors, but with a larger relative cross section for $\nu_e$ flavor since there are additional contributions from $W$ exchange.
Charged-current (CC) scattering on deuterons, $\nu_e + d \rightarrow e^{-} + p + p$, is only allowed for $\nu_e$ flavor as the kinematic thresholds for $\mu$ and $\tau$ are well-above the solar neutrino energies.
The neutral-current (NC) scattering on deuterons, $\nu_x + d \rightarrow \nu_x + p + n$ could occur for all flavors with equal cross section.
The NC channel measures the total neutrino flux and the CC channel measures only the $\nu_e$ flux, so a suppression in the CC channel relative to NC would be indicative of neutrino flavor transformation.

\begin{figure}
    \centering
    \includegraphics[width=0.75\linewidth]{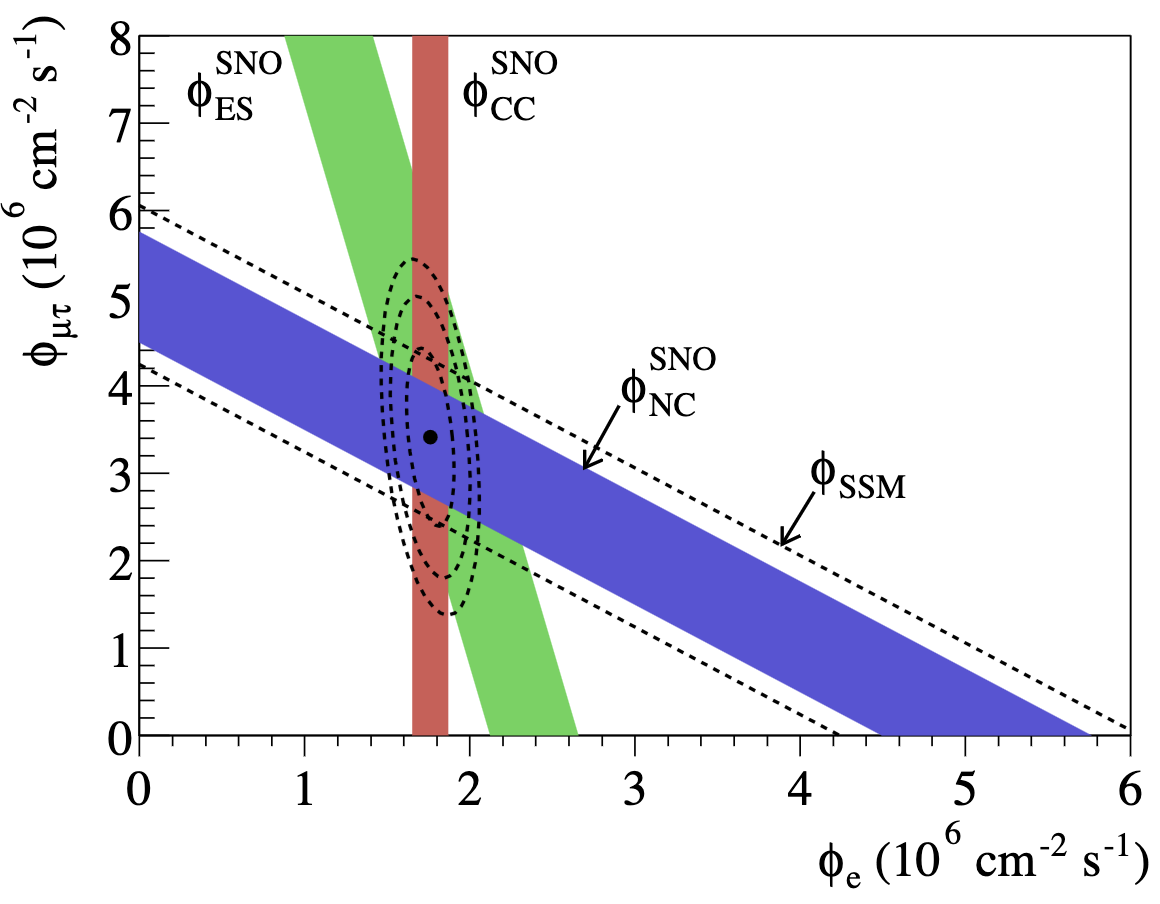}
    \caption{Measured fluxes of $^8$B solar neutrinos using charged-current scattering (red band), elastic scattering (green band), and neutral-current scattering (blue band) from SNO. The dashed band indicates the uncertainty from the SSM \cite{Bahcall:2000nu}. Figure from Ref. \cite{SNO:2002tuh}.}
    \label{fig:SNO_measurement}
\end{figure}

In 2002, SNO measured the $\nu_e$ flux using the CC interactions and found a flux of $\phi_{CC} = 1.76_{-0.05}^{+0.06}\,(\mathrm{stat.})_{-0.09}^{+0.09}\,(\mathrm{syst.})$ ${}\times 10^6~\mathrm{cm}^{-2}\,\mathrm{s}^{-1}$ and a NC-measured flux of $\phi_{NC} = 5.09_{-0.43}^{+0.44}\,(\mathrm{stat.})_{-0.43}^{+0.46}\,(\mathrm{syst.})$ ${}\times 10^6~\mathrm{cm}^{-2}\,\mathrm{s}^{-1}$~\cite{SNO:2002tuh}.
These measurements and their compatibility with each other are shown in \cref{fig:SNO_measurement}.
The NC-measured total flux was consistent with the SSM prediction \cite{Bahcall:2000nu}, demonstrating conclusively that the solar neutrino deficit arose from flavor transformation during propagation rather than mismodeling of the solar source.

However, the results of the solar neutrino experiments were consistent with several sets of mixing parameters \cite{Krastev:2001tv,Bahcall:2001zu,Bandyopadhyay:2002xj}.
The conclusive identification of the MSW effect with a large mixing angle as the solution to the solar neutrino deficit came from the results of the KamLAND experiment.
KamLAND was a liquid scintillator detector which observed $\bar{\nu}_e$ disappearance from nuclear reactors at an average baseline of $L \sim 180$ km \cite{KamLAND:2002uet, KamLAND:2004mhv}.
The KamLAND observation of $\bar{\nu}_e$ disappearance with a $\Delta m^2$ consistent with the large mixing angle solution to solar data, which itself was only viable when matter effects in the Sun were included \cite{Bahcall:2002ij,Fogli:2004zn}.
The combination of solar neutrino data with KamLAND established a complete solution to the solar neutrino problem.

\section{Atmospheric Neutrinos}\label{sec:atmospheric_neutrinos}
Atmospheric neutrinos are produced when cosmic rays interact in the upper atmosphere and subsequently create $\nu_\mu$ and $\nu_e$ with a wide range of energies and baselines spanning several orders of magnitude, making them a useful probe for neutrino oscillations across a wide range of $L/E$. 
When high-energy cosmic rays interact with nuclei in the Earth's atmosphere, they produce hadronic showers that primarily contain pions and kaons.
These mesons decay predominantly into muons and muon neutrinos.
The muons also decay and produce additional muon neutrinos and electron neutrinos, which gives a flavor ratio $\nu_\mu : \nu_e \approx 2:1$.
This well-defined flavor composition provides an experimental probe against which flavor-specific deficits (which could originate from neutrino oscillations) can be identified.
Atmospheric neutrinos were first observed by the Kolar Gold Fields experiment \cite{Achar:1965ova} and the experiment at East Rand Mine \cite{Reines:1965qk} in 1965, where horizontal neutrino-induced muon events were observed.
The IMB-1 detector reported the first hint of a deficit in 1986, measuring $26\pm3$\% $\mu$ events compared to a prediction of $31\pm1\%$, though the result was not statistically significant \cite{Haines:1986yf}.
In 1988, Kamiokande-II reported a more significant 59\% deficit in the observed muon-like events when compared to Monte Carlo predictions that could not be explained by detector systematic uncertainties, while the electron-like events matched predictions \cite{Kamiokande-II:1988sxn}.
A subsequent analysis with IMB-3 confirmed this deficit: a measurement of $36\pm2~(\mathrm{stat.})\pm2~(\mathrm{syst.})$\% muon events to a prediction of $51\pm1~(\mathrm{stat.})\pm5~(\mathrm{syst.})$\%, verifying the observation of Kamiokande-II \cite{Becker-Szendy:1992ory}.

\begin{figure}
    \centering
    \includegraphics[width=0.6\linewidth]{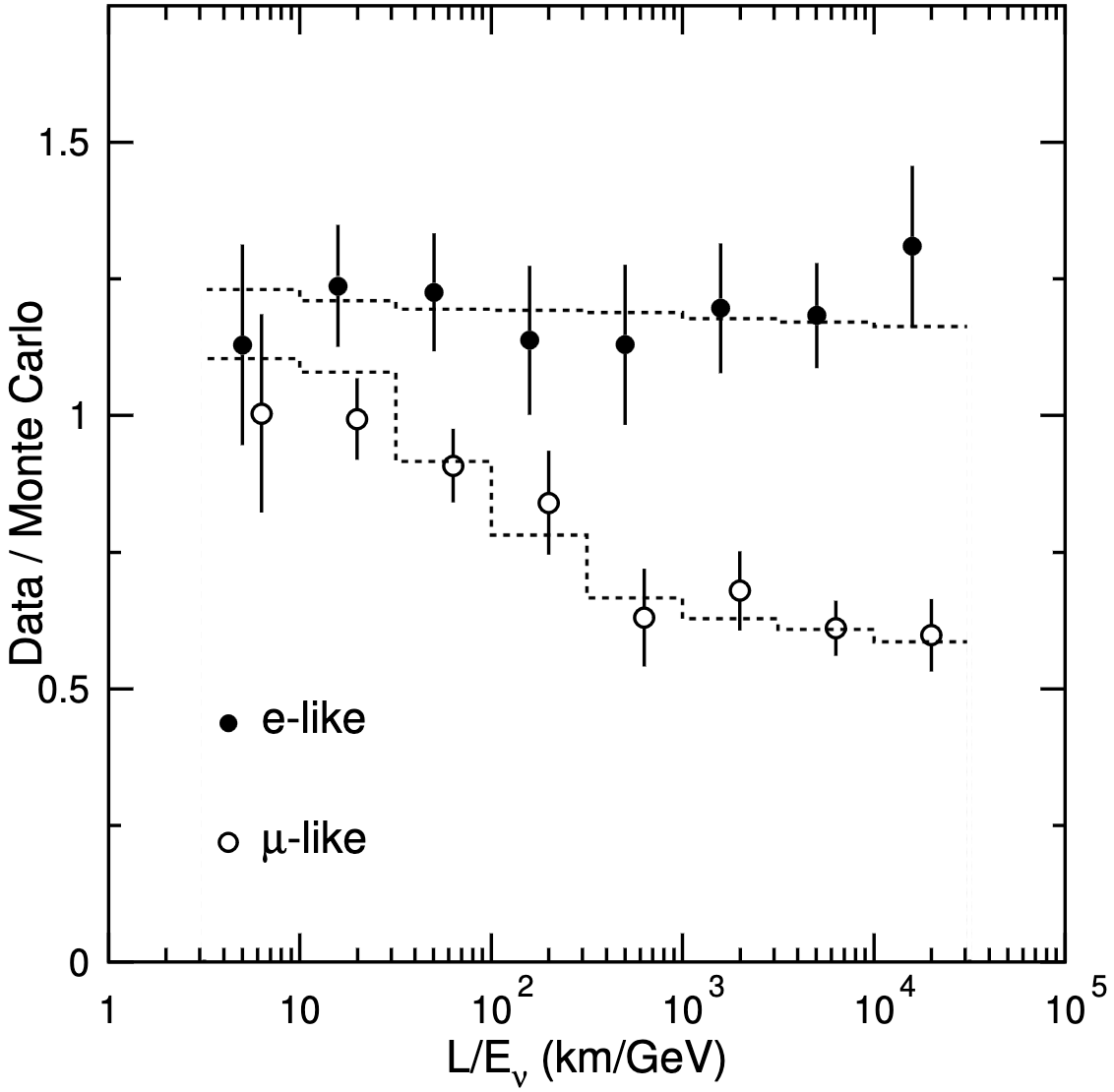}
    \caption{Ratio of data to Monte Carlo assuming no oscillations from SuperK as a function of $L/E$ demonstrating good agreement with the $\nu_\mu\rightarrow\nu_\tau$ oscillation hypothesis (dashed lines). Figure from Ref. \cite{Super-Kamiokande:1998kpq}.}
    \label{fig:superk_result}
\end{figure}

The resolution of the atmospheric neutrino deficit came with the result of the up-down asymmetry measurement by Super-Kamiokande (SuperK).
Since the neutrino oscillation probability depends on $L/E$, the zenith angles of atmospheric neutrinos provide a natural baseline lever-arm.
Down-going neutrinos travel $\mathcal{O}(10)$ km, while up-going neutrinos traverse a large fraction of the Earth up to $\mathcal{O}(10^4)$ km, producing an observable up-down asymmetry if oscillations are present.
SuperK measured the up-down asymmetry in the $\mu$-like events consistent with neutrino oscillations with a $\Delta m^2 \sim 10^{-3}$ eV$^2$ and a large mixing $\sin^2 (2\theta) > 0.86$ \cite{Super-Kamiokande:1998kpq}.
The absence of a corresponding asymmetry in the $e$-like channel disfavored $\nu_\mu\rightarrow\nu_e$ as the dominant oscillation channel, making the results more consistent with $\nu_\mu\rightarrow\nu_\tau$ oscillations as shown in \cref{fig:superk_result}.
The solar neutrino and atmospheric neutrino results combined established a clear three-neutrino oscillation picture that required non-zero neutrino masses.

\section{Short Baseline Experiments}\label{sec:short_baseline_experiments}
Early global fits to neutrino data indicated that the oscillations between $\nu_e$, $\nu_\mu$, and $\nu_\tau$ can be well-described by two mass splittings of $\approx2\times10^{-3}$ eV$^2$ and $7\times10^{-5}$ eV$^2$ \cite{Maltoni:2004ei}.
However, there were other hints that were indicative of oscillations at a much higher $\Delta m^2 \sim 1$ eV$^2$ from ``short-baseline'' experiments which correspond to searches with ranges of $L/E$ from about 0.1 to 10.
In this section, the experimental hints of oscillations that do not follow the same clear picture as the atmospheric and solar experiments will be presented.

\begin{figure}
    \centering
    \includegraphics[width=0.75\linewidth]{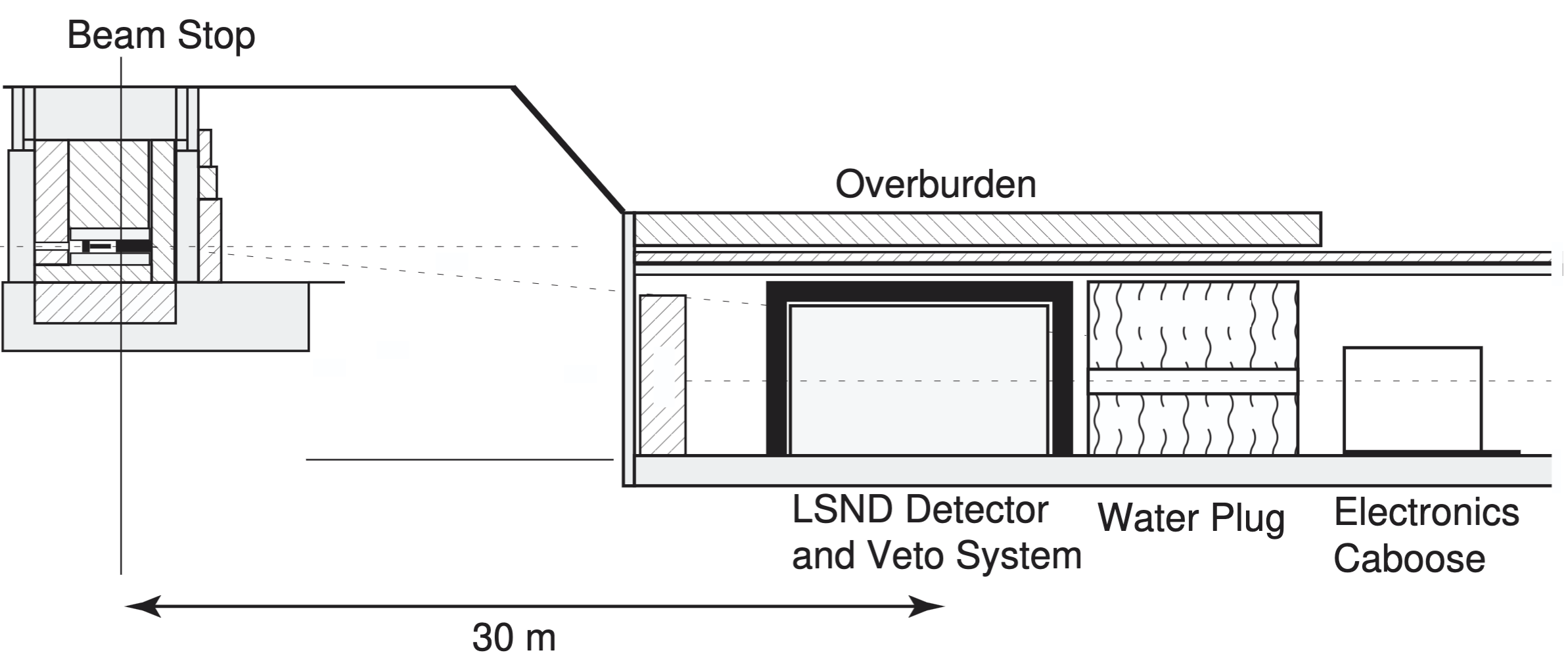}
    \caption{Schematic of the LSND experiment. Figure from  Ref. \cite{LSND:2001aii}.}
    \label{fig:lsnd_schematic}
\end{figure}

The first was observed by the Liquid Scintillator Neutrino Detector (LSND) at Los Alamos National Laboratory, which used an accelerator-based neutrino source with a close-by detector \cite{LSND:1996ubh}.
The detector was comprised of  167 tons of mineral oil with 1220 8-inch PMTs (approximately 20\% photocoverage) a distance of 30 meters from the neutrino source.
A schematic diagram of the LSND experimental setup is shown in \cref{fig:lsnd_schematic}.
The accelerator produced 800 MeV protons directed onto a target to produce pions, which were brought to rest before decaying via $\pi^+\rightarrow \mu^+ \nu_\mu$.
The muons decay through $\mu^+ \rightarrow e^+ \nu_e \bar{\nu}_\mu$ and the neutrinos propagate to the detector.
There was a small beam component to the $\bar{\nu}_e$ flux, originating from the $\pi^-$ decay chain, but it was significantly suppressed because 95\% of the $\pi^-$ are absorbed before decaying, and 88\% of the resulting $\mu^-$ of the surviving $\pi^-$ decays are captured as well.
Accounting for these factors, the $\bar{\nu}_e$ flux was then only 0.78\% of the $\bar{\nu}_\mu$ flux.

The detection mechanism for LSND was observing the Cherenkov light and scintillation light of the final-state $e^+$ of inverse beta decay $\bar{\nu}_e p \rightarrow e^+ n$, as well as the light signal of the subsequent neutron capture (2.2 MeV $\gamma$).
The two-component signal for $\bar{\nu}_e$ interactions allowed for the removal of the single-component $\nu_e$ interactions.
In 1996, LSND published a result showing an excess of $e^+$ events, measuring 22 events with an expected background of $4.6\pm0.6$ \cite{LSND:1996ubh}.
This excess in events was consistent with the appearance of $\bar{\nu}_e$ in a $\bar{\nu}_\mu$ beam that could not be explained by just the $\pi^-$ background for the reasons mentioned earlier.
If interpreted as neutrino oscillations, this provided favored regions with $\Delta m^2 > 0.1$ eV$^2$, much higher than the regions preferred by solar and atmospheric neutrino measurements.

\begin{figure}
    \centering
    \includegraphics[width=0.98\linewidth]{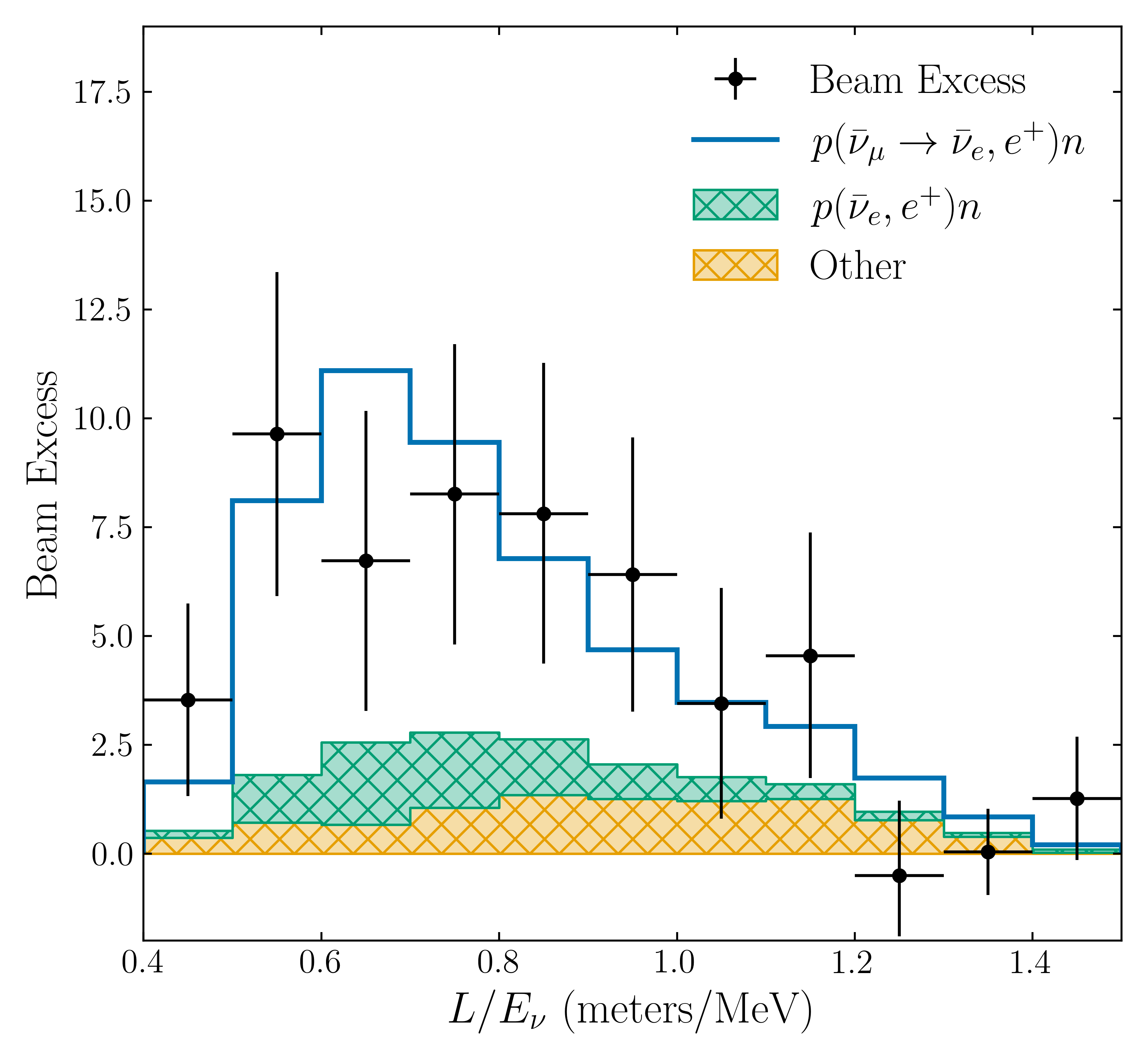}
    \caption{Measured beam excess of LSND as a function of baseline divided by neutrino energy ($L/E_\nu$). Data is shown in black. The stacked hashed histograms represent the expected background contamination from intrinsic $\bar{\nu}_e$ contamination from the beam (green) and other sources of background events (yellow). The signal model of $\bar{\nu}_\mu\rightarrow\bar{\nu}_e$ oscillations is shown in blue. This figure is a reproduction of Fig. 24 in Ref.~\cite{LSND:2001aii} for improved fidelity and readability.}
    \label{fig:LSND_data}
\end{figure}

To directly probe the excess observed by LSND, the MiniBooNE experiment was built at nearly the same $L/E$ to probe the oscillation hypothesis.
LSND used muon decay at rest with a 30m baseline, or roughly $L/E \approx 1$ m/MeV.
MiniBooNE was constructed in the Booster Neutrino Beam (BNB) at Fermilab, with a baseline of 500m and mean neutrino energy of 500 MeV such that the $L/E$ matched LSND \cite{MiniBooNE:2001sbk}.
At much higher neutrino energies, the physics for neutrino detection is different than that used for LSND.
The benefit is that any possible mismodeling of systematics from LSND would likely not be present in MiniBooNE.
The beam was operated in either neutrino or antineutrino mode with only a small contamination of wrong-sign neutrinos \cite{MiniBooNE:2008hfu}.
In neutrino mode, $\nu_\mu$ are produced from $\pi^+$ decay in flight and small, subleading contributions from $K^+$ decays.
$\nu_e$ can also be produced at $<1\%$ from the decay of the daughter muon from the $\pi^+$ decay, and are also rare products of the $K^+$ decays.

The detector was a spherical vessel lined with 1280 photomultiplier tubes with a total photocoverage of 11.3\% and filled with 818 tons of mineral oil \cite{MiniBooNE:2008paa}.
Neutrinos interacted via charged-current quasielastic scattering (CCQE) which produced either a final-state $e$ or $\mu$ depending on the neutrino flavor.
Observed events in the detector could be classified as $e$-like or $\mu$-like depending on the observed Cherenkov ring topology.
For $e$-like events, the Cherenkov ring is distorted (sometimes referred to as ``fuzzy'') due to radiative losses, whereas $\mu$-like events produce clean rings since they are minimum ionizing particles in this energy range.
A clean sample of $e$-like events was necessary to search for $\nu_\mu\rightarrow\nu_e$ oscillations.
Notably, $\pi^0\rightarrow\gamma\gamma$ was an important final state as it could be misidentified as $e$-like if one of the two Cherenkov rings was unidentifiable.
This could occur if one of the $\gamma$ was too low energy, or if one exited the detector before producing a measurable signal.
$\pi^0$ final states originate from neutral-current resonant scattering, where a nucleon is excited to a $\Delta$ and subsequently decays to $N+\pi^0$.
The other primary background to $e$-like events is $\Delta\rightarrow N\gamma$ decays, which produces a single electromagnetic shower.

\begin{figure}
    \centering
    \includegraphics[width=0.85\linewidth]{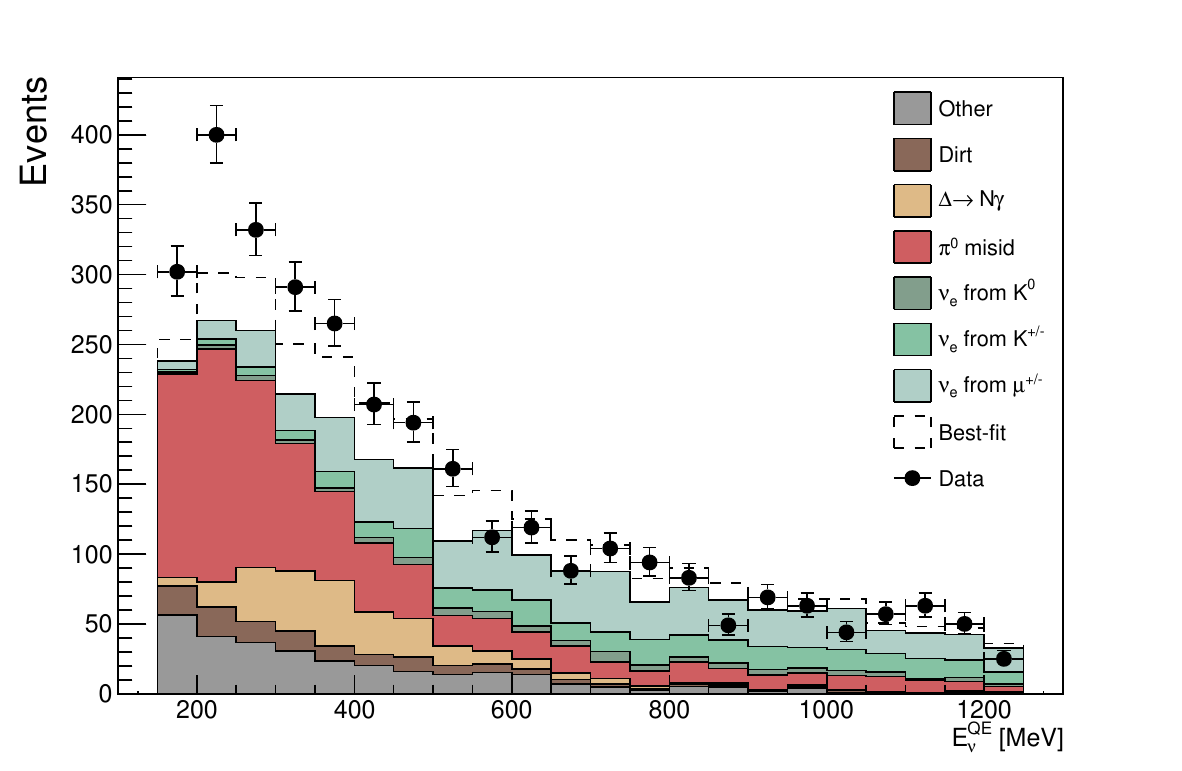}
    \caption{Excess events from the MiniBooNE experiment for 150 < $E_{\nu}^{QE}$ < 1250 MeV. Data points are marked as black dots, the stacked histograms indicate simulated background events, and the dashed line represents the best-fit 3+1 sterile neutrino expectation. Figure from Ref. \cite{MiniBooNE:2020pnu}.}
    \label{fig:miniboone_excess}
\end{figure}

MiniBooNE measured the energy and angular distributions of observed $e$-like events, finding a low-energy excess (\cref{fig:miniboone_excess}) consistent with neutrino oscillations in a similar region of parameter space as LSND at a significance of $4.8\sigma$ as of the latest oscillation analysis \cite{MiniBooNE:2018esg, MiniBooNE:2020pnu}.
This effect was observed in both neutrino and antineutrino modes \cite{MiniBooNE:2012maf}.
A three-fold enhancement of the $\Delta\rightarrow N\gamma$ process could accommodate the excess in the observed event distributions, but this is strongly disfavored by recent measurements by the MicroBooNE experiment \cite{MicroBooNE:2025ovj}.
MiniBooNE also searched for the disappearance of $\nu_\mu$ and $\bar{\nu}_\mu$ finding no significant signal \cite{MiniBooNE:2009ozf, MiniBooNE:2012meu, SciBooNE:2011qyf}.

The LSND and MiniBooNE anomalies found a strong preference for oscillations between $\nu_\mu\rightarrow\nu_e$ and $\bar{\nu}_\mu \rightarrow\bar{\nu}_e$ at high $\Delta m^2$ not consistent with solar or atmospheric neutrino oscillations.
However, these are not the only channels that observe anomalous oscillation-like behavior.
The gallium-based experiments that measured solar neutrinos found a large disappearance of $\nu_e$ while performing calibration runs with high-activity sources.
SAGE $^{51}$Cr source run yielded $r = \mathrm{Measured/Expected}=0.95\pm0.12$ \cite{SAGE:1998fvr} and $^{37}$Ar source run yielded $r = 0.79_{-0.10}^{+0.09}$ \cite{Abdurashitov:2005tb}.
GALLEX ran two $^{51}$Cr source runs and recalibrated after the solar neutrino data taking, finding $r = 0.953\pm0.11$ for source run one and $r = 0.812_{-0.11}^{+0.12}$ \cite{GALLEX:1997lja}.
The activity of the source used in the GALLEX calibration runs ($\sim 1.5$ MCi) was nearly $3\times$ higher than that of SAGE ($\sim 0.5$ MCi).

\begin{figure}
    \centering
    \includegraphics[width=0.85\linewidth]{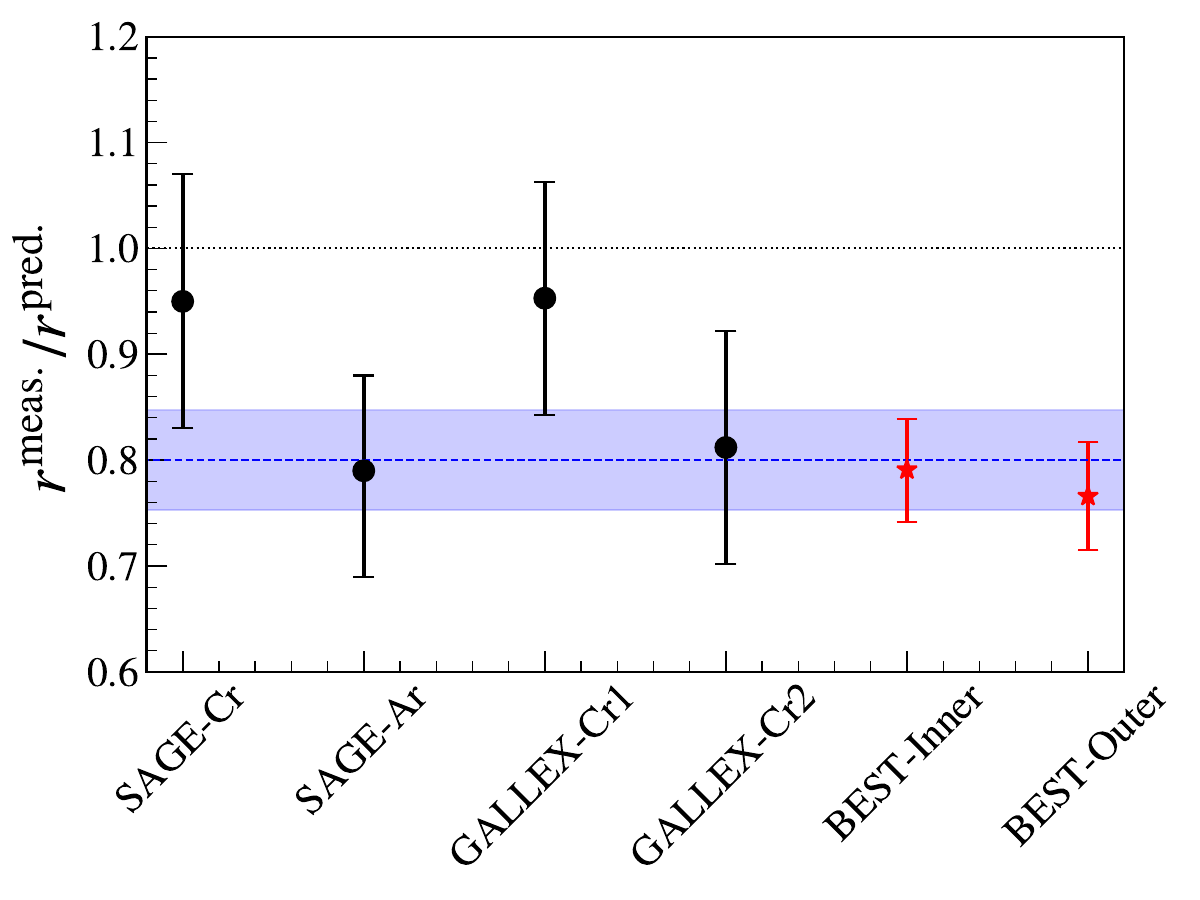}
    \caption{Ratio of measured events to expected for the SAGE, GALLEX source data analyses and the BEST experiment. The global average is shown as the dashed blue line with the band indicating its uncertainty. Figure from Ref. \cite{Elliott:2023cvh}.}
    \label{fig:gallium_exp_ratio}
\end{figure}

As a follow-up to the observed deficits in calibration source data, the BEST experiment was designed to use two gallium zones with a more intense (3 MCi) $^{51}$Cr source inserted into the center of the detector. 
With two zones, if the neutrino oscillation wavelength were on the order of 1 m, a difference between the measured events in each tank could observe it.
BEST observed $r = 0.79\pm0.05$ in the inner tank and $r = 0.77 \pm 0.05$ in the outer tank, confirming the observations made by the SAGE and GALLEX experiments with reduced uncertainties \cite{Barinov:2022wfh}.
Notably, there is an absence of a baseline effect as the ratios for both the inner and outer tank are the same within the quoted uncertainties.
If these results are to be interpreted under the oscillation hypothesis, this would indicate oscillations are occurring with a $\Delta m^2 \sim 1$ eV$^2$ and $\sin^2 (2\theta) \sim 0.4$ with a significance of 4$\sigma$.
Global analyses have shown that the total significance of SAGE, GALLEX, and BEST deficits exceed $5\sigma$ \cite{Brdar:2023cms}.
The ratios of the measured to predicted number of events of the gallium experiments are shown in \cref{fig:gallium_exp_ratio}.
To explain the so-called ``gallium anomaly'', thorough investigations of the different systematic effects have been performed \cite{Elliott:2023cvh}.
One of the primary suspects, the $^{71}$Ge decay half-life, has been studied from both the experimental and theoretical angles and has proven to be robust \cite{Elliott:2023xkb, Norman:2024hki, Collar:2023yew, Cai:2026twp}.
Currently, there are no sufficient explanations for the gallium anomaly.

The most significant signals suggesting neutrino oscillations with eV-scale mass splittings are the LSND, MiniBooNE, and gallium anomalies.
However, many other experiments have searched for these effects and found very weak preference or set exclusions.
Another class of experiments, which had their own anomalies, are the reactor-based experiments.
These experiments utilized nuclear reactors to search for the disappearance of $\bar{\nu}_e$.
Presently, the largest claimed signal was observed by the Neutrino-4 experiment at the SM-3 reactor in Dimitrovgrad, Russia \cite{Serebrov:2017ivv}.
The detector was designed to be able to probe multiple wavelengths of neutrino oscillations if $\Delta m^2 \sim 1$ eV$^2$, and the collaboration claimed to have observed $\bar{\nu}_e$ disappearance at a significance of $2.7\sigma$ \cite{Serebrov:2020kmd}. 
This result has been scrutinized by other experiments and theorists, which have criticized the statistical methodology and pointed out potential systematics that mimic an oscillation-like signature \cite{PROSPECT:2020raz, Giunti:2021iti}. 
Other reactor experiments, such as PROSPECT and STEREO, have not observed any signals and set strong exclusions on $\bar{\nu}_e$ disappearance \cite{STEREO:2022nzk, PROSPECT:2024gps}.
The DANSS experiment has published results finding no significant evidence for $\bar{\nu}_e$ disappearance \cite{DANSS:2018fnn}. 
More recent results indicate a slight preference for an oscillation signal at 2.1$\sigma$ \cite{Alekseev:2025cib}; the latest preliminary results have risen to a significance of 2.5$\sigma$, though these results have not yet appeared in a peer-reviewed publication \cite{Danilov:2025danss}.
The non-observation of $\bar{\nu}_e$ disappearance from reactor experiments is difficult to reconcile with the strong $\nu_e$ disappearance signals from the gallium experiments, as well as the LSND and MiniBooNE excesses.

Taken together, the LSND and MiniBooNE appearance signals and the gallium anomaly each exhibit oscillation-like behavior with $\Delta m^2 \sim 1$ eV$^2$, a scale incompatible with the two mass splittings established by solar and atmospheric neutrino measurements. Within the three-flavor oscillation framework, there is no mass eigenstate available to accommodate a splitting of this magnitude. The minimal extension that could explain these signals is the introduction of a fourth neutrino mass eigenstate ($m_4$). However, precision measurements of the $Z$ boson decay width at LEP constrained the number of light, weakly-interacting neutrino flavors to three \cite{ALEPH:2005ab}, requiring that any additional mass eigenstate have negligible coupling to the weak interaction. Such a state is referred to as a sterile neutrino.

\section{Sterile Neutrinos}\label{sec:sterile_intro}
\begin{figure}
    \centering
    \includegraphics[width=0.55\linewidth]{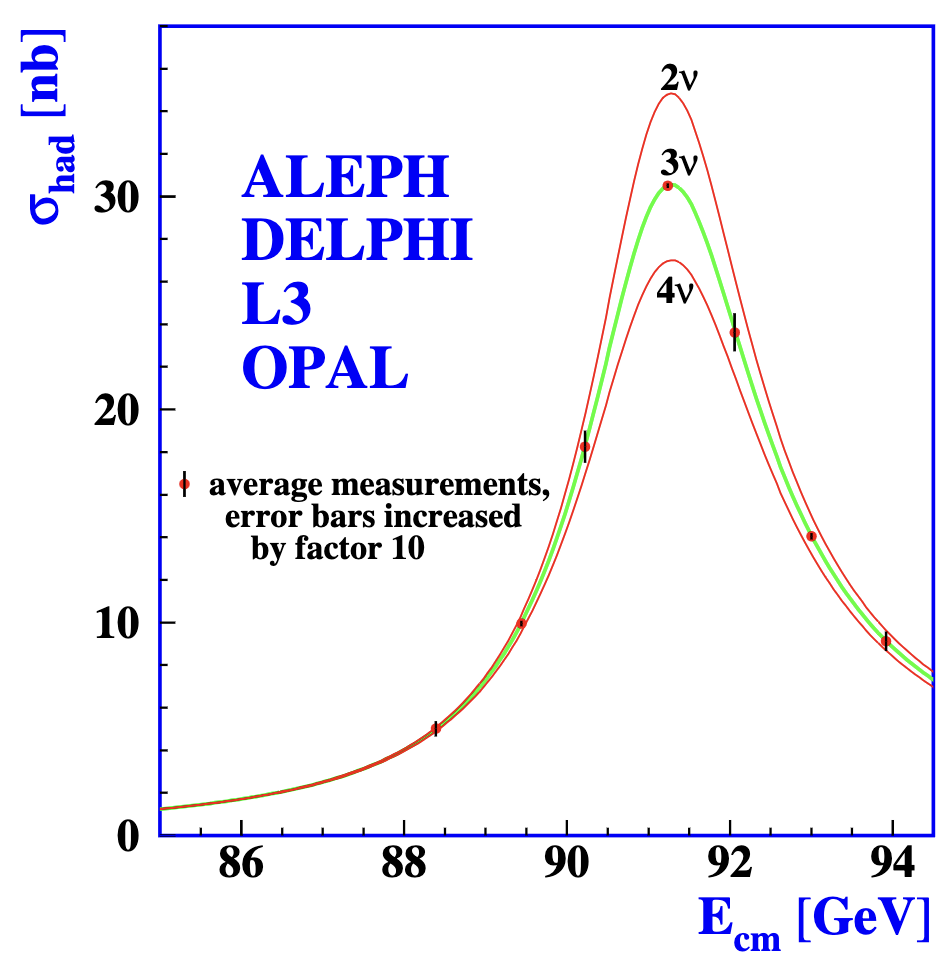}
    \caption{Hadronic production cross section at LEP as a function of center-of-mass energy $E_{\mathrm{cm}}$. Green line corresponds to number of neutrinos $N_\nu = 3$ and the red lines correspond to $N_\nu = 2$ and $N_\nu = 4$. Figure from Ref. \cite{ALEPH:2005ab}.}
    \label{fig:lep_z_width}
\end{figure}

There are complications with adding additional neutrino states. For example, one of the strongest constraints is from weak measurements at colliders, specifically the invisible decay width of the $Z$ boson,
\begin{equation}
    \Gamma_Z = \Gamma_{h} + \Gamma_e + \Gamma_\mu + \Gamma_\tau + N_{\nu}\Gamma_\nu
\end{equation}
where $N_\nu$ is the number of neutrinos and each of the $\Gamma_\alpha$ is the partial widths of species $\alpha$.
Note that this constraint only holds for neutrinos with $m_\nu < \frac{1}{2} m_{Z}$.
\cref{fig:lep_z_width} shows the variation in hadronic production cross section as a function of energy given $N_\nu = 2, 3, 4$.
The collider constraints on $N_\nu$ are very restrictive and exclude $N_\nu > 3$ \cite{ALEPH:1989ikb, ALEPH:2005ab}.
In order for the extension to neutrino oscillations to be viable, the new neutrino state(s) cannot interact via the weak interaction, or else it would appear in the invisible decay width of the Z boson.
These neutrinos are called ``sterile'', and the three neutrinos of the Standard Model are referred to as ``active''.
The simplest extension is to add one new sterile mass state $\nu_4$ and sterile flavor state $\nu_s$.
To describe neutrino flavor mixing with the new neutrino states, a row and column are added to the PMNS matrix which mixes the sterile state with the active flavors,
\begin{equation}
    U_{3+1} = \begin{pmatrix}
        U_{e1} & U_{e2} & U_{e3} & U_{e4} \\
        U_{\mu1} & U_{\mu2} & U_{\mu3} & U_{\mu4} \\
        U_{\tau1} & U_{\tau2} & U_{\tau3} & U_{\tau4} \\
        U_{s1} & U_{s2} & U_{s3} & U_{s4} \\
    \end{pmatrix}
\end{equation}

The results of sterile neutrino analyses typically look only at two-flavor oscillations.
This is a good approximation for most cases as typically neutrino sources are pure in one flavor and experiments search for the disappearance of that flavor, or the appearance of another.
For sterile neutrino searches, experiments typically report a single $\Delta m^2$ and $\sin^2(2\theta)$.
Most experiments are not sensitive to the oscillations with multiple sterile states and it becomes increasingly difficult to perform analyses searching for a higher number of sterile neutrinos.
Instead, the 3+1 model is used as a practical model for searching for the disappearance of the active neutrinos.
The relevant mixing parameters are then $\Delta m_{41}^2$ and typically only one active-sterile mixing angle $\theta$.

\begin{figure}
    \centering
    \includegraphics[width=0.9\linewidth]{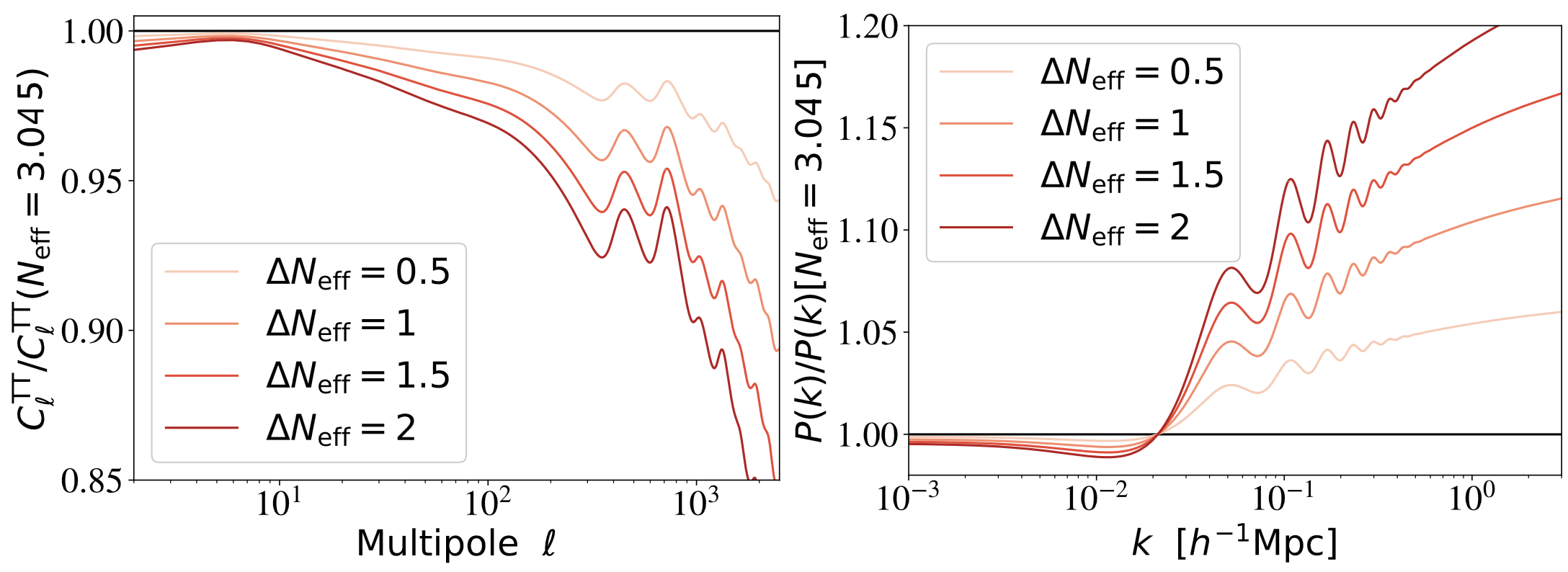}
    \caption{Left: The effect of changing $N_{\mathrm{eff}}$ on the CMB temperature spectrum compared to the Standard Model $N_{\mathrm{eff}} = 3.045$. Right: The effect of changing $N_{\mathrm{eff}}$ on the CMB power spectrum. Figure from Ref. \cite{ParticleDataGroup:2024cfk}.}
    \label{fig:neff_cosmology}
\end{figure}

In addition to the constraints on the number of neutrinos from collider data, there are also constraints that utilize cosmological measurements assuming $\Lambda$CDM.
The relevant cosmological parameter is the number of relativistic species $N_{\mathrm{eff}}$ which contribute to the energy density,
\begin{equation}
    \rho_{rad} = \rho_{\gamma} \left[ 1 + \frac{7}{8}\left(\frac{4}{11}\right)^{4/3} N_{\mathrm{eff}} \right].
\end{equation}
Light sterile neutrinos thermalize in the early universe through oscillations with the three active neutrinos before neutrino decoupling at $T \sim 2$--$3$ MeV.
Since the sterile neutrino adds to the radiation energy density, it increases the expansion rate ($H(z)$) during the radiation-dominated era of the universe.
This shifts the redshift of matter-radiation equality,
\begin{equation}
    1 + z_{\mathrm{eq}} = \frac{\omega_m}{\omega_\gamma} \frac{1}{1 + 0.227 N_{\mathrm{eff}}}
\end{equation}
where $\omega_m$ and $\omega_\gamma$ are the energy densities of matter and photons today.
$z_{\mathrm{eq}}$ is measured by taking the ratio of acoustic peaks of the CMB power spectrum, providing constraints on $N_{\mathrm{eff}}$ (see \cref{fig:neff_cosmology}) \cite{Abazajian:2012ys}. 
Other observables, such as the abundance of light elements, the sound horizon at recombination, and the high-$\ell$ tail of the power spectrum  are combined in fits which provide strong constraints on $N_{\mathrm{eff}}$.
The current measured value of $N_{\mathrm{eff}}$ using the latest data release of the Dark Energy Spectroscopic Instrument combined with CMB measurements from Planck and the Atacama Cosmology Telescope is $3.23_{-0.34}^{+0.35}$ which agrees with the Standard Model prediction \cite{Elbers:2025vlz}.
Additionally, the sum of neutrino masses $\sum m_\nu$ is constrained to be $< 0.0642$ eV at the 95\% C.L. with a prior constraining $\sum m_\nu > 0$ \cite{Elbers:2025vlz}.
Taking the limit of $m_1 \rightarrow 0$ in the normal mass ordering scheme ($m_1 < m_2 < m_3$), a lower bound on the sum of the neutrino masses can be determined from just the mass squared splittings $\sqrt{\vert\Delta m_{21}^2\vert} + \sqrt{\vert\Delta m_{31}^2\vert} \approx 0.059$ eV using the latest values from NuFit \cite{Esteban:2024eli}, near the limit from cosmology.
These measurements provide very strong constraints on eV-scale sterile neutrinos as they would have contributed to the early universe radiation density and altered $N_{\mathrm{eff}}$ without any additional interactions that modify thermalization.
The constraints on the sum of neutrino masses on the order of tens of meV rule out the allowed regions of the short-baseline experiments without additional physics that suppresses thermalization in the early universe.

\section{Unstable Sterile Neutrinos}\label{sec:unstable_sterile_neutrinos}
One method of alleviating the strong cosmology constraints is to introduce additional physics beyond oscillations to the 3+1 model.
A simple extension to the Standard Model introduces a scalar-neutrino interaction of the form \cite{Moss:2017pur},
\begin{equation}
    \mathcal{L} = \frac{g_{ij}^{s}}{2}\bar{\nu}_{i}^{c}\nu_{j}\phi + i \frac{g_{ij}^{p}}{2}\bar{\nu}_{i}^{c} \gamma_5 \nu_{j}\phi
\end{equation}
where $g_{ij}^{s}$ and $g_{ij}^{p}$ are scalar and pseudoscalar couplings between the neutrinos and a light scalar $\phi$. 
There are two decay processes: visible and invisible neutrino decay.
In the visible decay case, the daughter neutrino is an active flavor eigenstate and is detectable via weak interactions.
For invisible decays, the daughter is sterile and cannot be observed.
The introduction of a decay mechanism depletes the sterile component of neutrinos in the early universe, weakening the constraints that rely on sterile thermalization.
This model is the subject of the analysis presented in \cref{chapter:decay_analysis}.

\section{No Sterile Neutrinos?}
The experiments discussed above provide hints of oscillation-like signals at eV-scale $\Delta m^2$. However, a number of experiments probing the same parameter space have found no such evidence.
The KARMEN experiment was a similar $\bar{\nu}_\mu \rightarrow \bar{\nu}_e$ appearance experiment to LSND, but with a shorter mean baseline of 17.7 m \cite{Gemmeke:1990ix}.
KARMEN did not observe any hint of oscillations and excluded a substantial portion of the allowed regions from LSND, though a small region of the parameter space remained compatible with both results \cite{Armbruster:1998uk, Eitel:1999gt, KARMEN:2002zcm}.

As discussed in \cref{sec:short_baseline_experiments}, several other reactor-based neutrino oscillation experiments poised to probe the same region of parameter space as the gallium anomalies, LSND, and MiniBooNE anomalies did not find any signals.
The most relevant channel for the study presented in following chapters is the $\nu_\mu \rightarrow \nu_\mu$ disappearance channel.
The MINOS experiment and its upgrade MINOS+ were long-baseline accelerator-based neutrino experiments that searched for the disappearance of $\nu_\mu$ from the NuMI beam at Fermilab. 
Although $\nu_\mu$ disappearance does not probe the exact same signal as $\nu_e$ appearance in LSND and MiniBooNE, the two channels are related through unitarity in the 3+1 model.
The combined analysis found no evidence for oscillations, setting strong limits on $\nu_\mu$ disappearance for eV-scale sterile neutrinos \cite{MINOS:2017cae}. 

The MicroBooNE experiment was a 100 ton liquid argon time projection chamber situated in the Booster Neutrino Beam and off-axis from the NuMI beam at Fermilab \cite{MicroBooNE:2016pwy}.
It was designed as an experimental follow-up to the observation of the low-energy excess in MiniBooNE, leveraging the improved particle identification methods of liquid argon to separate electron and photon final states \cite{ArgoNeuT:2016wjb, MicroBooNE:2021gfj}.
The latest results of a 3+1 sterile neutrino search with MicroBooNE did not find any evidence for a low-energy excess \cite{MicroBooNE:2025nll}.
The experiment also explored the possible explanation that the MiniBooNE excess was due to an enhancement in $\Delta\rightarrow N\gamma$, but found no evidence for such an enhancement \cite{MicroBooNE:2021zai, MicroBooNE:2025ovj}.

\begin{figure}
    \centering
    \includegraphics[width=0.9\linewidth]{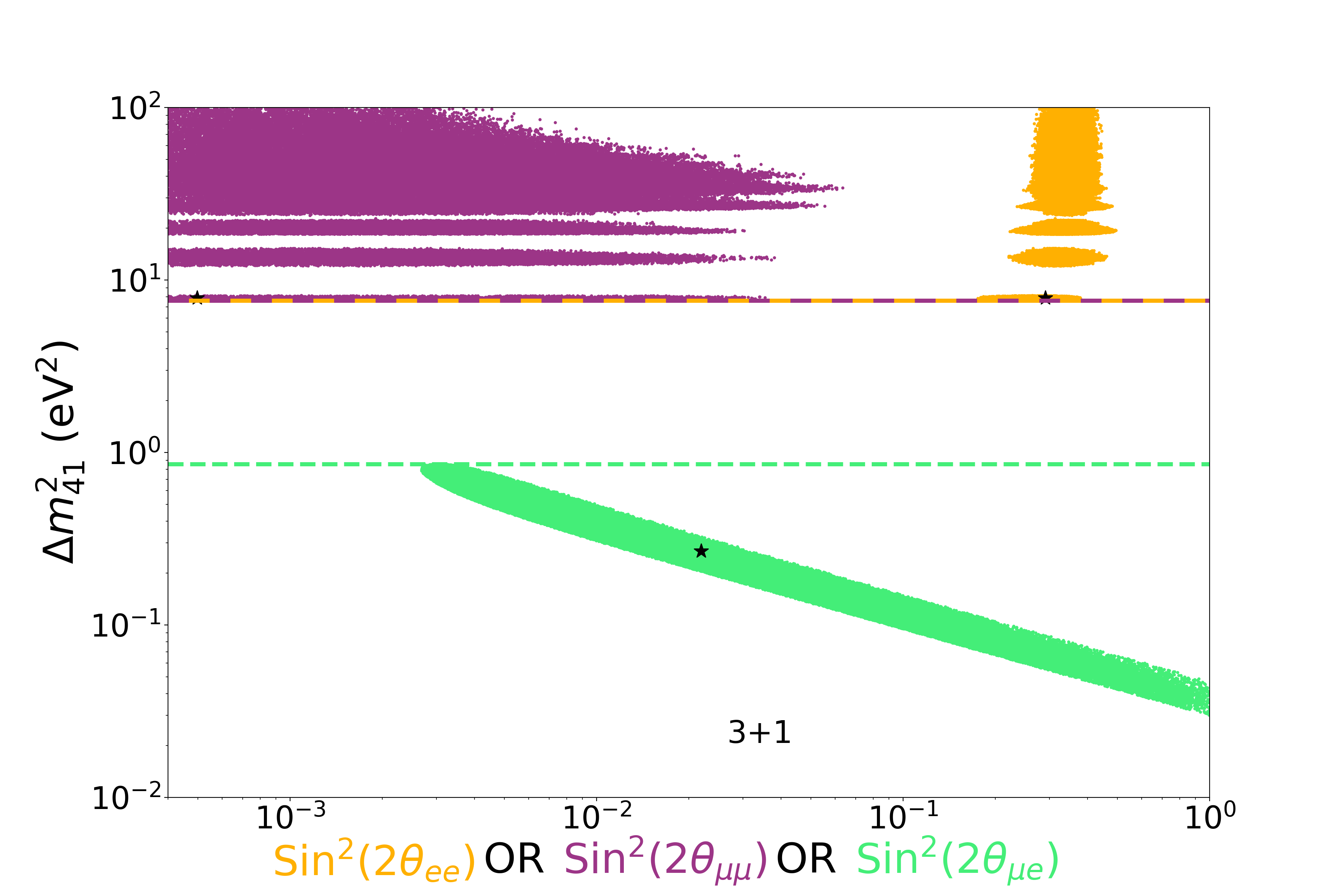}
    \caption{Tension in the 3+1 sterile neutrino landscape according to a global fit to experimental data from 2022. The x-axis shows mixing angles for electron-flavor disappearance (orange), electron-flavor appearance (green), muon-flavor disappearance and the y-axis indicates a shared $\Delta m_{41}^2$. The shaded regions correspond to the 95\% C.L. allowed regions obtained from fits to the corresponding categories of experiments. The dashed lines indicate the smallest/largest allowed $\Delta m_{41}^2$ and the non-alignment indicates tension between the different data sets. Figure from Ref. \cite{Hardin:2022muu}.}
    \label{fig:sterile_global_fit}
\end{figure}

The mix of observation and non-observation of sterile neutrino signals gives rise to a complicated experimental picture.
The combined statistical significance of the LSND, MiniBooNE, and gallium experiment signals is too large to be dismissed as fluctuations and there has not yet been a convincing argument for any systematic effects.
However, none of the experiments mentioned so far have observed any hint of $\nu_\mu$ disappearance at short baselines.
As other experiments and cosmological constraints have demonstrated, if there is some underlying new physics, then it is unlikely to be just a 3+1 sterile neutrino model without extra interactions.
Sterile neutrinos in the 3+1 framework cannot accommodate the signals and absence of signals measured in appearance and disappearance experiments as shown in \cref{fig:sterile_global_fit}.
This motivates searches for non-minimal sterile neutrino models which capture the oscillation-like behavior and alleviate tension within the experimental landscape.
In this thesis, a search for unstable sterile neutrinos using high-energy atmospheric neutrinos will be presented.

\chapter{Neutrino Deep-Inelastic Scattering Cross Sections}\label{chapter:cross_sections}

The neutrino-nucleon deep-inelastic scattering cross section is an important input to the study of atmospheric neutrinos in neutrino telescopes as they are required to predict the event rate.
The differential cross sections also provide information about the interaction kinematics that determine the outgoing momenta of final-state particles, which will be relevant for the studies described in \cref{chapter:sterile_signals} and \cref{chapter:meows2026}.
The following paper presents an updated calculation of the neutrino-nucleon and neutrino-nucleus deep-inelastic scattering (DIS) cross sections, with reduced uncertainties, updated parton distribution functions (PDFs), and improvements at both low and high energies.
This chapter consists of a brief overview of the calculation, followed by the published paper \cite{Weigel:2024gzh} that describes it in detail.

The baseline cross-section model used in most IceCube analyses is the calculation by Cooper-Sarkar, Mertsch, and Sarkar (CSMS) \cite{Cooper-Sarkar:2011jtt}.
This calculation uses next-to-leading order (NLO) QCD predictions for the neutrino DIS structure functions $F_{2}$, $F_{L}$, and $xF_{3}$.
At the time, the HERAPDF1.5 PDFs were the only set to include low-$x$ data from HERA, which is important for the study of high-energy neutrino interactions because of the accessible kinematic space in ($x$, $Q^2$).
The CSMS calculation has been in use for well over a decade and very successful in providing predictions for neutrino telescopes.
Since then, there have been substantial improvements in DGLAP evolution codes \cite{Bertone:2013vaa, Bertone:2017gds}, PDF fits \cite{AbdulKhalek:2019mzd,AbdulKhalek:2020yuc,NNPDF:2021njg,NNPDF:2024dpb,Hou:2019efy,PDF4LHCWorkingGroup:2022cjn,Bailey:2020ooq,McGowan:2022nag}, and nuclear PDF fits \cite{Eskola:2021nhw,Kovarik:2015cma,Duwentaster:2022kpv}.
These developments enable an updated cross-section calculation that uses PDF sets incorporating LHC data that extends to next-to-next-to-leading order (NNLO) in $\alpha_s$.
CT18ANNLO \cite{Hou:2019efy} was chosen as the baseline free-nucleon PDF set, and EPPS21 \cite{Eskola:2021nhw} was used for nuclear targets.
EPPS21 was chosen because it used CT18ANNLO as its free-nucleon baseline and parameterized nuclear modifications relative to it.
Uncertainties on the cross sections and inelasticity distributions are computed from the error sets provided by the fitting groups.

Beyond the updated PDF input, several improvements are made at both low and high energies. 
At low energies, the kinematically accessible $Q^2$ and $W^2$ shift to values where perturbative QCD is no longer reliable.
To extend the calculation into this regime, a phenomenological parameterization of the structure functions at low $Q^2$ was employed \cite{Jeong:2023hwe}.
At high energies, the top quark mass is included exactly as implemented in the FONLL-C mass scheme \cite{Forte:2010ta}, and the effects of small-x resummation are accounted for \cite{Bonvini:2016wki,Bonvini:2017ogt,Bonvini:2018xvt,Bonvini:2018iwt}.

The key features of the improved cross-section calculation are summarized below:
\begin{itemize}
    \item Free-nucleon DIS cross sections at NNLO
    \item Nuclear target DIS cross sections at NLO
    \item Final-state quark flavor separation of the cross sections
    \item Inclusion of the top quark mass
    \item Target mass corrections to the structure functions
    \item Extension to low $Q^2$ and low $W^2$ (``shallow inelastic scattering'')
    \item Quantified uncertainties on the inelasticity distribution
    \item Inclusion of final-state radiation
    \item Effects of small-x resummation
\end{itemize}
These improvements yield updated cross-section predictions with reduced uncertainties for IceCube analyses, particularly the differential inelasticity distributions that enable statistical separation of neutrinos and antineutrinos discussed in \cref{sec:inelasticity}.

\includepdf[pages=-]{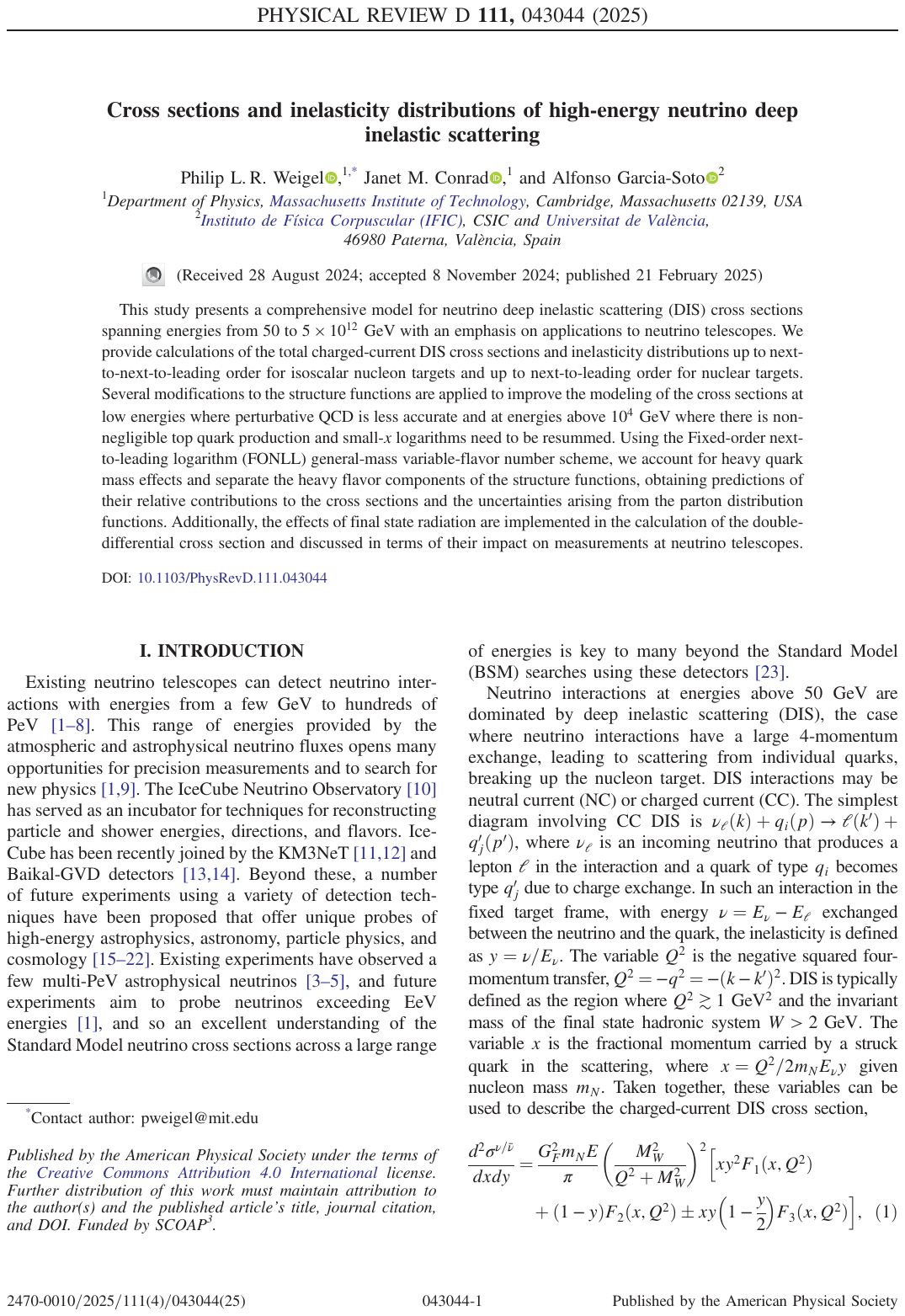}

\chapter{Events in the IceCube Neutrino Observatory}\label{chapter:icecube}
The IceCube Neutrino Observatory is a cubic-kilometer neutrino telescope at the geographic South Pole that detects atmospheric and astrophysical neutrinos across energies from $\sim$10 GeV to $>1$ PeV energies. 
This chapter describes the design of the detector, the understanding of the ice at the South Pole, and the types of signals observed that are relevant for the analysis presented in \cref{chapter:decay_analysis} and the development of an improved track event selection in \cref{chapter:meows2026}.

\section{The Active Detectors}\label{sec:active_detectors}
\begin{figure}[ht]
    \centering
    \includegraphics[width=0.85\linewidth]{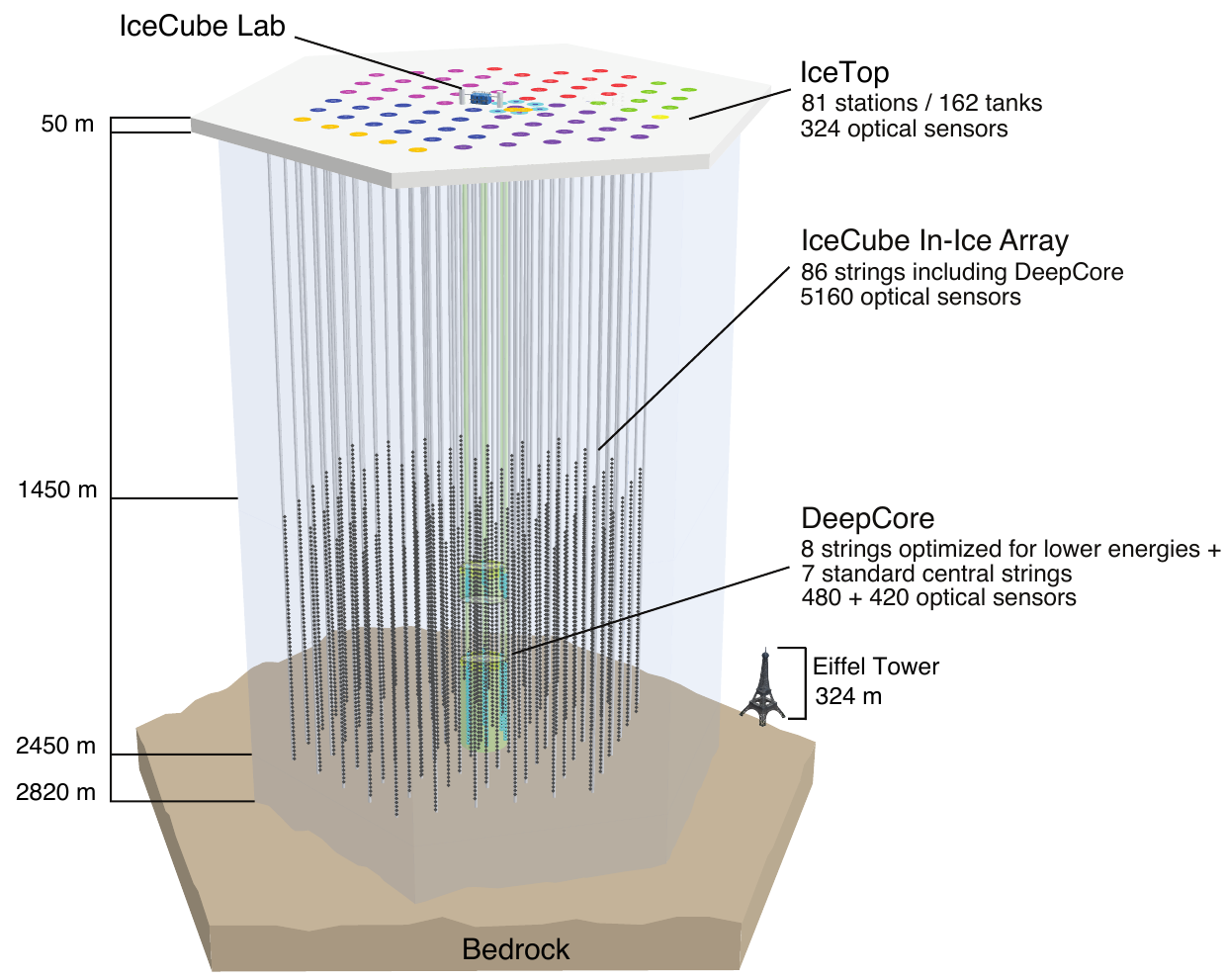}
    \caption{The IceCube detector. Figure from Ref. \cite{IceCube:2016zyt}.}
    \label{fig:icecube_detector}
\end{figure}

Highly detailed descriptions of the IceCube detector have been published elsewhere \cite{IceCube:2010dpc,IceCube:2008qbc,IceCube:2016zyt}.
This section serves as a brief overview.

IceCube is instrumented with 5,160 digital optical modules (DOMs) deployed in strings that detect Cherenkov light emitted by charged particles traversing the ice.
The DOMs shown in \cref{fig:icecube_detector} are located between 1450 and 2450 meters below the surface of the South Pole, completely shielded from surface light contamination.
Strings are spaced out on a hexagonal grid approximately 125 meters apart with a 17 meter vertical spacing between each DOM on a string.
DeepCore is a region of the detector with strings that have a more densely instrumented region near the center which is beneficial for the study of lower energy atmospheric neutrino oscillations.

\begin{figure}
    \centering
    \includegraphics[width=0.5\linewidth]{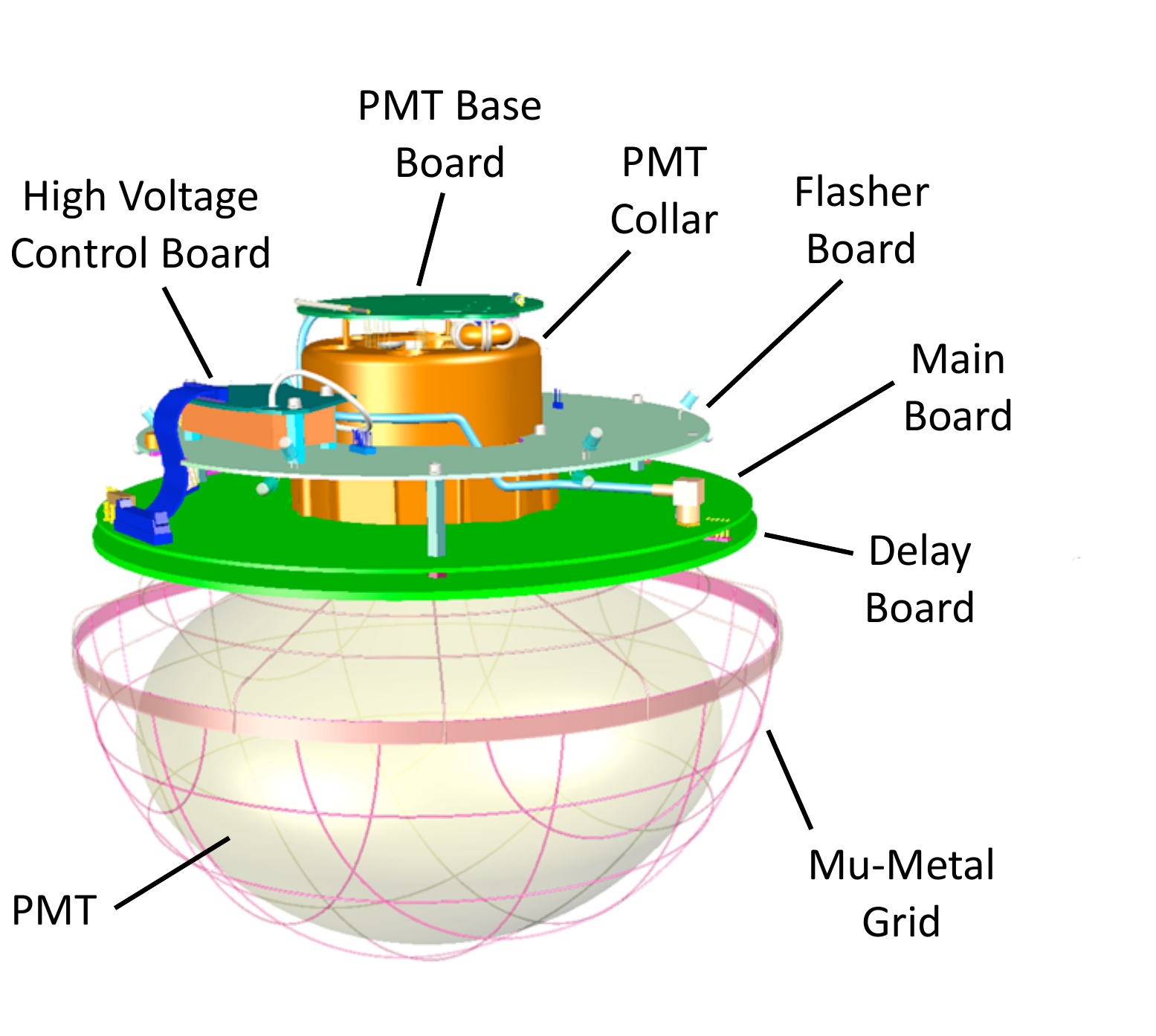}
    \caption{Model of the IceCube digital optical module. Figure from Ref.  \cite{IceCube:2016zyt}.}
    \label{fig:icecube_dom}
\end{figure}

The IceCube DOM is a spherical vessel containing a Hamamatsu R7081-02 photomultiplier tube (PMT) along with digitization hardware that performs basic signal processing before the signal is sent to the surface \cite{IceCube:2010dpc}.
The PMT is coupled to the glass pressure vessel using optical gel and contains magnetic shielding to prevent effects of the Earth's magnetic field on the propagation of electrons within the PMT.
The components of the DOM are shown in \cref{fig:icecube_dom}.
The quantum efficiency of the IceCube PMTs is $\sim25\%$ at a wavelength of 390 nm, but the full range of the spectral response spans from 300 nm to 650 nm \cite{IceCube:2020nwx}.
The dark rate, which is typically due to thermionic emission from the cathode, of the IceCube PMTs is approximately 300 Hz between -40 $^\circ$C and -20 $^\circ$C \cite{IceCube:2010dpc}.
Geothermal heat rising from the bedrock below the ice layers introduces a thermal gradient across the detector that spans this temperature range.

\section{Data Acquisition}\label{sec:data_acquisition}

\begin{figure}
    \centering
    \includegraphics[width=0.75\linewidth]{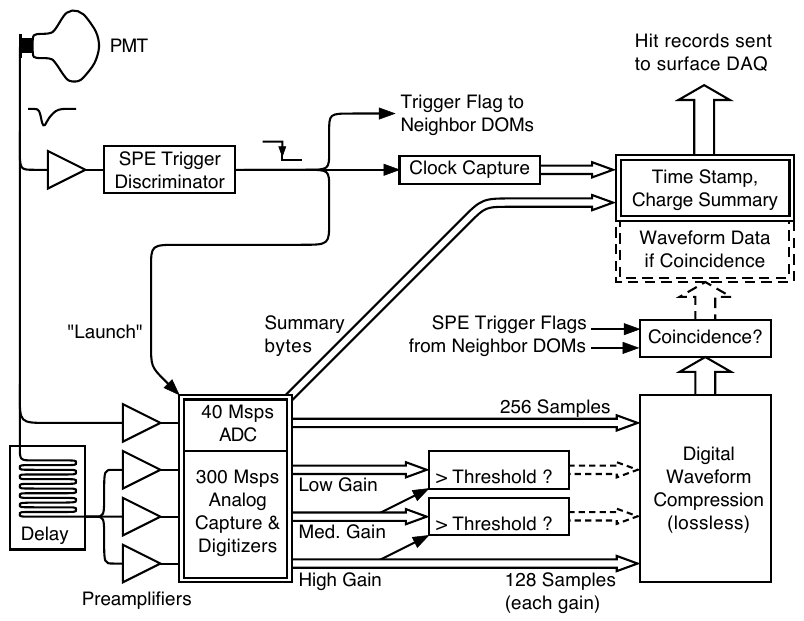}
    \caption{Diagram showing the components of the signal processing chain within an IceCube DOM. Figure from Ref. \cite{IceCube:2016zyt}.}
    \label{fig:icecube_dom_processing}
\end{figure}

An important aspect of the DOMs is the signal processing that occurs on-board prior to being sent to the surface.
A system diagram showing the components of the DOM that process the PMT signal is shown in \cref{fig:icecube_dom_processing}.
The charge signal produced by the PMT when a photon hits it is digitized by the DOM boards through several different paths.
The signal is fed into three amplifiers with gains of 0.25, 2, and 16 which is the input to the analog-to-digital converters (ADC).
The first of two ADCs is the Analog Transient Waveform Digitizer (ATWD) which takes in the amplified analog inputs and provides a dynamic range of 10-bit digitized values at a sampling rate of 300 MSPS.
The second ADC is the fast ADC (fADC), a slower 40 MSPS converter which captures longer time scale behavior of the signal.

The DOMs are calibrated using both on-board LED flashers, shown in \cref{fig:icecube_dom}, as a source of light with known position and intensity, as well as dark noise.
Each PMT is fit to a single photoelectron template, which is the distribution of charges observed for single photon detections.
This determines the gain setting of the PMT to establish a uniform response to light across the detector. 
Extensive studies of the DOM calibration efforts can be found in Refs. \cite{IceCube:2010dpc,IceCube:2016zyt,IceCube:2020nwx}.

Individual DOMs are triggered according to different conditions: hard local coincidence (HLC) or soft local coincidence (SLC).
An HLC is a state where one or more neighboring DOMs measure a signal and indicates that what was observed was unlikely to be induced by noise.
An SLC is when a DOM observes a signal without a neighboring DOM having observed anything.
If an HLC is observed, then the full waveforms of the ADCs are sent to the surface, and only the highest fADC value with adjacent bin values are sent if it is an SLC \cite{Axani:2019sbk}.
On the surface, the collected waveforms are aggregated into windows that represent an event using global triggers.
The SMT-8 trigger is one of the first levels of event processing and is a trigger that selects only events that contain eight or more HLC hits, known as the multiplicity, within a $5~\mu$s sliding window.
If the trigger condition is satisfied, then all hits within $-4~\mu$s and $+6~\mu$s of the trigger time are saved as an event \cite{IceCube:2016zyt}.
These events are saved and subject to additional levels of processing that form the basis for an analysis event selection.
The event rate of the SMT-8 trigger is approximately 3 kHz, which is almost entirely composed of down-going atmospheric muons \cite{IceCube:2015wro}.
Two levels of processing, called L1/online and L2/offline processing, are applied to every event which establish additional trigger conditions.
These different triggers are used to reduce rates to $\mathcal{O}(1-100)$ Hz for more specific physics analyses, such as the \texttt{CascadeFilter} which aims to select cascade events.
The selection of events after L2 processing is the starting point of the event selections discussed in \cref{sec:meows2023_event_selection} and \cref{sec:meows2026_event_selection}.

\section{The Ice Properties}\label{sec:ice_properties}

The optical properties of the ice govern the propagation of photons within IceCube, and strongly impact the number of photons that reach a given DOM and the timing distribution of their arrival.
Characterization of the ice properties is essential for reliable event reconstructions and is a significant source of systematic uncertainty in physics analyses.
The IceCube ice properties have been studied in great detail in previous studies \cite{bay2010south,IceCube:2013llx,Chirkin:2013lpu,IceCube:2023qua,IceCube:2024qxf}.
This section serves as an overview of the important ice properties relevant for the analysis presented in \cref{chapter:decay_analysis}.

\begin{figure}
    \centering
    \includegraphics[width=0.97\linewidth]{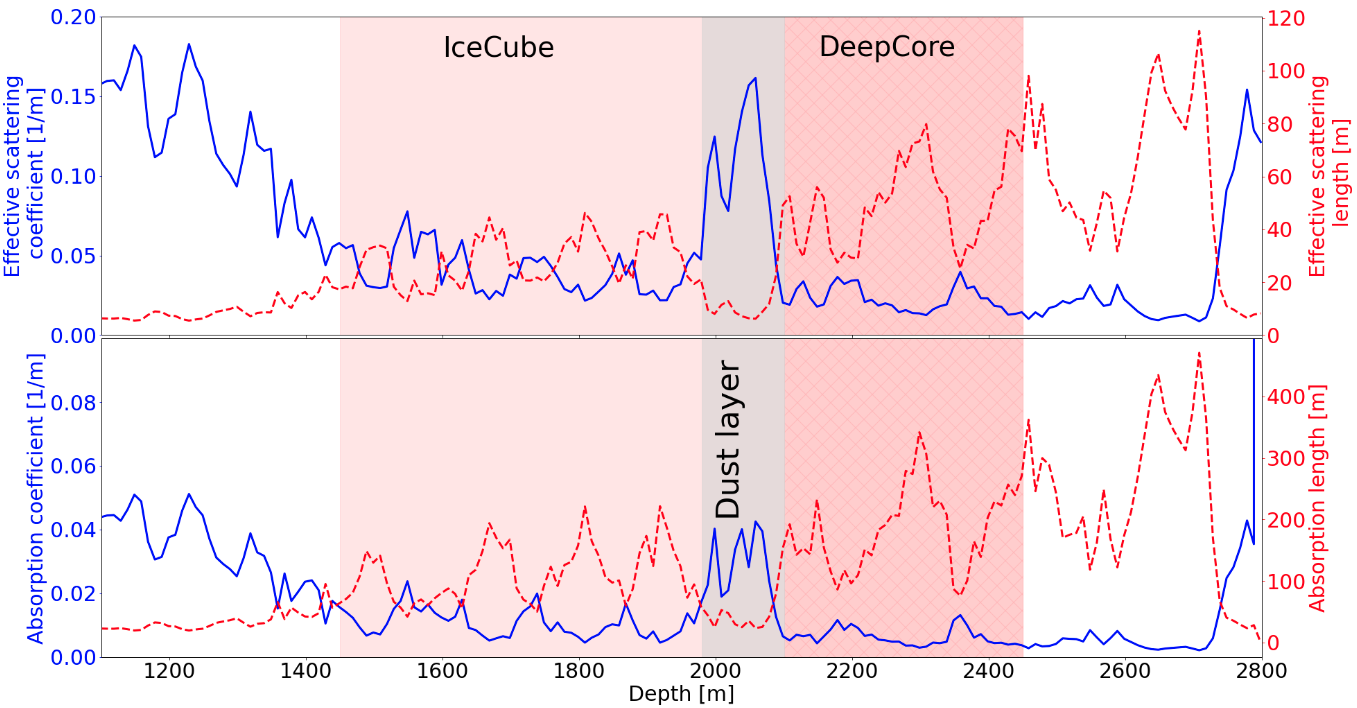}
    \caption{Top: Effective scattering coefficient (blue solid) and length (red dashed). The regions of the IceCube detector, the dust layer, and the DeepCore region are indicated by shaded bands. Bottom: Absorption coefficient (blue solid) and length (red dashed). Figure from Ref.  \cite{IceCube:2024qxf}.}
    \label{fig:icecube_ice}
\end{figure}

The bulk ice at the South Pole is very transparent at optical wavelengths, but the scattering and absorption effects vary with depth.
The photon scattering length, which is the mean free path of a photon between direction-changing scattering processes, and the absorption length depend on the local ice impurities and ice structure.
Impurities in the ice that decrease the scattering and absorption lengths are primarily dust particulates, salt grains, and trapped liquid acid droplets \cite{southpoleice_age_depth}.
As the ice was formed in layers over time, the description of the ice models typically report the effective scattering and absorption lengths as a function of depth.
The ice properties can be studied through several different means and have been measured using lasers, dust-loggers inserted during the string deployment, and the on-board LED flashers on the DOMs.
These measurements form the basis of the South Pole ice (SPICE) model, which provides the depth-dependent scattering and absorption coefficients used in IceCube simulation \cite{IceCube:2013llx, IceCube:2025tbm}.
\cref{fig:icecube_ice} shows the scattering and absorption lengths of light in the IceCube detector as a function of depth.

\begin{figure}[htbp]
    \centering
    \includegraphics[width=0.8\linewidth]{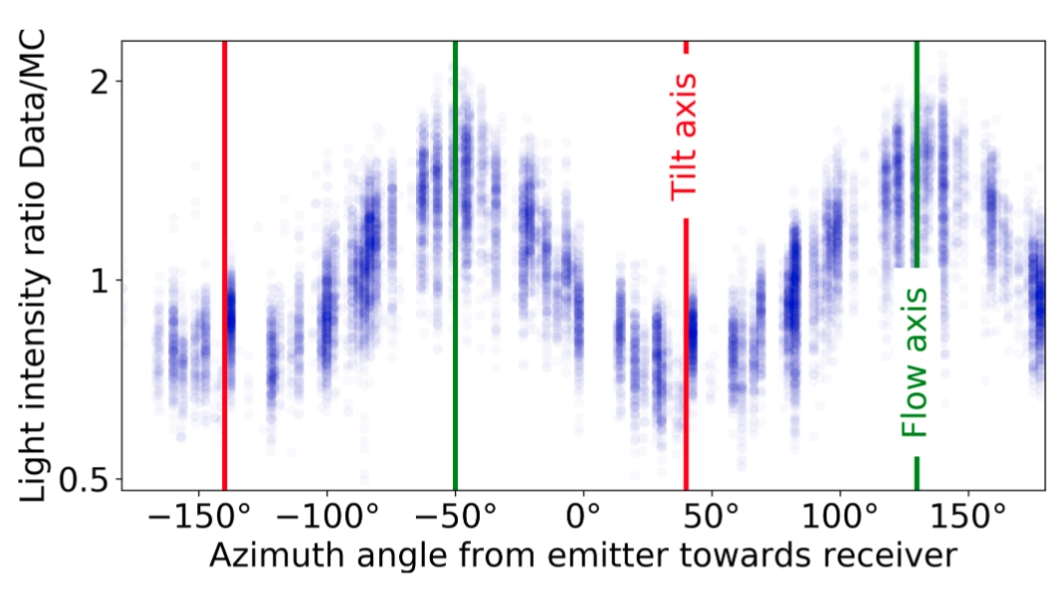}
    \caption{Ratio of measured light intensity using flashers compared to simulation assuming no ice tilt or ice flow as a function of azimuthal angle. Figure from Ref. \cite{IceCube:2024qxf}.}
    \label{fig:icecube_ice_anisotropy}
\end{figure}

In addition to the depth-dependent scattering and absorption processes, the ice exhibits additional geometric effects that complicate the modeling of photon propagation.
The ice layers are not perfectly flat.
There is a tilt to the ice layers which changes the height of the layers in the $x$-$y$ plane up to 60 meters across the full detector \cite{IceCube:2020nwx}.
An additional geometric effect is due to the flow direction of the ice which moves at a rate of about 10 meters per year \cite{southpole_ice_flow}, which produces a shear and introduces an azimuthal asymmetry.
A comparison of the light yield as a function of azimuthal angle is shown in \cref{fig:icecube_ice_anisotropy}, which highlights the directionality of the anisotropy with respect to the ice flow and tilt effects.

The IceCube strings were deployed into columns in the ice by melting it with hot-water drills, and the ice was refrozen once the strings were fully deployed \cite{IceCube:2016zyt}.
The resulting refrozen ice columns, known as ``hole ice'', have different optical properties from the surrounding bulk ice due to trapped air bubbles, which introduce additional scattering near the string \cite{IceCube:2013llx}.
The bubbles are concentrated in a 10 cm diameter inner column, while the full 60 cm diameter hole exhibits modified optical properties out to its edge \cite{IceCube:2013llx,IceCube:2021lmu}.
This effect can be modeled as a modification to the effective DOM angular acceptance and will be treated as such in \cref{chapter:decay_analysis}.

\clearpage

\section{Neutrino Fluxes at IceCube}\label{sec:neutrino_flux}

\subsection{Atmospheric Neutrino Flux}

\begin{figure}
    \centering
    \includegraphics[width=0.8\linewidth]{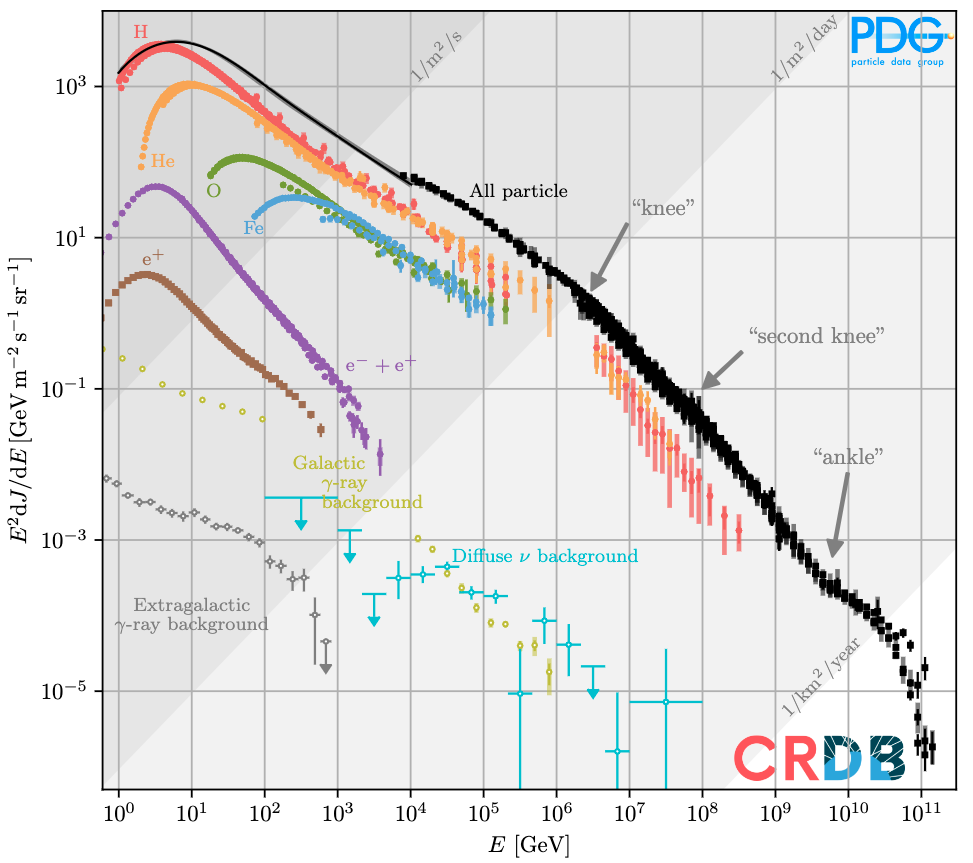}
    \caption{Cosmic ray spectrum separated by components from $10$ GeV to $10^{11}$ GeV. Figure from Ref. \cite{ParticleDataGroup:2024cfk}.}
    \label{fig:pdg_cosmic_rays}
\end{figure}

The dominant source of high-energy neutrinos at IceCube is the atmospheric neutrino flux (briefly described in \cref{sec:atmospheric_neutrinos}), which is generated by cosmic ray interactions in the atmosphere.
Cosmic rays are highly energetic particles produced in extreme astrophysical environments.
They are composed primarily of protons, helium nuclei, and other light elements, with smaller fractions of heavier nuclei \cite{ParticleDataGroup:2024cfk}.
Cosmic rays span many orders of magnitude in energy and arrive at Earth nearly isotropically.
The flux of cosmic rays is well-described by a power law for certain energy ranges,
\begin{equation}\label{eq:power_law}
    \Phi = \Phi_0 \left(\frac{E}{E_0}\right)^{\gamma}
\end{equation}
where $\gamma$ is the spectral index, $\Phi_0$ is the flux normalization, and $E_0$ is the reference energy.
Below a few PeV, the flux is well-described with $\gamma = -2.7$, but above those energies there is a spectral break known as the ``knee'' which steepens the spectral index to $\gamma \approx -3$, which is believed to be caused by changes in source and species composition \cite{LHAASO:2024knt,LHAASO:2025byy, ParticleDataGroup:2024cfk}.
At the very highest energies, there is another spectral break known as the ``ankle'' thought to mark the transition from galactic sources to extragalactic sources \cite{Aloisio:2012ba}.
The spectrum of cosmic rays as a function of energy is shown in \cref{fig:pdg_cosmic_rays}, with the species composition and the spectral breaks.

\begin{figure}
    \centering
    \includegraphics[width=0.95\linewidth]{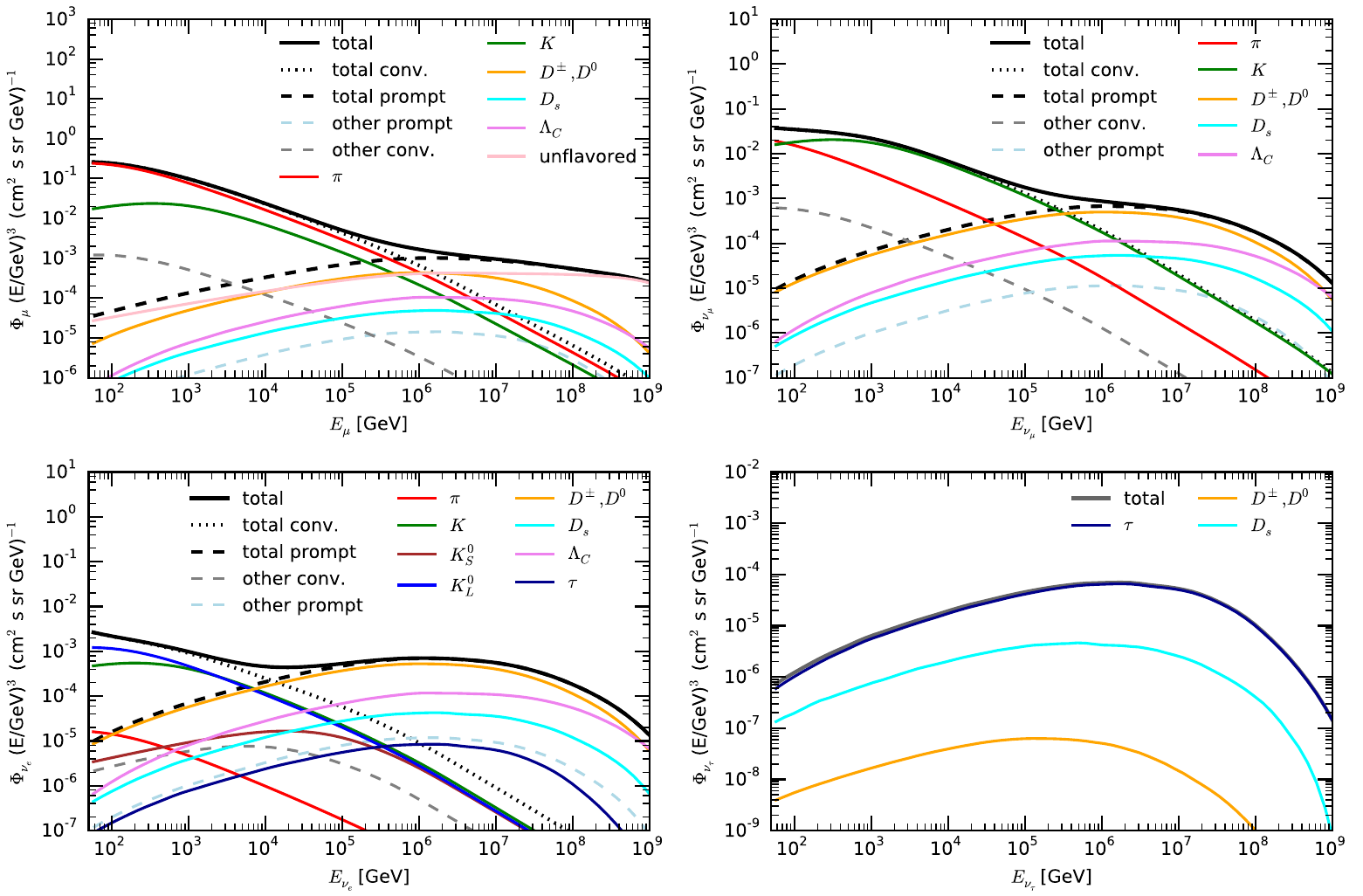}
    \caption{Atmospheric muon flux (top left), muon neutrino flux (top right), electron neutrino flux (bottom left), and tau neutrino flux (bottom right). Figure from Ref. \cite{Fedynitch:2015zma}.}
    \label{fig:mceq_fluxes}
\end{figure}

When a cosmic ray interacts in the upper atmosphere, it initiates a hadronic cascade that produces a population of mesons, predominantly pions and kaons.
Charged pions decay almost exclusively (99.988\% branching ratio \cite{ParticleDataGroup:2024cfk}) via $\pi^\pm \rightarrow \mu^\pm + \nu_\mu (\bar{\nu}_\mu)$ and the resulting muons subsequently decay via $\mu^\pm \rightarrow e^\pm + \nu_e(\bar{\nu}_e) + \bar{\nu}_\mu (\nu_\mu)$.
Kaons have many different leptonic, semi-leptonic, and purely hadronic decay modes that produce pions, muons, and electrons, the largest being $K^\pm \rightarrow \mu^\pm + \nu_\mu(\bar{\nu}_\mu)$ at 63.6\% branching ratio.
The hadronic interactions that produce these pions and kaons are described by various hadronic interaction models, such as SIBYLL~\cite{Ahn:2009wx, Riehn:2017mfm, Fedynitch:2018cbl, Riehn:2019jet}, DPMJET-III~\cite{Roesler:2000he}, EPOS-LHC~\cite{Pierog:2013ria}, and QGSJET-II~\cite{Ostapchenko:2010vb}.
In practice, the hadronic interaction model and primary cosmic ray flux model are used to numerically solve a set of cascade equations and produce a leptonic flux prediction using a code called \texttt{MCEq} \cite{Fedynitch:2015zbe, Fedynitch:2015zma}.
An additional component to the atmospheric neutrino flux originates from very short-lived particles such as $D$ mesons, $\Lambda$ baryons, and $\tau$ leptons \cite{Bhattacharya:2016jce}.
This is known as the prompt atmospheric neutrino flux and is the dominant component of the atmospheric flux above $\sim$ 100 TeV.
Despite this, the prompt flux has not been explicitly measured as it is subleading to the flux of neutrinos of astrophysical origin.
\cref{fig:mceq_fluxes} shows the resulting atmospheric muon and neutrino fluxes ($\nu_\mu$, $\nu_e$, and $\nu_\tau$) from each hadron species.

\clearpage

\subsection{Astrophysical Neutrino Flux}

Cosmic rays near astrophysical accelerators can interact and produce hadrons which subsequently produce very high energy neutrinos.
This flux is approximately isotropic and referred to as the diffuse astrophysical flux.
To date, several sources of astrophysical neutrinos have been identified by IceCube, such as TXS 0506+056~\cite{IceCube:2018cha}, NGC 1068~\cite{IceCube:2022der,IceCube:2025gdd}, and the galactic plane~\cite{IceCube:2023ame}.
The astrophysical neutrino flux is typically parameterized as a power law (\cref{eq:power_law}).
Measurements from IceCube~\cite{IceCube:2018pgc, IceCube:2020acn, IceCube:2020wum, Abbasi:2021qfz, IceCube:2024fxo} give a best-fit spectral index between $\gamma_{\mathrm{astro}} = 2.4$ and $2.9$ depending on the event sample and analysis, with an assumed equal flavor ratio at Earth and equal neutrino-to-antineutrino ratio.
A recent IceCube analysis found a preference for a spectral break in the astrophysical neutrino flux using a broken power law \cite{IceCube:2025tgp}, which will be the astrophysical flux model used in the analysis presented in \cref{chapter:decay_analysis}.

\section{Expected Neutrino Event Morphologies at High Energies}

\begin{figure}[ht]
    \centering
    \begin{subfigure}[b]{0.48\textwidth}
        \centering
        \includegraphics[width=\textwidth]{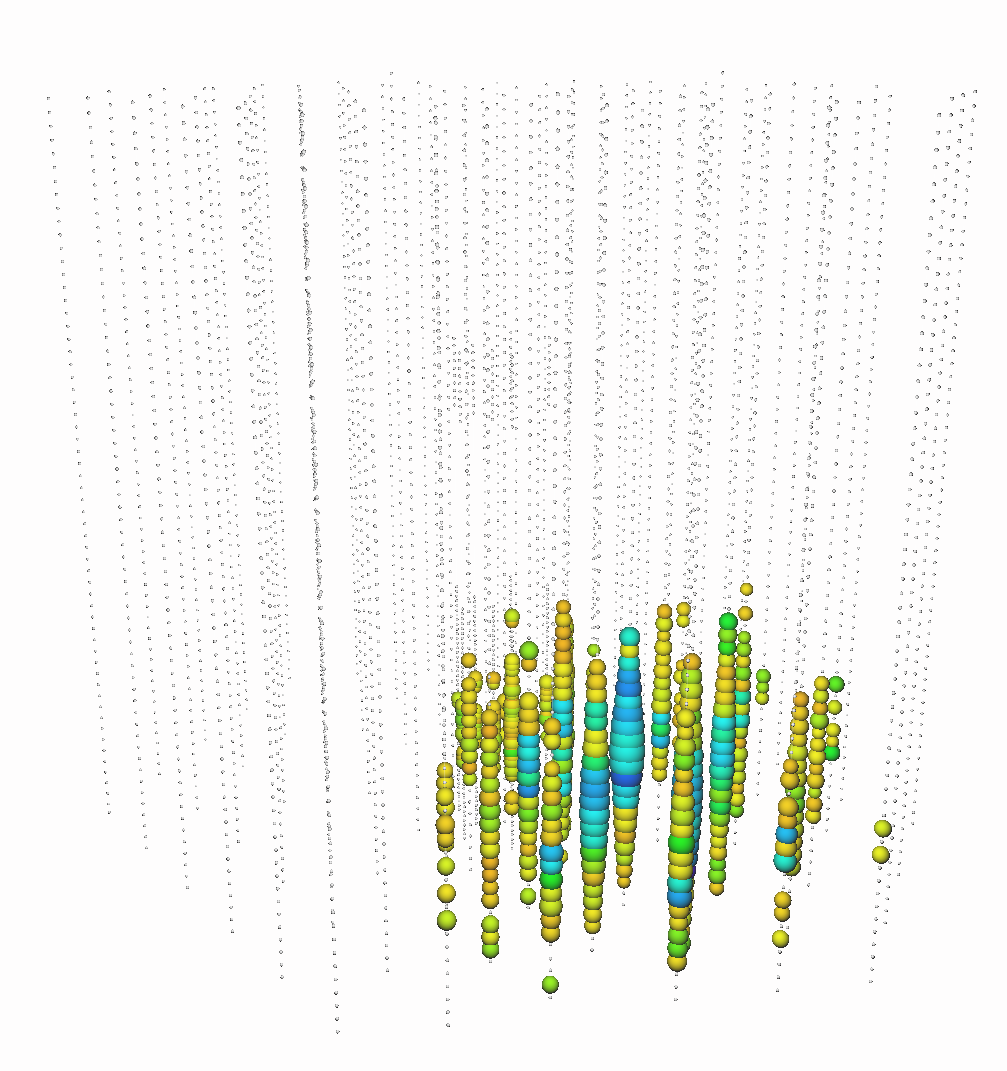}
        \caption{Cascade}
        \label{fig:icecube_cascade}
    \end{subfigure}
    \hfill
    \begin{subfigure}[b]{0.48\textwidth}
        \centering
        \includegraphics[width=\textwidth]{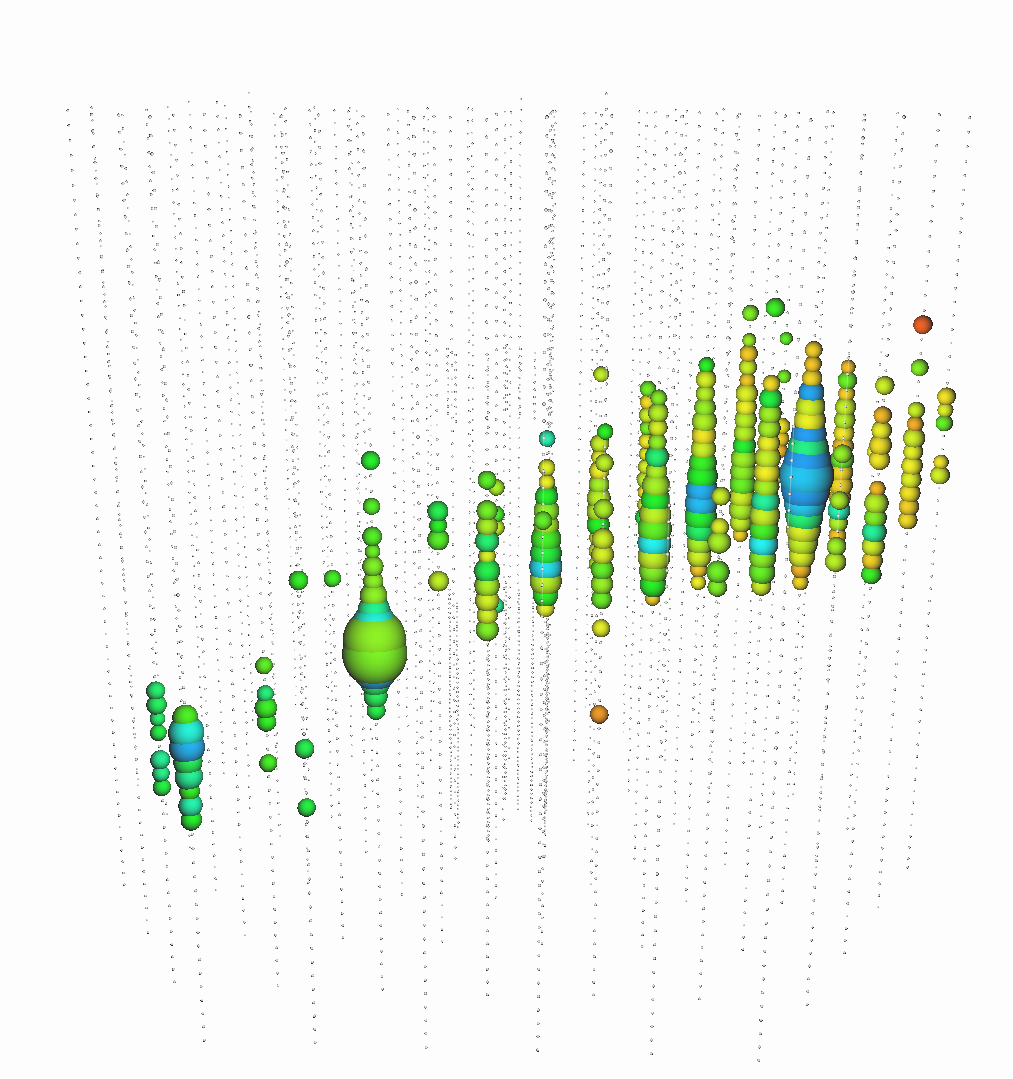}
        \caption{Starting track}
        \label{fig:icecube_starting_track}
    \end{subfigure}
    \caption{Event displays of a cascade event (left) and a starting track event (right). The size of the bubbles indicate the amount of charge observed and the color indicates the time that the charge was measured, where blue is early and red is late. Figure from Ref.  \cite{Kronmueller:2019jzh}.}
    \label{fig:icecube_event_displays}
\end{figure}

As the IceCube DOMs sparsely instrument the ice, the finer structure of events is washed out.
This means that IceCube can only differentiate between a few event morphologies.
The most common morphologies are tracks and cascades.
Cascades are near-spherical depositions of energy in the detector typically caused by electromagnetic or hadronic showers as shown in \cref{fig:icecube_cascade}.
Tracks, long and narrow depositions, are produced by muons which are able to traverse distances larger than the detector.
Neutrino events that are observed as tracks are almost entirely $\nu_\mu$ charged-current deep inelastic scattering, where the final state muon passes through the detector.
Charged-current $\nu_\tau$ events can also produce a track if the daughter $\tau$ decays to a muon, though this rate is small because of the 17 \% branching ratio and low $\nu_\tau$ flux at high-energies.
There are also compound morphologies, such as starting tracks.
A starting track is a cascade vertex with an outgoing track, produced by charged-current $\nu_\mu$/$\nu_\tau$ interactions occurring within the detector volume.
An example of a starting track is shown in \cref{fig:icecube_starting_track}.
Double cascades can be formed when a final-state $\tau$, from $\nu_\tau$ CC DIS, decays and gives an initial cascade from the DIS vertex separated by some distance before the $\tau$ decays.
The distance traveled by a $\tau$ in ice is approximately 50 m/PeV, making double cascade identification very difficult.

At higher energies, additional processes can occur.
Though these occur at a much smaller rate than the typical DIS events, it will be discussed here for completeness.
The Glashow resonance is the process $\bar{\nu}_{e} + e^{-} \rightarrow W^{-}$, which has a resonance energy of
\begin{equation}
    E_{res} = \frac{M_W^2}{2 m_e} = 6.3~\mathrm{PeV}.
\end{equation}
Neutrinos of this energy are rare, IceCube has observed only one Glashow resonance candidate event \cite{IceCube:2021rpz}.
$W$ bosons can also be produced on-shell through neutrino-nucleus scattering where a virtual photon is exchanged, $\nu_\ell + A \rightarrow \ell^- + W^+ + A^\prime$, where $A$ is the initial state nucleus and $A^\prime$ is the final state nucleus \cite{Zhou:2019vxt}.
Dimuon event morphologies are also possible through trident production, where two charged leptons are produced through neutrino-nucleus scattering, and $\nu_\mu$ CC DIS scattering that produces a charmed meson which subsequently decays to a muon.

\section{Energy Deposition and Measurement}\label{sec:energy_dep}

\begin{figure}
    \centering
    \includegraphics[width=0.8\linewidth]{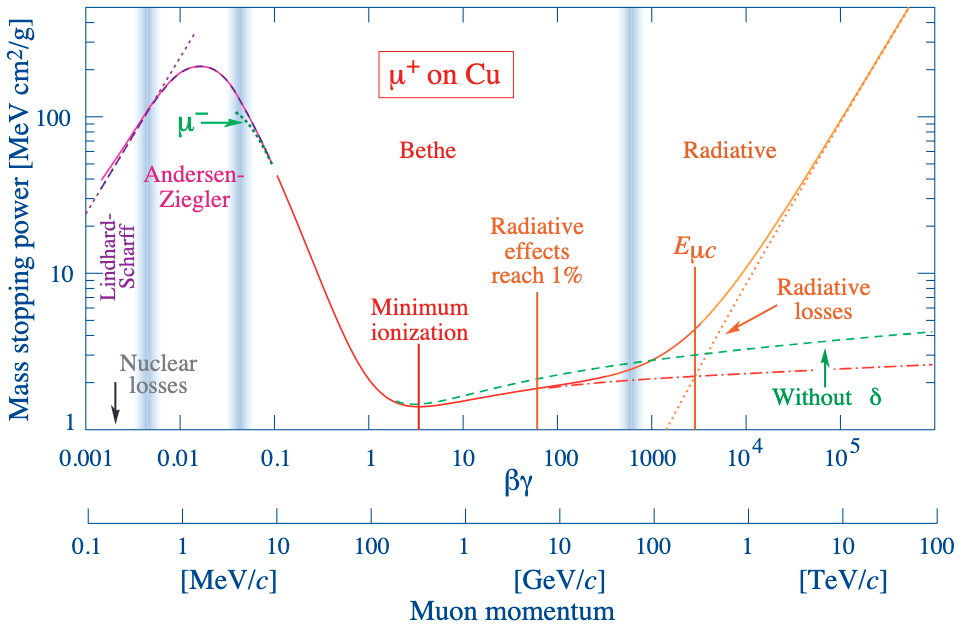}
    \caption{Muon energy loss on copper as a function of momentum. Figure from Ref. \cite{ParticleDataGroup:2024cfk}}
    \label{fig:muon_energy_loss_cu}
\end{figure}

Muons traveling through the ice lose energy through two processes.
At low energies ionization dominates. 
Muons interact electromagnetically with atoms and lose energy at a rate that is nearly independent of energy.
At higher energies, stochastic radiative processes become increasingly important: $e^{+}e^{-}$ pair production, bremsstrahlung, and photonuclear reactions produce electromagnetic cascades that deposit energy along the muon track.
The mean rate of energy loss is described by,
\begin{equation} \label{eq:energy_losses}
   - \left\langle \frac{dE}{dx} \right\rangle = a + b E
\end{equation}
where $a$ describes the losses due to ionization and $b$ describes the coefficient to the linearly-scaling radiative losses.
For muons in ice, $a = 0.259$ GeV mwe$^{-1}$ and $b = 3.63\times10^{-4}$ mwe$^{-1}$ \cite{Chirkin:2004hz}.
The critical energy is defined as $E_c = a / b \approx 700$ GeV where the energy losses of the two processes are equal.
Below this threshold, the deposited energy is weakly sensitive to the initial muon energy, making energy reconstruction from observed losses difficult.
Above the critical energy, the stochastic losses produce localized cascades along the muon track whose collective light yield provides a handle on the muon energy through the linear energy dependence in \cref{eq:energy_losses}.
The energy loss of muons propagating through a dense medium (copper) is shown in \cref{fig:muon_energy_loss_cu} to demonstrate the different regimes of losses.

The light produced by radiative losses originates from the secondary electromagnetic cascades initiated by the $e^{+}e^{-}$ pairs, bremsstrahlung photons, and the fragments of the photonuclear reactions.
The secondary particles travel faster than the speed of light in the medium, exceeding the Cherenkov threshold, and produce Cherenkov radiation that is detected by nearby DOMs.
The stochastic nature of the energy losses means that the deposited energy fluctuates event-by-event, introducing an irreducible uncertainty on the inferred muon energy even when the muon track is well-reconstructed.

An additional complication arises for muons whose parent neutrino interacted far outside of the instrumented detector volume.
In a charged-current deep inelastic scattering interaction with neutrino energy $E_\nu$, the outgoing hadronic shower carries energy $yE_\nu$ where $y$ is the inelasticity defined in \cref{chapter:cross_sections}.
The outgoing muon therefore leaves the interaction vertex with energy $(1-y)E_\nu$.
For a through-going track event, the hadronic shower at the vertex is not observed, so the full energy of the neutrino cannot be directly reconstructed.
The muon enters the detector having already lost an unknown amount of energy during propagation from the vertex, meaning that only the energy deposited within the instrumented volume is accessible.
Even if the muon energy could be determined exactly, there is an additional per-event uncertainty on the neutrino energy due to the unmeasurable inelasticity $y$.

Starting track events avoid both of these limitations.
When a neutrino interaction vertex occurs within the instrumented detector volume, the hadronic cascade at the vertex is observed and contributes to the overall deposited energy.
The muon energy can be inferred from the stochastic losses as it propagates, removing the ambiguity of how far it had traveled before being measured.
Starting track events therefore provide significantly better neutrino energy resolution compared to through-going track events.
However, this comes at a reduced number of events as through-going events are roughly three times as common as starting tracks in some high-energy IceCube data sets \cite{IceCubeCollaboration:2024dxk}.
The separation of starting tracks from through-going tracks in BSM physics searches, including the one presented in \cref{chapter:decay_analysis}, provides an increase in sensitivity due to the better energy resolution of the subsample.

\chapter{Oscillation Signals at Neutrino Telescopes}\label{chapter:sterile_signals}

This chapter expands on the oscillation signals introduced in \cref{chapter:exp_bkg} with a focus on the signatures expected at neutrino telescopes. 
Atmospheric neutrinos that traverse the Earth are subject to oscillations modified by matter effects analogous to the MSW effect in the Sun, with the strongest modifications occurring for trajectories that pass through the dense core of the Earth.
For sterile neutrino models with eV-scale mass splittings, the resonant disappearance feature appears at TeV energies and Earth-crossing baselines, which is accessible only to large-volume neutrino telescopes such as IceCube.

The search for a signal is made more complex by the limitations of neutrino telescopes.
First, the matter resonance enhances disappearance for only antineutrinos\footnote{Assuming $\theta_{24}<\pi/4$.}. 
Neutrino telescopes cannot distinguish $\nu$ from $\bar{\nu}$ on an event-by-event basis, however, so the resonance signature is partially diluted in the observed, combined $\nu+\bar{\nu}$ flux. 
Statistical separation through the inelasticity of charged-current interactions provides a handle on this composition and is discussed in \cref{sec:inelasticity}.
Second, neutrino telescopes have limited energy and angular resolution, which can obscure rapid oscillation features.
The treatment in this chapter looks at the truth-level energies and zenith angles with the impact of reconstruction effects deferred to later chapters.

The chapter is organized as follows. \cref{sec:vacuum_osc,sec:matter_effects,sec:sterile_msw} present the formalism for vacuum oscillations, matter effects in the active sector, and matter effects in the sterile sector. 
\cref{sec:earth_density} describes the Earth density profile that determines the matter potential along Earth-crossing trajectories. 
\cref{sec:neutrino_telescopes} translates these results into the energy and zenith-angle phase space accessible at a neutrino telescope, and the production-height correction is treated analytically in the following section. 
\Cref{sec:decay_signal} extends the formalism to unstable sterile neutrinos, the model targeted by the analysis in \cref{chapter:decay_analysis}. 
\cref{sec:inelasticity} discusses the inelasticity-based separation of neutrinos and antineutrinos that motivates new experimental directions presented in \cref{chapter:ml_recos,chapter:meows2026}.

\section{Vacuum Oscillations}\label{sec:vacuum_osc}
The most basic formalism for modeling sterile neutrino oscillations is to consider only vacuum oscillations. The neutrino flavor eigenstates can be expressed as linear combinations of the elements of the mixing matrix $U$ and mass eigenstates:
\begin{equation}
    \vert \nu_\ell \rangle = \sum_{m} U_{\ell m} \vert \nu_m \rangle.
\end{equation}
A flavor state propagating with momentum $p$ acquires a plane-wave factor $e^{ipx}$,
\begin{equation}
    \vert \nu_\ell(x,t=0) \rangle = \sum_m U_{\ell m} \vert \nu_m \rangle e^{ipx}.
\end{equation}
and time evolution is introduced through the energy eigenvalue $E_m$ of each mass eigenstate,
\begin{equation}
    \vert \nu_\ell(x,t) \rangle = \sum_{m} U_{\ell m} \vert \nu_m \rangle e^{ipx} e^{-iE_{m}t}.
\end{equation}

\begin{figure}
    \centering
    \includegraphics[width=0.9\linewidth]{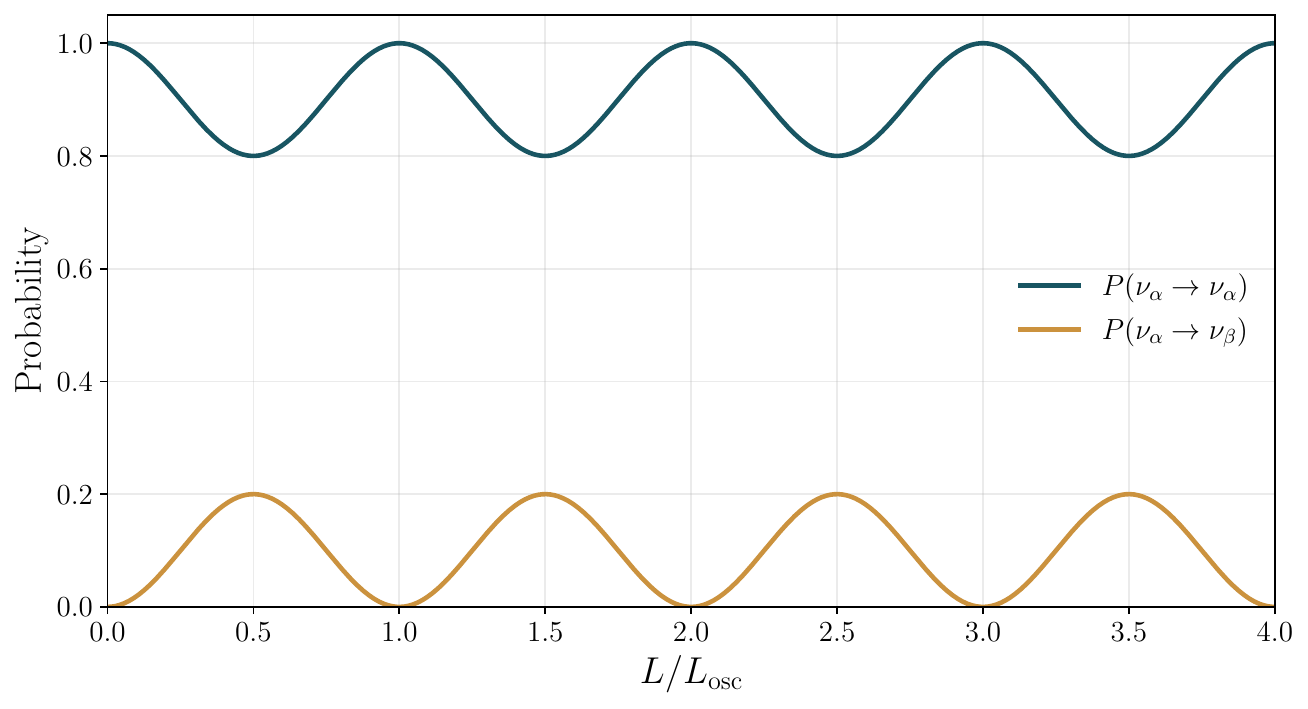}
    \caption{Two-neutrino oscillation probabilities over multiple oscillation lengths with an oscillation amplitude of $\sin^2 (2\theta) = 0.2$.}
    \label{fig:two_nu_osc}
\end{figure}

In the relativistic limit $p\gg m$, the energy of mass eigenstate $\nu_m$ can be approximated as $E_{m} \approx E + \frac{m^2}{2E}$.
Similarly, since the neutrinos propagate at $v\approx c$, setting $t = x$ yields:
\begin{equation}
    \vert \nu_\ell (x,t=x) \rangle = \sum_m U_{\ell m} \vert \nu_m \rangle e^{-i m^2 x / (2E)}.
\end{equation}
The probability of observing a neutrino of initial flavor $\ell$ at a later time is most easily illustrated in the two-neutrino model, where the mixing matrix is described by a single angle $\theta$:
\begin{equation}
    U = \begin{pmatrix}
        \cos \theta & \sin \theta \\
        -\sin \theta & \cos \theta
    \end{pmatrix}
\end{equation}
Substituting this mixing matrix into the propagated flavor state gives
\begin{equation}
    \vert \nu_\alpha \rangle = \cos \theta \vert \nu_1 \rangle e^{-i m_{1}^2 x / (2E)} + \sin \theta \vert \nu_2 \rangle e^{-i m_{2}^{2} x / (2E)}.
\end{equation}
The probability of the initial neutrino $\nu_\alpha$ being observed in the same flavor state at a distance $x$ is obtained by taking the inner product with $\vert \nu_\alpha (0) \rangle$,
\begin{equation}
    \langle \nu_\alpha (x) \vert \nu_\alpha (x=0) \rangle = \cos^2 \theta e^{-im_{1}^2 x / (2E)} + \sin^2 \theta e^{-im_{2}^2 x / (2E)}
\end{equation}
and then squaring and simplifying to compute the survival probability:
\begin{align}
    \vert \langle \nu_\alpha (x) \vert \nu_\alpha (0) \rangle \vert^2  & = \cos^4 \theta + \sin^4 \theta + 2 \cos^2 \theta \sin^2 \theta \cos\left( \frac{m_{2}^2 x}{2E} - \frac{m_{1}^2 x}{2E} \right) \\
    & = 1 - \sin^2(2\theta)\sin^2\left(\frac{\Delta m^2 x}{4 E}\right),
    \label{eq:two_nu_osc}
\end{align}
with $\Delta m^2 = m_{2}^2 - m_{1}^2$. $\Delta m^2$ is referred to as the squared mass splitting and it determines the oscillation wavelength $L_{\mathrm{osc}} = \frac{4\pi E}{\Delta m^2}$. 
The $\sin^2 (2\theta)$ term describes the maximum amplitude of the oscillations which occurs at $L_{\mathrm{osc}}/2$.
\cref{fig:two_nu_osc} demonstrates a simple two-neutrino oscillation case using \cref{eq:two_nu_osc} with $\sin^2 (2\theta) = 0.2$, showing both the survival probability $P(\nu_\alpha\rightarrow\nu_\alpha)$ and appearance probability $P(\nu_\alpha\rightarrow\nu_\beta) = 1 - P(\nu_\alpha\rightarrow\nu_\alpha)$.

\section{Matter Effects}\label{sec:matter_effects}
When neutrinos propagate through matter, an additional term must be added to the Hamiltonian to account for scattering contributions on the electron background.
The two-neutrino vacuum Hamiltonian in the flavor basis is,
\begin{equation}
    H_{v} = \frac{1}{2E} U \begin{pmatrix}
        m_{1}^2 & 0 \\
        0 & m_{2}^2
    \end{pmatrix}
    U^\dagger.
\end{equation}

In matter, the Hamiltonian must be modified to include the effect of matter density\footnote{There is also a contribution from the nucleon density, but it is proportional to the identity matrix and does not affect the oscillation dynamics of this system.} arising from coherent forward scattering of $\nu_e$ on electrons. The potential contribution can be expressed in terms of the electron number density and the forward scattering amplitude,
\begin{equation}
    V_e = \frac{N_{e}}{2E}\langle \mathcal{M}_{CC}^{\textrm{forward}} \rangle
\end{equation}
In the forward-scattering limit, the amplitude is expressed as a Fermi contact interaction averaged over electron spins (assuming an unpolarized medium),
\begin{equation}
    \langle \mathcal{M}_{CC}^{\textrm{forward}} \rangle = -\frac{G_F}{\sqrt{2}} \bar{u}_\nu \gamma^\mu (1 - \gamma^5) u_\nu \langle \bar{u}_{e} \gamma_\mu (1 - \gamma^5) u_e \rangle.
\end{equation}
Computing the amplitude gives $\langle \mathcal{M}_{CC}^{\textrm{forward}} \rangle = 2\sqrt{2}G_F E$, yielding the matter potential,
\begin{equation}
    V_e = \sqrt{2}G_{F}\begin{pmatrix}
        N_{e} & 0 \\
        0 & 0
    \end{pmatrix}.
\end{equation}
Combining the matter potential with the vacuum Hamiltonian,
\begin{equation}
    H_{m} = \frac{1}{4E} \begin{pmatrix}
        -\Delta m^2 \cos(2\theta) + 2\sqrt{2}G_{F}N_{e}E & \Delta m^2 \sin(2\theta) \\
        \Delta m^2 \sin(2\theta) & \Delta m^2 \cos(2\theta) - 2\sqrt{2}G_{F} N_{e}E
    \end{pmatrix}.
\end{equation}
Diagonalizing the matter-modified Hamiltonian yields a new effective mixing angle $\theta_m$ given by,
\begin{equation}
    \tan(2\theta_m) = \frac{\Delta m^2 \sin (2\theta) }{\Delta m^2 \cos(2\theta) - 2\sqrt{2}G_{F} N_{e}E }.
\end{equation}
The resulting eigenvalues provide the matter-modified effective $\Delta m^2$ as well,
\begin{equation}\label{eq:modified_dm2}
    \Delta m_{m}^2 = \Delta m^2 \sqrt{\sin^2(2\theta) + \left(\cos(2\theta) - \frac{2\sqrt{2}G_{F} N_{e}E}{\Delta m^2} \right)^2 }.
\end{equation}
In the limit of vanishing electron density, $N_e \rightarrow 0$, the matter-modified terms return to their vacuum equivalents: $\theta_m\rightarrow\theta$ and $\Delta m_m^2 \rightarrow \Delta m^2$ as expected. 
An important feature of the matter-modified Hamiltonian is the resonant energy, which is given by,
\begin{equation}\label{eq:msw_resonant_energy}
    E_{res} = \frac{\Delta m^2 \cos(2\theta)}{2\sqrt{2}G_F N_e}.
\end{equation}
Notably, at $E = E_{res}$, the matter-modified mixing angle satisfies $\tan(2\theta_m)\rightarrow \infty$, so $\theta_m = \pi/4$ and mixing becomes maximal even if the vacuum mixing angle is small.
This is the MSW effect \cite{Wolfenstein:1977ue,Mikheyev:1985zog} introduced in \cref{sec:solar_problem} in the context of solar neutrinos.
The expression for $E_{res}$ in \cref{eq:msw_resonant_energy} assumes the matter potential for $\nu_e$, which is positive for charged-current scattering on electrons.
For antineutrinos, the sign of the matter potential is reversed, yielding a negative $E_{res}$ that is unphysical.
Antineutrinos therefore do not undergo the MSW resonance in ordinary matter.
Alternatively, if $\cos(2\theta) < 0$ (i.e. $\theta > \pi/4$), then the signs are reversed and antineutrinos exhibit the resonance while neutrinos do not.

\section{Matter Effects with Sterile Neutrinos}\label{sec:sterile_msw}
Unlike the active flavors ($\nu_e$, $\nu_\mu$, $\nu_\tau$), the sterile flavor $\nu_s$ is not affected by either the charged-current or neutral-current matter potential because it does not interact weakly.
The charged-current and neutral-current potentials that impact the active flavors therefore generate a relative matter potential between active and sterile states, which enters the 3+1 matter Hamiltonian.
The corresponding contribution to the matter potential from neutral-current interactions with fermions is,
\begin{equation}
    V_{f} = \sqrt{2} G_{F} N_{f} T_{3}^{f}
\end{equation}
where $N_{f}$ is the fermion number density and $T_{3}^{f}$ is the weak isospin.
Assuming a neutral, isoscalar medium with $N_e = N_p = N_n$ simplifies the calculation and yields,
\begin{equation}
    V_{\textrm{NC}} = -\frac{\sqrt{2}}{2} G_{F} N_{n}.
\end{equation}
For the simple case of $\nu_\mu$ and $\nu_s$ oscillations, $V_{\mathrm{NC}}$ is the only contribution to the potential, since neither $\nu_\mu$ nor $\nu_s$ participates in charged-current scattering on electrons.
Following the same diagonalization procedure as in \cref{sec:matter_effects}, but with $V_{\mathrm{NC}}$ replacing $V_e$, yields the resonant energy,
\begin{equation}\label{eq:sterile_msw}
    E_{res} = -\frac{\Delta m^2 \cos(2\theta)}{\sqrt{2}G_{F} N_{n}}.
\end{equation}
For neutrinos, this expression is negative, which is unphysical.
The matter potential is reversed for antineutrinos, yielding a physical resonance with $E_{res}>0$.
Alternatively, two modified scenarios can produce the same physical effect:
\begin{itemize}
    \item $\Delta m^2 < 0$: 1+3 Sterile neutrino model
    \item $\theta > \frac{\pi}{4}$: Higher-octant sterile mixing
\end{itemize}

In both of these scenarios, the resonant energy is positive for neutrinos and negative for antineutrinos.
Presently, direct neutrino mass measurements and cosmological observations put stringent constraints on the allowed parameter space of 1+3 sterile neutrino models as discussed in \cref{sec:sterile_intro}. 
By contrast, the higher-octant scenario is poorly constrained by existing experiments. Distinguishing it from the lower-octant case requires probing matter effects at the TeV-scale energies and Earth-diameter baselines accessible only to large-volume neutrino telescopes such as IceCube, described in \cref{sec:neutrino_telescopes}.

\section{Earth Density Profile}\label{sec:earth_density}
The matter effects described in \cref{sec:sterile_msw} depend on the local density along the neutrino's path through the Earth.
For up-going atmospheric neutrinos, this path can traverse different layers of the Earth with varying densities spanning 3 g/cm$^3$ in the crust to 13 g/cm$^3$ in the inner core.
The standard Earth density model is the Preliminary Reference Earth Model (PREM) \cite{DZIEWONSKI1981297}, which provides the density as a function of radial distance.

\begin{figure}
    \centering
    \includegraphics[width=0.95\linewidth]{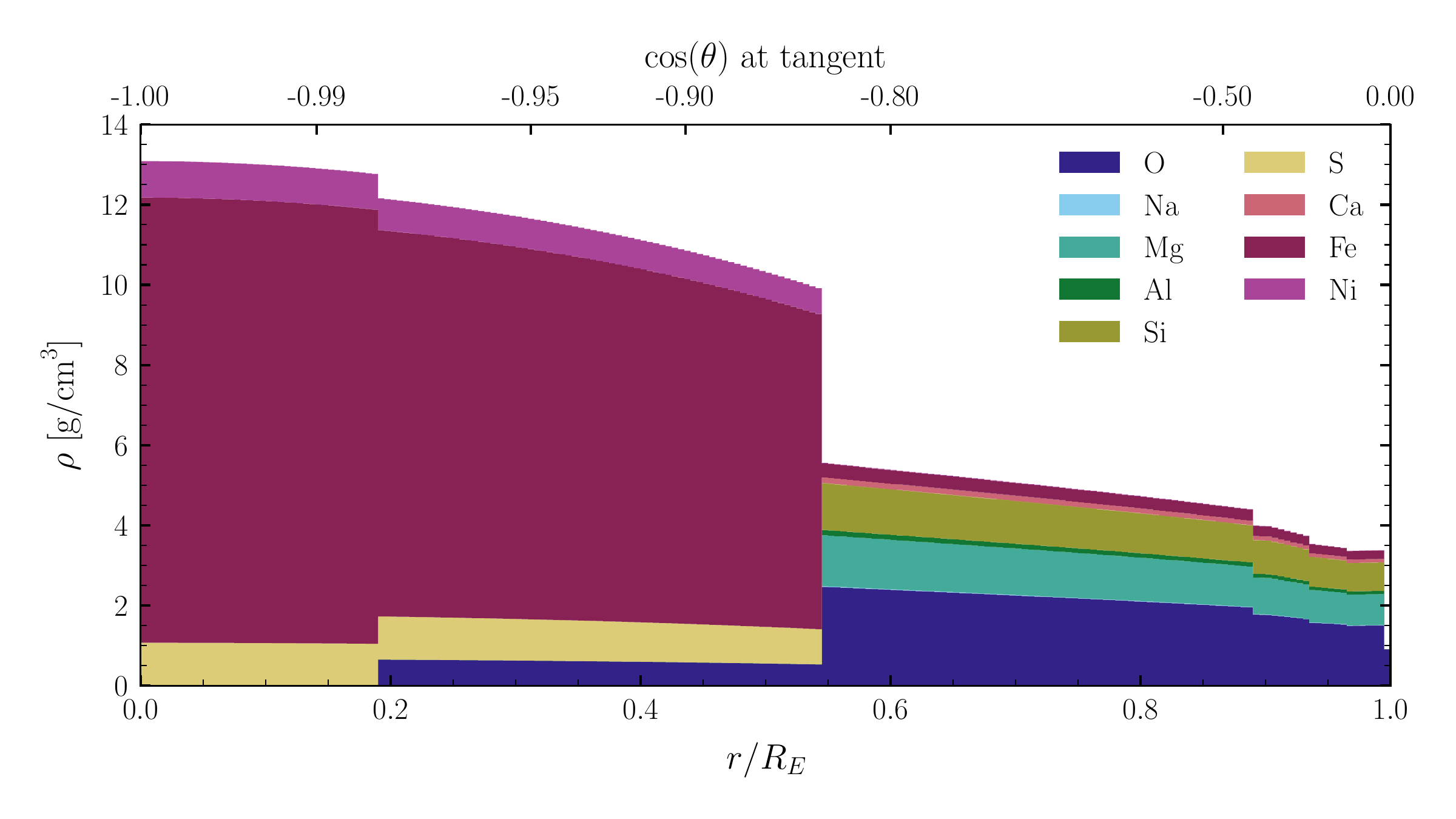}
    \caption{PREM density model of the Earth as a function of $r/R_E$ \cite{DZIEWONSKI1981297}. Earth composition data from Ref. \cite{Stacey_Davis_2008}.}
    \label{fig:earth_density_comp}
\end{figure}

\cref{fig:earth_density_comp} shows the PREM density profile along with the elemental composition from Ref. \cite{Stacey_Davis_2008} for each layer.
The density does not vary significantly within each layer, but has sharp discontinuities between layers.
Most notable is the core-mantle boundary at $r/R_E \approx 0.55$, where the density drops by approximately 50\% across the boundary.
The composition also changes significantly in each layer.
The core of the Earth is primarily composed of iron, whereas the mantle is primarily composed of oxygen, magnesium, and silicon.
For oscillations, the electron fraction $Y_e = N_e / (N_e + N_n)$ is typically used as a conversion factor between mass density $\rho$ and the electron number density $N_e$ which appears in the matter potential.
$Y_e \approx 0.5$ in the outer layers of the Earth, decreasing to $\approx0.466$ in the iron-rich core where the neutron number is slightly larger.

The density profile directly affects the location of the MSW resonance.
From \cref{eq:msw_resonant_energy}, the resonant energy is inversely proportional to $N_e$.
For an eV-scale sterile neutrino, $\Delta m_{41}^2 \sim 1~\mathrm{eV}^2$, the resonant energy is satisfied at different neutrino energies in different layers of the Earth.
Neutrinos crossing only the mantle encounter the resonance at higher energies than those that cross the core since $\rho_{\mathrm{core}} > \rho_{\mathrm{mantle}}$.
This causes the resonant disappearance effect to be broadened over a wider range of energies and zenith angles.

\section{Neutrino Telescopes}\label{sec:neutrino_telescopes}
\begin{figure}
    \centering
    \includegraphics[width=0.8\linewidth]{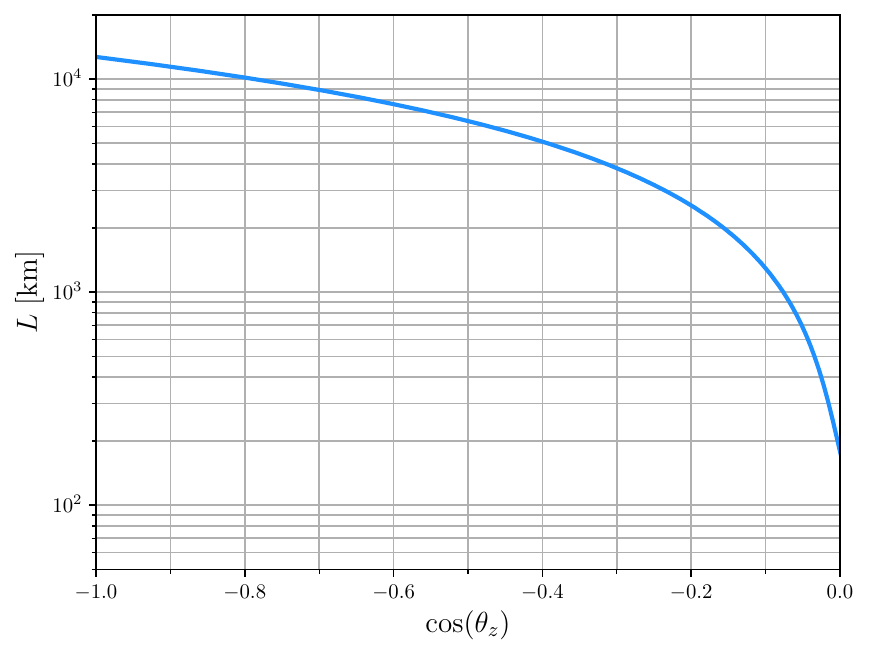}
    \caption{Baseline to the surface of the Earth for IceCube as a function of $\cos(\theta_z)$.}
    \label{fig:icecube_baseline_L_zenith}
\end{figure}

The observable signal of oscillations at a neutrino telescope is the modification of the atmospheric neutrino flux as a function of energy and zenith angle.
Two quantities determine the oscillation phase: the neutrino energy $E$ and the propagation baseline $L$ from the production point in the atmosphere to the detector.
$L$ is naturally parameterized by the zenith angle $\theta_z$ of the reconstructed neutrino direction, with $\cos(\theta_z) = 1$ corresponding to a vertical down-going neutrino and $\cos(\theta_z) = -1$ to vertical up-going neutrinos.
The up-going neutrinos traverse the Earth over long distances and can experience the matter effects discussed in \cref{sec:matter_effects,sec:sterile_msw}, while down-going neutrinos that only traverse a relatively short distance in the atmosphere do not and are well-described by vacuum oscillations.
The detector-surface baseline for IceCube is shown in \cref{fig:icecube_baseline_L_zenith} for up-going neutrinos, showing the large range of accessible $L$.

\subsection{Energy Regimes}
The energy range of atmospheric neutrinos determines the sensitivity of a neutrino telescope to oscillation signatures.
Standard atmospheric neutrino oscillations, as discussed in \cref{sec:atmospheric_neutrinos}, are driven by $\Delta m_{31}^2 \approx 2.5\times10^{-3}$ eV$^2$ producing an oscillation maximum of approximately 25 GeV for Earth-crossing baselines.
Above 100 GeV, the oscillation length dictated by $\Delta m_{31}^2$ exceeds the diameter of the Earth and the contribution to $\nu_\mu$ disappearance becomes negligible.

Matter-induced resonant disappearance effects from eV-scale sterile neutrinos occur at higher energies.
For $\Delta m_{41}^2 \sim 1$ eV$^2$, the resonance condition in \cref{eq:sterile_msw} is satisfied $E_{res}\sim1$ TeV can be met for neutrinos traversing the Earth's mantle or core.
The relevant neutrino energies for sterile neutrino searches at a neutrino telescope are then the up-going TeV-scale atmospheric neutrinos that experiments such as IceCube observe.

\subsection{Neutrino Baselines}\label{sec:neutrino_baselines}

For a detector at the surface, the neutrino baseline through the Earth is given by,
\begin{equation}\label{eq:surface_baseline}
    L = -2 R_{E} \cos(\theta_z),
\end{equation}
where $R_E$ is the radius of the Earth. 
This expression neglects the depth of the detector, which provides an additional contribution to the baseline near the horizon.
Existing neutrino telescopes are situated more than a kilometer below the surface.
For example, the instrumented region of IceCube lies between depths of 1.5 km and 2.5 km.
Including the detector depth $d$ the baseline expression becomes,
\begin{equation}\label{eq:depth_baseline}
    L = -(R_E - d) \cos(\theta_z) + \sqrt{R_{E}^2 - (R_E - d)^2 \sin^2 (\theta_z)}
\end{equation}
which reduces to \cref{eq:surface_baseline} for $d\rightarrow 0$.
Using the Earth's polar radius $R_p = 6357$ km, appropriate for a detector at the South Pole, and a depth $d = 2$ km, the baseline near the horizon is $L(\cos\theta_z = 0, d = 2~\text{km}) = 159.4~\text{km}$.
The baseline can also vary across the depth of the detector itself.
For IceCube, which spans 1 km vertically, the difference in baseline between events that traverse the top and bottom of the detector with $\cos(\theta_z) = 0$ is,
\begin{equation}
    \Delta L = 178.3~\mathrm{km} - 138.1~\mathrm{km} = 40.2~\mathrm{km}.
\end{equation}
The difference in baseline is largest for exactly $\cos(\theta_z) = 0$, but falls off as $\cos(\theta_z)$ decreases.
The variation is reduced to 7\% of the average baseline at $\cos(\theta_z) = -0.025$, and to 2\% at $\cos(\theta_z) = -0.05$.

\subsection{Structure of Matter Effects}

Near the horizon, $-0.2 \lesssim \cos(\theta_z) \lesssim 0$, neutrinos pass through the atmosphere and a short distance inside the Earth which is low density and do not produce matter effects.
This results in an oscillation pattern for both $\nu_\mu$ and $\bar{\nu}_\mu$ that reflects only vacuum oscillations.
The matter effects become significant for neutrinos propagating through the mantle with a zenith range $-0.8 \lesssim \cos(\theta_z) \lesssim -0.2$, inducing resonant $\bar{\nu}_\mu$ disappearance.
The strongest matter effects occur for paths that intersect the dense core of the Earth, $-1 < \cos(\theta_z) \lesssim -0.8$.
The broad resonance features across the $(E, \cos(\theta_z))$ plane are a consequence of the variation in matter density across the density layers, as discussed in \cref{sec:earth_density}.

\begin{figure}
    \centering
    \includegraphics[width=0.95\linewidth]{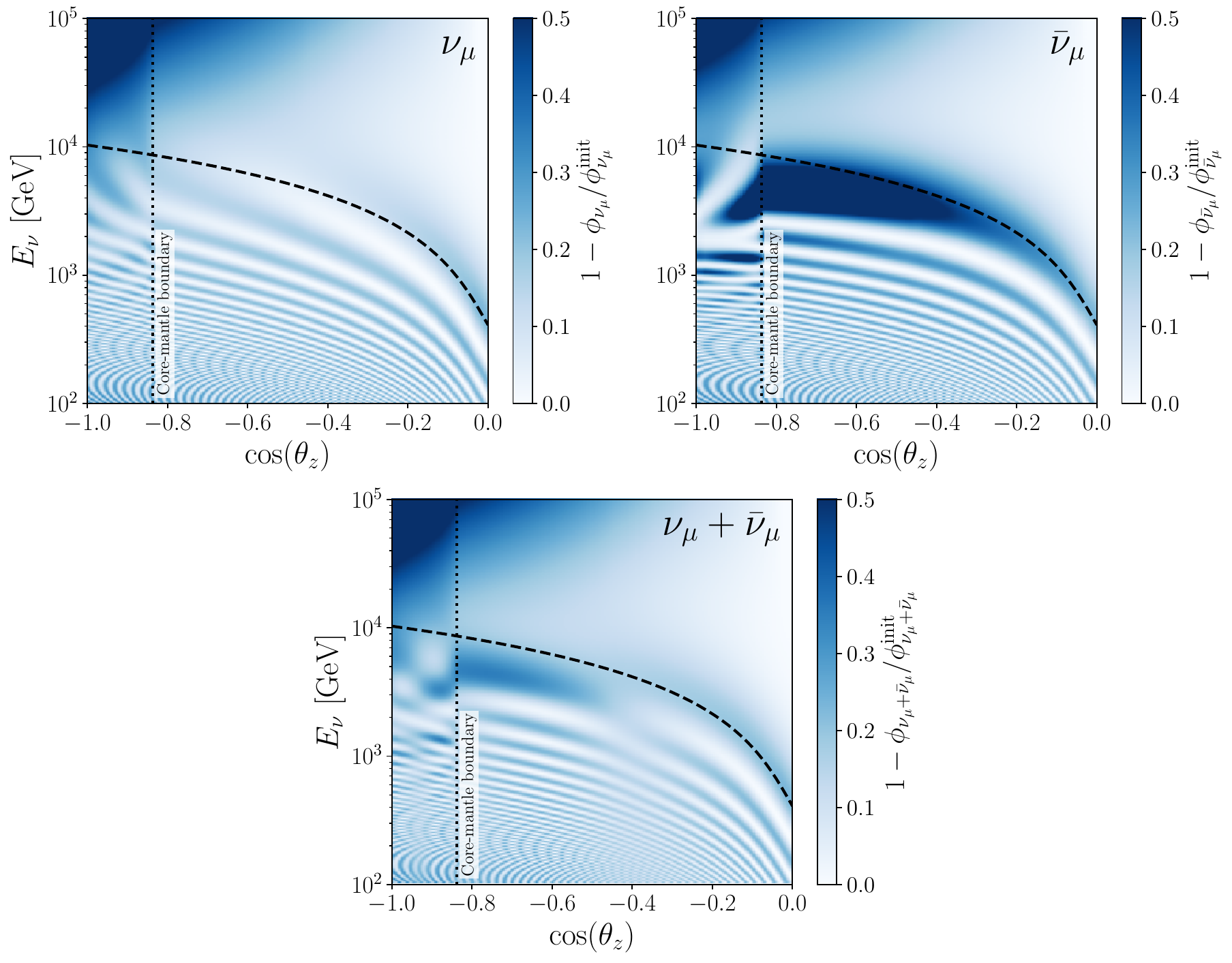}
    \caption{Oscillograms of 3+1 sterile neutrino oscillations with $\Delta m_{41}^2 = 1$ eV$^2$ and $\theta_{24} = 0.25$ for neutrinos (top left), antineutrinos (top right), and the combined flux (bottom). The core-mantle boundary is indicated by a dotted vertical line and the $L/E$ of the first vacuum oscillation maximum is indicated by a dashed line.}
    \label{fig:sterile_oscillogram}
\end{figure}

\cref{fig:sterile_oscillogram} shows an oscillogram of $\nu_\mu$, $\bar{\nu}_\mu$, and combined survival probabilities with respect to the initial flux across the relevant neutrino telescope energy and zenith ranges, illustrating these effects.
The dashed lines indicating the $L/E$ of the first oscillation maximum do not align exactly with the computed disappearance patterns, since matter effects modify the effective matter-modified $\Delta m_{41,m}^2$.
Analogous to \cref{eq:modified_dm2}, the effective $\Delta m_{41}^2$ is given by,
\begin{equation}
    \Delta m_{41,m}^2 = \Delta m_{41}^2 \sqrt{\sin^2(2\theta_{24}) + \left(\cos(2\theta_{24}) \pm \frac{\sqrt{2}G_{F} N_{n}E}{\Delta m_{41}^2} \right)^2 },
\end{equation}
which for neutrinos~(+ sign) increases with higher density and decreases for antineutrinos~(- sign).
Additionally, there is a region of disappearance at the highest energies and most up-going directions that corresponds to the absorption of neutrinos in the Earth from deep-inelastic scattering and does not originate from sterile mixing.

\clearpage

\section{Neutrino Production Height}
Atmospheric neutrinos are produced over a range of altitudes from approximately 10 km to 30 km, with the distribution depending on the neutrino energy and parent meson species \cite{Gaisser:2002jj, Barr:2004br, Honda:2015fha, Yanez:2023lsy}.
The baseline a neutrino travels for a given zenith angle depends on its production height.
The oscillation probability has to be averaged over the production height distribution since the production point of a single neutrino cannot be determined.
This section evaluates the impact of the neutrino production height profile using an analytic model and a Gaussian probability distribution for simplicity.
The first case that will be considered is the simple down-going neutrino case, and the second case will be the near-horizon region that is relevant for oscillation studies.
While this results in percent-level relative differences in the oscillation probability, these will be shown to be small compared to detector smearing effects in \cref{chapter:decay_analysis}.

\subsection{Production Height Distributions and Survival Probabilities}

Building on \cref{eq:depth_baseline}, the effect of the atmospheric neutrino production height $h$ can be added to the expression for the baseline,
\begin{equation}
    L = -(R_E - d) \cos(\theta_z) + \sqrt{ (R_E + h)^2 - (R_E - d)^2 \sin^2 (\theta_z)}.
\end{equation}
Consider a simple neutrino production height model that is Gaussian with mean production height $\bar{h}$ and width $\sigma$,
\begin{equation}
    p(h) = \frac{1}{\sqrt{2\pi\sigma^2}}e^{-(h-\bar{h})^2 / 2\sigma^2}.
\end{equation}
The average oscillation survival probability can be computed assuming a simplified two-neutrino scenario,
\begin{equation}\label{eq:avg_survival_prob}
    \langle P \rangle = 1 - \sin^2 (2\theta) \left\langle \sin^2 \left( \frac{\Delta m^2 L(h)}{4 E} \right)\right\rangle.
\end{equation}
The complexity of evaluating the average survival probability comes from the $L(h)$ expression and averaging over the probability distribution $p(h)$.
Different limits of zenith angles will be used to simplify this calculation.

\subsection{Down-going Neutrinos}
To begin characterizing the effects of neutrino production heights on the oscillation probabilities consider the case of vertically down-going neutrinos with $\cos(\theta_z) = 1$.
The baseline reduces to a simple form $L = h + d$ and the height-averaged oscillation factor becomes:
\begin{equation}
    \left\langle \sin^2 \left( \frac{\Delta m^2 L(h)}{4 E} \right)\right\rangle = \int_{0}^{\infty} p(h) \sin^2 \left(\frac{\Delta m^2 (h + d)}{4 E}\right) dh.
\end{equation}
Performing this integral with $\omega = \Delta m^2 / 4 E$ and extending the integral range to $(-\infty, \infty)$ which is valid for $\bar{h} \gg \sigma$, the integral evaluates to:
\begin{align}
    & = \frac{1}{2} - \frac{1}{2} \textrm{Re}\left\{ e^{2i\omega d} \int_{-\infty}^{\infty} \frac{1}{\sqrt{2\pi\sigma^2}}e^{-(h-\bar{h})^2/2\sigma^2}e^{2i\omega h} dh \right\} \\
    & = \frac{1}{2} - \frac{1}{2}\textrm{Re}\left\{ e^{2i\omega(d+\bar{h})} e^{-2\omega^2 \sigma^2} \right\} \\
    & = \frac{1}{2} - \frac{1}{2} e^{-2\omega^2\sigma^2} \cos\left( \frac{\Delta m^2 (d+\bar{h})}{2E}  \right).
\end{align}
This expression for the average can be substituted back into \cref{eq:avg_survival_prob} to obtain the height-averaged survival probability for $\cos(\theta_z) = 1$:
\begin{equation}
    \langle P \rangle = 1 - \frac{1}{2}\sin^2 (2\theta) \left(1 - \mathcal{D} \cos\left( \frac{\Delta m^2 (d+\bar{h})}{2 E}\right)\right)
\end{equation}
where a Gaussian damping factor $\mathcal{D}$ is defined as,
\begin{equation}
    \mathcal{D} = e^{-2\omega^2\sigma^2} = e^{-\frac{1}{2}\left( \frac{\Delta m^2 \sigma}{2 E} \right)^2}.
\end{equation}

\subsection{Near-horizon Expansion}

Similar to the detector depth discussion in \cref{sec:neutrino_baselines}, the zenith range of interest for oscillations is just below the horizon.
Take $\cos(\theta_z)$ to be small and negative such that $\cos(\theta_z) \approx -\epsilon < 0$ and $\sin^2 (\theta_z) \approx 1 - \epsilon^2$. 
Defining $r = R_E - d$ and assuming both $h\ll R_E$ and $d \ll R_E$,
\begin{equation}\label{eq:baseline_approx_eps}
    L(h) \approx r \epsilon + \sqrt{\ell(\epsilon)^2 + 2 R_E h}
\end{equation}
where $\ell(\epsilon) = 2 R_E d + (r\epsilon)^2$.
This expression for $L$ can be expanded to the second order in terms of $\Delta h = h - \bar{h}$,
\begin{equation}
    L(h) = \alpha + \beta \Delta h + \frac{1}{2}\gamma (\Delta h)^2 + \mathcal{O}(\Delta h^3).
\end{equation}
The three expansion coefficients can be determined from \cref{eq:baseline_approx_eps},
\begin{align}
    \alpha & = r \epsilon + \sqrt{\ell(\epsilon)^2 + 2 R_E \bar{h}} = r\epsilon + \Lambda \\
    \beta & = \eval{\frac{\partial L}{\partial h}}_{\bar{h}} =  \frac{R_E}{ \sqrt{\ell(\epsilon)^2 + 2 R_E \bar{h}}} = \frac{R_E}{\Lambda} \\
    \gamma & = \eval{\frac{\partial^2 L}{\partial h^2}}_{\bar{h}} = -\frac{R_E^2}{\Lambda^3}
\end{align}
where $\Lambda = \sqrt{\ell(\epsilon)^2 + 2 R_{E}\bar{h}}$ is defined for convenience.
\begin{equation}
    \langle e^{2i\omega L} \rangle = e^{2i\omega\alpha} \int \frac{1}{\sqrt{2\pi\sigma^2}} e^{- (\Delta h)^2/2\sigma^2} e^{2i\omega \beta \Delta h} e^{i\omega \gamma (\Delta h)^2} dh
\end{equation}
Defining an effective complex width $\sigma_{\mathrm{eff}}$ via,
\begin{equation}
    \frac{1}{\sigma_{\textrm{eff}}^2} = \frac{1}{\sigma^2} - 2i\omega\gamma.
\end{equation}
The integral becomes a single Gaussian,
\begin{equation}\label{eq:exp_val_post_integral}
    \langle e^{2 i \omega L} \rangle = \frac{\sigma_{\textrm{eff}}}{\sigma} e^{2i\omega \alpha - 2 \omega^2 \beta^2 \sigma_{\textrm{eff}}^2}.
\end{equation}
To separate $\sigma_{\mathrm{eff}}^2$ into a magnitude and a phase, the polar form is given by, 
\begin{equation}
    \frac{1}{\sigma_{\textrm{eff}}^2} = \frac{1}{\sigma^2} \left(1 - i \eta \right) = \frac{1}{\sigma^2} \left(\sqrt{1 + \eta^2} e^{i \arctan(\eta)} \right)
\end{equation}
where $\eta = 2 \omega \gamma \sigma^2$.
Substituting into \cref{eq:exp_val_post_integral} gives,
\begin{align}
    \langle e^{2i\omega L}\rangle & = \frac{1}{\left( 1 + \eta^2 \right)^{1/4}}e^{-\frac{2\omega^2 \beta^2 \sigma^2}{1 + \eta^2}} e^{i\left(2\omega\alpha + \frac{1}{2}\arctan(\eta) - \frac{2\omega^2\beta^2\sigma^2\eta}{1+\eta^2}\right)}. \\
\end{align}
This can be expressed in terms of a damping term $\mathcal{D}$ and a phase $\phi$, then substituted into the height-averaged survival probability using $\langle \sin^2 (\omega L) \rangle = \frac{1}{2} (1 - \textrm{Re}\left\{\langle e^{2i\omega L}\rangle\right\}) = \frac{1}{2}(1 - \mathcal{D} \cos(\phi))$,
\begin{equation}
    \langle P \rangle = 1 - \frac{1}{2}\sin^2 (2\theta) \left(1 - \mathcal{D} \cos \phi \right)
\end{equation}
where the damping term $\mathcal{D}$ and phase term $\phi$ have been defined as,
\begin{align}
    \mathcal{D} & = \frac{1}{\left( 1 + \eta^2 \right)^{1/4}}e^{-\frac{2\omega^2 \beta^2 \sigma^2}{1 + \eta^2}} \\
    \phi & = 2\omega\alpha + \frac{1}{2}\arctan(\eta) - \frac{2\omega^2\beta^2\sigma^2\eta}{1+\eta^2}.
\end{align}

\begin{figure}
    \centering
    \includegraphics[width=0.98\linewidth]{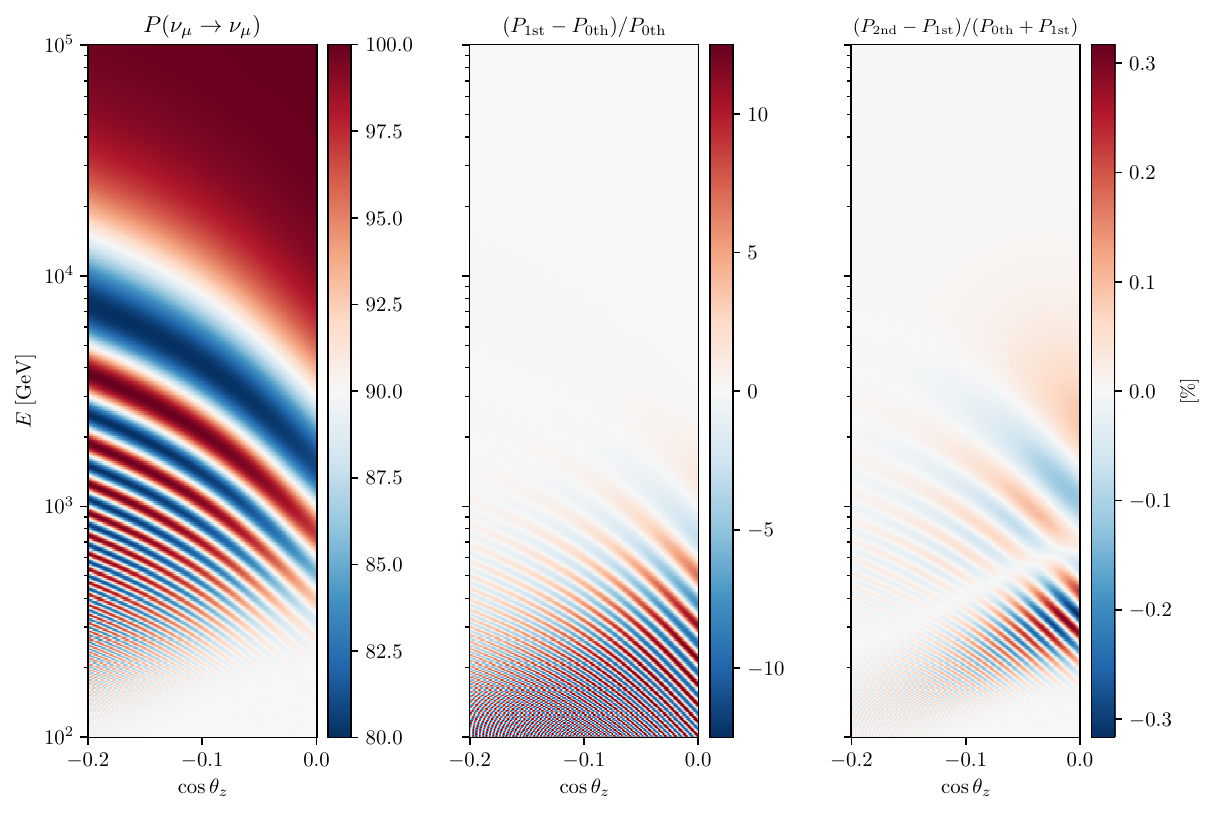}
    \caption{Left: $\nu_\mu$ survival probability given $\Delta m_{41}^2 = 3.5$ eV$^2$ and $\sin^2 (2\theta_{24}) = 0.2$ assuming neutrino production at exactly 20 km. Middle: Relative percent corrections to the zeroth-order calculation using the first-order correction with $\sigma = 6$ km. Right: Relative percent corrections to the first-order approximation using the second-order calculation.}
    \label{fig:second_order_horizon_oscillogram}
\end{figure}

\begin{figure}
    \centering
    \includegraphics[width=0.95\linewidth]{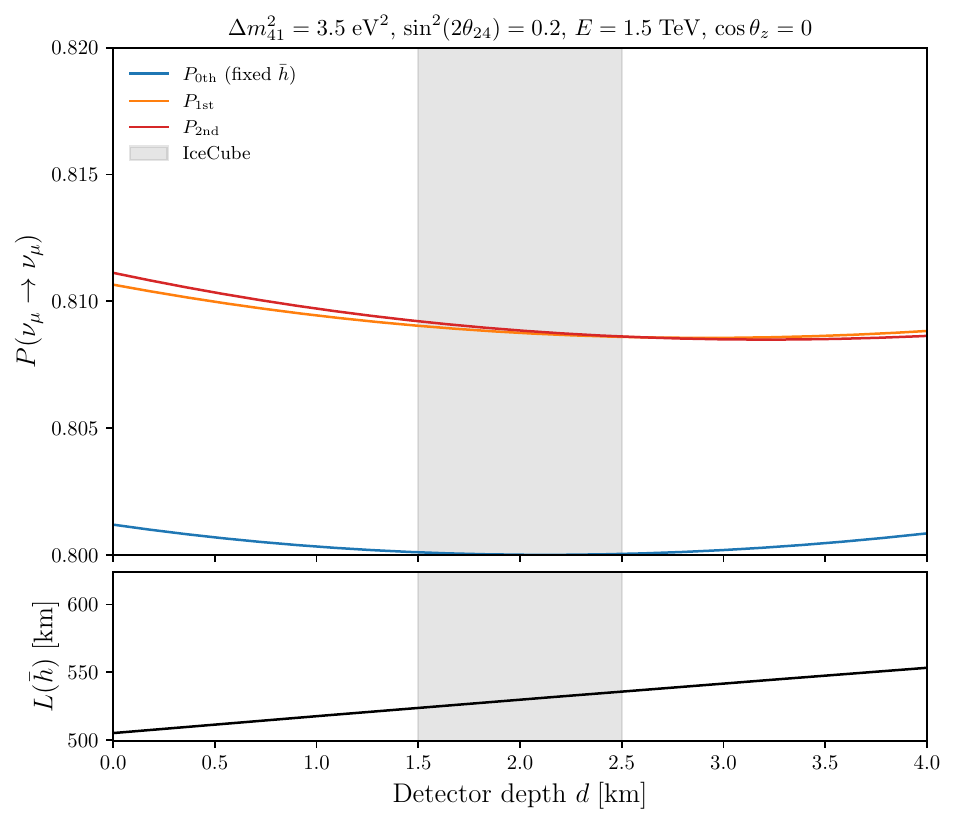}
    \caption{Top: $\nu_\mu$ survival probability given the zeroth- (blue), first- (orange), and second- (red) order calculations as a function of detector depth at a fixed energy and zenith angle. This calculation assumes a sterile neutrino model with $\Delta m_{41}^2 = 3.5$ eV$^2$ and $\sin^2(2\theta_{24})=0.2$ and a neutrino energy of 1.5 TeV and zenith angle of $\cos(\theta_z) = 0$. Neutrino production is assumed to be Gaussian distributed with $\bar{h} = 20$ km and $\sigma = 6$ km. Bottom: neutrino baseline from the mean production height to the detector. The approximate detector depth range of IceCube is shown as a gray band.}
    \label{fig:depth_scan}
\end{figure}

\cref{fig:second_order_horizon_oscillogram} shows the oscillation probabilities $P(\nu_\mu \rightarrow \nu_\mu)$ with the differences between the fixed-height calculation and the first- and second-order corrections assuming $\bar{h} = 20$ km and $\sigma = 6$ km.
This demonstrates that with this model, the first-order correction is at most a few percent and the second-order correction is negligible.
This effect is also quantified in \cref{fig:depth_scan}, which shows the probabilities over a range of detector depth at a fixed energy and zenith angle assuming the same neutrino production profile.

\clearpage

\section{Unstable Sterile Neutrinos}\label{sec:decay_signal}

The concept of unstable, or decaying, sterile neutrinos was introduced in \cref{sec:unstable_sterile_neutrinos} as a mechanism to evade cosmological constraints.
This section serves as a brief overview of the observable signature of this model in a neutrino telescope.

The inclusion of decay from $\nu_4 \rightarrow \phi \psi$, where $\phi$ and $\psi$ are invisible particles, depletes the $\nu_4$ component of the atmospheric neutrino flux as it propagates.
The decay rate $\Gamma = g^2 m_{4} / (16\pi)$ \cite{Moss:2017pur} is governed by $m_4 \approx \sqrt{\Delta m_{41}^2}$ with coupling constant $g$. 
This decay enters the Hamiltonian as an anti-Hermitian term,
\begin{equation}
    H \rightarrow H - i \Gamma/2.
\end{equation}
The stable case, which is the vanilla 3+1 sterile neutrino model, is recovered in the $g \rightarrow 0$ limit.

\begin{figure}
    \centering
    \includegraphics[width=0.95\linewidth]{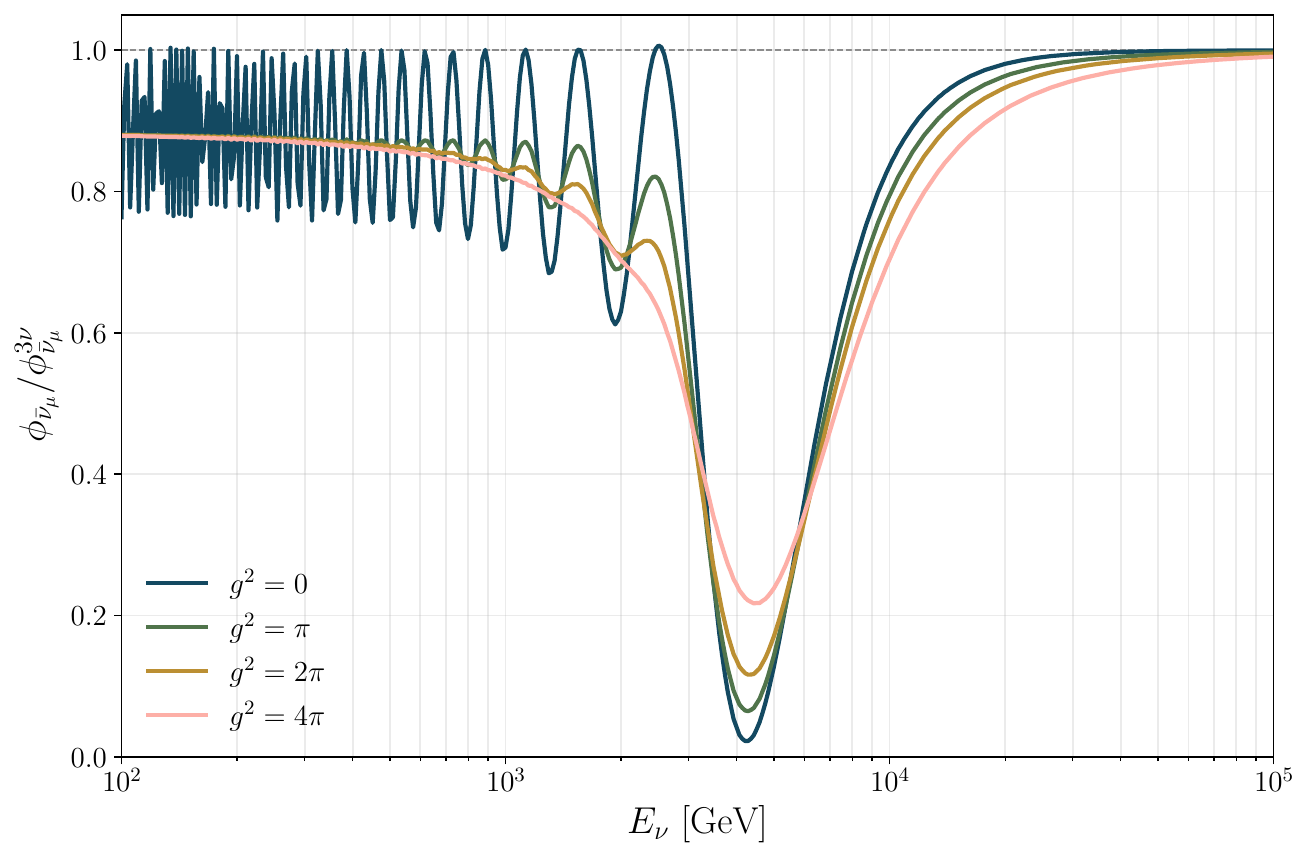}
    \caption{Ratio of fluxes for $\cos(\theta_z) = -0.8$, $\Delta m_{41}^2 = 1.0$ eV$^2$ and $\theta_{24} = 0.25$ compared to the $3\nu$ model over a range of energies given different coupling constants $g^2$.}
    \label{fig:1d_decay_oscillograms}
\end{figure}

\begin{figure}
    \centering
    \includegraphics[width=0.95\linewidth]{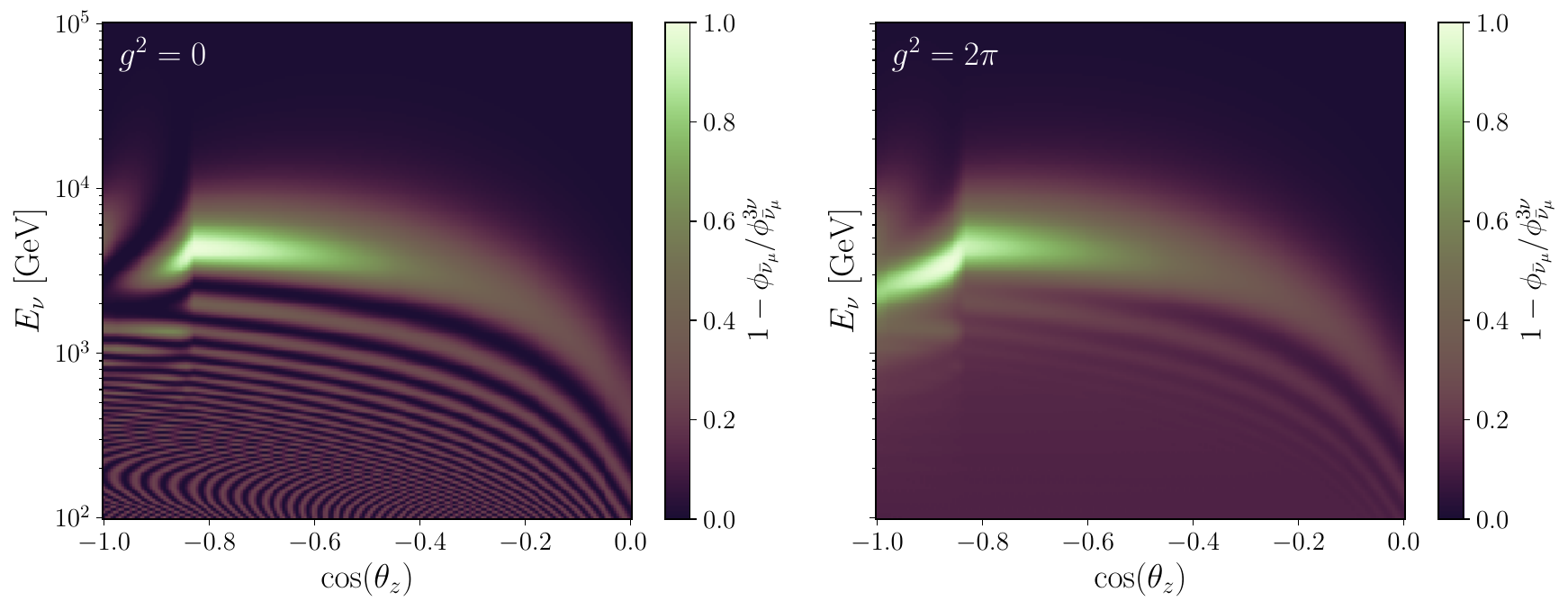}
    \caption{Left: $\bar{\nu}_\mu$ oscillogram comparing $\Delta m_{41}^2 = 1.0$ eV$^2$, $\theta_{24} = 0.25$, and $g^2=0$ to the $3\nu$ model. Right: Same as left, but with coupling constant $g^2 = 2\pi$.}
    \label{fig:decay_oscillograms}
\end{figure}

The decay introduces an additional damping of the oscillation amplitude \cite{Moss:2017pur, Hardin:2022muu}.
This is due to the $\nu_4$ depletion as it propagates, which no longer contributes to the flavor oscillations.
\cref{fig:1d_decay_oscillograms} demonstrates this effect on the neutrino oscillation probabilities at a fixed $\cos(\theta_z)$ for a range of $g^2 = 0,\pi,2\pi,4\pi$.
The resonant disappearance feature at $E \approx 4$ TeV broadens and shallows relative to the $g^2 = 0$ case as $g^2$ increases.
At lower energies, the standard oscillation features are increasingly damped as $g^2$ is increased.
This is demonstrated clearly in \cref{fig:decay_oscillograms} which compares the expected $\bar{\nu}_\mu$ flux over the energy and zenith space.
The resonance shifts to lower $\cos(\theta_z)$ and smears over a larger range of energies, while the vacuum oscillations are effectively averaged out to a normalization.
In \cref{chapter:decay_analysis}, an IceCube analysis searching for this damped oscillation signature will be presented.

\section{Inelasticity-based Separation of \texorpdfstring{$\nu$ and $\bar{\nu}$}{ν and ν̄}}\label{sec:inelasticity}

A difficulty in disentangling the sterile neutrino disappearance signature in neutrino telescopes is that the atmospheric flux contains more neutrinos than antineutrinos \cite{Gaisser:2002jj,Honda:2015fha,Yanez:2023lsy}.
The antineutrino disappearance signature is therefore diluted in the measured combined $\nu_\mu + \bar{\nu}_\mu$ flux.

Statistical separation can be achieved through the inelasticity of charged-current deep-inelastic scattering, $y = E_{\mathrm{had}} / E_\nu$ where $E_{\mathrm{had}}$ is the final-state hadronic shower energy and $E_\nu$ is the neutrino energy.
The V-A structure of the weak interaction produces different inelasticity distributions for $\nu$ DIS and $\bar{\nu}$ DIS at TeV energies.
For neutrinos, the $y$ distribution is approximately flat whereas the antineutrino distribution is peaked near $y = 0$ with a $(1-y^2)$ functional form as discussed in \cref{chapter:cross_sections}.

\begin{figure}
    \centering
    \includegraphics[width=0.98\linewidth]{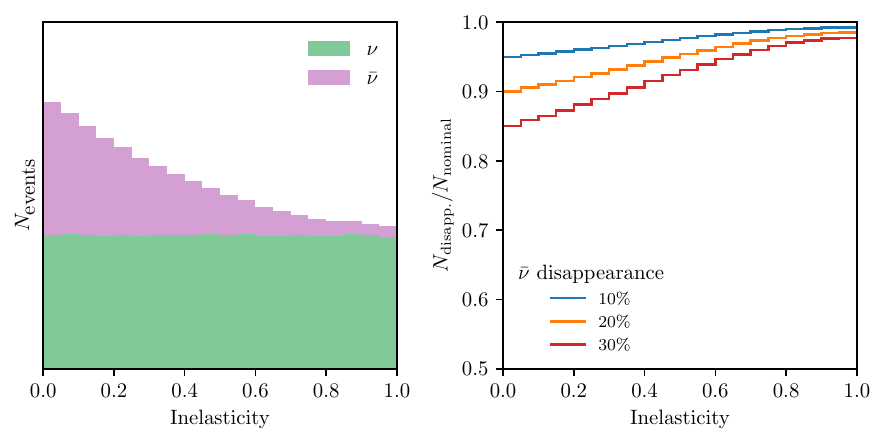}
    \caption{Left: Stacked inelasticity distributions of neutrinos (green) and antineutrinos (magenta) with toy MC. Right: Ratio of events under a sterile neutrino hypothesis that produces 10\%, 20\%, and 30\% disappearance of antineutrinos to the nominal expectation assuming no disappearance as a function of inelasticity.}
    \label{fig:example_inelasticity_disappearance}
\end{figure}

\begin{figure}
    \centering
    \includegraphics[width=0.98\linewidth]{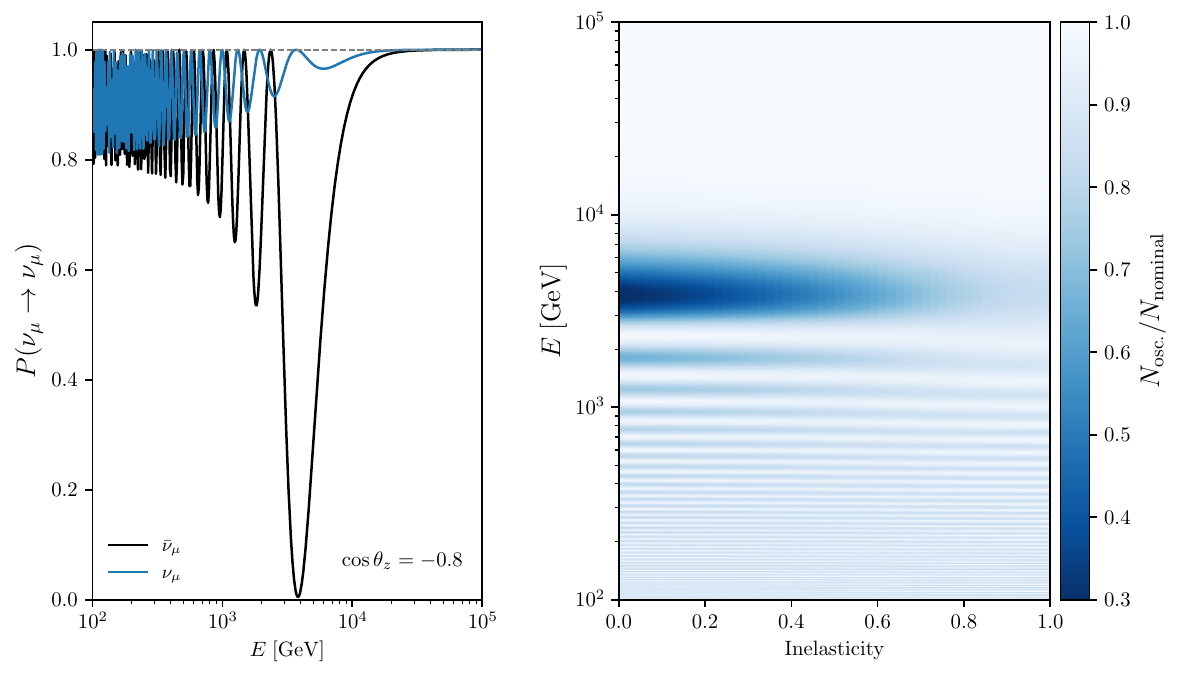}
    \caption{Left: Toy study of $\nu_\mu$ and $\bar{\nu}_\mu$ survival probabilities as a function of energy at $\cos(\theta_z) = -0.8$ given $\Delta m_{41}^2 = 1.0$ eV$^2$ and $\sin^2(2\theta_{24}) = 0.2$. Right: Ratio of events as a function of energy and inelasticity between the oscillation scenario and the no-oscillation (i.e. nominal) scenario.}
    \label{fig:inelasticity_oscillogram}
\end{figure}

The reconstructed inelasticity distribution therefore provides a handle on $\bar{\nu}_\mu$ disappearance: a sterile-induced deficit would appear predominantly at low $y$, where antineutrinos are concentrated.
This effect falls off at energies above 100 TeV because the sea quark contributions to DIS become important, making the $y$ distributions of neutrinos and antineutrinos nearly identical.
\cref{fig:example_inelasticity_disappearance} demonstrates a toy study with a neutrino and antineutrino sample.
The inclusion of antineutrino disappearance leads to a relative event deficit at lower inelasticity.
\cref{fig:inelasticity_oscillogram} shows this effect targeting the resonance region in zenith and energy.
Vacuum oscillations contribute negligibly to $\nu_\mu$ disappearance near the $\bar{\nu}_\mu$ resonance, so the matter resonance produces an observable $y$-dependent deficit.
This additional sterile signature motivates the development of new reconstructions targeting inelasticity, discussed in \cref{chapter:ml_recos}, and the development of a new IceCube event selection presented in \cref{chapter:meows2026}.

\chapter{An Improved Search for Unstable Sterile Neutrinos}\label{chapter:decay_analysis}

This chapter will present an analysis of IceCube data searching for an unstable sterile neutrino.
The unstable sterile neutrino model, also known as 3+1+Decay, is an extension of the vanilla 3+1 sterile neutrino model with the additional decay mechanism introduced in \cref{sec:unstable_sterile_neutrinos} and discussed further in \cref{sec:decay_signal}.
The model considered here assumes the invisible decay of $\nu_4$ into two daughter particles, the light scalar $\phi$ and neutrino state $\psi$, with a coupling constant $g$.
The lifetime of $\nu_4$ is given by \cite{Moss:2017pur, IceCubeCollaboration:2022tso},
\begin{equation}
    \tau = \frac{16\pi}{g^2 m_4}.
\end{equation}
For the invisible decay mode, this means that the final neutrino $\psi$ is either a right-handed neutrino or a left-handed antineutrino, which by construction does not couple to the Standard Model through the weak interaction.

\begin{figure}[htbp]
    \centering
    \includegraphics[width=0.75\linewidth]{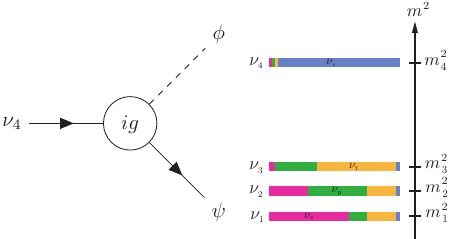}
    \caption{Left: Feynman diagram showing the decay of $\nu_4$ to two invisible particles $\phi$ and $\psi$. Right: Schematic of the neutrino mass ordering scheme with an eV-scale $\nu_4$ mass state and the relative flavor compositions (colored bars). Figure from Ref.~\cite{IceCube:2025wrv}.}
    \label{fig:decay_feynman}
\end{figure}

The model is described by three parameters: the active-sterile mixing angle $\theta_{24}$, the squared mass splitting $\Delta m_{41}^2$, and the coupling constant $g$ in the range $0 < g^2 < 4\pi$ that governs the decay rate\footnote{For the rest of this chapter, the square of the coupling constant ($g^2$) will be used.}.
As $m_4 \gg m_{1,2,3}$ for eV-scale sterile neutrinos, only one squared mass splitting is required to describe the oscillations.
It is assumed that the other mixing angles $\theta_{14}$ and $\theta_{34}$ are zero and do not impact this analysis since it is not sensitive to $\nu_e$ or $\nu_\tau$ appearance, only $\nu_\mu$ and $\bar{\nu}_\mu$ disappearance.
The observable effect of a non-zero coupling constant $g$ appears in the oscillation pattern as a damping effect as discussed in \cref{chapter:sterile_signals}.
\cref{fig:decay_feynman} schematically shows the invisible decay process considered in this analysis alongside the assumed mass ordering with the eV-scale $\nu_4$ state and corresponding flavor composition of each mass eigenstate.

\begin{figure}[htbp]
    \centering
    \includegraphics[width=0.85\linewidth]{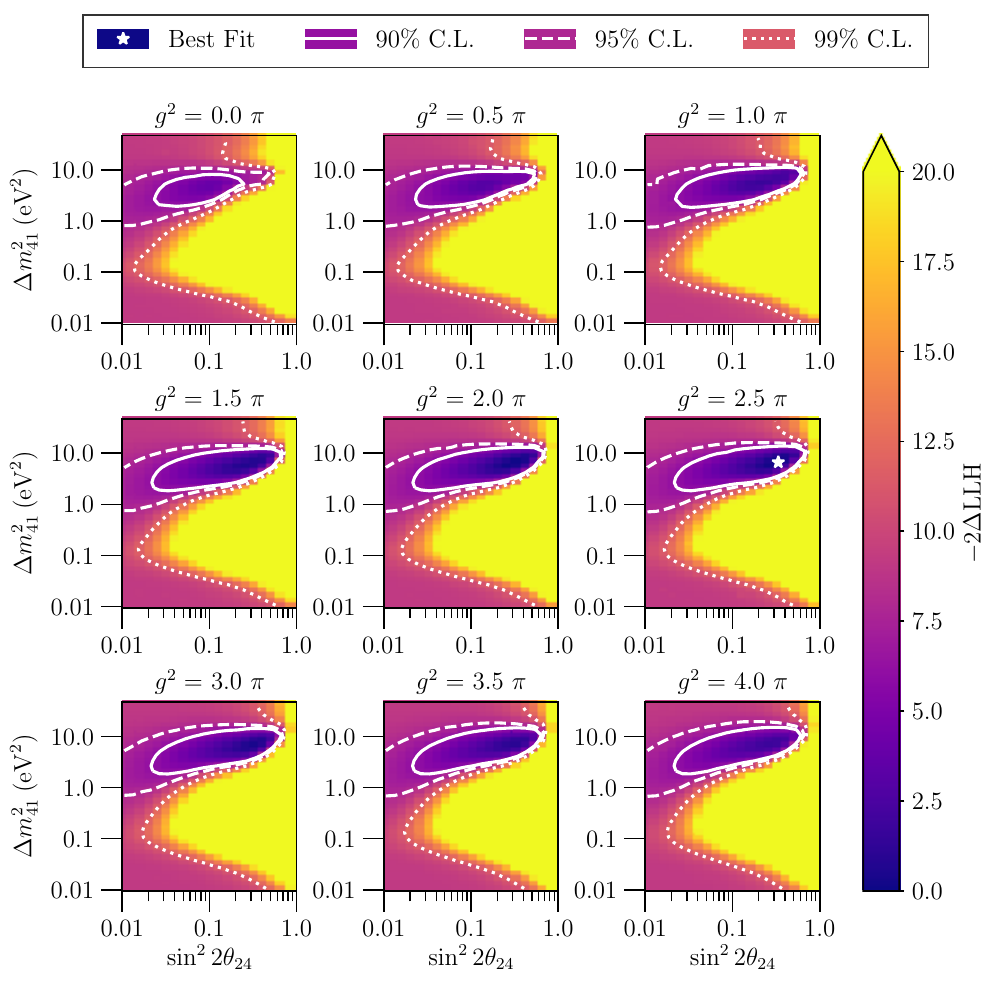}
    \caption{Results of the first search for unstable sterile neutrinos using the MEOWS-2020 event selection. The best-fit was found to be at $g^2 = 2.5\pi$. Figure from Ref. \cite{IceCubeCollaboration:2022tso}.}
    \label{fig:marjon_analysis_results}
\end{figure}

The high-energy sterile neutrino analyses with IceCube have proceeded as iterations based on a common event selection known as the ``Matter-Enhanced Oscillations With Steriles'' (MEOWS) event selection \cite{IceCube:2020phf,IceCube:2020tka}.
The approach is to begin with the vanilla 3+1 sterile neutrino analysis and then expand the analysis to more complex models.
The unstable sterile decay model was explored for the first time using the MEOWS-2020 event selection \cite{IceCubeCollaboration:2022tso}.
The result found allowed regions (closed contours) at 90\% C.L., as seen in \cref{fig:marjon_analysis_results}, where each subplot represents a bin in $g^2$.
This slight preference for non-zero $g^2$ motivates further studies reported in this thesis.

The MEOWS-2023 event selection incorporated significant improvements over the previous iteration.
New machine-learning reconstructions and BDT-based background rejection techniques improved signal efficiency and allowed for a separated sub-sample of starting track events with better energy resolution \cite{IceCubeCollaboration:2024dxk}.
The details of the MEOWS-2023 event selection will be discussed in \cref{sec:meows2023_event_selection}.

\begin{figure}[htbp]
    \centering
    \includegraphics[width=0.7\linewidth]{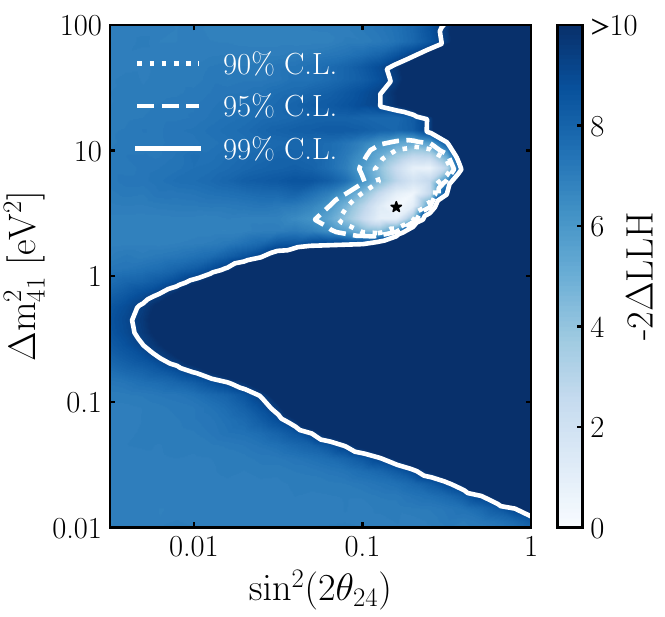}
    \caption{Result of the IceCube 3+1 sterile neutrino analysis using the MEOWS-2023 event selection. The color indicates the test statistic ($\mathrm{TS} = -2\Delta \mathrm{LLH}$). The best-fit point is shown as a black star with the 90\% (dotted), 95\% (dashed), and 99\% (solid) C.L. contours. Figure from Ref. \cite{IceCubeCollaboration:2024dxk}.}
    \label{fig:alfonso_analysis_result}
\end{figure}

The first application of MEOWS-2023 was the updated search for the vanilla 3+1 sterile neutrino \cite{IceCubeCollaboration:2024nle}.
That result found a best-fit favoring 3+1 over null at 2.2$\sigma$ \cite{IceCubeCollaboration:2024nle}, as seen in \cref{fig:alfonso_analysis_result}.
The questions that then followed were:
\begin{enumerate}
    \item Does expanding to the unstable 3+1 model provide a better fit to the data?
    \item Would an updated analysis be consistent with the previous MEOWS-2020 unstable sterile neutrino analysis?
\end{enumerate}
At the same time, global fits were being performed with accelerator-based oscillation data. An allowed region was found that preferred the decay scenario over the vanilla 3+1 model \cite{Hardin:2022muu}. A third question follows from this result:
\begin{enumerate}[start=3]
    \item Would the results of a MEOWS-2023 unstable sterile neutrino analysis be compatible with the global fit results?
\end{enumerate}

This chapter is organized as follows.
\cref{sec:meows2023_simulation} describes the methodology for simulating neutrino events and performing the first levels of processing to obtain a realistic Monte Carlo sample.
\cref{sec:meows2023_event_selection} explains the cuts and reconstructions that are applied to both simulation and data to form the MEOWS-2023 event selection with minimal background contamination.
\cref{sec:fit_procedure} presents the statistical formulation of the analysis and likelihood fit procedure.
\cref{sec:meows2023_systematics} details the systematic uncertainties in the analysis that are represented as nuisance parameters in the fit.
The results of this analysis are presented in \cref{sec:meows2023_results} and their interpretation in the context of the experimental landscape described in \cref{chapter:exp_bkg}.
Lastly, post-unblinding checks were performed to probe the stability of the analysis with respect to different splits of the dataset, the results of which are shown in \cref{sec:meows2023_postunblinding}.

\section{Simulation}\label{sec:meows2023_simulation}

\subsection{Event Generation}
The first step in simulated event production is sampling truth-level event parameters.
Neutrino directions are sampled uniformly in $\cos(\theta_z)$, where $\theta_z$ is the zenith angle, and in azimuthal angle $\phi$.
Each event is assigned an initial weight, and an event can be reweighted according to a flux model later to obtain a physically accurate rate prediction.
The energies of the simulated neutrino events are sampled from an $E^{-2}$ power law.
This is a harder spectrum than the actual spectral index of atmospheric neutrinos in this energy range ($\gamma \approx -3.7$).
Setting $\gamma = -2$ allows for more high-energy neutrino events to be generated to reduce uncertainty from Monte Carlo statistics, which can be reweighted later to reflect the physical rates.

The kinematic properties and interaction point of the neutrino interactions are more nuanced. 
The \texttt{LeptonInjector} code \cite{IceCube:2020tcq} is used to sample the interaction points of neutrino events near the detector, which takes into account the possible final-state muon range to increase the simulation efficiency of MC events that will produce a visible track.
The kinematic parameters, namely Bjorken-$x$ and inelasticity $y$ are sampled from the double-differential cross section $\frac{d^2\sigma}{dx dy}$ at a given neutrino energy $E$.
If the sampled $x$ or $y$ falls outside of the allowed kinematic phase space determined by the threshold: $Q^2 = 2 m E x y > Q_{\mathrm{min}}^2 = 1~\mathrm{GeV}^2$ where $m$ is the nucleon mass, then it is resampled.

\subsection{Particle and Light Propagation} \label{sec:particle_light_prop}
The outputs of the event sampling are the initial conditions for simulating the final-state lepton and hadronic shower of the neutrino DIS interaction.
For $\nu_\mu$ CC DIS interactions, where the final state is composed of a $\mu$ and hadronic shower $h$, two different propagation methods are used.
In the case of the $\mu$, the energy losses along the track are computed using the descriptions of the processes in \cref{sec:energy_dep}.
The energy losses at these energies are almost entirely cascade-like stochastic/radiative losses rather than ionizing continuous energy losses.
The position and energies of the losses ($\delta$ electrons, bremsstrahlung, pair production) along the muon track are calculated using \texttt{PROPOSAL} \cite{koehne2013proposal}.
The light signal produced by the muon is the sum of the muon's Cherenkov light and the Cherenkov light produced by charged particles in the stochastic losses.
Above the critical energy, the stochastic losses dominate the total light yield.
In the hadronic shower stemming from the DIS interaction vertex, an approximation is used to compute the relative light yield rather than simulating the complicated final-state dynamics.
The equation for the light yield of a hadronic shower given shower energy $E_h$ is,
\begin{equation}
    F = 1 - (1 - f_0) \left( \frac{E_h}{E_0} \right)^{-m}
\end{equation}
where $f_0$, $E_0$, and $m$ are determined from simulation  \cite{raedel2012cherenkov}.
The energy deposited by the shower that is visible is then $F E_h$, and to simulate it a package called \texttt{CMC} \cite{Voigt:2008zz} is used to generate photons.

With a series of energy losses and positions $\{\vec{r}_{i}, E_{i} \}$, the Cherenkov photon yield is computed and the photons are propagated through the detector using the photon propagation code \texttt{clsim} \cite{Schwanekamp:2022ybd}. For reference, the wavelength-dependent photon emission from the muon track follows the Frank-Tamm formula \cite{Frank:1937fk},
\begin{equation}
    \frac{d^2 N}{dx d\lambda} = \frac{2\pi\alpha}{\lambda^2} \left( 1 - \frac{1}{n(\lambda)^2 \beta^2} \right).
\end{equation}
The angle $\theta_c$ in which the Cherenkov photons are emitted with respect to the direction of the lepton is given by,
\begin{equation}
    \cos(\theta_{c}) = \frac{1}{n(\lambda)\beta},
\end{equation}
where $n(\lambda)$ is the wavelength-dependent index of refraction and $\beta = v/c$.

In ice, the absorption length of light is much larger than the scattering length, meaning photons scatter many times before being absorbed.
This means that photons are scattered significantly and typically do not point back to the source, so the Cherenkov cone cannot be reconstructed like typical water Cherenkov detectors. 
The photon propagation accounts for the different optical properties of each layer of ice in the detector. 
At each step of the photon propagation, the scattering and absorption properties are recomputed. 
Additionally, the directional properties of the ice discussed in \cref{sec:ice_properties} are accounted for, namely the ice tilt and anisotropic absorption properties due to the ice flow. 
Once a photon reaches the column of refrozen ice near the DOM, a parameterized model is used to simulate its angular dependence. The relative change in angular sensitivity $\epsilon$ is given by the forward hole ice model which is parameterized as \cite{IceCube:2020tka},
\begin{equation}
    \frac{\epsilon}{\epsilon_{0}} = 0.34\left(1+\frac{3}{2} x - \frac{1}{2}x^3\right)+p_{1} x(x^2 - 1)^3 + p_2 e^{10(x - 1.2)},
\end{equation}
where $x = \cos(\eta)$ with $\eta$ being the angle from the -$z$ axis (direction of the DOM). Both $p_1$ and $p_2$ are free parameters that alter the angular sensitivity in the backward and forward regions respectively. The nominal values are set to $p_1 = 0.3$ and $p_2 = -1$ which were determined by fits to flasher data assuming the nominal bulk ice model.

\subsection{Detector Response Simulation}

Events are initially simulated using a higher-than-nominal DOM efficiency which produces a larger sample of simulated hits on the DOMs than the nominal DOM efficiency would.
These hits are then downsampled according to the target DOM efficiency.
This downsampling technique is used to produce multiple MC sets that can be used to model the effect of the DOM efficiency systematic uncertainty.
Once the angular acceptance from the hole ice model and DOM efficiency are applied, the Monte Carlo photoelectrons (MCPEs) are then generated according to a template.
The single photoelectron (SPE) template is used to sample the resulting MCPE charges from a single photon hitting the photocathode.
A number of effects that produce extra photoelectrons are also taken into account, such as late pulses and afterpulses which are PMT effects.
The MCPE series is then processed into a ``pulse series'' which represents the digitized data saved by each DOM and recorded for the event.
This stage of the simulation results in MC events that match what the detector will actually read out.
Each stage of event processing after this treats MC and data identically.

\section{Event Selection} \label{sec:meows2023_event_selection}

The IceCube detector records approximately 3~kHz of triggered events, the majority of which are detector noise or atmospheric muons that must be rejected to obtain a pure neutrino sample.
The two stages of the event selection used in this analysis begin after collaboration-wide filtering levels L1 and L2 mentioned in \cref{sec:data_acquisition}. 
These stages apply generic filtering and reconstruction algorithms that provide analyzers with quantities to begin selecting high-quality events.

The MEOWS-2023 event selection begins with selecting events that pass the \texttt{MuonFilter}.
The \texttt{MuonFilter} selects events within certain regimes of $\cos(\theta_z^{reco})$ requiring varying levels of fit quality and number of hit DOMs ($\mathrm{N}_{\mathrm{chan}}$). 
These cuts remove a large number of events that are low charge and unlikely to be well-reconstructed. The rate of events that pass these cuts is approximately 30 Hz which are still overwhelmingly atmospheric muons. 
To select the neutrino events, more computationally expensive reconstructions and aggressive cuts must be performed.

In the first level of MEOWS processing after L2, a precut is applied to the events which further restricts events reconstructed above the horizon.
While only tracks below the horizon ($\cos(\theta_z^{true}) < 0$) are wanted, preserving high-quality tracks at this stage is beneficial since the basic direction reconstruction applied in L1/L2 could be inaccurate compared to the more expensive reconstructions.
Above $\cos(\theta_{z}^{reco}) > 0$, only events with 100 photoelectrons and a charge-weighted distance less than 200m pass the precut.
Below that, events pass the selection if: $\log_{10} Q_{tot} \geq 0.6 (\cos(\theta_z^{reco}) - 0.5) + 2.5$, where $Q_{tot}$ is the total observed charge for the event.
This simple precut, which utilizes the reconstructed zenith angle from the L1 and L2 reconstructions (precomputed) and simple charge variables reduces the event rate to $\sim$few Hz.
At this rate, it becomes computationally feasible to run more computationally expensive reconstructions on the events.

The next step in the event selection is to apply a topological event splitter, which takes an event (which could contain coincident atmospheric muons) and splits it into multiple events based on the topology.
After the event splitting, additional pulse cleaning is applied to remove any possible PMT afterpulses which are not well-modeled and could bias reconstructions.
A sequence of likelihood-based track reconstructions is performed on each event.
Each track reconstruction is fed into the successive one as a seed to improve fit stability and ensure a good final fit, assessed by the reduced log-likelihood.
The angular uncertainties in zenith and azimuth are computed from the track best fit by scanning the local region of the best-fit direction.
An additional track fit is performed with a Bayesian zenith prior that encourages the fitter to assume the track is coming from the Southern sky, or above the horizon.
A likelihood ratio test is formed to compare the previous track best fit to the Bayesian down-going track fit and events that are statistically consistent with a down-going hypothesis are rejected.
Regions with a poor angular resolution and fit quality are also cut. 
The final step in this level of processing is an estimation of the track energy, called ``MuEx'' which assumes an infinite muon track hypothesis and fits to the observed energy losses \cite{IceCube:2013dkx}.
The event rate at this level of processing is several tens of mHz.
The majority of events at this stage are atmospheric muon backgrounds that are mistakenly reconstructed as up-going.

\begin{figure}[htbp]
    \centering
    \includegraphics[width=0.85\linewidth]{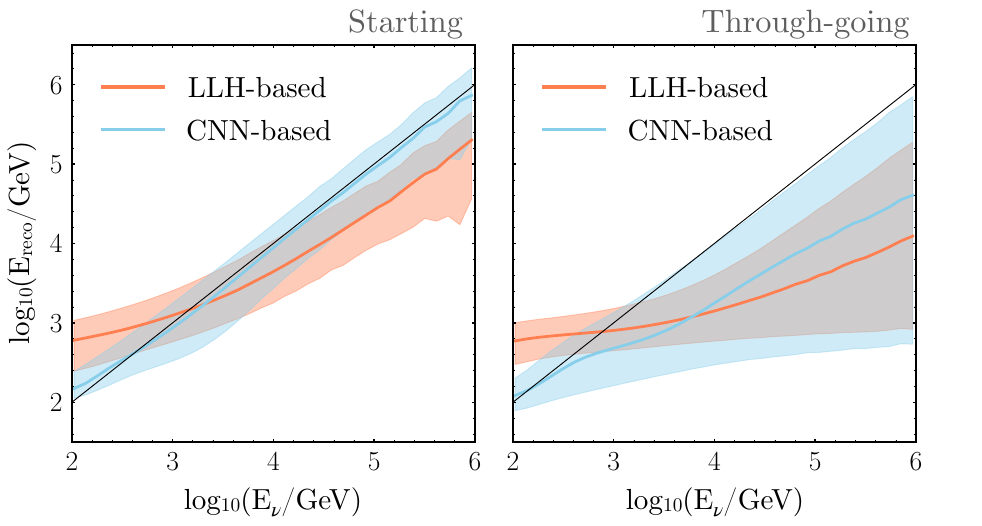}
    \caption{Reconstructed energy for starting tracks (left) and through-going tracks (right) as a function of true energy for LLH-based reconstruction used in MEOWS-2020 (orange) and the CNN-based reconstruction used in MEOWS-2023 (blue). The median is shown as a solid line and the band represents the 90\% uncertainty. Figure from Ref. \cite{IceCubeCollaboration:2024dxk}.}
    \label{fig:meows2023_energy}
\end{figure}

The final stage of the event selection aims to reduce the atmospheric muon background to zero by making very restrictive cuts.
The first step is to apply two convolutional neural network (CNN) reconstructions on the events that passed the previous cuts.
The first neural network classifies the events as either: through-going track, starting track, skimming track, stopping track, or cascade \cite{Glauch:2021kke}.
Events with a starting track score $>0.99$ are classified as starting tracks and are used as a subsample with excellent energy resolution since the energy of the neutrino interaction vertex is measured.
The second CNN estimates the combined visible energy of the event, which includes contributions from both the muon and hadronic shower if present.
In contrast to the likelihood-based MuEx energy estimator used in MEOWS-2020, there is no explicit event hypothesis assumed by the neural network and it is trained to infer the deposited energy by all particles in an event.
The reconstruction performance and uncertainty comparing both the LLH-based (MuEx) and the DNN-based reconstructions are shown in \cref{fig:meows2023_energy}.

\begin{figure}[htbp]
    \centering
    \includegraphics[width=0.75\linewidth]{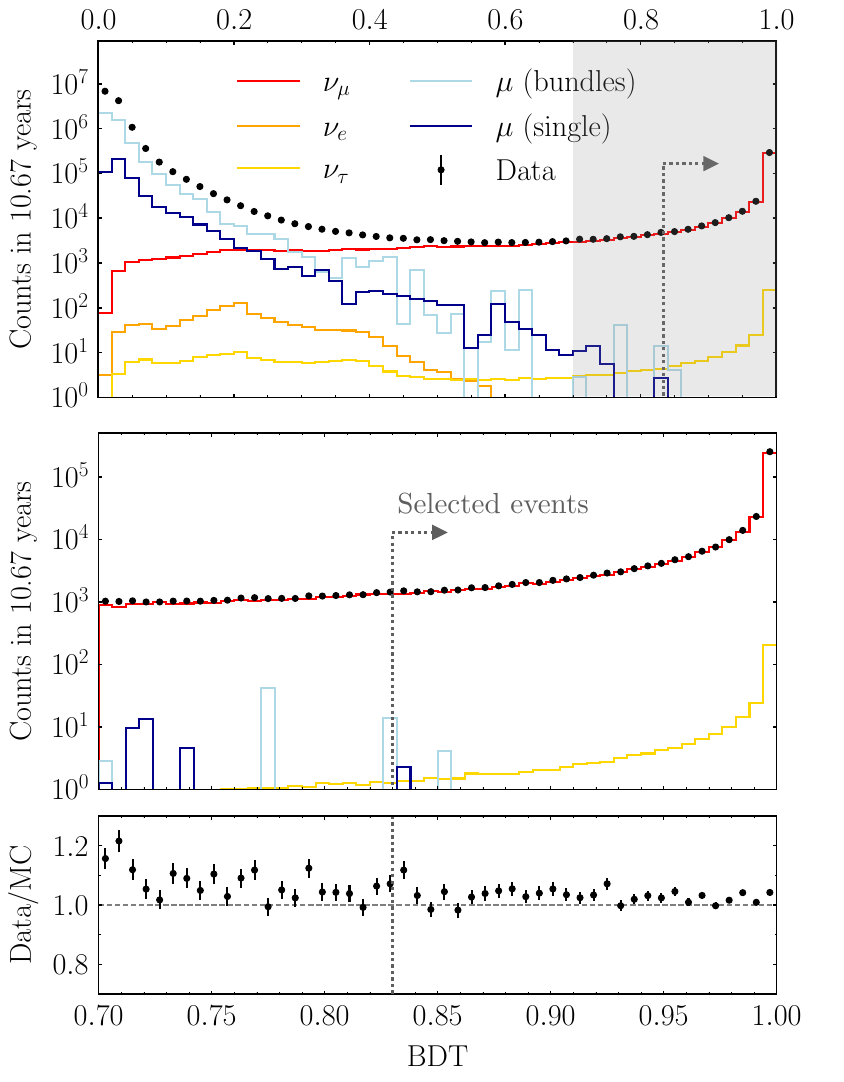}
    \caption{Top: BDT scores for data (black dots) compared against $\nu_e$ (orange), $\nu_\mu$ (red), $\nu_\tau$ (yellow), single atmospheric $\mu$ (light blue), and atmospheric $\mu$ bundles (dark blue) simulation. The dashed line and arrow indicates the BDT score cut used for the analysis at 0.83. Middle: BDT scores zoomed in to the analysis selection region. Bottom: Data/MC comparison in the analysis selection region. Figure from Ref. \cite{IceCubeCollaboration:2024dxk}.}
    \label{fig:meows2023_bdt_scores}
\end{figure}

The final selection of events is performed by a boosted decision tree (BDT).
The BDT was trained on event features that capture morphology, energy, direction, track fit quality metrics, as well as low-level charge variables.
A variety of background and signal MC samples were used to ensure good Data/MC agreement with the simple labels of 0 = background (atmospheric $\mu$ and cascades) and 1 = signal ($\nu_\mu$ tracks).
The background sample was comprised of 300k CORSIKA \cite{Heck:1998vt} air shower events and 400,000 single muon events, and the signal sample contained nearly 4 million simulated muon neutrino events.
A significant initial set of background MC was required since few background events pass through the filtering to the BDT stage.
The $\nu_\mu$ rate estimated from MC passing the BDT cut, before energy and zenith cuts, is $2.1$ mHz, with an estimated background rate of $0.101$ $\mu$Hz.
The BDT scores of the signal and background MC compared to data are shown in \cref{fig:meows2023_bdt_scores}.
With a zenith cut of $\cos(\theta_z^{reco}) < 0$ and an energy cut $500~\mathrm{GeV} < E_{reco} < 100~\mathrm{TeV}$, the signal rate is $961$ $\mu$Hz and the background rate is $0.019$ $\mu$Hz.
A small fraction of astrophysical $\nu_\tau$ events interact outside the detector and produce up-going muon tracks indistinguishable from $\nu_\mu$ tracks, contributing an additional $0.994~\mu$Hz under the assumption of a nominal astrophysical flux with 1:1:1 flavor composition at Earth.
Summing over all contributions, this results in an overall $\nu_\mu$ purity of 99.9\%.

\begin{figure}[htbp]
    \centering
    \includegraphics[width=0.95\linewidth]{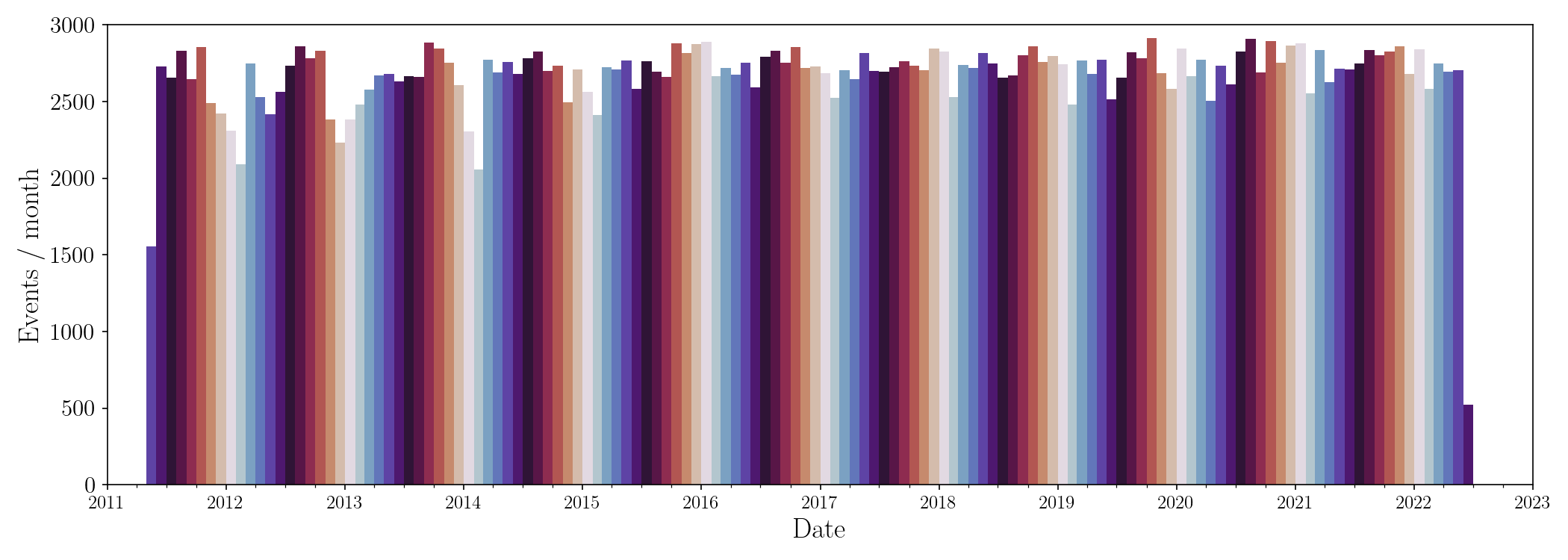}
    \caption{Monthly event rate of the MEOWS-2023 event selection from 2011 to 2022.}
    \label{fig:meows2023_livetime}
\end{figure}

Events were processed with the full processing chain using 10.67 years of collected data, resulting in a total of 93,529 starting tracks and 274,309 through-going tracks \cite{IceCubeCollaboration:2024dxk}.
The monthly event rate over the processed detector live time is shown in \cref{fig:meows2023_livetime}.

\section{Fitting Methodology} \label{sec:fit_procedure}

This analysis used a binned maximum-likelihood fit over energy and zenith for starting and through-going tracks to extract the sterile neutrino parameters from the data.
The fit was implemented in the $\texttt{GollumFit}$ fitting framework \cite{IceCube:2025yvq}.

The analysis is designed to use a likelihood ratio test, where the test statistic (TS) is given by the ratio of likelihoods, or the difference in log-likelihood (LLH),
\begin{equation}\label{eq:test_statistic}
    \mathrm{TS}(\theta) = -2\Delta\mathrm{LLH} = 2(\textrm{LLH}(\theta) - \textrm{LLH}_{\textrm{min}}).
\end{equation}

The LLH is determined at each point $\theta$ in the physics parameter space by profiling over the nuisance parameters $\eta$. The simplest likelihood for counting events in a binned analysis is the Poisson likelihood given by,
\begin{equation}\label{eq:poisson_likelihood}
    \mathcal{L}_{\mathrm{Poisson}}(\theta\vert k) = \frac{\lambda(\theta)^{k} e^{-\lambda(\theta)}}{k!}
\end{equation}
where $\lambda(\theta)$ is the expectation for a given physics hypothesis $\theta$ and $k$ is the observed count. 
In this analysis $\lambda(\theta)$ is the expected number of events in each analysis bin, computed by summing the weights of MC events falling in that bin under the physics hypothesis $\theta$.
The Monte Carlo method introduces an intrinsic statistical uncertainty because the simulated sample is finite and only converges to the true expectation in the limit of infinite statistics.
The uncertainty introduced from statistical variations in the MC is given by the square of the event weights,
\begin{equation}\label{eq:mc_stat_unc}
    \sigma_{\mathrm{MC}}^2 = \sum_{i} w_{i}(\theta)^2.
\end{equation}
To fully account for the uncertainties introduced from MC sampling, the Schneider-Arg\"uelles-Yuan (SAY) likelihood \cite{Arguelles:2019izp} is used,
\begin{equation}\label{eq:SAY_likelihood}
    \mathcal{L}_{\mathrm{SAY}} (\theta \vert k) = \left(\frac{\mu}{\sigma^2}\right)^{\frac{\mu^2}{\sigma^2} + 1} \Gamma\left( k + \frac{\mu^2}{\sigma^2} + 1 \right) \left[ k!\left( 1 + \frac{\mu}{\sigma^2} \right)^{k+\frac{\mu^2}{\sigma^2}+1} \Gamma \left( \frac{\mu^2}{\sigma^2} + 1 \right) \right]^{-1}.
\end{equation}
This is expressed in terms of gamma functions $\Gamma$ and includes the MC statistical uncertainties defined in \cref{eq:mc_stat_unc}.
In the limit of no MC statistical uncertainty, $\sigma^2 \rightarrow 0$, the SAY likelihood reduces to the Poisson form of \cref{eq:poisson_likelihood}.

Evaluating the likelihood at each point in the physics parameter space requires reweighting the full MC sample, which is computationally prohibitive.
To address this, simulated events are compressed using the FastMC technique, which groups MC events that are similar in phase space and combines their weights \cite{IceCube:2025yvq}.
This compression reduces the number of events requiring reweighting by approximately a factor of ten, with a corresponding reduction in fit run time, while preserving the accuracy of the binned likelihood to within the MC statistical uncertainty.
Because the physics parameters $\theta$ affect the true neutrino flux and therefore the event weights prior to compression, a separate FastMC set must be generated for each point in the physics parameter space.
For each point in the parameter space, each of the flux components is computed and used to calculate their contributions to the physical event rate.

With a FastMC for each point in the parameter space, the likelihood is minimized over the nuisance parameters using the L-BFGS algorithm \cite{liu1989limited} with box constraints (L-BFGS-B) \cite{byrd1995limited}.
This minimizer constructs an approximate inverse Hessian at each step that captures the local curvature of the nuisance parameter space to find the minimum.
The box constraints impose hard boundaries on the physics systematics, preventing numerical instability and out-of-bounds evaluations.
Because the full systematic treatment discussed in \cref{sec:meows2023_systematics} contains 36 nuisance parameters, the minimizer can become stuck in local minima.
To overcome this, each fit to the nuisance parameters at each point in physics parameter space is run using ten seeded values of the systematics.

\section{Systematic Treatment}\label{sec:meows2023_systematics}

\subsection{Atmospheric Neutrino Flux}

Atmospheric neutrinos are produced in the atmosphere by the decays of mesons produced by cosmic ray interactions with the air as discussed in \cref{sec:neutrino_flux}.
There are two main contributions to the uncertainty of the neutrino flux, which come from the uncertainty in the cosmic ray flux and the uncertainties stemming from the hadronic interaction model.
In this analysis, the \texttt{daemonflux} atmospheric neutrino flux model \cite{Yanez:2023lsy} is used.
The neutrino flux is calibrated using observed atmospheric muon data from a number of experiments \cite{Haino:2004nq, CMS:2010yju, L3:2004sed, Allkofer:1985ey, Matsuno:1984kq, MINOS:2007laz, OPERA:2014blf}.

\begin{figure}[htbp]
    \centering
    \includegraphics[width=0.65\linewidth]{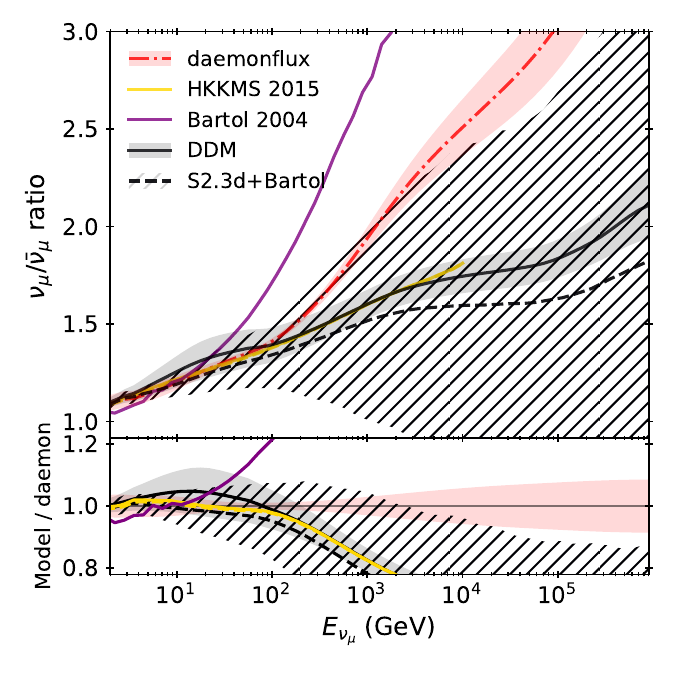}
    \caption{Ratio of the atmospheric $\nu_\mu$ to $\bar{\nu}_\mu$ fluxes for different models as a function of true neutrino energy. Figure from Ref.  \cite{Yanez:2023lsy}.}
    \label{fig:daemonflux_nunubar}
\end{figure}

\texttt{daemonflux} combines the global spline fit (GSF) model  \cite{Dembinski:2017zsh} for the primary cosmic ray fluxes with the data-driven model (DDM) \cite{Fedynitch:2022vty} of non-perturbative hadronic interactions that produce neutrinos.
The number of parameters used by GSF is reduced to six using principal component analysis to select only the largest contributors to the uncertainty.
Observed atmospheric muon data from various experiments are then used to calibrate the parameters of the model.
The resulting flux model provides a significantly improved atmospheric neutrino flux model with reduced uncertainties across the entire analysis energy range over previous models (e.g. the Bartol model \cite{Barr:2006it}).
Notably, the fraction of antineutrinos in the conventional atmospheric flux is smaller than that of previous analyses \cite{IceCube:2020tka, IceCubeCollaboration:2022tso}.
The $\nu_\mu$/$\bar{\nu}_\mu$ ratio for different flux models with respect to \texttt{daemonflux} is shown in \cref{fig:daemonflux_nunubar}.

In this analysis, the six cosmic-ray PCA parameters (denoted GSF1 to GSF6) and ten of the DDM hadronic parameters are used.
This reduced set of hadronic parameters was determined by observing the variations in expected muon neutrinos and selecting only the parameters with an observable effect.
The hadronic parameters varied in the fit are: $K_{+}^{158G}$, $K_{-}^{158G}$, $\pi_{+}^{20T}$, $\pi_{-}^{20T}$, $K_{+}^{2P}$, $K_{-}^{2P}$, $\pi_{+}^{2P}$, $\pi_{-}^{2P}$, $p_{2P}$, $n_{2P}$.
The definitions of these parameters can be found in Ref. \cite{Yanez:2023lsy}.
In this analysis, each parameter is perturbed by $1\sigma$ and the atmospheric fluxes are recomputed at each point in the sterile neutrino parameter space.
Each MC event is then assigned an additional weight that corresponds to the variation of the event rate with respect to that flux parameter.
The fluxes for each component are propagated through the Earth to the detector using $\texttt{nuSQuIDS}$ \cite{Delgado:2014lyt, Arguelles:2021twb}, which numerically evolves the neutrino ensemble in matter for a given set of sterile neutrino parameters.

\subsection{Atmospheric Uncertainties}
Kaons produced in cosmic ray air showers interact with nuclei in the atmosphere as they propagate, losing energy before they decay to produce neutrinos.
The result of this a shift in the spectral shape of high-energy atmospheric $\nu_\mu$.
This depends on the kaon-nucleon cross section, which has not been measured at center-of-mass energies relevant for air showers \cite{ParticleDataGroup:2024cfk}.
However, the total hadron-hadron cross section can be expressed as \cite{Halzen:2011xc},
\begin{equation}
    \sigma = Z_{ab} + B \log^2 \left( \frac{s}{s_{0}^{ab}} \right)
\end{equation}
where $B$ is a universal constant for all hadron-hadron scattering at high energies, $Z_{ab}$ is a projectile-dependent constant, and $s_0^{ab}$ is a scale factor.
The values for these constants for $Kp$ scattering have been determined from fits to data \cite{Halzen:2011xc}.
A conservative uncertainty of 7.5\% is applied to $\sigma_{K-\mathrm{Air}}$ to cover the $\sigma_{Kp}$ uncertainty and any unaccounted-for nuclear effects that could be relevant for $O$ or $N$ nuclei \cite{IceCube:2020tka}.
In the fit, this is implemented as a shape effect that modifies the neutrino flux as a function of true energy and zenith angle.

The effective scattering length of mesons in the atmosphere depends on the local density.
The uncertainty on the atmospheric density was obtained in a previous analysis \cite{IceCube:2020tka} by simulating the effects of shifting the density within allowed range determined by temperature data from the NASA Atmospheric InfraRed Sounder (AIRS) satellite \cite{airs_data}.
The neutrino flux under atmospheric density assumptions could be computed with a primary cosmic ray model and hadronic interaction model then propagated to the surface.
Similar to the kaon-air cross section, the implementation of this systematic uncertainty is a shift in the event weights in true energy and zenith as a function of a density parameter.

\subsection{Astrophysical and Prompt Atmospheric Neutrino Flux}

\begin{figure}[htbp]
    \centering
    \includegraphics[width=0.8\linewidth]{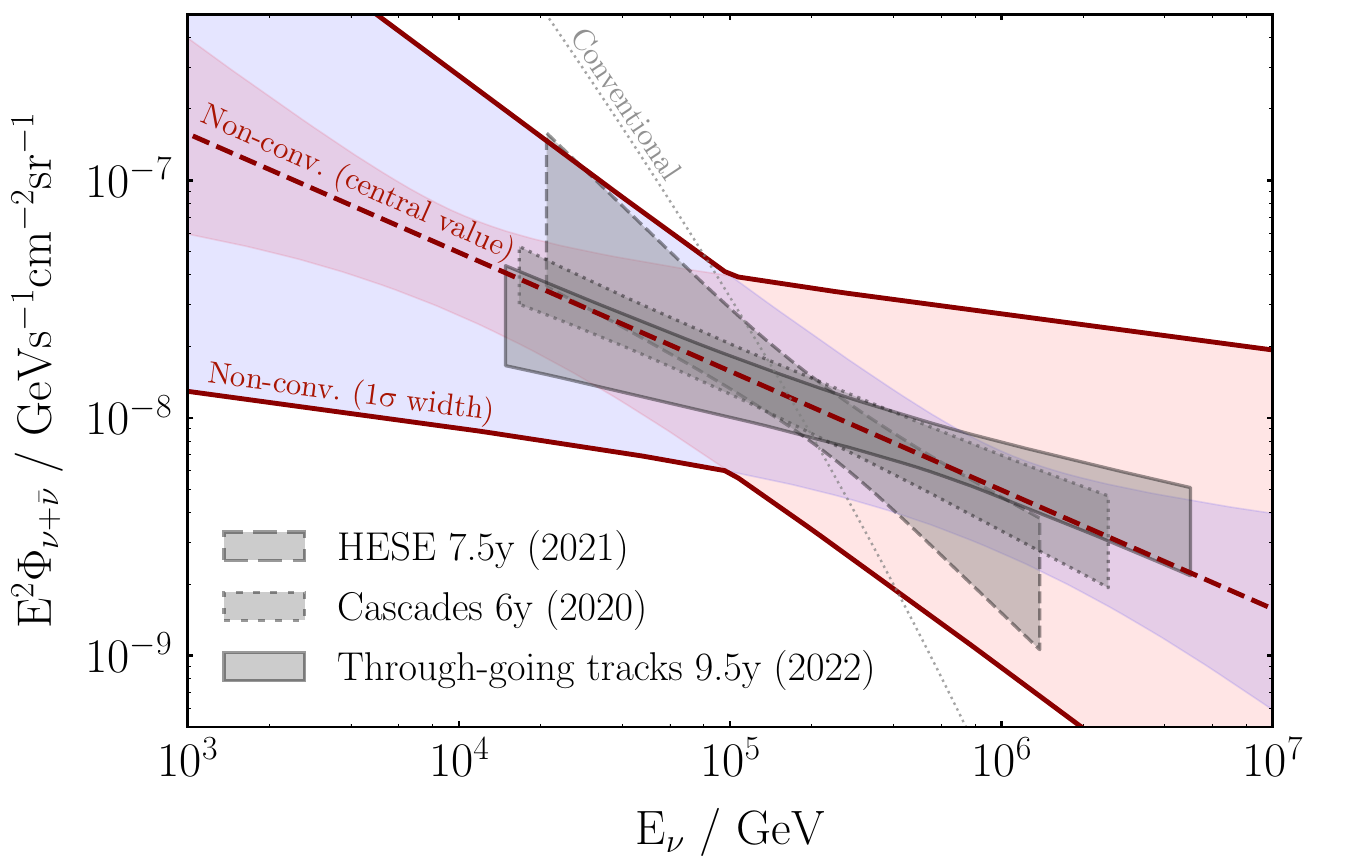}
    \caption{The range of the astrophysical flux prior used in the analysis. Results of previous IceCube measurements of the astrophysical neutrino flux are shown in gray bands \cite{IceCube:2020wum, Abbasi:2021qfz, IceCube:2020acn}. Figure from Ref. \cite{IceCubeCollaboration:2024dxk}.}
    \label{fig:meows2023_nonconv}
\end{figure}

In MEOWS-2023, the upper energy limit of the neutrino events in the selection extends up to 100 TeV.
This means that the flux component from the astrophysical and prompt atmospheric fluxes are not negligible.
The prompt atmospheric flux component used in this analysis is the prediction from the Sibyll2.3c model \cite{Riehn:2017mfm}.
For the astrophysical neutrino flux, a conservative broken power law model rather than a single power law is used to accommodate a potential break point in the astrophysical flux spectrum,
\begin{equation}
    \Phi(E) = \Phi_{0} 
    \begin{cases}
    \left(\dfrac{E}{E_{b}}\right)^{-\gamma_{1}} & E \leq E_{b} \\[6pt]
    \left(\dfrac{E}{E_{b}}\right)^{-\gamma_{2}} & E > E_{b}
    \end{cases}
\end{equation}
Below $E_b$, the $\gamma_1$ spectral index is used and above $E_b$, $\gamma_2$ is used with a shared normalization $\Phi_0$.
The flavor ratio of the astrophysical flux is assumed to be $1:1:1$ with equal portions of neutrinos and antineutrinos.
The nominal astrophysical flux and the prior shape are shown in \cref{fig:meows2023_nonconv} along with recent measurements of the astrophysical flux from IceCube \cite{IceCube:2020wum, Abbasi:2021qfz, IceCube:2020acn}.

\subsection{Neutrino Attenuation}

At energies above $10$ TeV, neutrinos that pass through the Earth's core start to be absorbed through neutrino deep inelastic interactions.
The baseline cross-section model for IceCube analyses is the CSMS model \cite{Cooper-Sarkar:2011jtt}. 
While the cross sections for scattering off of protons and neutrons are treated separately, nuclear effects are not accounted for.
To fully span the range of mismodeling of the neutrino-nucleon cross-sections and possible nuclear modifications, an overall $10\%$ uncertainty is applied to the neutrino and antineutrino cross sections.
A spline that modifies the MC weights is created by propagating neutrinos through the Earth with these modified cross sections, for neutrinos and antineutrinos separately.

\subsection{Detector Systematics}

There are several sources of uncertainty that stem from an imperfect understanding of the IceCube detector.
The main contributors are the models of the bulk ice and hole ice, with an additional contribution from the overall efficiency of the DOMs.

The uncertainties in the bulk ice are modeled using the Snowstorm method \cite{IceCube:2019lxi}.
In the Snowstorm method, the absorption and scattering profile of the ice is perturbed using a Fourier expansion that allows for expressive representations of the ice model uncertainty.
The amplitudes and phases of the Fourier components are constrained by fits to flasher data, which define the relative uncertainties and correlations.
In total, five amplitudes ($\mathrm{A}_{0}$ to $\mathrm{A}_{4}$) and four phases ($\mathrm{Phs}_{1}$ to $\mathrm{Phs}_{4})$ are used. Systematic MC sets were produced with variations in these parameters to capture the variations in reconstructed quantities.
For the hole ice systematic uncertainty, MC sets were produced with discrete variations in the hole ice model equal to the total MC statistics of the full nominal MC set.
At the final level, splines were created to capture the changes in event rates in each bin as a function of the hole ice $p_2$ parameter as defined in \cref{sec:particle_light_prop}.
The splines provide a smoothly varying prediction that captures the impact of the hole ice uncertainty with well-defined gradients for the minimizer.

\FloatBarrier

\section{Analysis Sensitivity}\label{sec:meows2023_sensitivity}
The sensitivity of the unstable sterile neutrino decay analysis was determined through Asimov data studies assuming Wilks' Theorem \cite{Wilks:1938dza}.
An Asimov data set is defined as the MC prediction under the nominal systematics and without Poisson statistical fluctuations.
This provides an event expectation for each bin in energy and zenith given each set of sterile parameters or null (no-sterile) hypothesis.
The atmospheric and prompt fluxes are calculated for each point in the sterile parameter space, along with variations in the conventional atmospheric flux with respect to each \texttt{daemonflux} systematic parameter.
To compute the sensitivity, the data is taken to be the expectation under the null hypothesis and nominal systematics and fit assuming the sterile hypothesis at each point in the parameter space.

\begin{figure}[h!]
    \centering
    \includegraphics[width=0.9\linewidth]{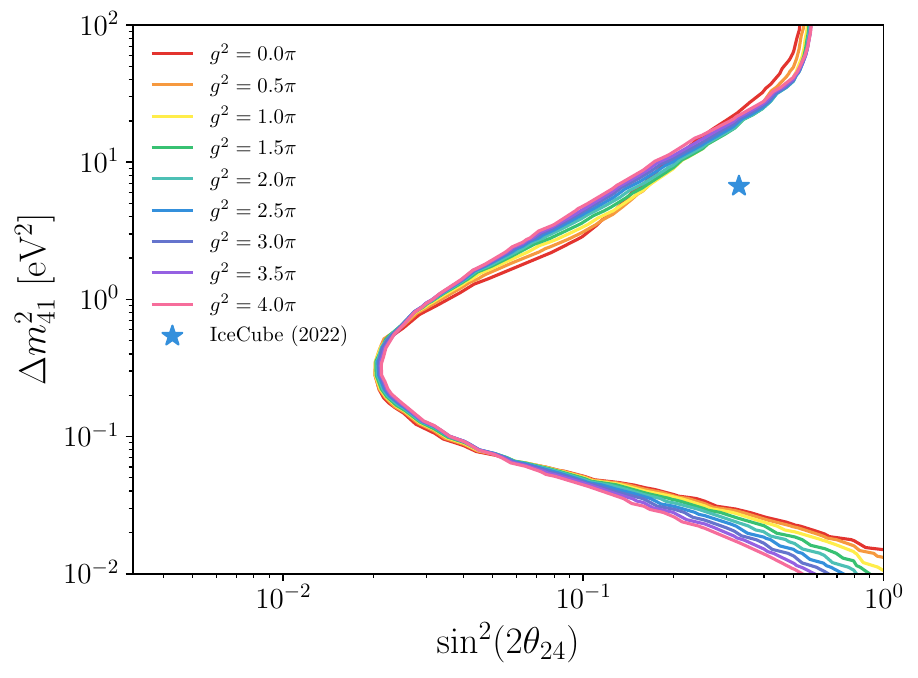}
    \caption{90\% C.L. Asimov sensitivities of the improved unstable sterile neutrino analysis. Each color corresponds to different values of $g^2$ in increments of 0.5$\pi$ from 0 to 4$\pi$. The best-fit point of the MEOWS-2020 3+1+Decay analysis \cite{IceCubeCollaboration:2022tso} is shown as a blue star at $g^2 = 2.5\pi$.}
    \label{fig:meows_decay_asimov_90}
\end{figure}

The analysis was set up to fit events in the region $500~\mathrm{GeV} < E_{reco} < 100~\mathrm{TeV}$ and $-1 < \cos(\theta_z^{reco}) < 0$.
There are 22 bins logarithmically spaced in energy and 20 linearly spaced bins in $\cos(\theta_z^{reco})$.
Additionally, the starting tracks are separated from through-going tracks by selecting events with a starting score $>0.99$.
In total, there are 880 bins in the analysis, 440 for starting and through-going tracks each.
While the signals of eV-scale sterile neutrinos appear primarily above 1 TeV, the inclusion of the low-energy region helps constrain the systematic uncertainties in the fit.
The 90\% Asimov sensitivity is shown in \cref{fig:meows_decay_asimov_90}, where each curve represents one of the nine values of $g^2$ in the fit, compared to the best-fit point of the previous IceCube unstable sterile neutrino analysis.

\begin{figure}[htbp]
    \centering
    \includegraphics[width=0.9\linewidth]{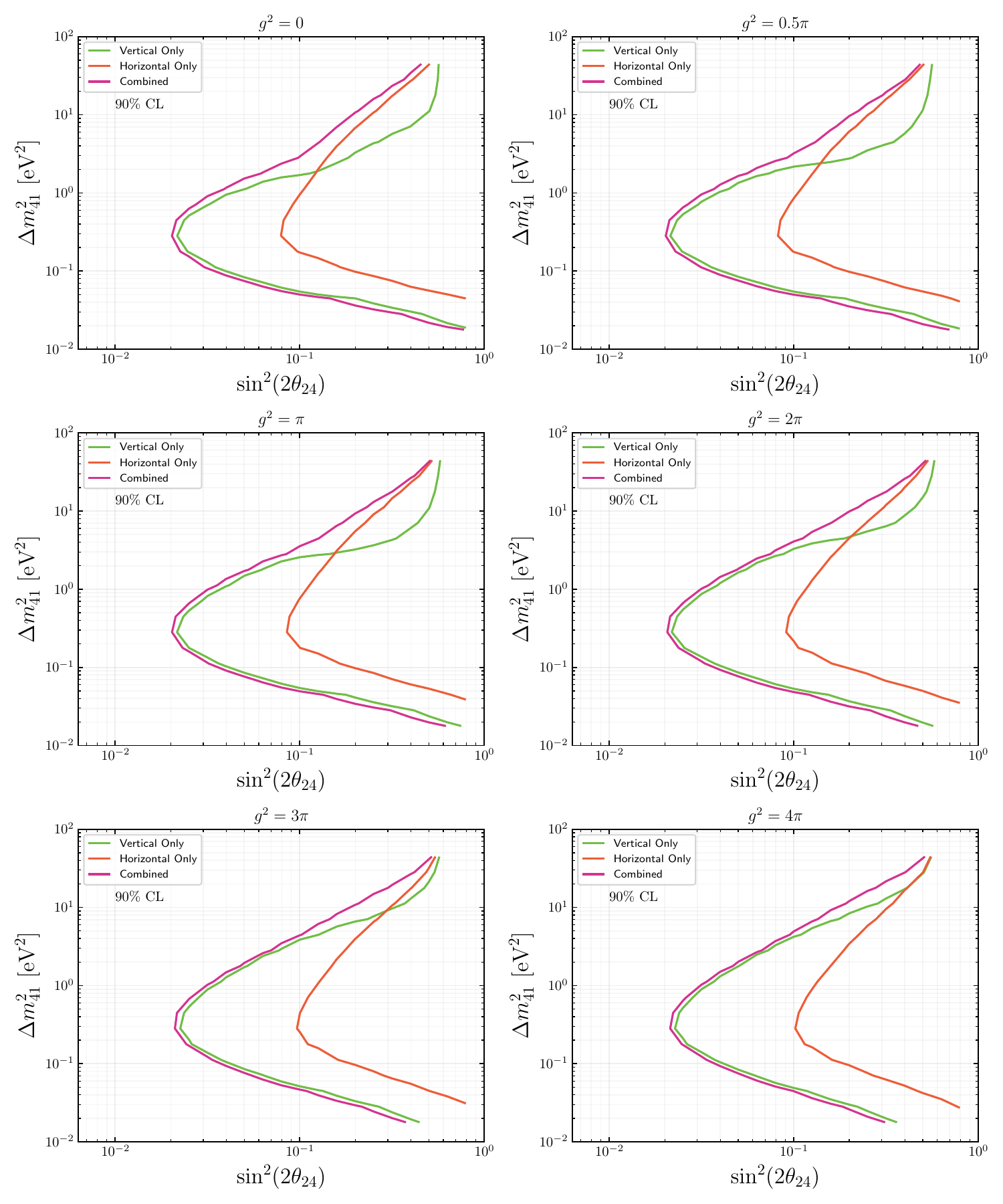}
    \caption{90\% C.L. Asimov sensitivities of the 3+1+Decay analysis for the horizon and vertical regions separated for $g^2=0,0.5\pi,\pi,2\pi,3\pi,4\pi$.}
    \label{fig:meows_decay_asimov_hv}
\end{figure}

\begin{figure}[htbp]
    \centering
    \includegraphics[width=0.9\linewidth]{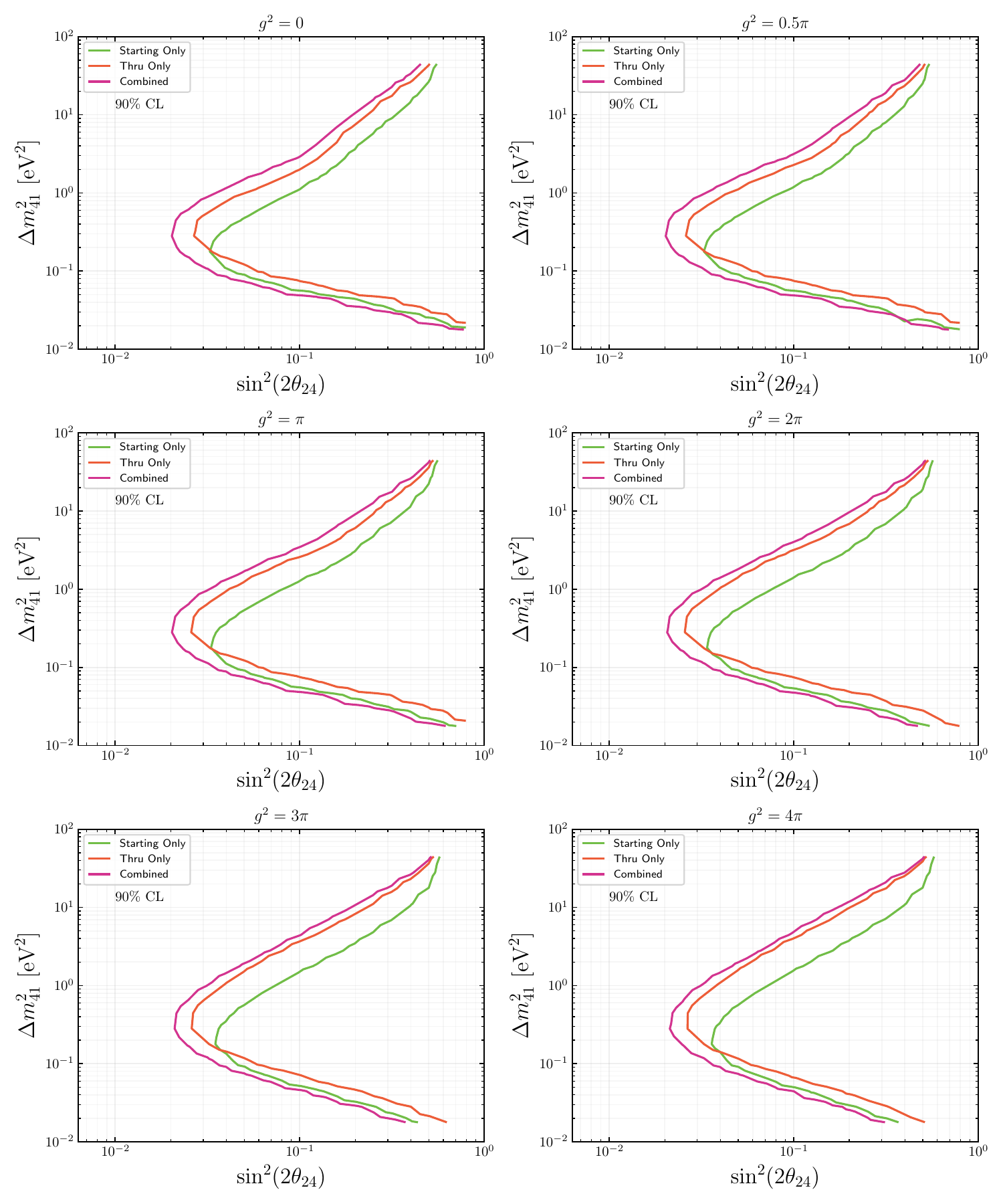}
    \caption{90\% C.L. Asimov sensitivities of the 3+1+Decay analysis for starting and through-going tracks separated for $g^2=0,0.5\pi,\pi,2\pi,3\pi,4\pi$.}
    \label{fig:meows_decay_asimov_ts}
\end{figure}

Several checks were performed to evaluate the contribution of certain components of the data to the overall sensitivity.
\cref{fig:meows_decay_asimov_hv} demonstrates the combined effects of the vertical ($-1<\cos(\theta_z)<-0.2$) and horizon ($-0.2<\cos(\theta_z)<0$) regions. The vertical region is required to probe the resonance region which produces the largest signal.
In the horizon region, the signal is almost entirely defined by the vacuum $L/E$-type oscillations.
For higher $g^2$, the contributions to the sensitivity from the horizon region are less dramatic because of oscillation damping effects.
This means that the sensitivity is driven mostly by the vertical region from the resonance as the smaller components of the signal get washed out.
The starting track subsample, with its enhanced energy resolution, primarily improves sensitivity to low $\Delta m_{41}^2$, as demonstrated in \cref{fig:meows_decay_asimov_ts}.

\FloatBarrier

\section{Blind Fits and Results} \label{sec:meows2023_results}
Prior to unblinding the analysis, a sequence of pre-unblinding checks were performed to ensure the fit was successful and the data was being modeled correctly. The steps of the blind fits are outlined in \cref{tab:blind_fit_plan}.

\begin{table}[htbp]
    \centering
    \begin{tabular}{ |c|c| } 
        \hline
        Step & Stop Condition \\
        \hline
        Blind fit over 3D space & \\
        \hline
        Calculate BF systematics pulls & Any systematic pulls $>3\sigma$ \\
        \hline 
        Ensemble test with BF point & BF TS has a p-value $< 5\%$ \\
                                    & compared to ensemble test \\
        \hline
        Check BF TS vs. ensemble & 6 or more bins in each \\
        test bin-wise            & sample have p-value $>3\sigma$ \\
        \hline
        Calculate 1D data pulls in & 3 or more bins pull \\
        zenith and energy          & $>3\sigma$ in any sample \\
        \hline
        Plot and show 1D zenith & $\chi^2$ p-value $< 5\%$\\
        and energy distributions& \\
        \hline
    \end{tabular}
    \caption{Blind fit plan of the MEOWS-2023 unstable sterile neutrino analysis.}
    \label{tab:blind_fit_plan}
\end{table}

\begin{figure}[htbp]
    \centering
    \includegraphics[width=0.98\linewidth]{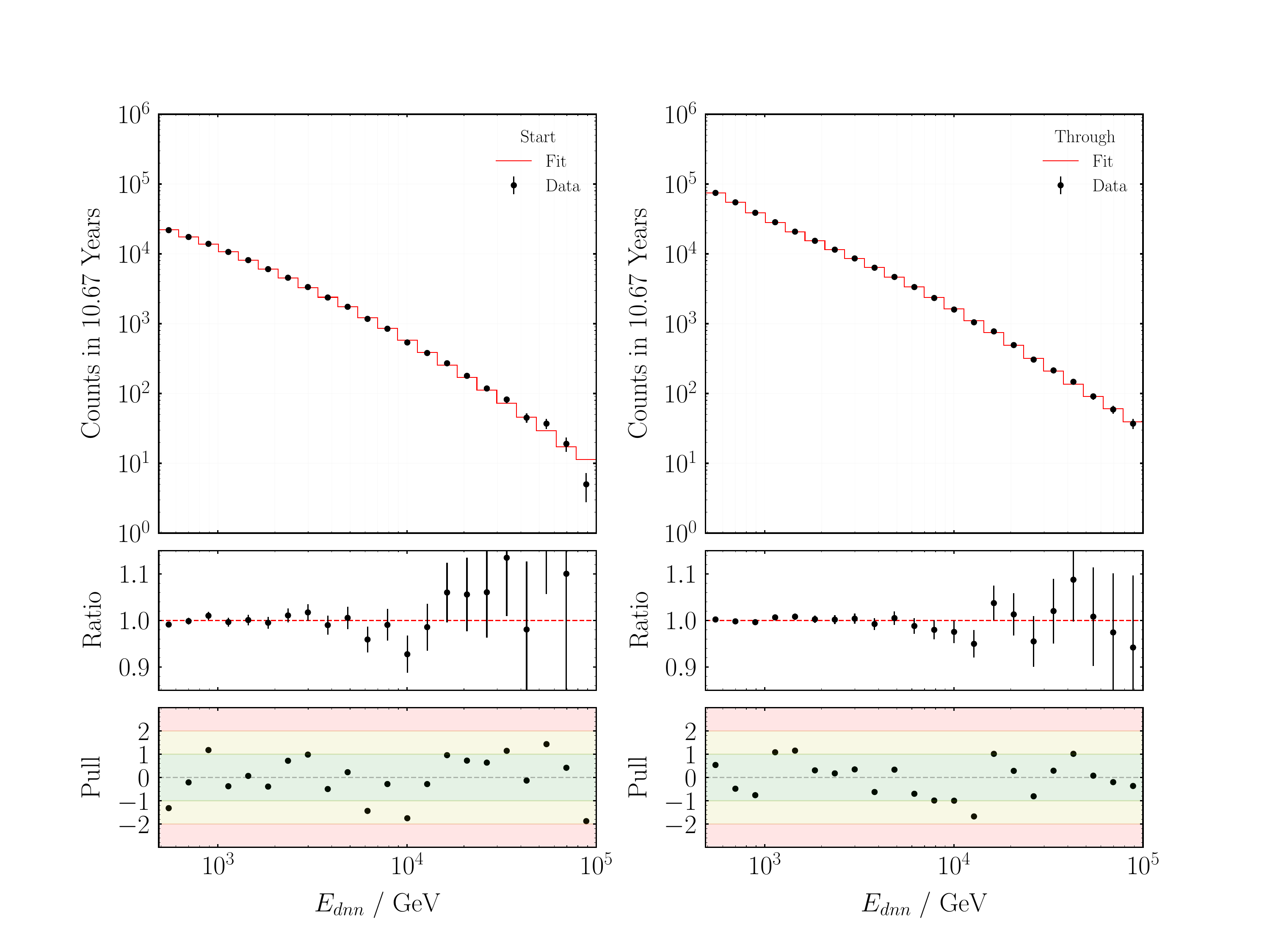}
    \caption{Top: One dimensional distributions of reconstructed $E_{reco}$ in GeV for starting tracks (left) and through-going tracks (right) of the MC expectation (red) and observed data (black). Middle: ratio of data to MC expectation. Bottom: Pulls in terms of $\sigma$ of data compared to MC expectation.}
    \label{fig:meows_decay_unblinding_1d_energy}
\end{figure}

\begin{figure}[htbp]
    \centering
    \includegraphics[width=0.98\linewidth]{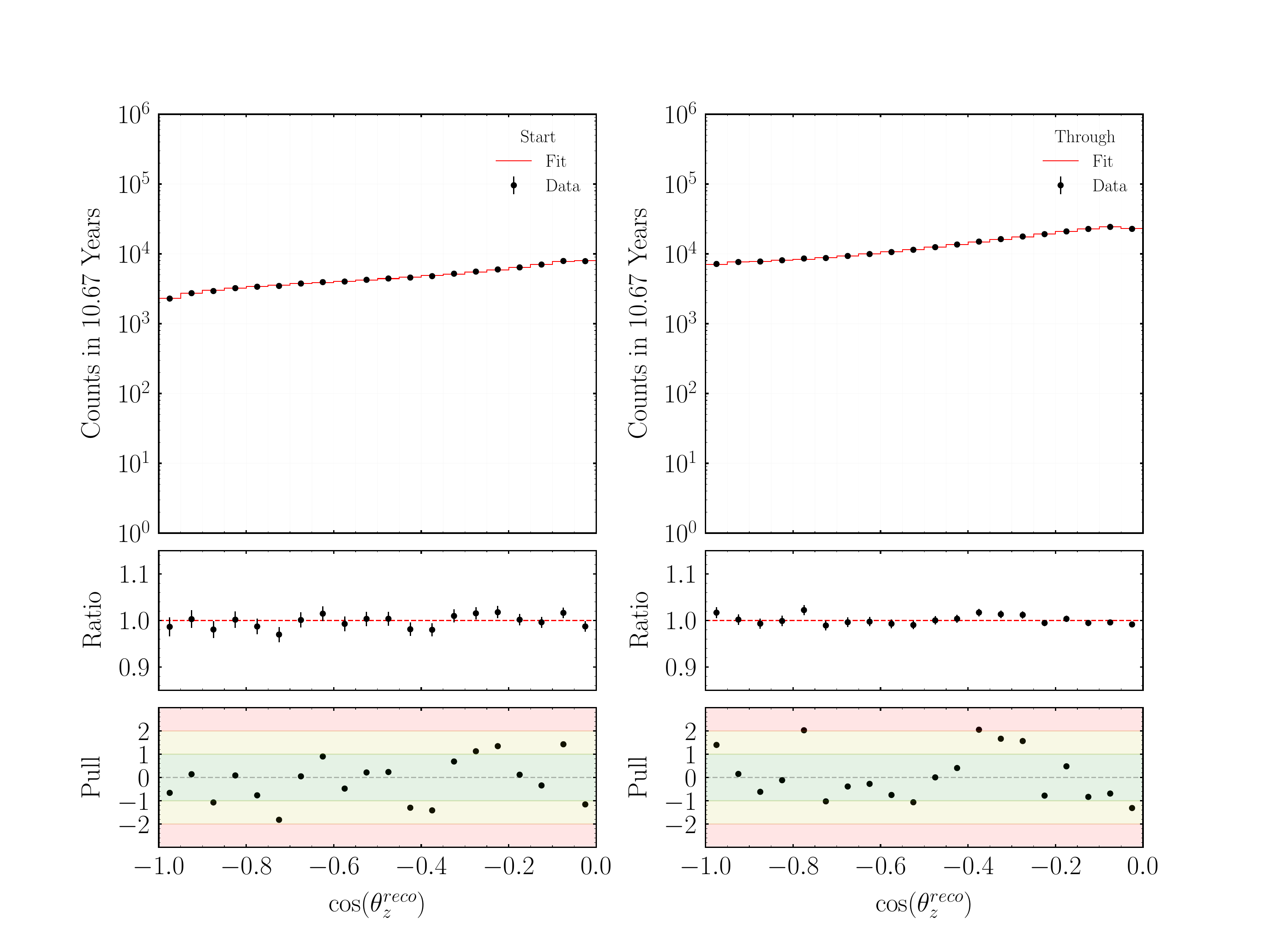}
    \caption{Top: One dimensional distributions of reconstructed $\cos(\theta_z)$ for starting tracks (left) and through-going tracks (right) of the MC expectation (red) and observed data (black). Middle: ratio of data to MC expectation. Bottom: Pulls in terms of $\sigma$ of data compared to MC expectation.}
    \label{fig:meows_decay_unblinding_1d_zenith}
\end{figure}

\begin{figure}[htbp]
    \centering
    \includegraphics[width=0.98\linewidth]{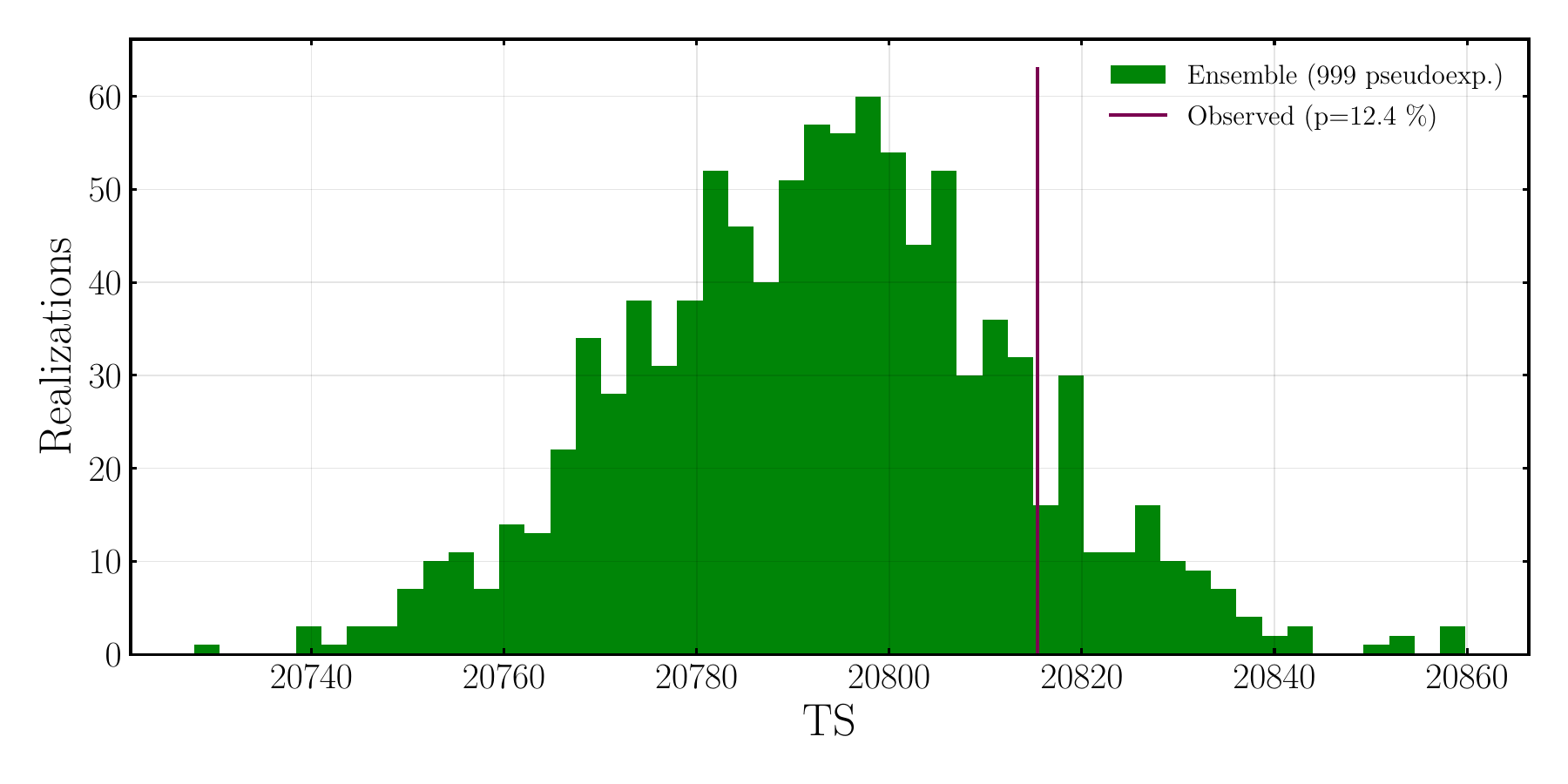}
    \caption{Likelihood values for 999 pseudoexperiments generated from the nominal expectation (green) overlaid with the likelihood of the observed data (purple line).}
    \label{fig:meows_decay_unblinding_gof}
\end{figure}

\begin{figure}[htbp]
    \centering
    \includegraphics[width=0.65\linewidth]{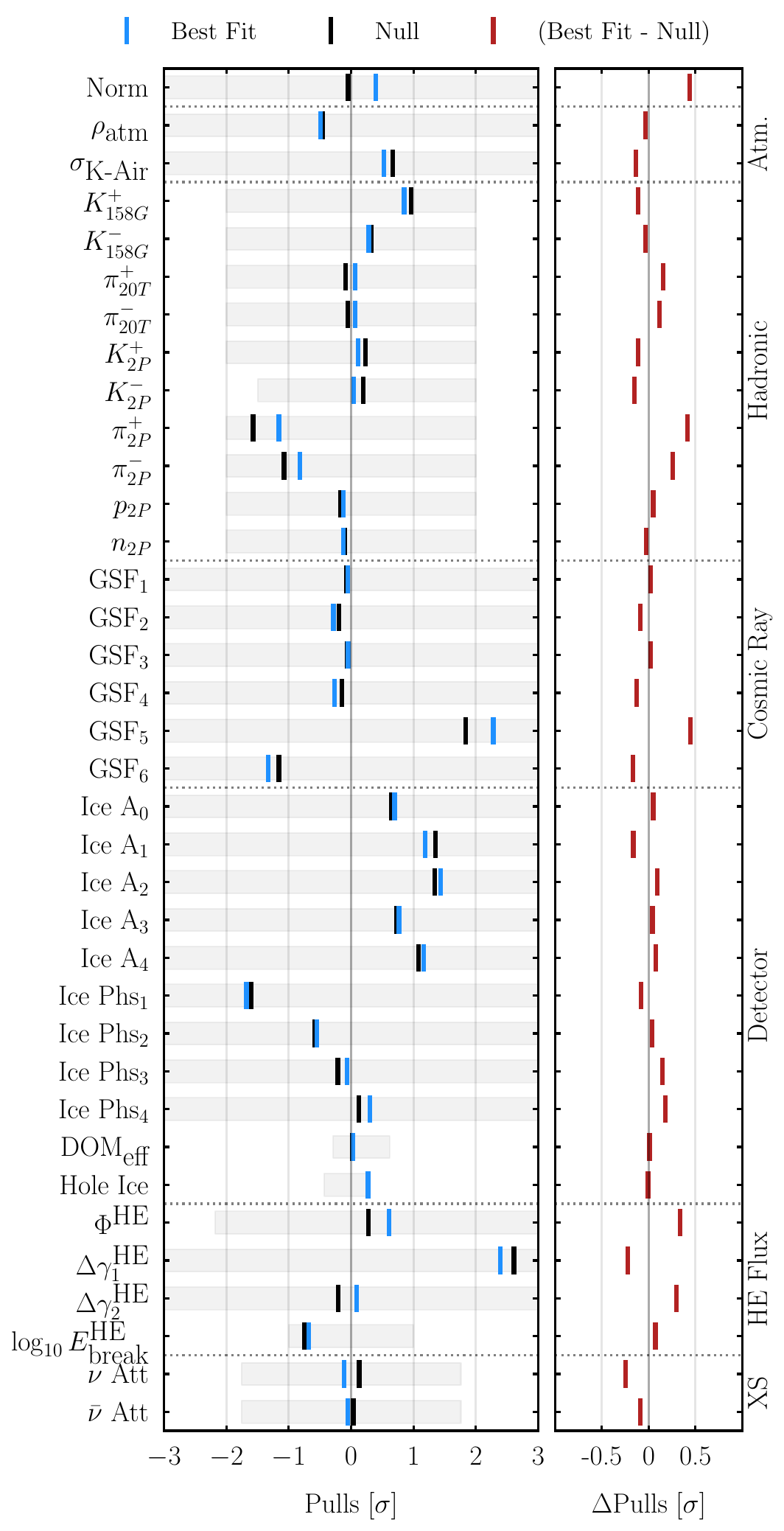}
    \caption{Systematic pulls of the best-fit (blue) and null hypothesis (black) points in the unstable sterile neutrino analysis. The differences between best-fit and null are shown in the right sub-plot. Each systematic parameter is described in detail in Ref. \cite{IceCubeCollaboration:2024dxk}. Figure from Ref. \cite{IceCube:2025wrv}.}
    \label{fig:meows_decay_pulls}
\end{figure}

First, the fit to data is performed over the entire parameter space in $\Delta m_{41}^2$, $\sin^2(2\theta_{24})$, and  $g^2$.
The best-fit point was identified and the systematic pulls were checked to see if any parameter pulled greater than $3\sigma$.
The pull distributions are shown in \cref{fig:meows_decay_pulls} with no systematics pulling beyond the stopping criterion.
Then, an ensemble of pseudoexperiments was generated at the best-fit point and the data TS is compared to the ensemble, as shown in \cref{fig:meows_decay_unblinding_gof}.
If the best-fit data TS p-value was less than 5\%, this could indicate a possible mismodeling of the MC.
After the ensemble p-value check, the bin-wise pulls were checked to see if six or more bins in each of the starting/through-going samples pulled greater than $3\sigma$.
The one-dimensional event distributions in energy and zenith with the data pulls are shown in \cref{fig:meows_decay_unblinding_1d_energy} and \cref{fig:meows_decay_unblinding_1d_zenith} respectively.

\begin{figure}[htbp]
    \centering
    \includegraphics[width=0.5\linewidth]{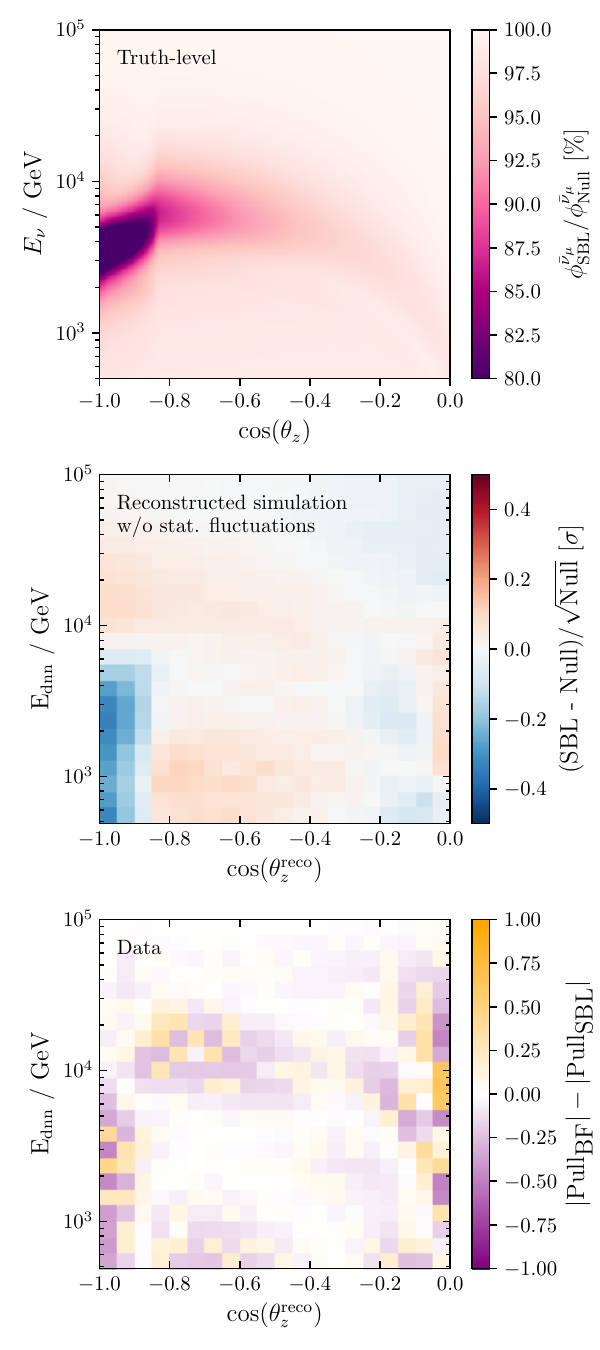}
    \caption{Top: $\bar{\nu}_\mu$ flux deficit under the best-fit model of the SBL global fit compared to the no-sterile hypothesis \cite{Hardin:2022muu}. Middle: Event count pulls comparing SBL best-fit to null in reconstructed space. Bottom: Difference in data pulls between the best-fit of this analysis and the best-fit of the SBL global fit. Purple indicates better agreement with the best-fit parameters of this analysis, orange indicates better agreement with the SBL best-fit parameters. Figure from Ref.~\cite{IceCube:2025wrv}.}
    \label{fig:meows_decay_osc_exp_pulls}
\end{figure}

The main result was unblinded after the conditions of the blind fit tests were met.
The best-fit point of the analysis was found to be $\Delta m_{41}^2 = 3.5~\mathrm{eV}^2$, $\sin^2(2\theta_{24})=0.16$, and $g^2 = 0$, the same point as the no-decay sterile neutrino analysis \cite{IceCube:2024dlz, IceCubeCollaboration:2024dxk}.
The significance of this point with respect to the null hypothesis (no sterile) is at $1.8\sigma$ assuming Wilks' theorem with three degrees of freedom, with a p-value of $7.4\%$.
The region of parameter space preferred by the short baseline experiments, determined by a global fit to these experiments \cite{Hardin:2022muu}, was almost entirely excluded at 95\% C.L.
Only a small region in the $g^2 = 4\pi$ region lies outside of the 90\% C.L. contour. 
In \cref{fig:meows_decay_osc_exp_pulls}, the expected signal of the SBL best-fit is shown with respect to the null/no-sterile neutrino model in true (top) and reconstructed (middle) variables.
The bottom subplot shows the comparison between the data pulls under the best-fit hypothesis of this analysis to the SBL best-fit hypothesis.

\begin{figure}[htbp]
    \centering
    \includegraphics[width=0.7\linewidth]{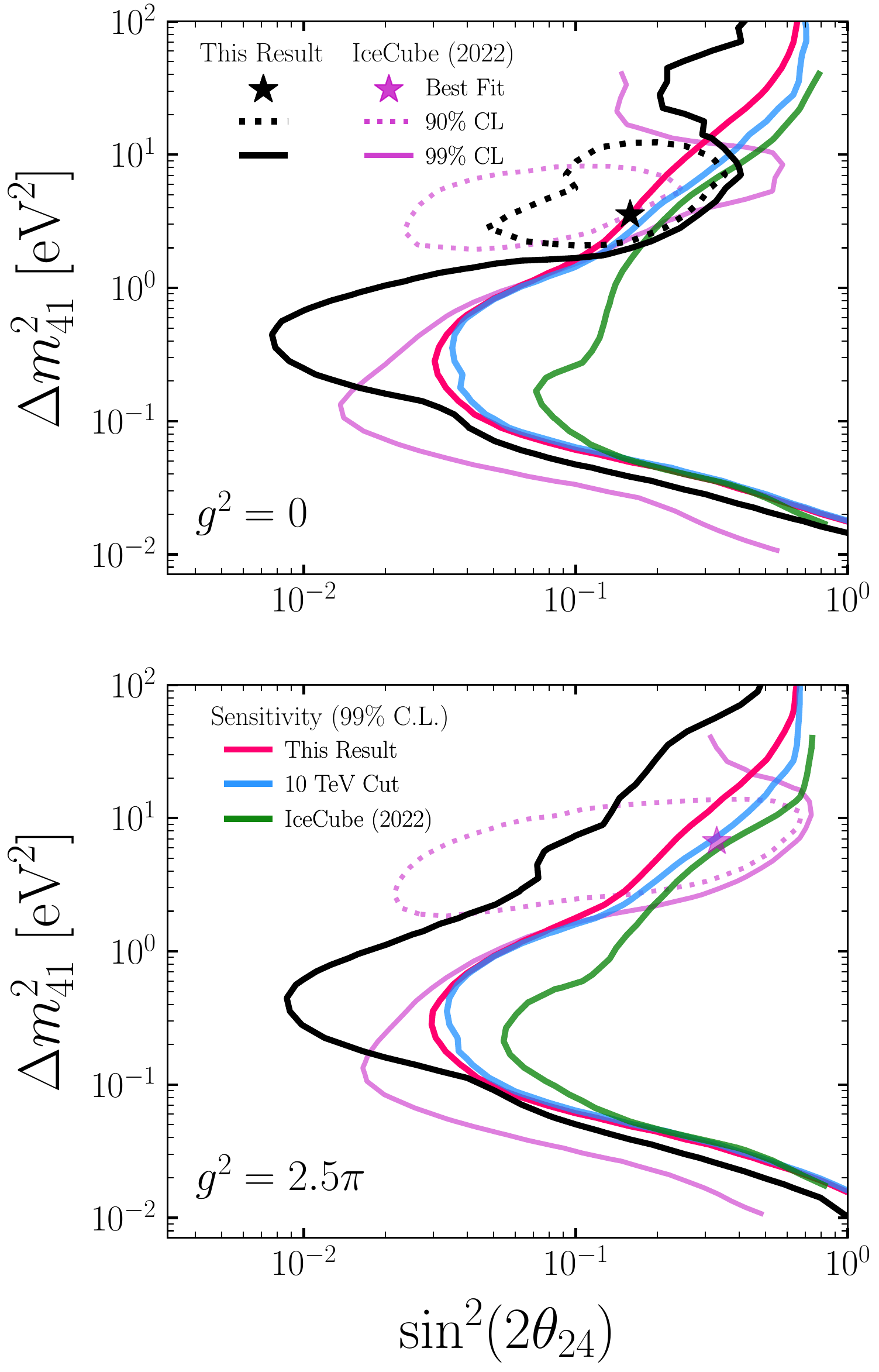}
    \caption{Comparison of results. The result of the improved unstable sterile neutrino analysis is shown in black, and the result of the 2022 analysis in pink. Overlaid are the 99\% C.L. Asimov sensitivities of this analysis (magenta), this analysis with a 10 TeV upper energy cut (blue), and the previous analysis \cite{IceCubeCollaboration:2022tso} (green). Figure from Ref.~\cite{IceCube:2025wrv}.}
    \label{fig:meows_decay_compat}
\end{figure}

Though this analysis excludes a large fraction of the allowed region for $g^2 > 0$ from the previous iteration of this analysis
\cite{IceCubeCollaboration:2022tso}, it should be noted that the likelihood landscape is shallow with a large overlap in the $g^2 = 0$ bin indicating no significant tension between the results.
The comparison between these results is shown in \cref{fig:meows_decay_compat}.
This result indicates that the unstable sterile neutrino hypothesis is unlikely to be an explanation for the short baseline anomalies.

\begin{figure}[htbp]
    \centering
    \includegraphics[width=0.95\linewidth]{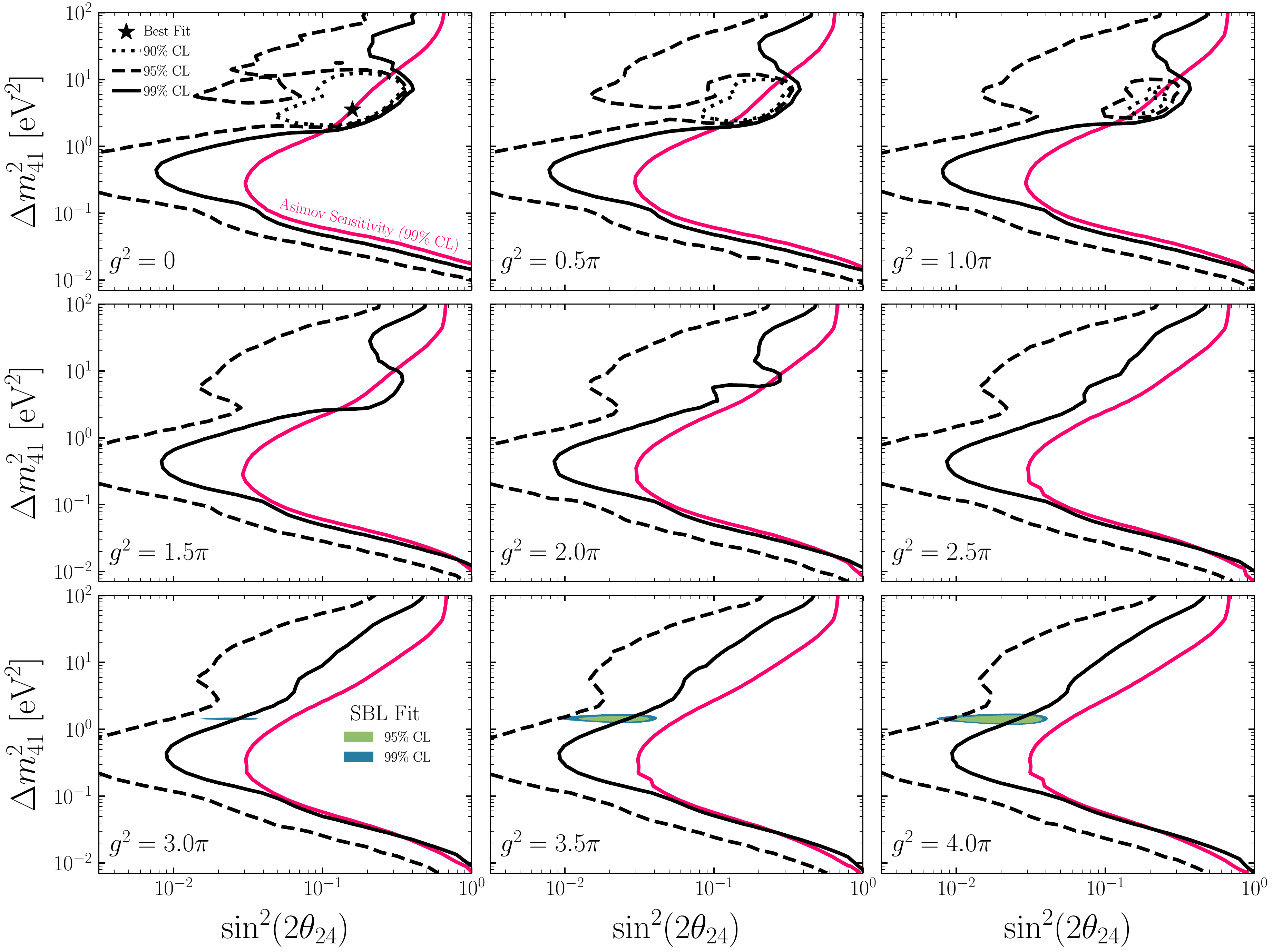}
    \caption{Results of the unstable sterile neutrino analysis divided into slices of the parameter space in $g^2$. The best-fit is indicated by a black star at $g^2=0$ with the 90\% (black dotted), 95\% (black dashed), and 99\% (black solid) C.L. contours, with the 99\% Asimov sensitivity (pink). The 95\% (green) and 99\% (blue) preferred regions from the global fits \cite{Hardin:2022muu} are shown for $g^2> 3\pi$. Figure from Ref.~\cite{IceCube:2025wrv}.}
    \label{fig:meows_decay_result}
\end{figure}

\section{Post-unblinding Checks} \label{sec:meows2023_postunblinding}
After the unblinding of the unstable sterile neutrino analysis, several post-unblinding studies were performed to probe this result, specifically the $g^2 = 0$ slice of the parameter space that also corresponds to the vanilla 3+1 sterile neutrino analysis.

\subsection{Split Tests}
To test whether the signal is being driven by mismodeled systematics, the fits to split datasets were performed and the results of the splits were compared to each other.
For example, to test if the signal was being driven by a mismodeling in the horizon region, the event selection was split into two subsamples: one for $-1<\cos(\theta_z^{reco}) < -0.2$ and the other for $-0.2 < \cos(\theta_z) < 0$.
The number of events in each subsample is approximately the same for this zenith split value meaning that the statistical uncertainty is the same.
For the split tests, this exact zenith cut is important because it does two things: it isolates a region of the reconstructed parameter space that probes only the vacuum oscillations, and it also analyzes the horizon region which other IceCube analyses have found to be problematic in terms of Data/MC agreement.
The result of this test showed no major disagreement in the two samples with no unexpected differences in systematic pulls.
Some systematics are expected to pull differently as each subsample has varying sensitivity to them.
This zenith split test was performed for varying zenith split values and is detailed in \cref{tab:zenith_splits}.
All fits, with the exception of the -0.05 value, fit to approximately the same region of $\Delta m_{41}^2$ and $\sin^2 (2\theta_{24})$.
The horizon sample of the -0.05 split value fits to a higher $\Delta m_{41}^2$ but the likelihood profile is very shallow over much of the parameter space since it has a much smaller fraction of the event selection and is not very sensitive.
These results indicate that there is no significant mismodeling near the horizon region that leads to a spurious sterile neutrino signal.

\begin{table}[htbp]
\begin{center}
\begin{tabular}{ |l|c|c|c| } \hline
$\cos (\theta_z^{reco}) $ split value & BF $\Delta m_{41}^2$ [eV$^2$] & BF $\sin^2 (2\theta_{24})$ & Null rejection ($\sigma$)\\ \hline
\multicolumn{4} {|c|} {Upgoing Sample}\\ \hline
-0.3 & 2.8 & 0.23 & 3.1 \\
-0.25 & 2.8 & 0.23 & 2.9 \\
-0.20 & 2.8 & 0.23 & 2.9 \\
-0.15 & 2.8 & 0.23 & 2.1 \\
-0.10 & 7.1 & 0.30 & 2.3 \\
-0.05 & 7.1 & 0.26 & 2.1 \\ \hline
\multicolumn{4} {|c|} {Horizon Sample}\\ \hline
-0.3 & 4.5 & 0.20 & 1.8 \\
-0.25 & 4.5 & 0.20 & 1.9 \\
-0.20 & 5.6 & 0.20 & 1.7 \\
-0.15 & 7.1 & 0.20 & 1.4 \\
-0.10 & 7.1 & 0.20 & 0.9 \\
-0.05 & 17.8 & 0.18 & 0.3 \\ \hline
\multicolumn{4} {|c|} {3+1 Full Fit}\\ \hline
  & 3.5 & 0.16 & 2.2 \\ \hline
\end{tabular}
\end{center}
\caption{Results of the split dataset fits. For the upgoing sample, the region is defined by $-1 < \cos(\theta_z^{reco}) < \mathrm{cut}$ and for the horizon sample it is $\mathrm{cut} < \cos(\theta_z^{reco}) < 0$. The null rejection $\sigma$ is computed for the 3+1 model with two degrees of freedom.}
\label{tab:zenith_splits}
\end{table}

\subsection{Randomness and Subsignal Tests}
Another method of verifying that the result is not being negatively affected by a mismodeled systematic is to analyze the bin-by-bin variations with respect to the best-fit expectation.
The behavior of the bin fluctuations was validated during the blind fit steps in \cref{sec:meows2023_results}.
The tests used to probe this use a simple suite of checks developed by the National Institute of Standards and Technology \cite{rukhin2001statistical} to verify that the bin-wise data pulls/fluctuations match that of a true random process, known as the ``NIST tests''.
These are run in two modes: \textit{frequency} mode or \textit{run} mode.
In \textit{frequency} mode, the check validates the frequency that 1's and 0's appear (which are defined by bins pulling positive or negative, respectively).
In \textit{run} mode, it checks that the lengths of sequences match the expectation of random fluctuations.
The columns of bins in $\cos(\theta_z^{reco})$ were checked with these tests, with only one column (the range $-0.75<\cos(\theta_z^{reco})<-0.7$) found to have $p=0.3\%$ in the frequency test for starting tracks.
For the runs test, one other column in starting tracks was found with $p=0.5\%$ in the $-0.55 < \cos(\theta_z^{reco})<-0.5$ range.
Though 2 in 80 columns have a small p-value, there is no indication that they are driving the result since they are in the lower statistics starting track sample and do not encompass a major fraction of the vacuum oscillation signal prediction near the horizon or the resonant disappearance prediction.

\begin{figure}[htbp]
    \centering
    \includegraphics[width=0.9\linewidth]{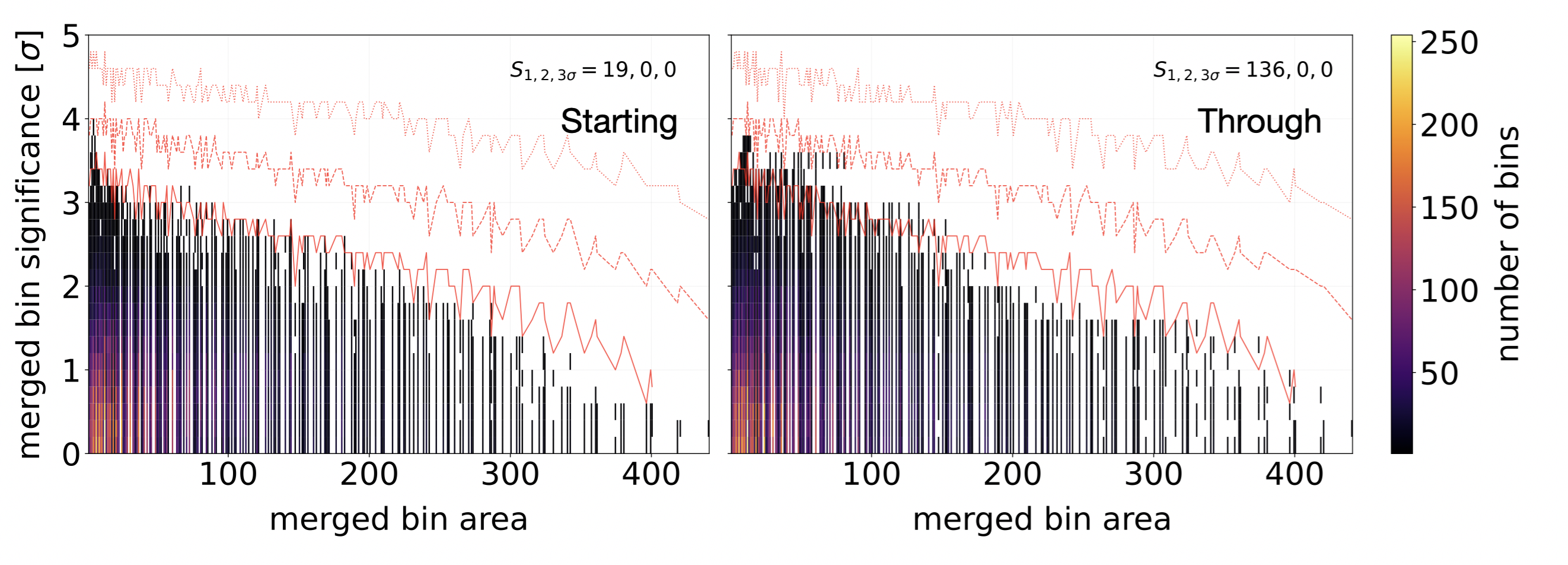}
    \caption{Results of the subsignal test separated by starting tracks and through-going tracks. Figure courtesy of A. Garcia.}
    \label{fig:xsigma_test}
\end{figure}

A second method of probing systematic issues with the event selection is the subsignal test.
This model-agnostic method approaches the problem by testing different scales in the reconstructed histogram that may encompass nearby bin correlations by testing clusters of bins.
Pseudoexperiments are generated from Poisson fluctuations of the nominal event expectation histogram and every combination of merged neighboring bins is compared to the nominal expectation.
This generates predictions for $1\sigma$, $2\sigma$, and $3\sigma$ statistical fluctuations given the area of the merged bins.
The same bin merging scheme is then applied to the real experimental data and compared to the predictions.
The test was performed and found no merged-bin combinations that exceeded the $2\sigma$ prediction, and all other combinations fell near the $1\sigma$ expectation or smaller indicating good agreement.
\cref{fig:xsigma_test} shows the merged bin significance as a function of merged bin areas.

\section{Analysis Conclusions}

The updated analysis searching for unstable sterile neutrinos with the MEOWS-2023 event selection did not find a preference for decay $(g^2 > 0)$, and it recovered the same best-fit point as the 3+1 analysis at $\Delta m_{41}^2 = 3.5$ eV$^2$ and $\sin^2 (2\theta_{24}) = 0.16$ \cite{IceCubeCollaboration:2024nle}.
Answering the three questions posed in the introduction chapter: 
\begin{enumerate}
    \item The unstable 3+1 model does not provide a better fit to IceCube data.
    \item This analysis excludes the best-fit point and much of the preferred regions of the MEOWS-2020 unstable sterile neutrino analysis, but the results are compatible due to the significant overlap in the allowed regions for $g^2 = 0 $ as demonstrated by \cref{fig:meows_decay_compat}.
    \item The global fit allowed regions indicated by the global fits to accelerator data are largely excluded at 95\% C.L.
\end{enumerate}
The results of this analysis indicate that unstable sterile neutrinos are not compatible with IceCube data in the parameter space preferred by short-baseline experiments.

\chapter{New Reconstruction Techniques}\label{chapter:ml_recos}
The reconstruction of neutrino properties from raw detector signals is a central challenge in neutrino telescope physics. 
In IceCube, the observable quantities available for any given event are the arrival times and integrated charges of Cherenkov photons recorded at each DOM.
The charge readings can be thought of as a sparse point cloud of data in a four-dimensional space $(\vec{x}, t)$.
Using this information, a reconstruction or classification machine-learning (ML) model can be trained to infer neutrino flavor, position of the interaction vertex, energy, direction, and other kinematic variables such as inelasticity.
Traditional reconstruction approaches address this problem through likelihood maximization, which is computationally expensive and relies on an assumed event hypothesis \cite{Aartsen:2013bfa, IceCube:2013dkx}.
Machine-learning methods offer an alternative by learning a direct mapping from raw detector data to reconstructed quantities, reducing rigid model dependence.

A fundamental challenge in applying machine learning to neutrino telescope data is sparsity. 
For a typical IceCube event, only a small fraction of the DOMs record any signal. 
Rather than processing the full detector volume, the most recently used architectures within IceCube represent only the active DOMs and their recorded pulse series as inputs.
This naturally motivates sequence- and graph-based architectures that operate on the set of observed pulses directly, without requiring a dense spatial representation of the detector.

Transformer architectures, originally developed for natural language processing, provide a mechanism for modeling long-range dependencies in sequential data through the attention operation. 
Applied to neutrino telescope reconstruction, a transformer can, in principle, operate on the entire sequence of pulses across active DOMs, capturing spatio-temporal correlations that are inaccessible to methods that require compression of the pulse series into summary statistics such as CNNs. 
Graph transformer architectures extend this further by incorporating topological information from the detector geometry into the attention computation, enabling the model to exploit the known spatial relationships between DOMs as an inductive bias \cite{Ju2023ACS}. 
State-space models (SSMs), and in particular the MAMBA architecture \cite{mamba, mamba2}, offer a complementary approach to sequence modeling with linear rather than quadratic computational complexity in sequence length $L$, making them attractive for the long pulse sequences characteristic of high-energy IceCube events.

The performance of these architectures is evaluated in two complementary ways.  
First, the transformer and MAMBA architectures developed for this thesis are applied to IceCube Monte Carlo simulation, and the transformer-based inelasticity reconstruction is validated against experimental data within the event selection used for the sterile neutrino analysis.
This check verifies that there is no significant mismodeling in the MC that would cause the model to misreconstruct events in real data.
Second, as part of the NuBench project \cite{Orsoe:2025yru} which is an open-source benchmarking framework developed to enable reproducible, cross-experiment comparisons of neutrino telescope reconstruction methods.
The Graph Inductive bias Transformer (GRIT) \cite{Ma2023GraphIB} is benchmarked against other architectures used by experiments using a set of standardized simulated datasets with different detector geometries.

This chapter describes each of these architectures and their application to IceCube reconstruction. 
\cref{sec:data_formatting} discusses the formatting of IceCube pulse data as input to the models. \cref{sec:transformers} introduces the transformer architecture. \cref{sec:grit} describes the GRIT graph transformer. \cref{sec:mamba} presents the MAMBA state-space model.
In \cref{sec:applied_ml}, the transformer and MAMBA architectures are applied to IceCube reconstruction tasks including energy and inelasticity reconstruction, and the transformer-based inelasticity reconstruction is validated against MEOWS-2023 data.
The GRIT architecture is evaluated separately in the NuBench benchmark study discussed in \cref{sec:nubench}, rather than in the IceCube-specific applications of this section.

\section{Data Representation}\label{sec:data_formatting}
In a neutrino telescope, charges $q$ are measured at times $t$ to form a pulse series $(q_i, t_i)$ for each DOM or similar detector module.
The formatting of these pulse series to match the expected input shape of different deep neural networks can be implemented in many ways.
Convolutional neural networks (CNNs) were one of the first neural network model architectures used in IceCube to perform event reconstructions \cite{Kronmueller:2019jzh, Abbasi:2021ryj}.
CNNs worked by performing successive convolutions between learned kernels and the input data, treating it essentially as an image.
To encode the input data in a way that can be processed by a CNN, summary statistics of the observed charge distributions were computed for each DOM. 
That is, for a given pulse series, certain statistics were computed to capture its properties.
A minimal example of summary statistics would be to compute the following properties:
\begin{itemize}
    \item $Q_{tot}$: Total charge on DOM
    \item $t_{\textrm{first}}$: Time of the first pulse
    \item $t_{\textrm{last}}$: Time of last pulse
\end{itemize}
More specific properties could be used to increase the model dependence on finer pulse series details:
\begin{itemize}
    \item $Q_{\textrm{first}}$: Charge of the first pulse
    \item $Q_{\textrm{10 ns}}$: Charge integrated over the first 10 ns
    \item $t_{25\%}$: Time at which 25\% of the charge was collected 
    \item $t_{50\%}$: Time at which 50\% of the charge was collected 
    \item $t_{75\%}$: Time at which 75\% of the charge was collected 
\end{itemize}
By increasing the specificity of the summary statistics, ML-based reconstructions become more precise as they can leverage fine-grained details about the event.
This is particularly important for reconstruction and classification algorithms that rely on the spatio-temporal development of an event, such as morphology.
An issue with summary statistics is that it smooths over fine-grained details of the event signature that could be leveraged in a reconstruction.
In the next sections, transformer and state-space models will be introduced that can overcome this limitation.

\section{Transformer Models}\label{sec:transformers}
A limitation of CNNs for neutrino telescope reconstruction is their reliance on local receptive fields.
Each convolutional layer can only combine information from spatially adjacent ``pixels'' in the input, so capturing correlations between widely separated regions of an event, such as the first and last pulse of a track, requires many successive layers where information is downsampled and reweighted \cite{cnn_receptive_fields, Vaswani:2017lxt}.
This makes it difficult to preserve positional information of individual pulses across the network depth.

An event in a neutrino telescope can be expressed as a point cloud.
Each observed pulse at position $\vec{x}$ and time $t$ can be expressed by a four-dimensional point in $(\vec{x}, t)$ and observed charge $q$.
Rather than enforcing some sort of geometry, such as a regular grid in a traditional CNN, transformers operate directly on the set of pulses and model correlations between all pairs of pulses simultaneously through the attention mechanism \cite{Bahdanau2014NeuralMT, Vaswani:2017lxt}.
The core operation of a transformer is scaled dot-product attention,
\begin{equation}\label{eq:attention}
    \mathrm{Attention}(Q, K, V) = \mathrm{softmax}\left(\frac{Q K^T}{\sqrt{d_k}}\right)V.
\end{equation}
$Q$, $K$, and $V$ are input matrices called queries, keys, and values, and $d_k$ is the dimensionality of the keys. 
In a variant known as self-attention, $Q$, $K$, and $V$ are each obtained by multiplying the input $\vec{x}$ by a separate learnable weight matrix (for example, $\vec{q}_i = W_{Q} \vec{x}_i$).
In most machine-learning models that utilize attention, each layer actually performs attention multiple times in parallel with independent sets of learned projection matrices, known as multi-head attention.
The softmax operation normalizes its input vector to produce a set of attention scores that sum to one \cite{Bahdanau2014NeuralMT},
\begin{equation}
    \mathrm{softmax}(\vec{y})_i = \frac{e^{y_i}}{\sum_j e^{y_j}}.
\end{equation}
A transformer network consists of many layers of successive attention operations followed by a feed-forward network that acts to aggregate the attention outputs.
In reconstruction tasks, the last layer of the network projects the final output of the network to the output dimensionality.
For example, two output dimensions for a direction reconstruction ($\theta$, $\phi$) or one for an energy reconstruction.

The computational cost of the attention operation scales as $L^2$, where $L$ is the length of the input sequence, because the attention weight matrix $Q K^T$ has to be computed for $L\times L$ elements.
For high-energy events in a neutrino telescope, an individual event may contain thousands of pulses, presenting a significant computational challenge due to the quadratic scaling.
Alternative architectures that alleviate this will be discussed in \cref{sec:grit} and \cref{sec:mamba}.

A further consideration specific to transformer models is that the self-attention mechanism in \cref{eq:attention} is invariant to permutations of the input sequence \cite{Qi2016PointNetDL}.
Reordering the input elements produces a reordered output, but does not change its values.
In natural language processing, where the ordering of words does matter, input embeddings are applied to explicitly encode sequence position information and break this symmetry \cite{Vaswani:2017lxt, Shaw2018SelfAttentionWR, Su2021RoFormerET}.
For neutrino telescope pulse data, the $(\vec{x}, t)$ coordinates can be added directly as input features, providing the positional information without additional positional encoding techniques.
Some techniques apply a space-time encoding to the input sequence to improve temporal correlations for reconstruction tasks, improving the performance \cite{Bukhari:2023ezc}.

\section{Graph Transformer Models}\label{sec:grit}
CNNs benefit strongly from regular grid data (such as images constructed from pixels), but capturing non-regular behavior requires significant architectural changes and remains only an approximation \cite{Huennefeld:2019rrf, Abbasi:2021ryj}.
For neutrino telescope data, detector geometry provides valuable domain knowledge as the spatial relationships between DOMs are useful for capturing local context in the event reconstruction.
It is more efficient to start by looking at the signals of nearby DOMs than those that are far away or those that do not observe anything.

Several architectures of graph transformer models have been developed to expand upon the traditional transformer architecture and integrate more topological information into the models \cite{zhang2020graph, dwivedi2021generalization, graphormer, Ju2023ACS}.
In the point cloud case discussed in \cref{sec:transformers}, the full pulse series information can be leveraged along with additional topological information, such as the distance between DOMs.
In a simple event graph, the observed pulses (or their summary statistics) are treated as nodes with edges defined by their spatial separation.
An example of edge information is the Euclidean distance between DOMs, i.e., $e_{i,j} =  \vert\vert x_j - x_i\vert\vert$.

In the GRIT architecture \cite{Ma2023GraphIB}, the edges are additional inputs and outputs of the attention operation. 
This means that both the nodes and edges are being transformed as they pass through each layer of the graph transformer model. 
The following formalism of this graph transformer network follows the derivations from Ref. \cite{Ma2023GraphIB}.
The first component of the GRIT architecture is the computation of context scores,
\begin{equation}
    c_{i,j} = \left( \mathbf{W}_{Q} x_i + \mathbf{W}_{K} x_j \right) \odot \mathbf{W}_{Ew} e_{i,j}.
\end{equation}
In this step, the queries and keys are different nodes and are combined with learnable weights $\mathbf{W}_{Q}$ and $\mathbf{W}_{K}$ before being combined with the weighted edges $\mathbf{W}_{Ew} e_{i,j}$ via element-wise multiplication. The edges are then updated according to,
\begin{equation}
    \hat{e}_{i,j} = \textrm{ReLU}\left(\rho(c_{i,j}) + \mathbf{W}_{Eb} e_{i,j} \right).
\end{equation}
The context scores are rescaled using the signed square root operation $\rho(x) = \textrm{sign}(x)\sqrt{\vert x \vert}$ for stability and combined with the edge biases. The edges do not undergo any further modification, and the next step is to update the nodes of the graph.
The attention scores are computed using the updated edges and an additional learnable weight matrix $\mathbf{W}_{A}$,
\begin{equation}
    \alpha_{i,j} = \textrm{softmax}(\mathbf{W} \hat{e}_{i,j})
\end{equation}
These attention scores are then used to compute the output nodes $\hat{x}_i$ by performing a weighted sum over the $j$ nodes,
\begin{equation}
    \hat{x}_{i} = \sum_{j} \alpha_{i,j} \left(\mathbf{W}_{V} x_j + \mathbf{W}_{Ev} \hat{e}_{i,j} \right).
\end{equation}
The output of the GRIT attention operation is an updated graph $\hat{G} = (\hat{x}, \hat{e})$ from the original graph $G = (x, e)$.
These operations can be repeated a number of times before being fed into an aggregation mechanism (i.e. pooling) and then mapped to the output dimension with an MLP.

\begin{figure}
    \centering
    \includegraphics[width=0.9\linewidth]{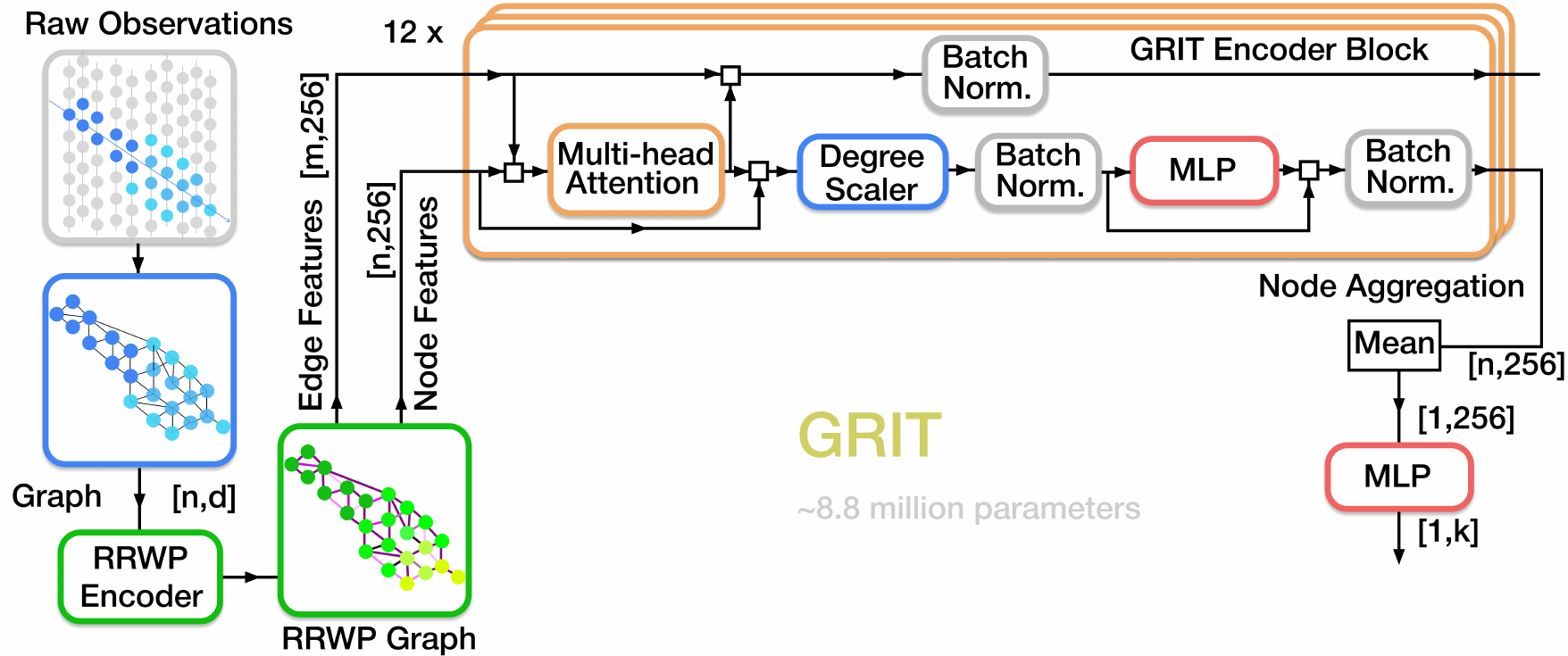}
    \caption{Architecture of the GRIT graph transformer model adapted to neutrino telescope reconstruction/classification tasks. Observed pulse data is encoded into a graph using the RRWP method. Graph node and edge information is updated at each stage of the GRIT encoder. Node information is aggregated and fed through a multilayer perceptron to provide an output. Figure from Ref. \cite{Orsoe:2025yru}.}
    \label{fig:grit_architecture}
\end{figure}

\begin{figure}[ht]
    \centering
    \begin{subfigure}[b]{0.24\textwidth}
        \centering
        \includegraphics[width=\textwidth]{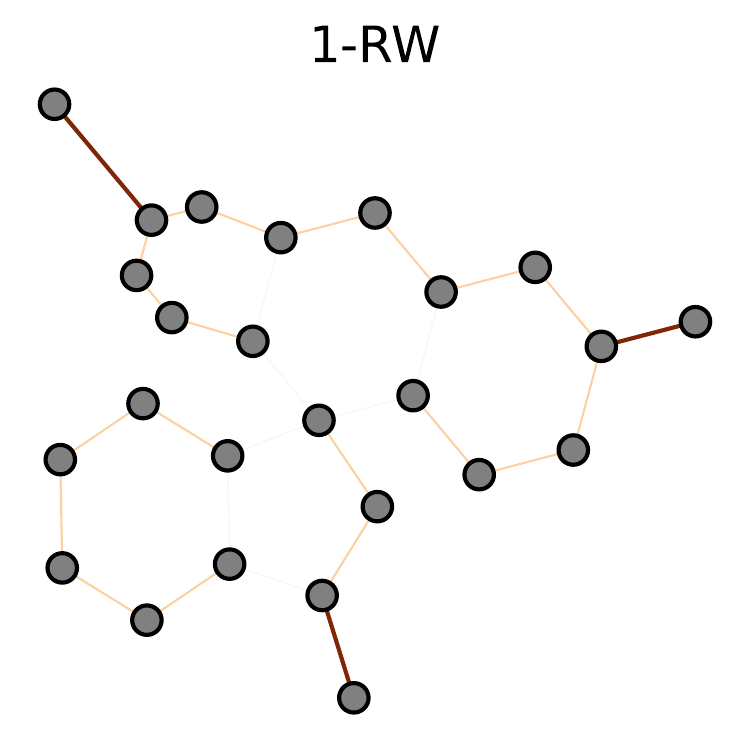}
        \caption{}
    \end{subfigure}
    \hfill
    \begin{subfigure}[b]{0.24\textwidth}
        \centering
        \includegraphics[width=\textwidth]{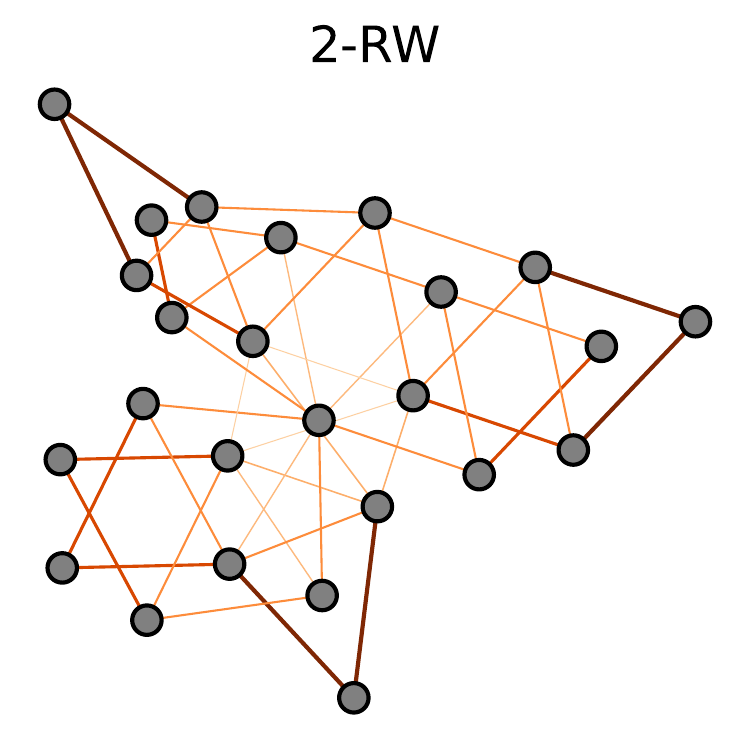}
        \caption{}
    \end{subfigure}
    \begin{subfigure}[b]{0.24\textwidth}
        \centering
        \includegraphics[width=\textwidth]{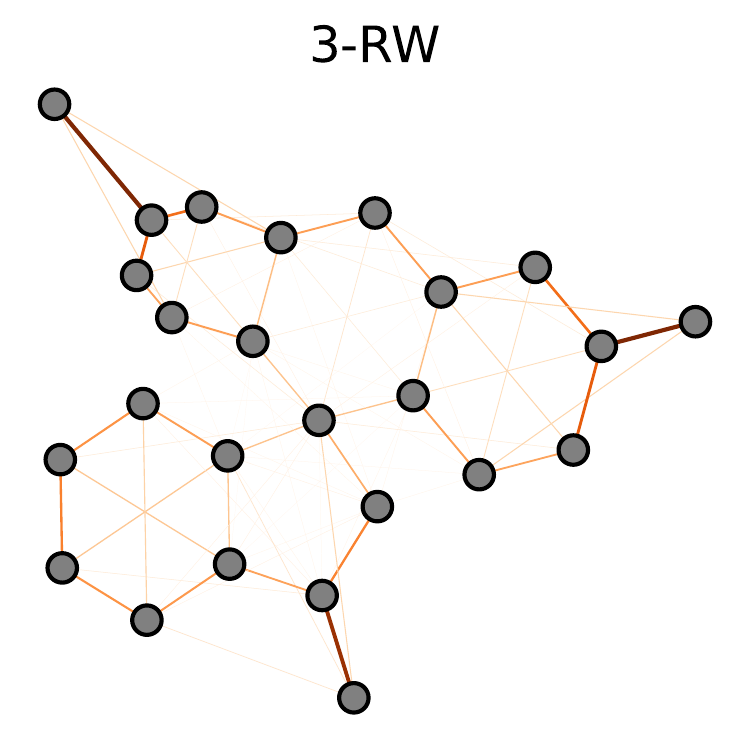}
        \caption{}
    \end{subfigure}
    \begin{subfigure}[b]{0.24\textwidth}
        \centering
        \includegraphics[width=\textwidth]{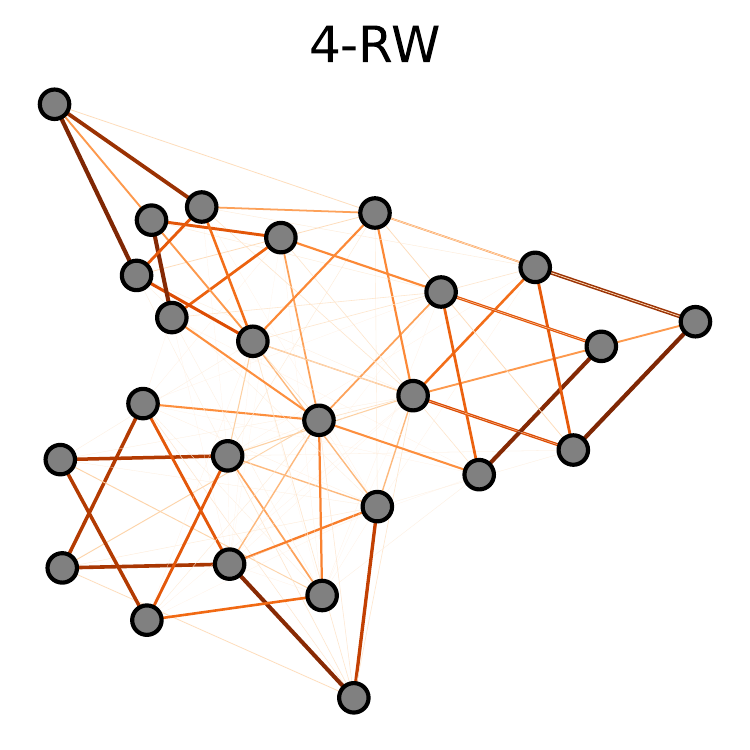}
        \caption{}
    \end{subfigure}
    \caption{Example of applying the walk probability matrix $M$ multiple times on a fluorescein molecule. Edge colors indicate the magnitude of $P_{i,j}$ between nodes, where darker indicates a larger value. Figure from Ref. \cite{Ma2023GraphIB}.}
    \label{fig:grit_rrwp}
\end{figure}

An important aspect of the GRIT graph transformer model is the inclusion of relative random walk probability (RRWP) encoding as an inductive bias \cite{Ma2023GraphIB, Ju2023ACS}.
A walk probability matrix is defined as $M = D^{-1} A$, where $A$ is the graph adjacency matrix and $D$ is a diagonal matrix with elements $D_{ii}$ corresponding to the degree of node $i$.
The element $M_{i,j}$ is treated as the probability of randomly walking from node $i$ to $j$.
Successive powers of the $M$ matrix form an increasingly dense matrix, as demonstrated by \cref{fig:grit_rrwp}.

An implementation of this model was developed for the GraphNeT machine-learning framework \cite{Sogaard:2022qgg} and is used in a benchmarking study performed on open data in \cref{sec:nubench}.

\section{MAMBA}\label{sec:mamba}
The MAMBA architecture \cite{mamba} takes a fundamentally different approach from the attention-based architectures mentioned previously.
It belongs to a class of models called ``state-space models''. 
A continuous-time state-space model is governed by the system of equations,
\begin{equation}\label{eq:state_space_one}
    \vec{h}^\prime (t) = \textrm{A} \vec{h}(t) + \textrm{B} \vec{x}(t)
\end{equation}
\begin{equation}\label{eq:state_space_two}
    \vec{y}(t) = \textrm{C} \vec{h}(t)
\end{equation}
In this model, the input $x(t)$ is used in both the equations for the internal/hidden state $h(t)$ and the output $y(t)$. Notably, \cref{eq:state_space_one} uses the current input and hidden state to compute the change of the hidden state with respect to time.
A contribution of $+\mathrm{D} \vec{x}(t)$ can be added to \cref{eq:state_space_two} to create a skip/residual connection.

To be used as a machine-learning model, where there is no continuous time but rather a discrete sequence of inputs, there is a discretized analog of the model above using recurrence \cite{mamba}.
\begin{equation}\label{eq:discrete_state_space}
    \vec{h}[n] = \bar{\mathrm{A}} \vec{h}[n-1] + \bar{\mathrm{B}} \vec{x}[n]
\end{equation}
\begin{equation}
    \vec{y}[n] = \mathrm{C} \vec{h}[n]
\end{equation}
The resulting $\bar{\mathrm{A}}$ and $\bar{\mathrm{B}}$ matrices are a result of discretizing the continuous-time version of the state-space model. This is derived by first starting with the analytic solution for $h(t)$,
\begin{equation}
    \vec{h}(t) = e^{\mathrm{A} t} \vec{h}(0) + \int_{0}^{t} \mathrm{B} e^{\mathrm{A}(t-t^\prime)} x(t^\prime) dt^\prime 
\end{equation}
Defining $\vec{h}[n] = \vec{h}(n\Delta)$ and $\vec{x}[n] = \vec{x}(n\Delta)$, where $\Delta$ is the time step and $n$ is the step number, and integrating, the expression for $\vec{h}[n]$ can be expressed in terms of $\vec{h}[n-1]$ and $\vec{x}[n]$,
\begin{equation}
    \vec{h}[n] = e^{\Delta \mathrm{A}} \vec{h}[n-1] + (\Delta \mathrm{A})^{-1} (\mathrm{exp}(\Delta \mathrm{A}) - \mathrm{I}) (\Delta \mathrm{B}) \vec{x}[n].
\end{equation}
\cref{eq:discrete_state_space} is obtained by defining the discretized matrices as,
\begin{equation}
    \bar{\mathrm{A}} = e^{\Delta \mathrm{A}}
\end{equation}
\begin{equation}
    \bar{\mathrm{B}} = (\Delta \mathrm{A})^{-1} (\mathrm{exp}(\Delta \mathrm{A} )- \mathrm{I}) \Delta \mathrm{B}.
\end{equation}
State-space models have a unique feature that allows them to be expressed in two ways: the recursive equation (as shown above) and a convolution equation,
\begin{equation}
    \vec{y} = \bar{\mathrm{K}}\otimes\vec{x}.
\end{equation}
The convolution kernel at step $n$ is defined as,
\begin{equation}
    \bar{\mathrm{K}} = (C\bar{B}, C\bar{A}\bar{B}, ..., C\bar{A}^n \bar{B}).
\end{equation}
So far, the only effect that the input has on the hidden state and output is through the addition of $\bar{\mathrm{B}} \vec{x}[n]$. 
The MAMBA architecture allows for dynamic matrices $\bar{\mathrm{B}}$, $\bar{\mathrm{C}}$, and step size $\Delta$ to enable input-dependent sequence interactions necessary for expressive deep neural networks.
This is referred to as the ``selection mechanism''.
In MAMBA, these quantities are all learnable, and $\Delta[n]$ dynamically controls how much the state changes at each step.
The important feature of the MAMBA architecture is how its computational complexity scales with the sequence length.
Using hardware-efficient algorithms for computing recurrence relations, MAMBA scales as $\mathcal{O}(L)$ in terms of the sequence length $L$ \cite{mamba, mamba2}.
This is a significant reduction in complexity compared to the vanilla transformer, making the MAMBA architecture a desirable option for modeling long-sequence data.
The application of MAMBA to IceCube reconstruction tasks is presented in \cref{sec:applied_ml}.

\begin{figure}
    \centering
    \includegraphics[width=0.98\linewidth]{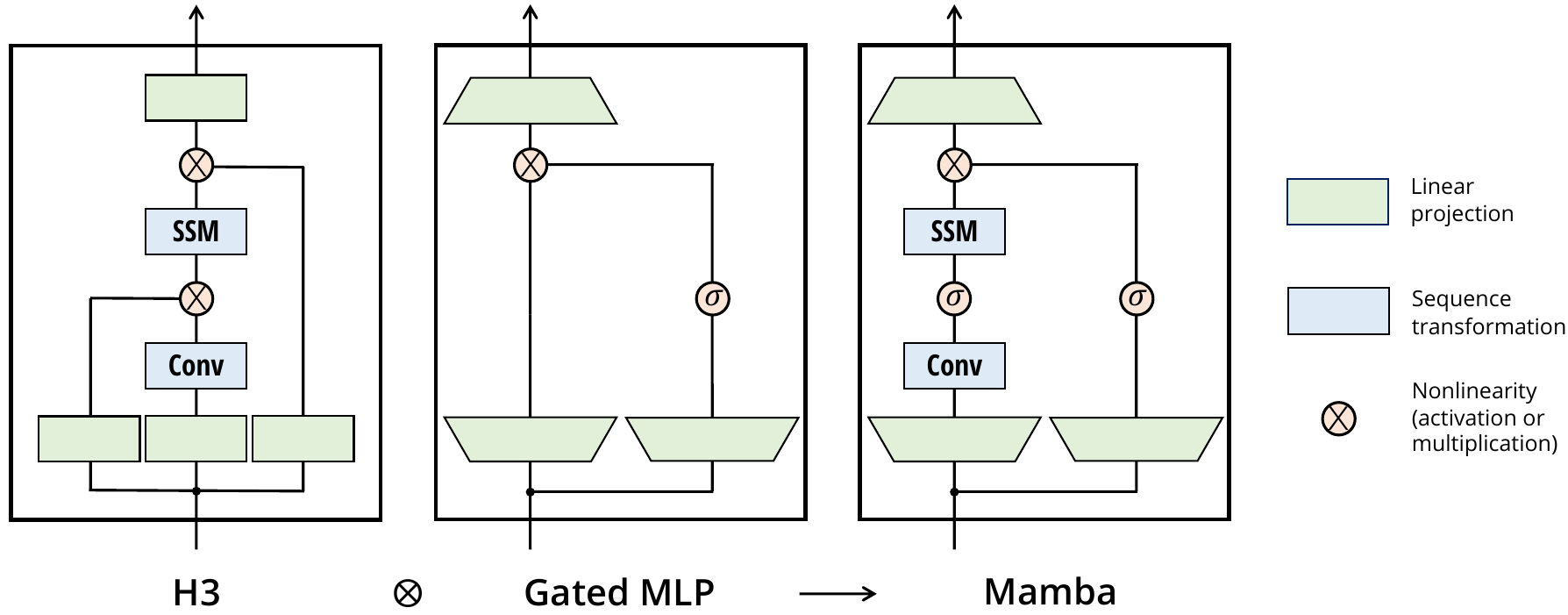}
    \caption{Overview of the MAMBA architecture. The MAMBA block combines the features of the H3 block from Ref. \cite{fu2023hungry} as the SSM backbone and a gated multilayer perceptron (MLP). Figure from Ref.  \cite{mamba}.}
    \label{fig:mamba_arch}
\end{figure}

\section{Applied Reconstructions in IceCube}\label{sec:applied_ml}

In this section, the transformer and MAMBA deep neural network architectures are applied to IceCube reconstruction tasks and their performance is compared to other models.

\begin{figure}
    \centering
    \includegraphics[width=0.9\linewidth]{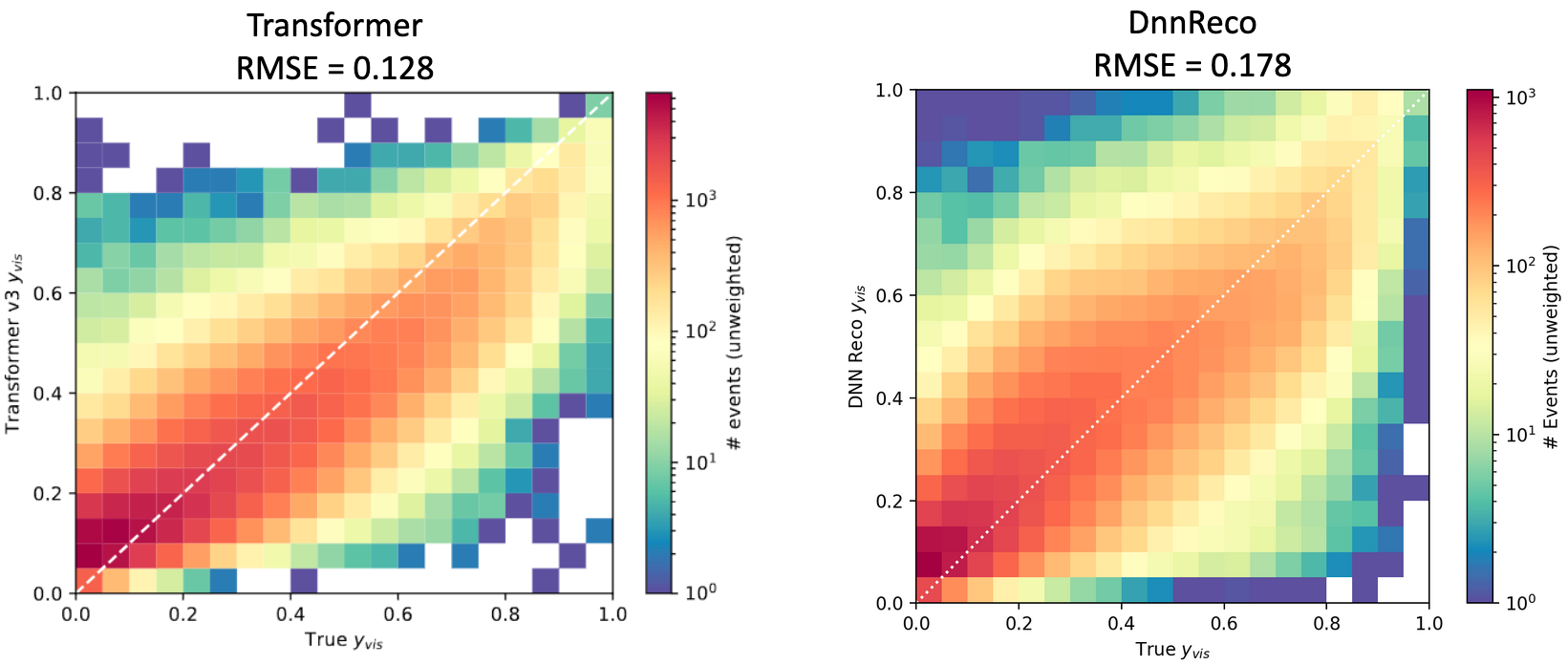}
    \caption{Comparison between a transformer model trained on the full pulse series (left) to a convolutional neural network model called DnnReco \cite{Abbasi:2021ryj} (right) trained on summary statistics for an inelasticity reconstruction task. The performance is measured in terms of the root mean square error shown above each subplot.}
    \label{fig:transformer_cnn_comparison}
\end{figure}

\begin{figure}
    \centering
    \includegraphics[width=0.7\linewidth]{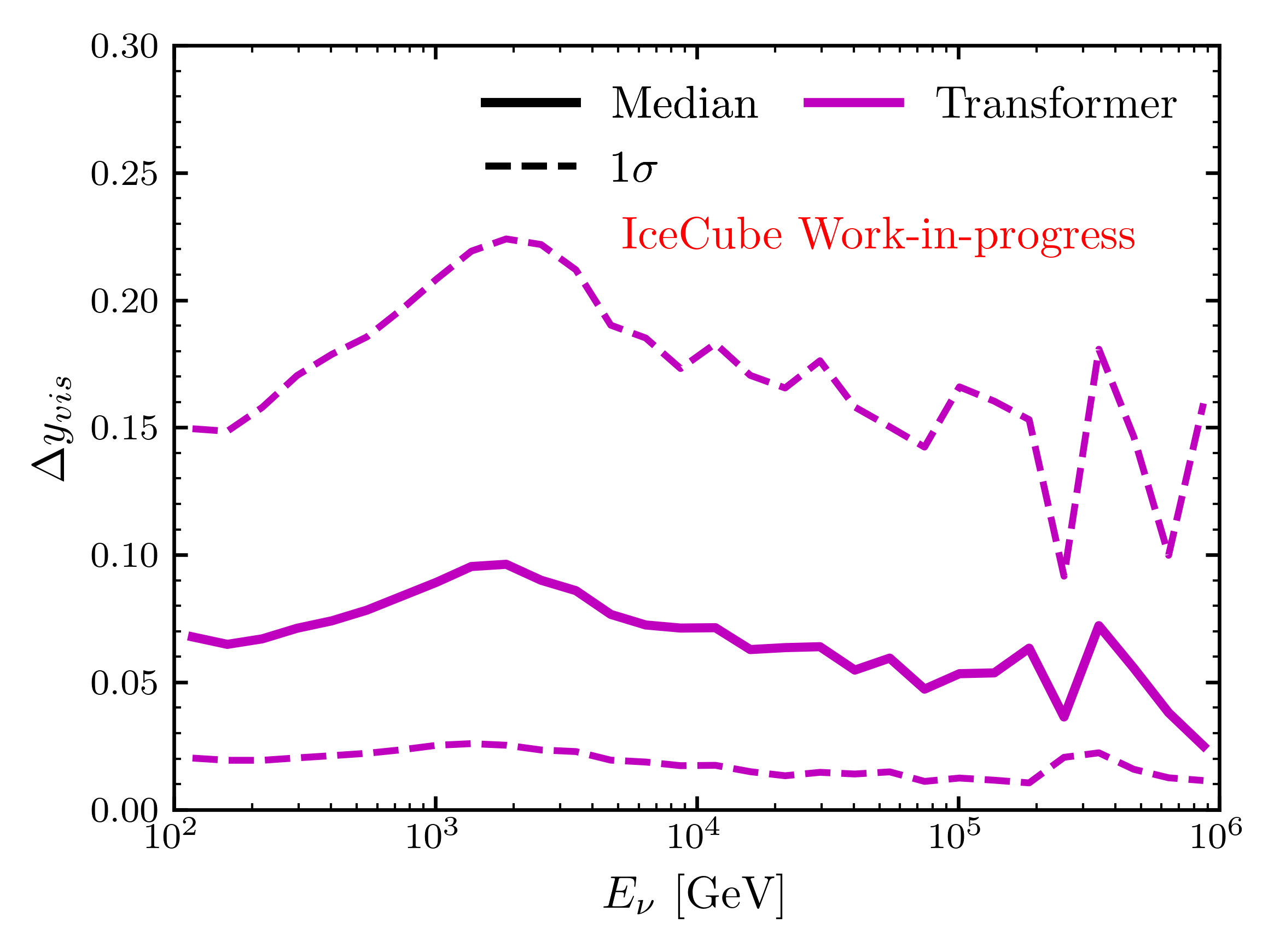}
    \caption{Absolute error of the transformer-based inelasticity reconstruction $\Delta y_{vis} = \vert y_{true} - y_{reco} \vert$ as a function of neutrino energy $E_\nu$. At high energies, $E_\nu > 10^5$ GeV, the MC statistics are low leading to noisy measurements.}
    \label{fig:transformer_cnn_inelasticity_energy}
\end{figure}

\cref{fig:transformer_cnn_comparison} compares the results of an inelasticity reconstruction task using a transformer model and a convolutional neural network.
The transformer network was based on the IceMix architecture from Ref. \cite{Bukhari:2023ezc} and trained on pulse series data of starting $\nu_\mu$ events, but the pulse series is truncated to 256 pulses when longer, which affects approximately 10\% of the training sample.
The convolutional network model was based on a model used for IceCube cascade reconstruction \cite{Abbasi:2021ryj} and used summary statistics of the pulse series as input.
The root mean square error (RMSE) is calculated over all reconstructed events as a performance metric.
The transformer RMSE was 0.128 while the CNN RMSE was 0.178, which is a substantial improvement in reconstruction quality.
The median and $\pm1\sigma$ bands of the reconstruction error as a function of neutrino energy using the transformer model are shown in 
\cref{fig:transformer_cnn_inelasticity_energy}.

The MAMBA architecture introduced in \cref{sec:mamba} was adapted to perform reconstruction and classification tasks with pulse series data.
Matching the typical information flow of a transformer network, the MAMBA blocks were arranged sequentially, with interwoven MLPs.
Compared to the transformer architecture, this MAMBA model could be trained on events up to a maximum pulse series length of 4096 with no measurable slowdown to training speed.
Using an Nvidia A100 GPU, the largest MAMBA models could perform event inference (including pre-processing) at $\sim$800 Hz compared to $\sim$50 Hz with the transformer. 
Additionally, the memory footprint of the MAMBA model was 20\% that of the transformer model.

\begin{figure}
    \centering
    \includegraphics[width=0.8\linewidth]{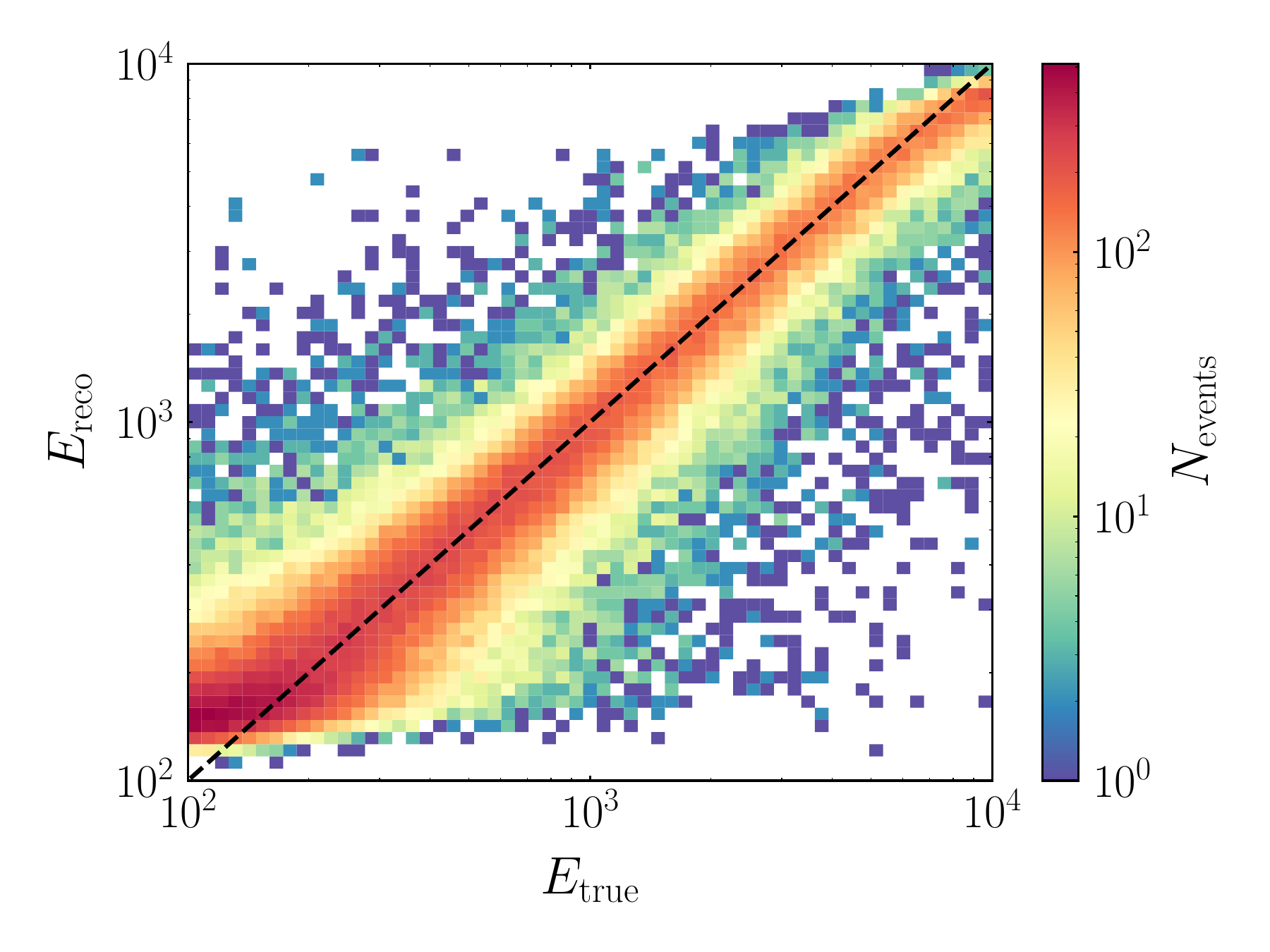}
    \caption{Energy reconstruction task performance using the MAMBA model trained on starting $\nu_\mu$ CC events. The black line indicates $E_{true} = E_{reco}$.}
    \label{fig:mamba_energy_reco}
\end{figure}

Several versions of the MAMBA model were trained for different tasks, in particular inelasticity and energy reconstructions.
The energy reconstruction performance is shown in \cref{fig:mamba_energy_reco}, which shows strong alignment with the $E_{true} = E_{reco}$ line, indicating a low-bias reconstruction.
The training dataset spanned $E_\nu \in [10^2, 10^4]$ GeV, so events near these boundaries are reconstructed with reduced accuracy because the model has less statistical coverage at the edges.

\begin{figure}[h]
    \centering
    \includegraphics[width=0.7\linewidth]{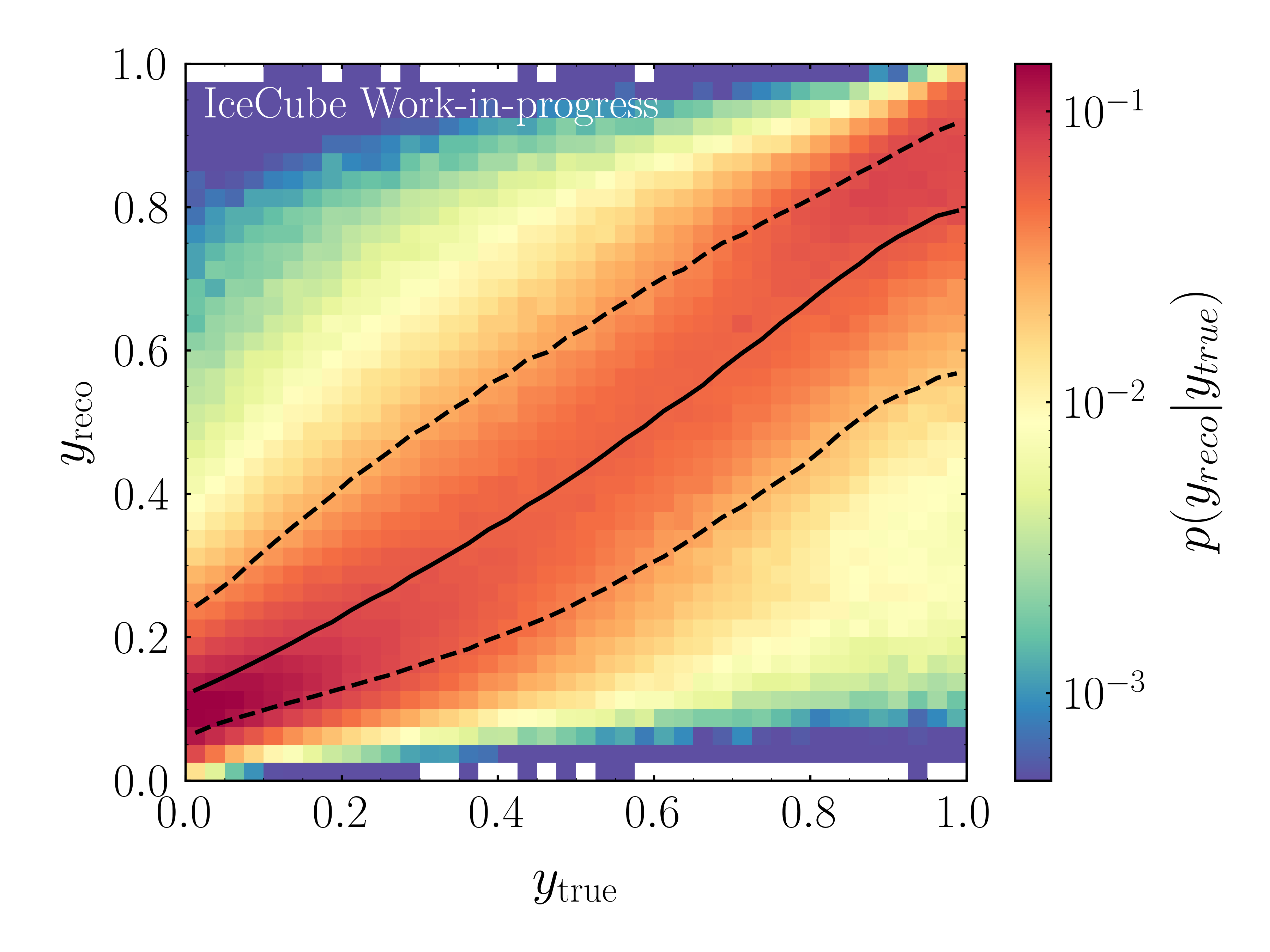}
    \caption{True vs. reconstructed inelasticity for the MAMBA-based inelasticity reconstruction. The color indicates the probability of reconstructing an event as $y_{reco}$ given $y_{true}$. The median is shown as the solid black line with the dashed lines spanning the $1\sigma$ band.}
    \label{fig:mamba_inelasticity}
\end{figure}

\begin{figure}[h]
    \centering
    \includegraphics[width=0.7\linewidth]{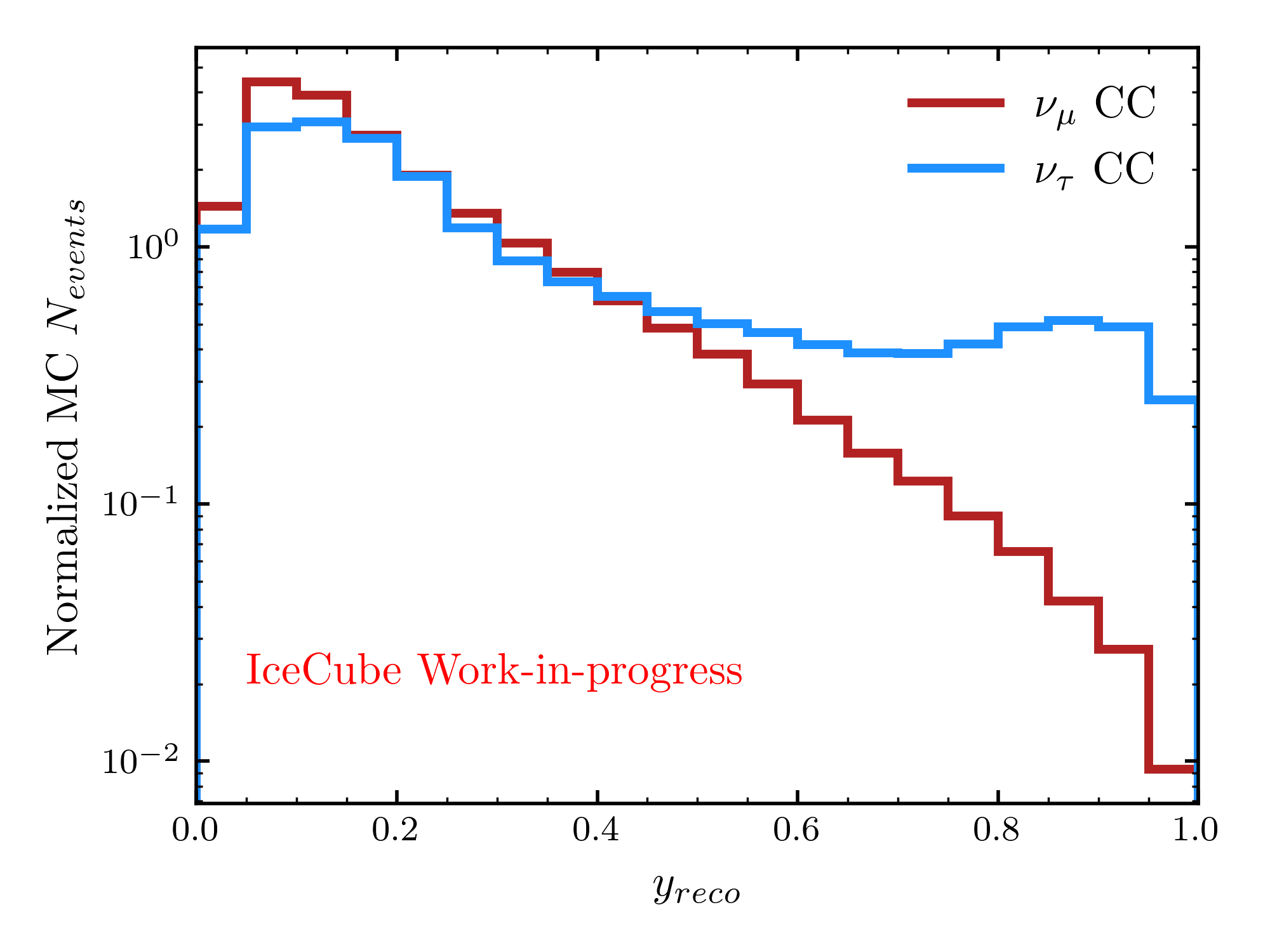}
    \caption{The MAMBA-based inelasticity reconstruction applied to a sample of unweighted $\nu_\mu$ (red) and $\nu_\tau$ (blue) MC events. Each MC set is normalized to unity.}
    \label{fig:mamba_inelasticity_tau}
\end{figure}

The inelasticity reconstruction was trained in the same manner as the transformer reconstruction mentioned above, and the reconstruction performance is shown in \cref{fig:mamba_inelasticity}.
The RMSE of this reconstruction was $0.13$, which is comparable to that of the transformer model, but the computational cost is substantially lower for MAMBA due to its $\mathcal{O}(L)$ scaling.
As a test, the model was applied to a sample of $\nu_\tau$ CC simulation in which the final-state $\tau$ decays to a muon through $\tau^{-} \rightarrow \mu^{-} \bar{\nu}_\mu \nu_\tau$ and energy is carried away invisibly by the neutrinos.
Consequently, the visible muon energy is lower than the final-state lepton energy in a $\nu_\mu$ CC event at the same neutrino energy, while the hadronic shower energy is unchanged.
The resulting visible inelasticity of the interaction is shifted to higher values.
The results of this test are shown in \cref{fig:mamba_inelasticity_tau}, demonstrating a clear bump in the inelasticity at high-$y_{reco}$.
To validate that this approach could be used for a $\nu_\tau$ search, further studies of systematic effects are required, which is beyond the scope of this study.

\begin{figure}
    \centering
    \includegraphics[width=0.8\linewidth]{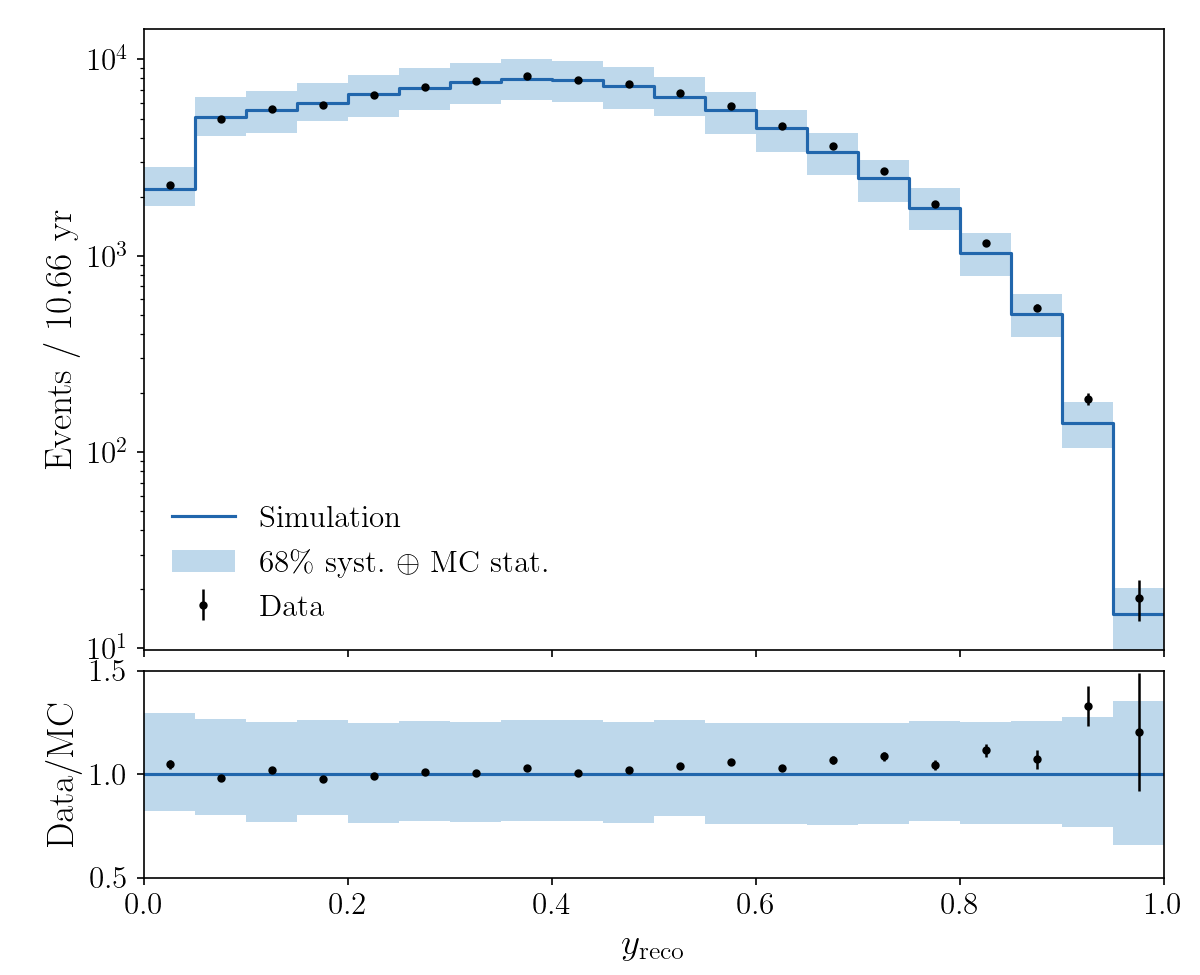}
    \caption{Comparison of the reconstructed inelasticity of data to MC using the MEOWS-2023 event selection using starting tracks only. The uncertainty bands on the MC combine the MC statistical uncertainty with the systematic uncertainty in quadrature with the exception of hole ice and DOM efficiency, which were not included.}
    \label{fig:meows2023_inelasticity}
\end{figure}

As a final validation, the transformer-based inelasticity reconstruction was applied to both the Monte Carlo simulation and experimental data of the MEOWS-2023 event selection used in the analysis presented in \cref{chapter:decay_analysis}.
The purpose of this study was to evaluate the data/MC agreement of the reconstruction.
Data/MC disagreement would indicate possible mismodeling of the physics processes in the simulation that is being learned by the model and could bias the reconstruction.
The MC was reweighted according to the nominal systematics, and no fit was performed.
The results of this test are shown in \cref{fig:meows2023_inelasticity}, showing good agreement between data and MC.
The total uncertainty was obtained by combining statistical and systematic uncertainties, except the hole ice and DOM efficiency effects which required separate systematic MC sets.

\section{Benchmarking Reconstructions}\label{sec:nubench}

A critical component of evaluating the performance of neural networks on certain reconstruction tasks is a consistent benchmarking dataset.
The NuBench project was designed to address the lack of a consistent, open-source dataset for neutrino telescope reconstruction studies.
Developing such a dataset presents several significant technical challenges.
First, it requires a software suite to simulate the propagation of charged particles in water or ice and then subsequent Cherenkov photons.
The readout of photons by a sensor (e.g. a DOM) requires a simulation of device hardware that is experiment-specific, and necessary calibration information is not typically public.
In this work, the sensors were modeled as 30 cm spherical PMTs with a quantum efficiency of 20\%, as the framework could not accommodate multi-PMT sensors.
The signals observed by the sensors then need to be digitized and noise must be added, which varies greatly experiment-to-experiment.

\begin{figure}
    \centering
    \includegraphics[width=0.95\linewidth]{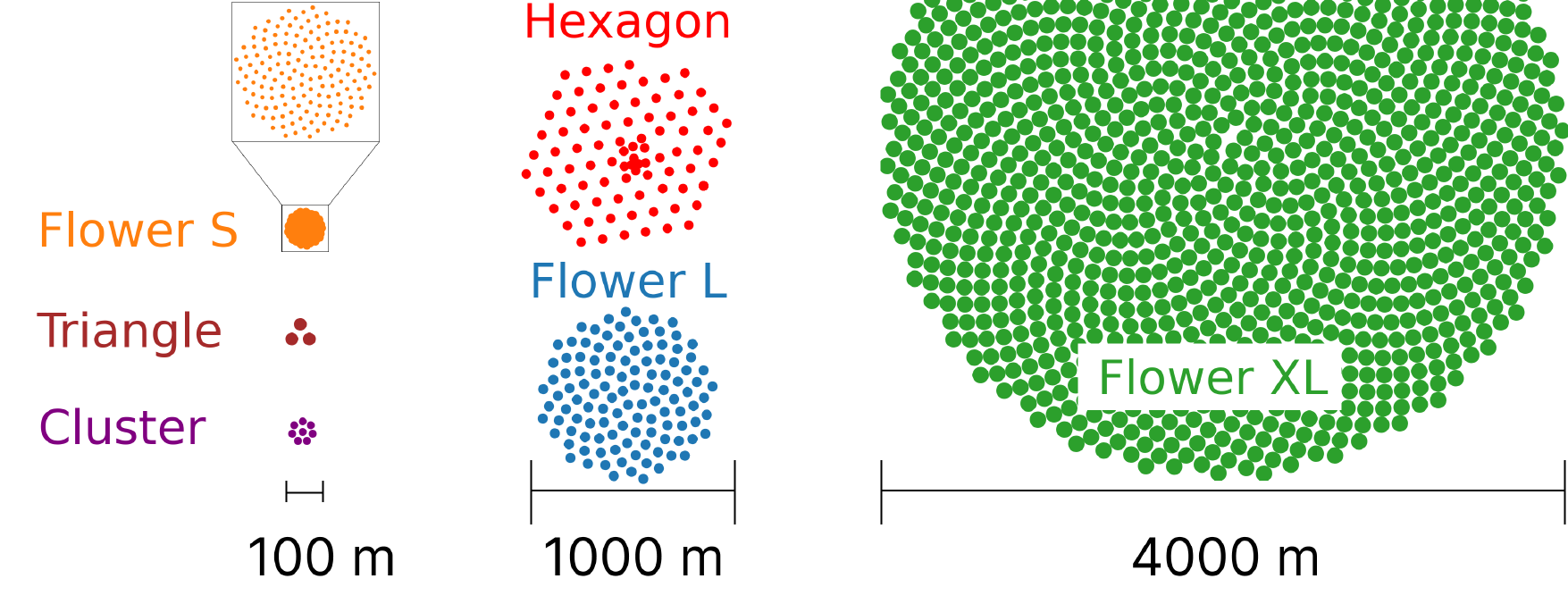}
    \caption{The different geometries that were used in simulation production for NuBench. All geometries were used for water-based simulation sets, and an additional simulation set was produced using low-energy neutrinos in ice for the ``Hexagon'' geometry. Figure from Ref. \cite{Orsoe:2025yru}.}
    \label{fig:nubench_geometries}
\end{figure}

For NuBench, events were generated using \texttt{PROMETHEUS} \cite{Lazar:2023rol} which simulates neutrino events and the light propagation for any detector geometry for either water or ice.
Event properties are first sampled from realistic kinematic distributions using \texttt{LeptonInjector} \cite{IceCube:2020tcq}.
\texttt{PROPOSAL} \cite{koehne2013proposal} is used to propagate the leptons through the medium and track the energy losses.
\texttt{PROMETHEUS} uses the custom \texttt{fennel} and \texttt{hyperion} packages for photon propagation code in water, and \texttt{PPC} \cite{chirkin_ppc_2023} for ice.
Once the photons reach the optical modules, the PMT quantum efficiency is applied, effectively downsampling the detected photons.
The remaining photons are then binned according to a timing window representing the sampling speed, and the assigned charge per pulse is drawn from a Gaussian distribution.
A sample of track and cascade events was generated for several geometries representative of real neutrino telescope designs.

\begin{figure}
    \centering
    \includegraphics[width=0.9\linewidth]{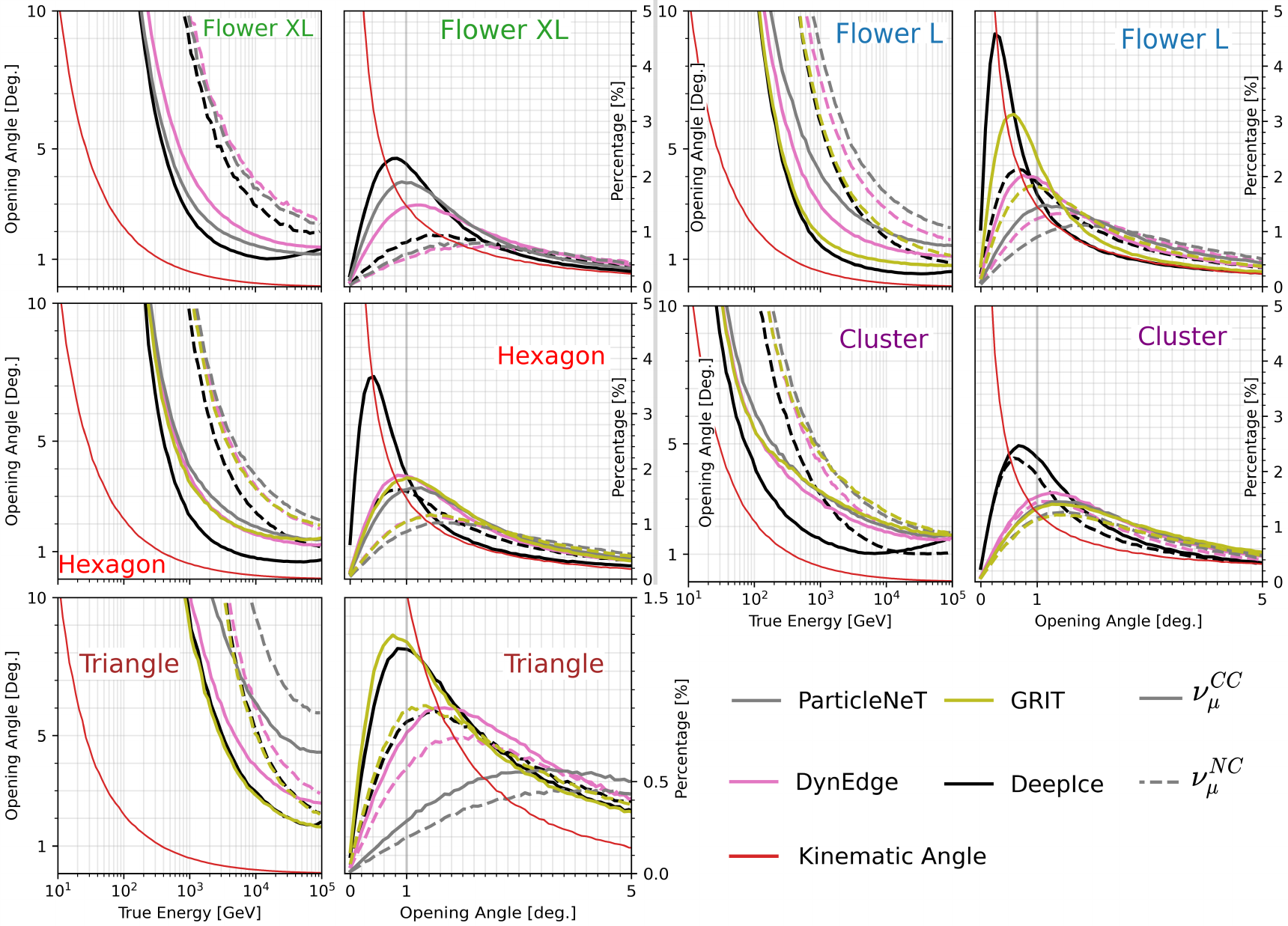}
    \caption{Result of the direction benchmark study using ParticleNet (gray), GRIT (gold), DynEdge (pink), and DeepIce (black). The median opening angle as a function of true energy and opening angle distributions are shown for each model. The kinematic angle (red) is the true angular separation between the outgoing lepton and the incoming neutrino direction. Charged-current $\nu_\mu$ events (tracks) are shown as solid lines and neutral-current $\nu_\mu$ events (cascades) are shown as dashed lines. Figure from Ref. \cite{Orsoe:2025yru}.}
    \label{fig:nubench_direction}
\end{figure}

Several neural network architectures were employed in the NuBench studies, primarily ones that are actively used in IceCube and KM3NeT analyses.
ParticleNet is a graph convolutional neural network originally developed for jet identification and subsequently adapted to neutrino telescope data formats \cite{Qu:2019gqs}.
DynEdge is another graph convolutional neural network developed by the IceCube collaboration, originally for the reconstruction of low-energy events in the DeepCore subarray \cite{Abbasi:2022ypr}.
The GRIT architecture used here is the same as discussed in \cref{sec:grit}.
Finally, the DeepIce transformer architecture was developed by competitors in an open data challenge where monetary prizes were awarded for the best-performing direction reconstructions on simulated data \cite{Bukhari:2023ezc}.

No single architecture performed best across all metrics; different architectures excelled at different combinations of tasks and geometries.
A selection of results from the direction reconstruction tasks is shown in \cref{fig:nubench_direction}.
Several trends were observed in the results of the different reconstruction tasks.
First, for direction reconstruction tasks, transformer models performed better than the convolution-based models.
This is consistent with the expectation that capturing large-scale correlations, such as the direction of a long track, requires global context across the full event.
In the vertex reconstruction tasks, the models that leverage local convolutions outperformed the transformer models.
Denser detector geometries perform better for energy, vertex, and inelasticity reconstruction, while larger sparse geometries perform better for direction reconstruction \cite{Orsoe:2025yru}.
\chapter{Towards the Next Sterile Neutrino Analysis}\label{chapter:meows2026}

The MEOWS-2023 analysis presented in \cref{chapter:decay_analysis} relied on simulation, event selection, and fitting methodology largely inherited from prior IceCube sterile neutrino searches. 
This chapter presents improvements to each of these components, developed for the next generation of IceCube sterile neutrino searches.
The most important addition to the event selection is the inclusion of reconstructed inelasticity, which was shown to be feasible in \cref{sec:applied_ml}.
As discussed in \cref{sec:inelasticity}, the inelasticity distribution could provide a way to statistically separate $\nu_\mu$ from $\bar{\nu}_\mu$ and improve sensitivity to the sterile-induced resonant disappearance of $\bar{\nu}_\mu$.
The cross-section calculation and inelasticity distributions presented in \cref{chapter:cross_sections} provide an improved prediction for the inelasticity distributions and uncertainties for this type of sterile neutrino search.

Three categories of improvements are presented. 
Updates to the Monte Carlo simulation include a more complete treatment of tau polarization and decay, the double-differential neutrino cross sections of \cref{chapter:cross_sections}, and revised detector modeling (\cref{sec:meows2026_simulation}). 
A new event selection leveraging the transformer-based reconstructions of \cref{chapter:ml_recos} achieves a factor of 2.3 increased signal efficiency with negligible atmospheric muon contamination (\cref{sec:meows2026_event_selection}). 
Analysis framework improvements address computational limitations of MEOWS-2023 analyses through GPU-accelerated likelihood evaluation, adjoint-mode gradient computation, systematic spline refits, and minimizer warm-starting (\cref{sec:meows2026_fitting}).

\section{Monte Carlo Simulation}\label{sec:meows2026_simulation}

\begin{figure}
    \centering
    \includegraphics[width=0.98\linewidth]{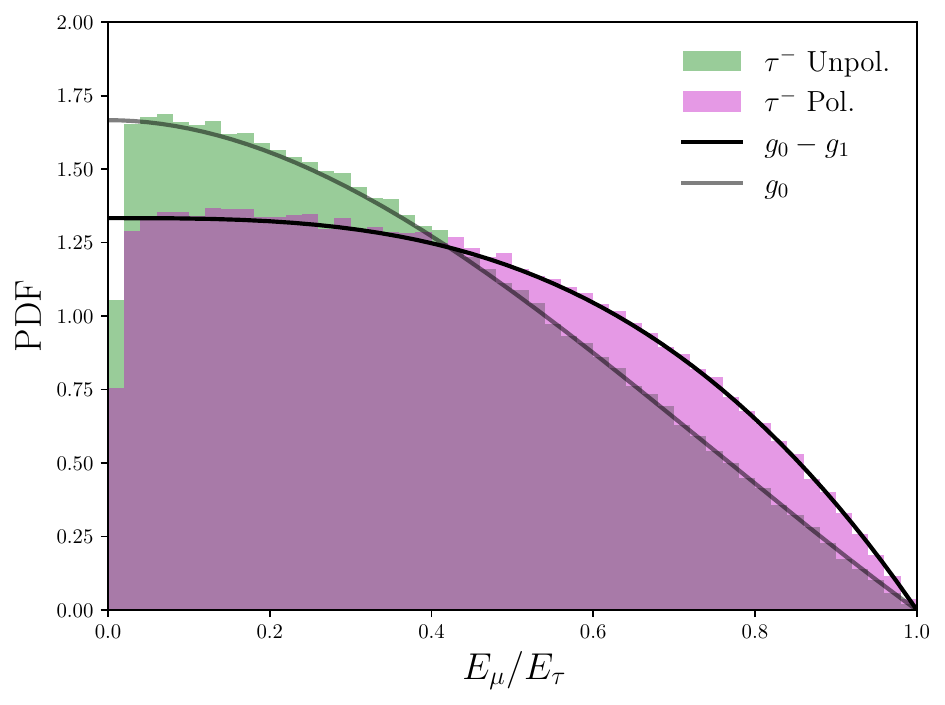}
    \caption{Probability density functions of $z = E_\mu / E_\tau$ for $\tau$ decays. The histograms show the results of a Monte Carlo study for polarized (magenta) and unpolarized (green) taus using TAUOLA. This is compared to the analytic approximations for polarized (black) and unpolarized (gray) decays.}
    \label{fig:tau_polarization}
\end{figure}

One of the improvements from the IceCube 3+1 sterile neutrino analysis with non-zero $\theta_{34}$ \cite{IceCube:2024pky} was the incorporation of tau polarization effects using an analytic method.
Since this event selection aims to be applicable to a broader range of analyses, a more complete treatment of $\tau$ decay effects was implemented using the \texttt{TAUOLA} library \cite{Was:2000st,Chrzaszcz:2016fte}, which simulates the correct branching ratios to all accessible final states with a full treatment of lepton polarization.
For this work, the important decay modes are the decays to a muon: $\tau^- \rightarrow \mu^- \bar{\nu}_{\mu} \nu_\tau$ and $\tau^+ \rightarrow \mu^+ \nu_\mu \bar{\nu}_\tau$. 
These processes occur with a branching ratio of approximately 17\% \cite{ParticleDataGroup:2024cfk}.
The partitioning of energy to the outgoing muon $z = \frac{E_\mu}{E_\tau}$ depends on the polarization of the initial $\tau$.
The probability of a decay with energy ratio $z$ can be computed analytically and is $p(z) = g_0$ for the unpolarized case and $p(z) = g_0 - g_1$ for the fully polarized case, where $g_0$ and $g_1$ are defined as \cite{Bhattacharya:2016jce, Arguelles:2022bma},
\begin{equation}
    g_0 = \frac{5}{3} - 3z^2 + \frac{4}{3} z^3
\end{equation}
\begin{equation}
    g_1 = \frac{1}{3} - 3 z^2 + \frac{8}{3} z^3
\end{equation}
A sample of $\nu_\tau$ MC events was generated and the interactions were simulated using TAUOLA.
The distributions of the outgoing muon energy of the MC events compared to the analytical expressions are shown in \cref{fig:tau_polarization}, validating the TAUOLA approach.

The new cross sections developed in \cref{chapter:cross_sections} were implemented in the simulation code to sample events from the double-differential distributions.
This was done by creating splines of the total cross section $\sigma(E)$ and double-differential cross section $\frac{d^2 \sigma}{dxdy}$ using points sampled across grids of $E$, $x$, and $y$ with high density to form an accurate interpolator.
In contrast to earlier versions of this simulation, the cross section could now be sampled at low $Q^2$ and $W^2$ which had previously been cut as no cross section information was available for those ranges.
While at high energies, there are negligible contributions from the low-$Q^2$ region, though having a more correct model of low energies will correctly capture events that are misreconstructed to higher energies.

Significant improvements to detector modeling were incorporated.
First, a newer model of the bulk ice was utilized which accommodates for birefringence effects of the ice and an improved ice layer tilt treatment \cite{IceCube:2024qxf}.
Second, an improved model of the DOM single photoelectron (SPE) template was used for converting MC photoelectrons to pulses.
This new SPE template is formed by fitting calibration data to a three-Gaussian model which provides a better fit than previous iterations of the SPE template.

\section{New Event Selection}\label{sec:meows2026_event_selection}
The event selection improvements are aimed at replacing the MEOWS-2023 XLevel and CNN/BDT-based reconstructions and classifiers by leveraging the new transformer-based models discussed in \cref{chapter:ml_recos} to improve statistics and provide better reconstructions. The event selection is divided into three levels:
\begin{itemize}
    \item M1 Level: Precuts and basic reconstructions
    \item M2 Level: BDT cut
    \item M3 Level: Reconstruction and classification
\end{itemize}

For M1 level, many of the cuts that were performed in XLevel were replaced with a DNN-based background classification score cut and cuts on the reconstructed zenith and angular uncertainty from the DNN-based angular reconstruction.
The first step of M1 level requires each event to pass at least one of the following filters: the \texttt{MuonFilter}, \texttt{CascadeFilter}, \texttt{HighQFilter} (high charge), or the \texttt{HESEFilter} (high energy starting events).
The addition of the \texttt{CascadeFilter} is specifically included to preserve high-inelasticity starting track events, which otherwise could be misreconstructed as a cascades early in the processing and subsequently cut.
The \texttt{HighQFilter} and \texttt{HESEFilter} are included to retain high energy events.
Prior to reconstruction, a series of pulse cleaning steps are performed on the pulse series.
A procedure called ``IceHive'' \cite{Zoll:2016wvz} is applied, which uses a clustering algorithm to split a pulse series into multiple pulse series if there is a clear temporal or topological difference between clusters while also removing pulses that are most likely noise.
The second step of pulse cleaning is to apply a time window cut on each DOM from its trigger time to remove afterpulsing effects which are not well-modeled.

The first transformer model that is applied is the ``micro'' background classifier.
This classifier is a small transformer-based binary classifier trained on pulse statistics that outputs a score from 0 to 1 indicating the probability of being a background event (i.e. an atmospheric muon).
This network uses summary statistics and has a much smaller model size of about 1 million parameters than other machine-learning reconstructions to perform these classifications when the data rate is still high.
Immediately after the classifier is run, any event with a background classifier score $>0.1$ is cut.
This first precut removes a significant fraction of all background events while preserving neutrino events, which is important for early stages of the event selection where per-event computation time must be kept to a minimum.
The resulting rate is approximately 1.5 Hz from an initial 60 Hz, removing a significant fraction of background atmospheric muons.
The fast DNN-based precut allows for more computationally expensive reconstructions to be run during the earlier processing levels, as rates above tens of Hz would utilize significant computational resources when processing experimental data.

The next precut that is applied in M1 level is a cut on an angular uncertainty metric.
A machine-learning model called Neptune \cite{Yu:2025etb}, which is a transformer-based network developed from the models in \cref{sec:transformers}, was trained on tracks and cascades to perform a direction reconstruction.
The loss function used for training was the von-Mises Fisher loss function \cite{scott2021mises} based on the probability density function,
\begin{equation}
    p(\vec{x}, \vec{\mu}, \kappa) = \frac{\kappa^{n/2 - 1}}{(2\pi)^{n/2} I_{n/2 - 1}(\kappa)} \exp\left(\kappa \vec{\mu}^T \vec{x}\right)
\end{equation}
where $\kappa$ is a concentration parameter, $n$ is the dimensionality of the vector space, $\vec{\mu}$ is a normalized unit vector, and $I_{v}$ is the modified Bessel function of the first kind at order $v$.
As a loss function, $\vec{\mu} = \frac{\vec{x}_{true}}{\vert\vert \vec{x}_{true}\vert\vert}$ which determines the direction of minimal loss.
The $\kappa$ parameter is a learned output of the model and is correlated to the angular uncertainty. 
This is not treated exactly as an angular uncertainty, as it is unlikely to provide the correct statistical coverage, but can still be used to quantify reconstruction quality.
It is unlikely that events with a small $\kappa$ will be well-constructed further in the event processing chain, so events with $\kappa < 30$ are cut.
This reduces the atmospheric muon background rate to approximately 350 mHz and the atmospheric neutrino rate to 5.8 mHz.
Overall, the precuts after applying the filter conditions reduced the background rate to 0.8\% of the post-filter rate and the neutrino rate to 85.7\%.
The selection with the different filters, the background score cut, and the $\kappa$ cut constitute the precuts of this event selection.

In the previous iteration of the event selection, the reduced log-likelihood (rlogl) value from the likelihood-based direction reconstruction proved to be a powerful variable for removing background events.
After the precuts, a track hypothesis $(x, y, z, t, \theta, \phi)$ is created by first running a fast likelihood-based track reconstruction called LineFit \cite{linefit_dumand} to obtain a track anchor point $(x, y, z, t)$, and then combining it with the zenith and azimuth from the Neptune direction reconstruction.
This track hypothesis is then used as a seed to MPEFit \cite{AMANDA:2003vtt}, a track direction reconstruction, which provides rlogl.
After this reconstruction, charge-based variables are computed using the cleaned pulse series to be used as BDT input features in the next level of processing.
The last step of the M1 level processing is running a Neptune-based morphology classifier to provide probabilities of the event being either: starting track, through-going track, cascade, or a muon bundle.

At M2 level, the starting rate is at a reasonable level where expensive DNN-based reconstructions can be run on a CPU and finish in a reasonable time, or the reconstructions can be run with a GPU and not require an excessive number of resources.
To further reduce this, the BDT cut can be applied as the first step in this processing level as all of the required input variables were obtained in M1 level.
The BDT was trained using \texttt{XGBoost} \cite{chen2016xgboost} with 2,773,270 CORSIKA MC events spanning several datasets that passed the M1 level cuts and 10,252,696 neutrino MC events which includes both tracks and CC and NC cascades. 
90\% of this sample is used to train the BDT and the other 10\% is used to validate to ensure there is no overfitting. 
The selected features for the BDT are largely high-level reconstruction/classification values as well as some low-level charge distribution information:
\begin{itemize}
    \item \texttt{neptune\_kappa}: Angular uncertainty from transformer reconstruction
    \item \texttt{neptune\_zenith}: Reconstructed zenith angle from transformer reconstruction
    \item \texttt{MuonFilter}: Passes the MuonFilter
    \item \texttt{CascadeFilter}: Passes the CascadeFilter
    \item \texttt{background\_score}: Transformer-based binary classifier background score
    \item \texttt{cascade\_score}: Cascade probability from morphology classifier
    \item \texttt{thru\_track\_score}: Through-going probability from morphology classifier
    \item \texttt{starting\_track\_score}: Starting track probability from morphology classifier
    \item \texttt{bundle\_score}: Muon bundle probability from morphology classifier
    \item \texttt{MPEFit\_neptune\_rlogl}: Reduced LLH from MPEFit with neptune seed
    \item \texttt{NoDC\_Qtot}: Total charge without DeepCore (DC)
    \item \texttt{NoDC\_NChan}: Total number of hit DOMs without DC
    \item \texttt{NoDC\_AvgDomDistQTotDom}: Average charge-weighted distance without DC
    \item \texttt{NoDC\_TrackHitsDistributionSmoothness}: Track hit smoothness without DC
    \item \texttt{NoDC\_TrackHitsSeparationLength}: Hit separation distance without DC
\end{itemize}

\begin{figure}[H]
    \centering
    \includegraphics[width=\linewidth]{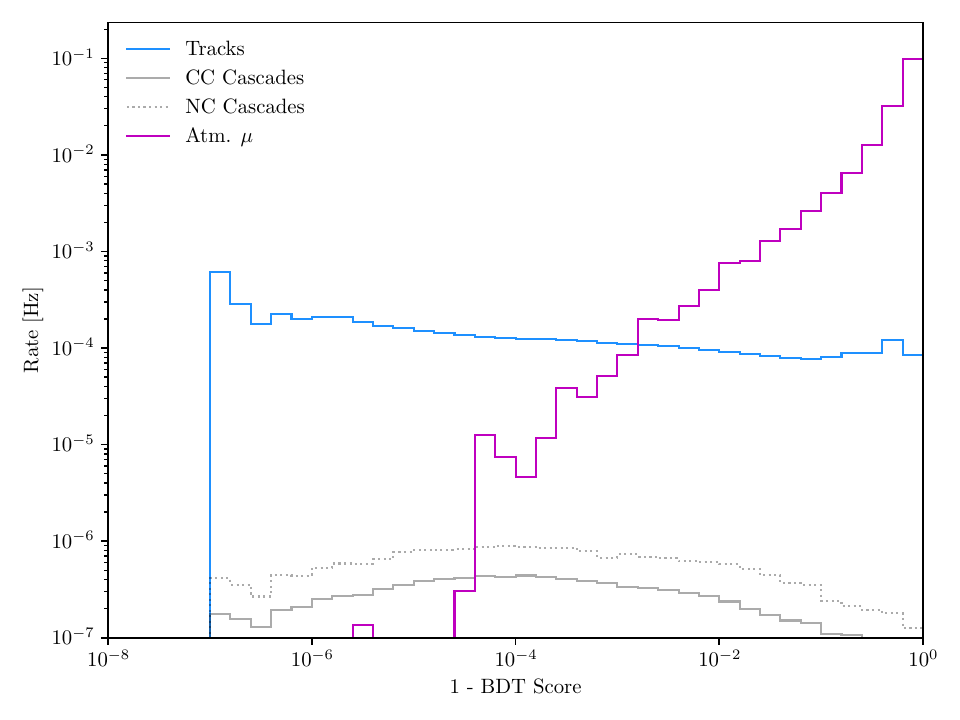}
    \caption{Scores from the final boosted decision tree for tracks (blue), CC cascades (gray solid), NC cascades (gray dashed), and background atmospheric muons (magenta). Note that multiple overlapping CORSIKA MC sets were used to construct the histogram of atmospheric muons and so the rates are an overestimation of the true atmospheric muon rate.}
    \label{fig:meows_2025_bdt_scores}
\end{figure}

A BDT score cut of 0.99999 was selected to reduce the background to negligible rates while maintaining a very large sample of neutrino events.
The distribution of BDT scores separated by tracks, CC cascades, NC cascades, and CORSIKA atmospheric muon events are shown in \cref{fig:meows_2025_bdt_scores}.
Only a single atmospheric muon MC event passes the BDT cut with a weight corresponding to an event rate of 4 per year, or 52 atmospheric muons over 13 years of data.

After the BDT cut, three large Neptune models are applied at M3 level: visible energy reconstruction, visible inelasticity reconstruction, and a starting event binary classifier (starting vs. through-going tracks).
With 13 years of livetime, the total number of expected events that pass the event selection is $1.02\times10^6$.
This means that the expected yearly rate is about $7.7\times10^4$ events/year.
Compared to the old event selection rate of $3.4\times10^4$ events/year, this is a factor of 2.3 increase in selection efficiency.
Of the $1.02\times10^6$ events in the new event selection, $2.85\times10^5$ are classified as starting tracks and $7.38\times10^5$ are classified through-going tracks.
This means that approximately 28\% of the events are starting tracks, a 3\% increase over the 25\% from MEOWS-2023.
Comparing these numbers to the expected rate of background atmospheric muons, it is expected that there will be a negligible contamination of $0.005\%$.
Without any further cuts, the contamination from cascades is higher: approximately 5.5 $\mu$Hz for NC cascades and 2.5 $\mu$Hz for CC cascades.
This sums to $3.2\times10^3$ cascades over 13 years, which could be further reduced by an additional cut on the cascade score from the morphology classifier but this is beyond the scope of this work.

\section{Analysis Framework Improvements}\label{sec:meows2026_fitting}
During the post-unblinding checks of this analysis, several aspects of the fits were found to be excessively computationally intensive.
In particular, a single fit at a given point in parameter space, profiling over the nuisance parameters, takes on average four hours.
For the full parameter space of the unstable sterile neutrino analysis and a single seed, this sums to nearly 40,000 CPU-hours or 4.5 CPU-years.
This computation cost makes it infeasible to perform certain analysis stability tests and fits to pseudoexperiments even with large computing clusters.
The most computationally intensive components of the fits are driven by event reweighting, which relies on slow\footnote{Relative to the other reweighting methods.} spline evaluations and gradient calculations.

The step of reweighting events is structured as a for-loop over events, meaning that each event can be weighted independently from others.
This allows event weights to be computed in parallel across multiple CPU cores or GPUs.
However, significant architectural changes must be made to the \texttt{GollumFit} code to accommodate GPU workflows.
These architectural changes are the subject of \cref{sec:gpu_fits}.

\subsection{Systematic Splines}
\begin{figure}[H]
    \centering
    \includegraphics[width=0.82\linewidth]{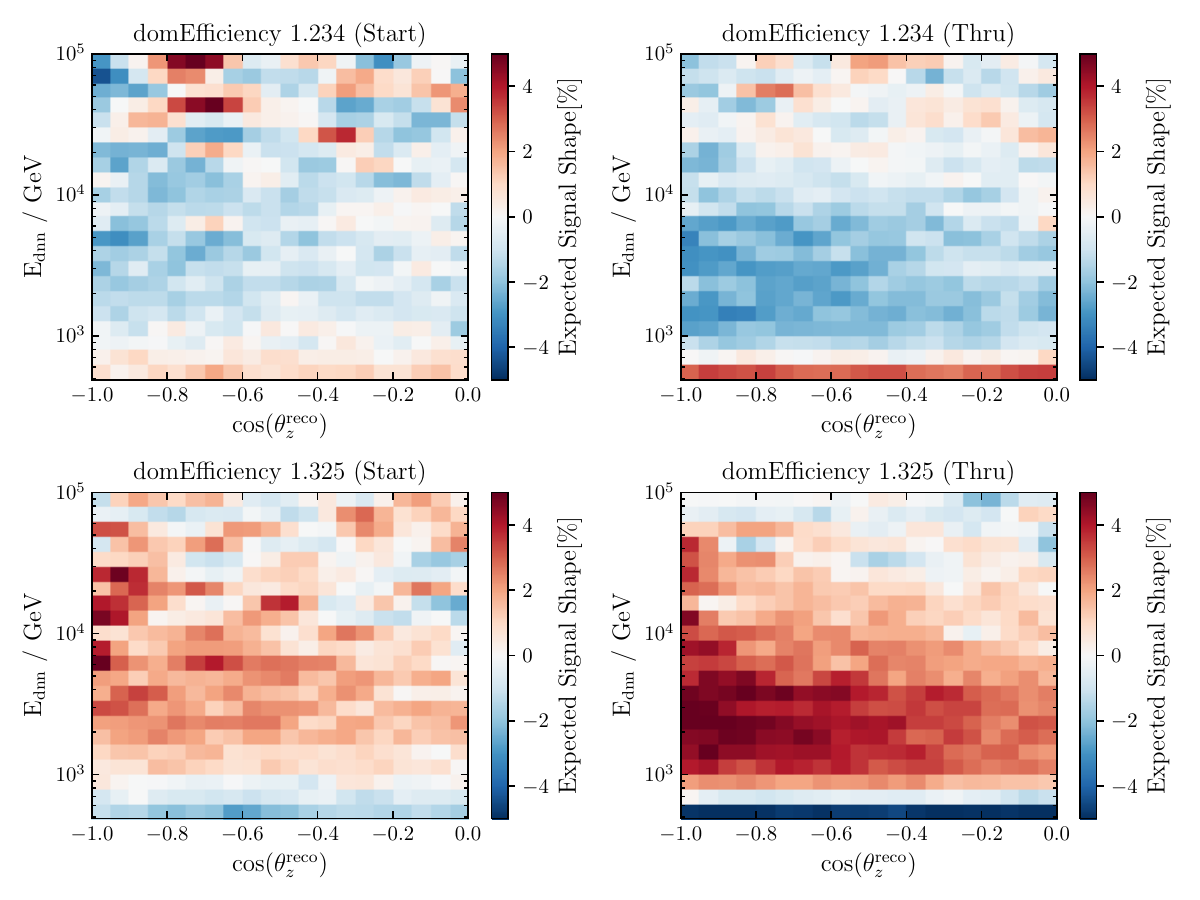}
    \caption{DOM efficiency systematic spline prior to the refit with $s=10^{-6}$.}
    \label{fig:domeff_spline_old}
\end{figure}

\begin{figure}[H]
    \centering
    \includegraphics[width=0.82\linewidth]{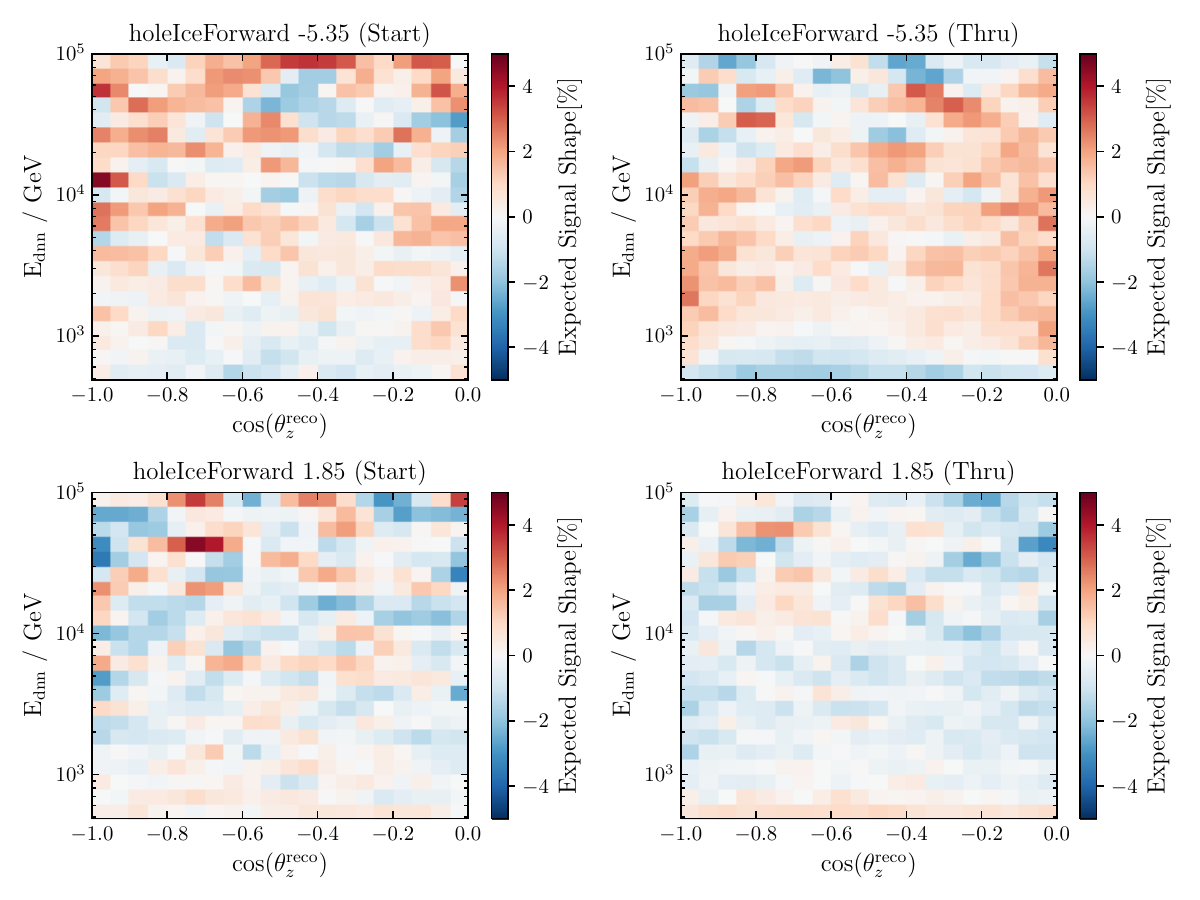}
    \caption{Hole ice systematic spline prior to the refit with $s=10^{-6}$.}
    \label{fig:holeice_spline_old}
\end{figure}

As described in \cref{sec:meows2023_systematics}, some of the systematic uncertainties are implemented by generating discrete MC sets and building splines.
The systematic splines for the hole ice and DOM efficiency were found to be noisy.
The noise was caused by an overly restrictive smoothing parameter ($s$) during the spline fits.
This caused the shapes in reconstructed space to follow MC statistical fluctuations as demonstrated by \cref{fig:domeff_spline_old} and \cref{fig:holeice_spline_old}.
The resulting noise increased the number of minimizer steps required to find the minimum and slowed down the physics fit. 

\begin{figure}[H]
    \centering
    \includegraphics[width=0.82\linewidth]{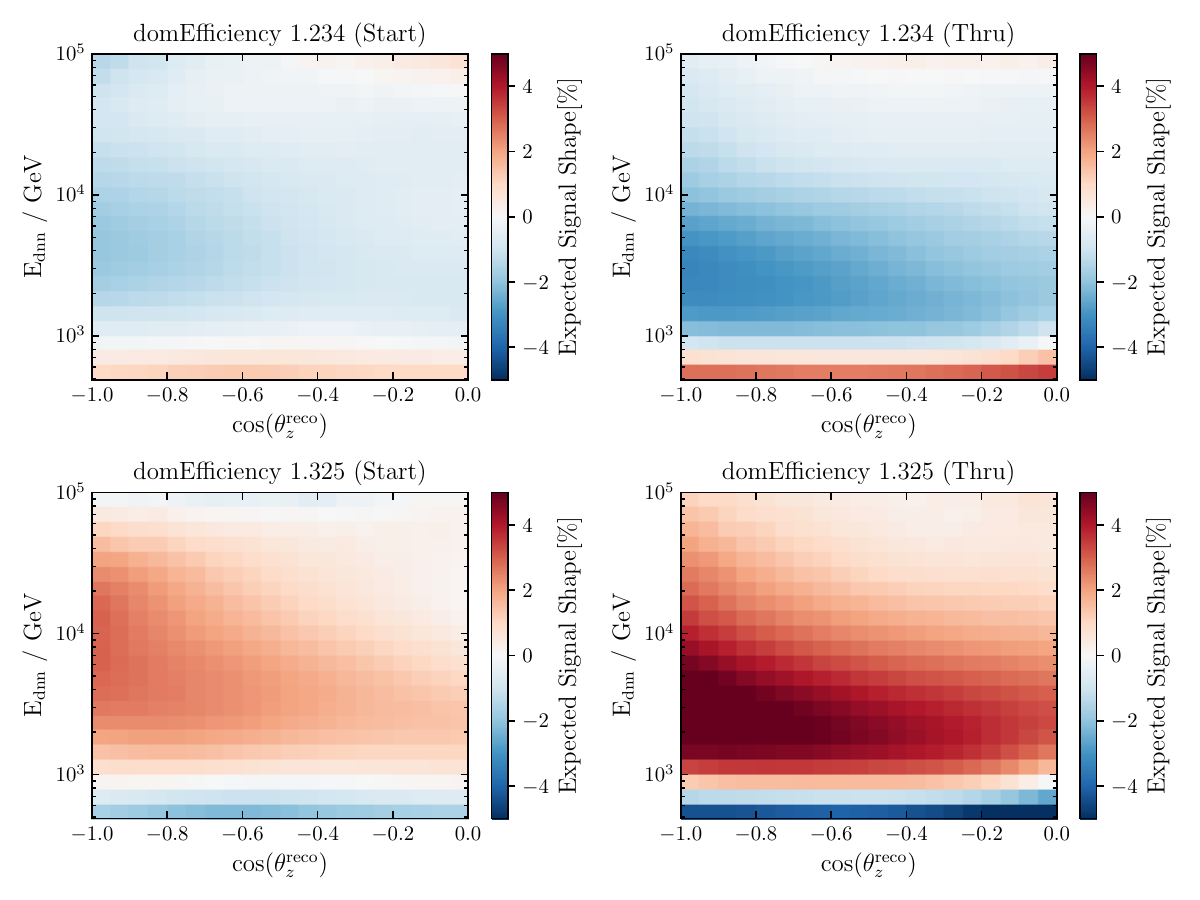}
    \caption{DOM efficiency systematic spline after refitting with $s=10^{-3}$.}
    \label{fig:domeff_spline_new}
\end{figure}

\begin{figure}[H]
    \centering
    \includegraphics[width=0.82\linewidth]{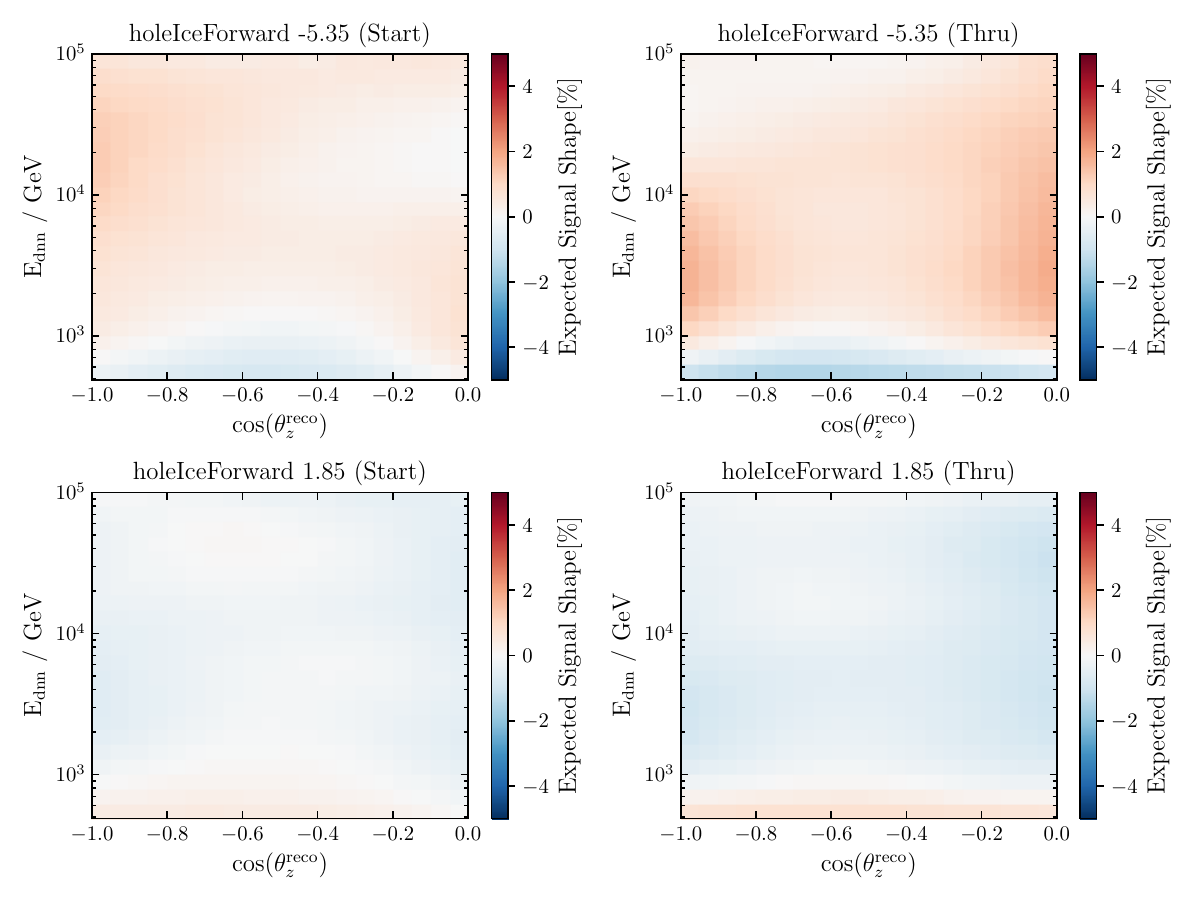}
    \caption{Hole ice systematic spline after refitting with $s=10^{-3}$.}
    \label{fig:holeice_spline_new}
\end{figure}

To address this, the splines were refit with a range of smoothing parameter values, and the results were analyzed to identify the smoothing parameter that captured the systematic trend without following statistical fluctuations.
The selected smoothing parameter was $s = 10^{-3}$, whereas the previous fit used $s = 10^{-6}$.
The final systematic splines with the improved fit are shown in \cref{fig:domeff_spline_new} and \cref{fig:holeice_spline_new} for DOM efficiency and hole ice respectively.
The analysis fits were rerun with the new splines and reproduced the results of the original fits, with negligible shifts in the systematic pulls or change in likelihood values.
This verifies that the analysis was not significantly impacted by the noisy splines and that the improvements were primarily in fit time and minimizer iteration count.


\clearpage

\subsection{GPU-Accelerated Fits}\label{sec:gpu_fits}

The likelihood evaluation in \texttt{GollumFit} is dominated by a per-event reweighting step.
For each Monte Carlo event, a weight must be computed as a function of 38 nuisance parameters by evaluating several \texttt{photospline} B-spline corrections (DOM efficiency, hole ice, and attenuation for each flux component), multiplying by ice gradient and atmospheric density factors, applying 16 linear modifiers for hadronic and cosmic-ray uncertainties, and applying a broken power law tilt reweighting factor for the astrophysical flux.
The resulting weights are accumulated into an 880-bin histogram (22 energy bins $\times$ 20 zenith bins $\times$ 2 topology bins), and the SAY log-likelihood \cite{Arguelles:2019izp} is summed over bins with penalty terms for the priors.
On a CPU with single-threaded evaluation, one likelihood call takes about 4 seconds, so a full profile-likelihood scan over a grid of physics parameters can consume thousands of CPU-hours.
The per-event reweighting is highly parallelizable which is ideal for a GPU workload.

The per-event weight is computed using a weighting equation that combines the three flux components with detector and atmospheric corrections:
\begin{equation}
    w = \mathrm{norm}\times\mathrm{ice}\times(\mathrm{conv}+\mathrm{prompt}+\mathrm{astro})
\end{equation}
where each flux term incorporates spline-based corrections for DOM efficiency, bulk ice, hole ice, and neutrino attenuation, along with atmospheric and astrophysical flux modifiers. 
These terms can be expanded in terms of the cached event weights $w_{\mathrm{cached}}$ and the systematic parameter weight modifiers as follows:
\begin{equation}
    \mathrm{conv} = \Phi_{\mathrm{c}} \times \mathrm{DOM}_{\mathrm{eff}}^{\mathrm{c}} \times \mathrm{HI}^{\mathrm{c}} \times \mathrm{Att}^{\mathrm{c}} \times w_{\mathrm{atm}}
\end{equation}
\begin{equation}
    \mathrm{prompt} = N_{\mathrm{p}} \times w_{\mathrm{cached}}^{\mathrm{p}} \times \mathrm{DOM}_{\mathrm{eff}}^{\mathrm{p}} \times \mathrm{HI}^{\mathrm{p}} \times \mathrm{Att}^{\mathrm{p}}
\end{equation}
\begin{equation}
    \mathrm{astro} = N_{\mathrm{a}}\times w_{\mathrm{cached}}^{\mathrm{a}} \times \mathrm{DOM}_{\mathrm{eff}}^{\mathrm{a}} \times \mathrm{HI}^{\mathrm{a}} \times \mathrm{Att}^{\mathrm{a}} \times w_{\nu/\bar{\nu}} \times w_{\mathrm{tilt}}
\end{equation}
\begin{equation}
    \mathrm{ice} = \prod_{i=0}^{8} \left( 1 + p_{i} c_{i} \right)
\end{equation}
The superscripts $c$, $p$, and $a$ correspond to the conventional atmospheric, prompt atmospheric, and astrophysical neutrino flux components.
Importantly, the $\mathrm{DOM}_{\mathrm{eff}}$, $\mathrm{HI}$, and $\mathrm{Att}$ terms are implemented as B-splines which will be discussed later.
These terms are unique for each neutrino flavor, type, and source (conventional, prompt, astrophysics) since they are constructed using propagated neutrino fluxes.
The bulk ice term is the product of contributions from the Snowstorm gradients.
In the default \texttt{GollumFit} implementation, gradients with respect to the 38 nuisance parameters\footnote{In the analysis presented in \cref{chapter:decay_analysis}, only 36 of the 38 parameters are fit.} are computed using forward-mode automatic differentiation \cite{IceCube:2025yvq, phystools}.

Modern NVIDIA GPUs split their computational resources into streaming multiprocessors (SMs), each of which executes groups of 32 threads at a time known as warps \cite{kirk2016programming, nvidia_cuda_cpp_guide}.
Each SM has limited fast on-chip memory, which is divided into registers, shared memory, and an L1 cache.
Programs running on a GPU are implemented as kernels, which are functions that can be launched on many threads simultaneously.
Threads are grouped into thread blocks, which are the units of allocation to SMs, while warps are units of scheduling.
Occupancy measures the number of active warps on an SM relative to the maximum supported, and high occupancy hides memory latency by allowing the SM to switch to other ready warps while some wait on memory.
Memory coalescing refers to the GPU's ability to service a warp's memory request in a single transaction when consecutive threads access consecutive addresses. 
Scattered accesses require multiple transactions and hence reduce bandwidth efficiency.
These GPU program design considerations motivate different data representations and kernel implementations which will be integrated into the fitting framework.

In \texttt{GollumFit}, a new GPU fit accelerator was implemented that mirrors the CPU likelihood computation pipeline using CUDA kernels for event weight calculations, histogramming, and the likelihood evaluation.
Additional GPU data objects were implemented for storing B-spline parameterizations and various intermediate calculations.
Event data is transformed once from an array-of-structures data object used in the CPU pipeline to a structure-of-arrays layout so that a warp of 32 threads read 32 consecutive values of any field in a single 128-byte coalesced transaction, and fixed-reference spline values are precomputed once and cached per-event.
The per-event kernel replaces expensive numerical operations that are repeated often with computationally cheaper but mathematically equivalent operations (e.g. $10^x \rightarrow 2^{x \log_2 10}$) and uses explicit fused multiply-add intrinsics for the ice-gradient and flux-modifier products.
All kernels were numerically validated against the CPU reference to sub-ppm agreement.
For the value-only likelihood evaluation (no gradient), the likelihood evaluation on an NVIDIA A100 MIG 3g.20gb GPU \cite{nvidia_a100_whitepaper} yielded a 250 times speedup, with approximately 17 ms per likelihood evaluation, compared to the single-thread CPU baseline.

Accelerating the gradient calculation is more difficult because of the much larger memory footprint.
The forward-mode automatic differentiation method utilizes a templated dual-number type that carries a scalar value and 38 derivatives through every arithmetic operation \cite{IceCube:2025yvq, phystools}.
Each dual number occupies 312 bytes of memory and the gradient kernel consumes the hardware maximum of 255 registers per thread, spills 14.6 kB per thread to L1-cached local memory, and is limited to one 256-thread block per SM.
Profiling the gradient kernel showed that it became bounded by memory bandwidth at 67\% of peak DRAM while compute utilization sat at 4\% indicating that the SMs were waiting on the spilled derivative arrays.
To overcome this, a three-stage approach that exploits the structure of the computational quantities was implemented to replace the forward-mode gradient method for GPU fits.

The adjoint gradient method \cite{bartholomew2000automatic, baydin2018automatic, margossian2019review} splits the gradient computation into three stages: forward pass, bin sensitivities, and event gradients.
The forward pass is a single scalar propagation through the reweighting, binning, and SAY log-likelihood calculation.
The bin sensitivities are the derivatives of the log-likelihood with respect to the bin weight sum $w_b = \sum_{i\in b} w_i$ and squared weight sum $w_b^2 = \sum_{i \in b} \frac{w_i^2}{n_i}$ where $n_i$ is the number of MC events combined into a single event $i$ during FastMC compression ($n_i = 1$ without any compression).
The sensitivities are expressed as,
\begin{equation}
    \overline{w}_{b} = \frac{\partial \mathcal{L}}{\partial w_{b}},\quad\quad
    \overline{w_{b}^2} = \frac{\partial \mathcal{L}}{\partial w_{b}^2}.
\end{equation}
The likelihood can be expanded out into products of derivatives using the chain rule,
\begin{equation}
    \frac{\partial \mathcal{L}}{\partial w_{i}} = \frac{\partial \mathcal{L}}{\partial w_{b}}\frac{\partial w_b}{\partial w_{i}} + \frac{\partial \mathcal{L}}{\partial w_b^2}\frac{\partial w_b^2}{\partial w_i},
\end{equation}
and substituting $\frac{\partial w_b}{\partial w_i} = 1$ and $\frac{\partial w_b^2}{\partial w_i} = \frac{2 w_i}{n_i}$ yields the per-event adjoint scalar,
\begin{equation}
    \lambda_i = \overline{w}_b + \frac{2w_i}{n_i} \overline{w_b^2}
\end{equation}
Using the adjoint scalar terms, the derivative of the likelihood function can be computed with respect to each nuisance parameter $\eta$,
\begin{equation}
    \frac{\partial \mathcal{L}}{\partial \eta_j} = \sum_i \lambda_i \frac{\partial w_i}{\partial \eta_j}
\end{equation}
The per-parameter gradients $\frac{\partial w_i}{\partial \eta_j}$ take analytic forms determined by how each nuisance parameter enters the weight equation: a scale factor, linear flux modifier, multiplicative correction, etc.
For the spline-based systematics, the analytic form contains the derivative of the underlying B-spline, which can be provided by \texttt{photospline} with minimal cost since the basis-function derivatives share the same knot-interval lookup with the ordinary value evaluation.

The first stage of the gradient method was to compute the full weighting formula and cache intermediate per-event quantities.
The second stage differentiates the SAY likelihood formula per bin using only two gradients with respect to event weight $w$ and the squared weight $w^2$.
The third stage contracts these to a per-event adjoint $\lambda_i$ and then computes the 38 analytic partial derivatives from the cached intermediates.
The gradients are computed in groups of eight parameters at a time to relieve register pressure, with shared-memory atomics used across groups.
The register footprint drops to about 50, occupancy increases to 4 blocks per SM, and no stack spills to local memory occur.
Spline evaluations, which dominated most of the forward-mode evaluation time, are performed only once per forward pass, with cached values reused in the gradient calculation.

With the finalized GPU implementation and adjoint gradient method, the simple scalar likelihood evaluation yielded a 250 times speedup and the full gradient calculation a 26 times speedup compared to the original automatic differentiation scheme.
This significantly improves overall fit speed, since a single fit can require hundreds of likelihood and gradient evaluations.
One limitation is that the number of available GPUs on a computing cluster is typically much lower than the number of CPUs.
For the gradient-based likelihood minimizations, such as the frequentist fit in \cref{chapter:decay_analysis}, it may still be advantageous to run on CPUs if no other optimizations can be made.
The scalar-only evaluations benefit most from the GPU acceleration and are well-suited to Bayesian analyses that evaluate the likelihood many times without requiring gradients, such as those presented in Refs. \cite{IceCubeCollaboration:2022tso, IceCube:2024dlz}.

\subsection{Warm-starting}

A second major improvement to fitting speed was the incorporation of warm-starting, in which the L-BFGS-B minimizer is initialized using the results of previous fits.
Since the likelihood should not vary significantly between neighboring points in physics parameter space, neither should the systematic parameters $\vec{\eta}$.
This means that the minimizer can be seeded using the results of nearby points.
A significant portion of the minimization time is spent not on finding the best $\vec{\eta}$, but on constructing a good approximation of the inverse Hessian used for curvature information.
For an effective warm-start to the optimizer, both the parameters and the approximate inverse Hessian from nearby fits can be loaded as the starting point.

\begin{figure}
    \centering
    \includegraphics[width=\linewidth]{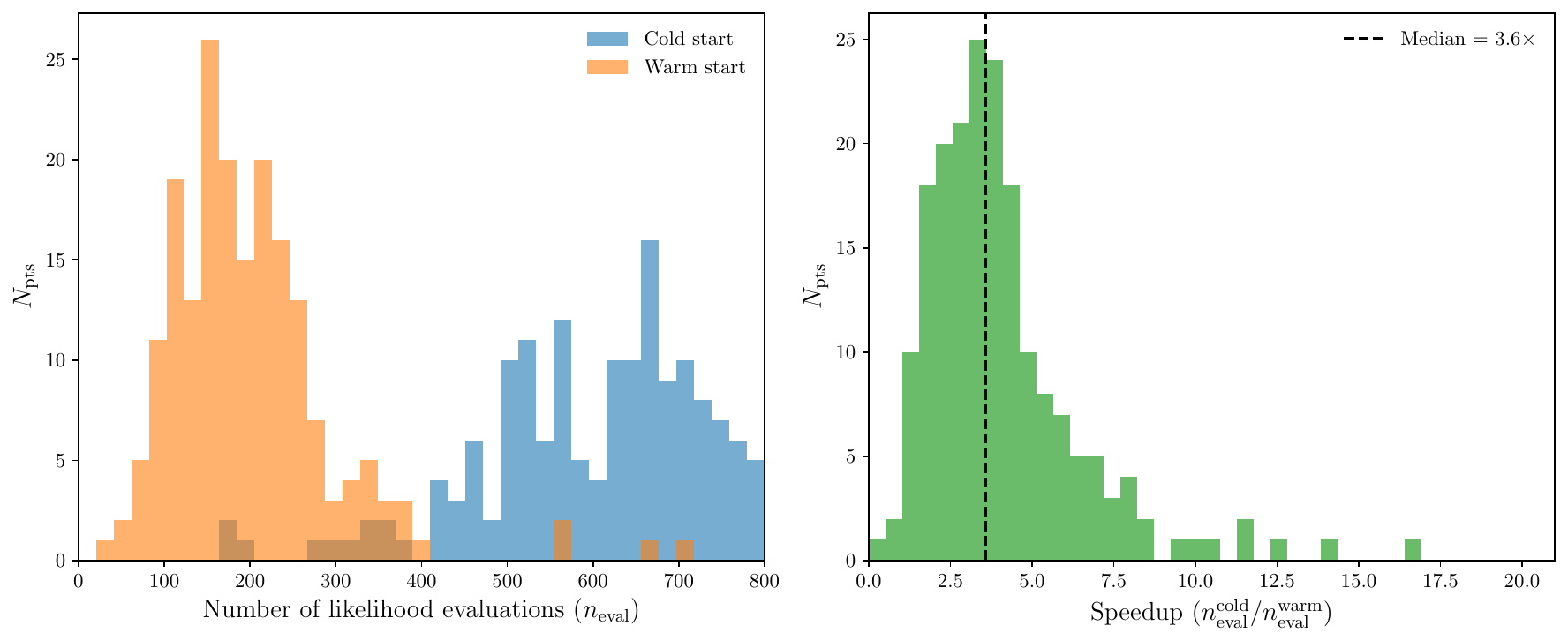}
    \caption{Left: Histogram of the number of likelihood evaluations to complete a fit for 191 points in the sterile parameter space using the cold-start/default method (blue) and the warm-start method (orange). Right: The ratio of cold-start likelihood evaluations to warm-start likelihood evaluations. The median speedup is 3.6$\times$ as indicated by the vertical black dashed line.}
    \label{fig:warmstart_neval_comparison}
\end{figure}

With the warm-start methods implemented into \texttt{GollumFit}, which loads the inverse Hessian and nuisance parameters into the minimizer, 191 points in the sterile physics parameter space were fit using both the cold- and warm-start methods.
The number of likelihood evaluations required to converge for the warm-start case was significantly lower, with a median speedup of 3.6 times.
Some warm-start fits took a similar number of evaluations as the cold-start fits.
Separately, a subset of the fits converged to a minimum that was 0.5 to 1 LLH units lower than the cold-start minimum\footnote{The warm-start fits find better minima given a single seed. In an analysis, multiple seeds are used to find the true minimum, which in these cases agrees with the warm-start minimum.}, whereas most of the other points agreed to within 0.05 LLH units.
This suggests that warm-started fits may require fewer seeds to converge to the true minimum.
These results are shown in \cref{fig:warmstart_neval_comparison}.

\chapter{Conclusions}\label{chapter:conclusions}

This thesis has presented an analysis of 10.67 years of IceCube data searching for the signature of unstable sterile neutrinos with up-going atmospheric neutrinos.
The analysis targeted an extension of the 3+1 sterile neutrino model in which the heavy mass eigenstate $\nu_4$ decays to two invisible particles.
This scenario preserves the oscillation phenomenology that can explain short-baseline anomalies while suppressing sterile thermalization in the early universe, alleviating tension from bounds set by cosmological measurements that restrict the vanilla 3+1 sterile neutrino model.
The analysis found no preference for sterile decay over the no-decay 3+1 hypothesis, and excluded most of the preferred parameter space from global fits to data from short-baseline experiments at 90\% C.L.
This result represents a strong constraint on the unstable sterile neutrino decay model, indicating that it is unlikely to be the explanation for the anomalies observed by short-baseline experiments.

The remainder of this thesis established a foundation for a next-generation sterile neutrino analysis with IceCube.
The matter-enhanced disappearance of $\bar{\nu}_\mu$ is a unique signal of eV-scale sterile neutrinos that neutrino telescopes can search for.
In the existing analyses, the expected $\bar{\nu}_\mu$ disappearance signal is partially diluted in the combined observable $\nu_\mu + \bar{\nu}_\mu$ flux because neutrino telescopes cannot distinguish $\nu_\mu$ from $\bar{\nu}_\mu$ on an event-by-event basis.
Developments in IceCube reconstruction techniques now make it possible to estimate the inelasticity of $\nu_\mu$ charged-current deep-inelastic scattering interactions within the IceCube detector.
The measured inelasticity and understanding of the inelasticity distributions presented in \cref{chapter:cross_sections} provide a way to perform a statistical separation of $\nu_\mu$ and $\bar{\nu}_\mu$.

Together, these contributions establish the infrastructure for an improved next-generation sterile neutrino search.
The updated event selection has better signal efficiency, providing larger samples of both starting and through-going tracks, with improved reconstruction quality.
The cross-section calculations and reconstruction tools developed in this thesis are not specific to the sterile neutrino search and have broad applications to other IceCube physics analyses.
The short-baseline anomalies are persistent oscillation-like signatures inconsistent with the three-flavor framework, in tension with both null searches at other experiments and cosmological constraints. 
This situation parallels the period that preceded the establishment of neutrino oscillations.
The solar and atmospheric neutrino anomalies persisted for decades before the convergence of independent measurements established a resolution.
While the question of whether the anomalies observed by LSND, MiniBooNE, and the gallium experiments reflect physics beyond the Standard Model is not resolved, the work presented here has narrowed the space of viable solutions and developed experimental methodology to advance the search toward an eventual resolution.



\appendix
\chapter{The IsoDAR Control System}

The IsoDAR experiment is a proposed accelerator-based $\bar{\nu}_e$ disappearance experiment that aims to search for $3+1$ neutrino oscillations at an average baseline of $L\sim17$ m \cite{Bungau:2012ys,Alonso:2021kyu,Alonso:2022mup}.
To do this, neutrinos will be produced by accelerating protons onto a $^7$Be target using a high-current cyclotron.
Neutrons produced in the target are then captured by a sleeve of $^7$Li, and the subsequent $^8$Li nuclei beta decay with a half-life of 839 ms, producing a large flux of $\bar{\nu}_e$ with a well-understood energy spectrum \cite{Bungau:2018spu}.
The technical difficulty in constructing such an experiment is the production of a 10 mA, 60 MeV proton beam with a compact accelerator.

A compact high-current cyclotron has been designed with a radiofrequency quadrupole pre-accelerator and buncher that directly injects $\mathrm{H}_{2}^+$ into the cyclotron \cite{Winklehner:2015cqd,Winklehner:2018kqi,Winklehner:2021gew,Koser:2021rcl}.
The $\mathrm{H}_{2}^+$ is accelerated and then stripped after exiting the cyclotron to produce protons.
The initial $\mathrm{H}_{2}^+$ beam is produced by a multicusp ion source.
A first design of the ion source, MIST-1, has been constructed and commissioned at MIT \cite{Winklehner:2018fbq,Winklehner:2025nqo}.
Part of this development included a control system for tuning the ion source parameters and recording measurements from the diagnostic instrumentation.
The following paper \cite{Weigel:2023kfw} was the culmination of work on a control system design by several undergraduate students and the author of this thesis.

\includepdf[pages=-]{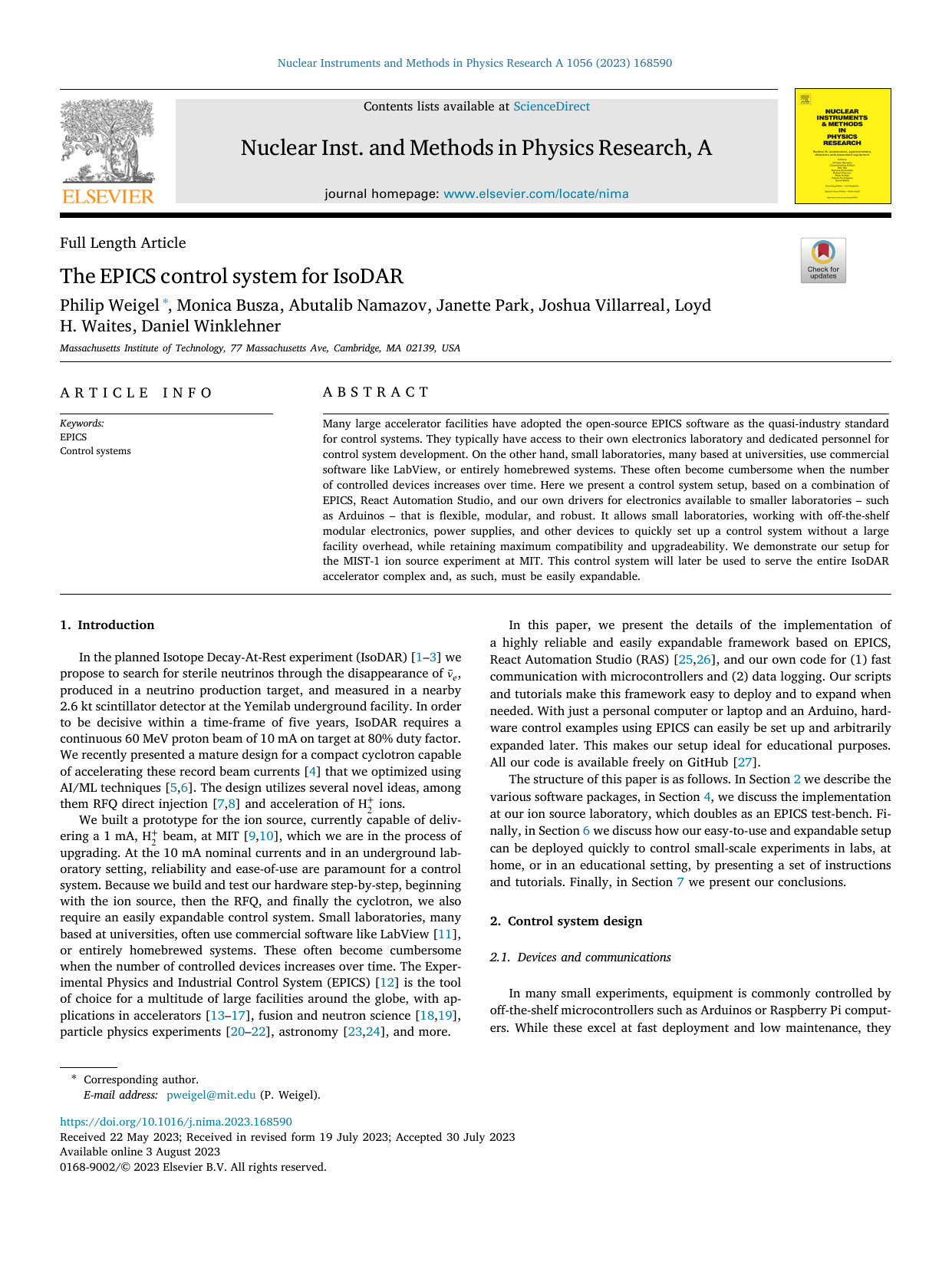}


\begin{singlespace}

\cleardoublepage
\phantomsection
\addcontentsline{toc}{chapter}{Bibliography}

\bibliography{main}
\bibliographystyle{unsrt}
\end{singlespace}




\end{document}